\documentclass{aa}  

\usepackage{graphicx}
\usepackage{xcolor}
\usepackage{soul}
\usepackage{mathrsfs} 
\usepackage{amsmath}
\usepackage[scr=rsfs,cal=boondox]{mathalfa} 
\usepackage{soul} 

\usepackage{txfonts}
\newcommand{\nmn}[1]{{\color{black} #1}}

\newcommand{\FIG}[1]{#1}

\begin{document}

   \title{Radiation-magnetohydrodynamics with \texttt{MPI-AMRVAC} using flux-limited diffusion}

   \subtitle{}

   \author{N. Narechania
          \inst{1}\fnmsep\thanks{corresponding author}
          \and
          R. Keppens \inst{1}
          \and
          A. ud-Doula \inst{2,3}
          \and
          N. Moens\inst{3}
          \and
          J. Sundqvist\inst{3}
          }

   \institute{Centre for mathematical Plasma Astrophysics, Department of Mathematics, KU Leuven, Celestijnenlaan 200B, 3001 Leuven, Belgium\\
              \email{nishant.narechania@ipp.mpg.de}
         \and
             Penn State Scranton, 120 Ridge View Drive, Dunmore, PA 18512, USA
         \and
             Instituut voor Sterrenkunde, KU Leuven, Celestijnenlaan, 200D, 3001 Leuven, Belgium
             }

   \date{July 18, 2024}

 
  \abstract
   {Radiation plays a significant role in solar and astrophysical environments as it may constitute a sizeable fraction of the energy density, momentum flux, and the total pressure. Modelling the dynamic interaction between radiation and magnetized plasmas in such environments is an intricate and computationally costly task.}
   {The goal of this work is to demonstrate the capabilities of the open-source parallel, block-adaptive computational framework \texttt{MPI-AMRVAC}, in solving equations of radiation-magnetohydrodynamics (RMHD), and to present benchmark test cases relevant for radiation-dominated magnetized plasmas.}
{The existing magnetohydrodynamics (MHD) and flux-limited diffusion (FLD) radiative-hydrodynamics physics modules are combined to solve the equations of radiation-magnetohydrodynamics (RMHD) on block-adaptive finite volume \nmn{Cartesian} meshes \nmn{in any dimensionality. 
}}
{We introduce and validate several benchmark test cases such as steady radiative MHD shocks, radiation-damped linear MHD waves, radiation-modified Riemann problems and a multi-dimensional radiative magnetoconvection case. We recall the basic governing Rankine-Hugoniot relations for shocks and the dispersion relation for linear MHD waves in the presence of optically thick radiation fields where the diffusion limit is reached. The RMHD system allows for 8 linear wave types, where the classical 7-wave MHD picture (entropy and three wave pairs for slow, Alfv\'en and fast) is augmented with a radiative diffusion mode.}
{The \texttt{MPI-AMRVAC} code now has the capability to perform multidimensional RMHD simulations with mesh adaptation making it well-suited for larger scientific applications to study magnetized matter-radiation interactions in solar and stellar interiors and atmospheres.
}

   \maketitle
%

\section{Introduction}
Radiation is a key driver in many solar and astrophysical processes and could be a determining factor in the overall dynamics of such flows. For example, radiation is necessary to prevent `overcooling' and to accurately predict star formation frequency in simulations of galaxy evolution \citep{emerick2018stellar}. The process of granulation in the Sun's convection zone is driven by radiative cooling \citep{stein1998simulations,stein2000realistic,jacoutot2008realistic}. Radiation is also important in the accretion processes around black holes \citep{igumenshchev2003three, 2019ApJ...880...67J} and stars \citep{tomida2010radiation}. 
The force exerted by radiation onto stellar gas is the driving force behind the winds of massive stars and determines the velocity profile and structure of their stellar winds \citep{1975ApJ...195..157C,cassinelli1979stellar,moens2022first,esseldeurs20233d,debnath20242d,2024arXiv240616517V}. The coupling between the plasma and the radiation field can be a truly non-local process, and may strongly depend on the radiation wavelength (or frequency) and the plasma density and temperature. Although in general the full radiative transfer equation dictating the change of intensity (at specific frequency) needs to be solved along all possible rays, it is customary to use moments of the radiative transfer equation and reformulate the matter-radiation interaction using coupled momentum and energy equations for plasma and radiation fields \citep{2024FrASS..1146812W}. These in essence reformulate the complex extinction and emission processes (where energy is taken away or added to the radiation beam) into angle and frequency averaged opacities and emission coefficients. Known limits are the diffusion limit (strong coupling between radiation and matter, such as in deep stellar interior layers), and the free-stream limit, adequate when the radiation and matter is uncoupled, i.e. the optically thin radiative regime.

For many applications, it suffices to solve radiation-hydrodynamic (RHD) equations where the included (angular and frequency integrated) moments of the transfer equation are the radiation energy density and the radiation flux vector. In the case of significant magnetic fields, the RHD equations are complemented with a description of the magnetic field to give the radiation-magnetohydrodynamics equations (RMHD). RMHD is necessary to study phenomena in solar contexts, where subphotosphere to coronal layers indeed represent a clear transition between optically thick and thin regimes, in the presence of dynamically important magnetic fields (which dominate in the solar corona, or in sunspots). The development of efficient and accurate numerical techniques to solve these systems is a daunting task. Over the years, researchers have developed several RHD and RMHD codes with a wide range of techniques. For example, \citet{commerccon2011radiation} used an RHD solver to study the fragmentation and collapse of prestellar dense cores. \citet{van2011crash} developed the block-adaptive framework \texttt{CRASH} to perform multi-dimensional RHD simulations of two-temperature plasmas. \citet{johnson2010numerical} performed RHD simulations of radiative acoustic waves and radiative diffusion waves. \citet{hayes2006simulating} studied acoustic waves and shocks in radiative media using \texttt{ZEUS-MP}, a parallel, adaptive, multiphysics computational framework. \citet{yang2012multidimensional} studied hydrodynamic radiating supercritical shocks using a shock-capturing HLLD Riemann solver. \citet{kim2017modeling} modelled the UV feedback from massive stars using the 3D RHD module of the Athena code. \citet{moens2022radiation} implemented radiation-hydrodynamics in the parallel, block-adaptive simulation framework \texttt{MPI-AMRVAC}, which is also at the base of this work, and used it to study the outflows of Wolf-Rayet-type stars \citep{moens2022first}. This code was later also used to study the coupled turbulent envelopes and outflows of O-type stars by \citet{debnath20242d}. \citet{gonzalez2015multigroup} performed 3D RHD simulations of protostellar collapse using the parallel, adaptive \texttt{RAMSES} code.

\citet{jiang2012godunov} developed an RMHD algorithm to study phenomena such as the photon bubble instability and radiative ablation of dense clouds. The Athena++ code was used to perform global 3D RMHD simulations of accretion disks surrounding supermassive black holes \citep{2019ApJ...880...67J,2019ApJ...885..144J,2020ApJ...900...25J}. \citet{flock2013radiation} used the \texttt{PLUTO} code to perform global 3D RMHD simulations of heating of protoplanetary discs by the magneto-rotational instability. \citet{jiang2014radiation} also explored the role of the magnetorotational instability in the formation of hot accretion disk coronae using 3D RMHD simulations. \citet{farcy2022radiation} performed RMHD simulations using the \texttt{RAMSES-RT} code to study the effect of cosmic rays on star formation efficiency in galaxies. 3D RMHD simulations of protostellar collapse have also been performed to study formation of circumstellar disks and protostellar cores \citep{tomida2012radiation,tomida2015radiation}. \citet{ohsuga2009global} performed 2D RMHD simulations of black hole accretion disks. \texttt{MURaM}, a code designed specifically to simulate stellar atmospheres, was recently used by \citet{panja20203d} to perform 3D RMHD simulations of starspots. \texttt{MURaM} was later used by \citet{przybylski2022chromospheric} to perform RMHD simulation of non-equilibrium hydrogen ionization effects on the chromosphere. The 3D RMHD code known as \texttt{Bifrost} has been used to study the transition from optically thin to thick media in the chromosphere \citep{carlsson2016publicly} and more recently ambipolar diffusion processes in solar atmospheres \citep{nobrega2020ambipolar}. \citet{iijima2017three} studied the formation of solar chromospheric jets through 3D RMHD simulations with the \texttt{RAMENS} code. \citet{khomenko2018three} used the \texttt{MANCHA3D} code \citep{modestov2024mancha3d} to perform 3D RMHD simulations of magnetoconvection incorporating non-ideal effects such as ambipolar diffusion.

To avoid the complexity of the full integro-differential radiative transfer equation, moment formulations use closure relations required for coupling the radiation equations with the equations of HD and MHD. One such approach is the Variable Eddington Tensor (VET) formalism \citep{stone1992zeus,hayes2003beyond,jiang2012godunov} where the zeroth and second angular moments are related by a variable Eddington tensor computed from a time-independent transfer equation at every time step. This VET approach requires the solution of the time-dependent radiation energy and radiation momentum equations. A similar approach is the M1-closure scheme \citep{gonzalez2007heracles,skinner2013two,bloch2021high}, which also requires the solution of the time-dependent radiation energy and radiation momentum equations, but the radiation stress tensor is closed by an assumed analytic function of the radiation energy density and radiation momentum. An even simpler approach is the flux-limited diffusion (FLD) approximation, which was first employed by \citet{alme1974numerical} and has been implemented by several researchers since then \citep{minerbo1978maximum,levermore1981flux,turner2001module,hayes2006simulating,van2011crash,yang2012multidimensional,moens2022radiation}. In this approach, only the radiation energy equation needs to be solved, whereas the radiation momentum is assumed proportional to the gradient of the radiation energy density, using a so-called empirical flux-limiter as a closure relation. In this approach, a time-independent radiation momentum equation is solved to obtain a closure for the Eddington tensor which becomes a function of the radiation energy density and its gradient. The FLD approach, although the simplest and computationally cheapest approach among all the above approaches, still fully recovers the optically thin and thick limits and retains a lot of the important physics. \nmn{However, FLD has several notable and potentially severe constraints that users must be mindful of. These include the inability to handle situations with distinct beams and shadows \citep{hayes2003beyond, davis2012radiation}, erroneous predictions in transition regions between optically thick and optically thin regimes \citep{boley2007three} and the inability to model radiation viscosity effects \citep{mihalas1984foundations, castor2004radiation, jiang2012godunov}, among others.} For an in-depth review of the various prescriptions used for modelling the radiation field, the reader is referred to the work by \citet{2024FrASS..1146812W}. In this work, we introduce the FLD approximation for an RMHD module as added to the open-source \texttt{MPI-AMRVAC} code.

The \texttt{MPI-AMRVAC} code uses Fortran 90 and MPI for solving hyperbolic and elliptic PDEs on parallel, block-adaptive grids. \texttt{MPI-AMRVAC} originally focused on special relativistic HD and MHD regimes \citep{keppens2012parallel}, but has been fine-tuned to simulate solar and non-relativistic astrophysical magnetized plasmas \citep{porth2014mpi,xia2018mpi,keppens2023mpi}. The magnetohydrodynamics module has been used for studying several solar phenomena in the solar corona such as flux rope formation \citep{xia2014three}, prominence formation \citep{xia2012simulations,xia2014simulating,xia2016formation,jenkins2022,dion2024}, prominence oscillations \citep{zhou2018three}, solar flares \citep{ruan2020,ruan2023,ruan2024,druett2024exploring} and coronal rain dynamics \citep{Li2022,Li2023,jercic24}. Over the years, \texttt{MPI-AMRVAC} has been upgraded through the efforts of several researchers to include more capabilities such as high order reconstruction, gas-dust coupling, multi-fluid modelling, Super-Time-Stepping (STS), Implicit-Explicit (IMEX) schemes among others \citep{keppens2023mpi}. Its modular structure with options for various numerical schemes and switches to include many different physics terms makes it relatively easy to integrate new modules into the existing code. Recently, \citet{moens2022radiation} implemented a two-temperature FLD approach in \texttt{MPI-AMRVAC} to handle coupling of the hydrodynamics module with radiation. We extend this FLD approach to be used along with the equations of magnetohydrodynamics to model the interaction of radiation with magnetized plasmas.

The organization of this paper is as follows:
Section~\ref{sec:rmhd_eqns} below describes the equations of radiative-magnetohydrodynamics and the flux-limited diffusion method used for obtaining the radiative force density source term for coupling between the radiation and MHD equations. Section~\ref{sec:num_imp} describes the numerical methods used to solve the governing equations of RMHD. Numerical tests used to benchmark the framework are detailed in Section~\ref{sec:tests}. Section~\ref{sec:conclusions} discusses our findings and provides general conclusions.

\section{RMHD equations}\label{sec:rmhd_eqns}
The equations of magnetohydrodynamics in conservative form, extended with the radiation source terms, are given by
\begin{equation}\label{eq:mhd_mass}
\frac{\partial {\rho}}{\partial {t}} + {\nabla} \cdot \left({\rho} {\mbox{\bf v}}\right) = 0,
\end{equation}
\begin{equation}\label{eq:mhd_mom}
\begin{split}
\frac{\partial (\rho{\mbox{\bf v}})}{\partial {t}} + {\nabla}\cdot\left(\rho \mbox{\bf v} \mbox{\bf v} - \mbox{\bf B} \mbox{\bf B} + {\left(p + \frac{\mbox{\bf B} \cdot \mbox{\bf B}}{2}\right)}\mbox{\bf I}\right) = \bf f_r,
\end{split}
\end{equation}
\begin{equation}\label{eq:mhd_energy}
\begin{split}
\frac{\partial e}{\partial t} + 
\nabla\cdot\left(\left(e + p + \frac{\mbox{\bf B}\cdot\mbox{\bf B}}{2}\right)\mbox{\bf v} - (\mbox{\bf B}\cdot \mbox{\bf v})\mbox{\bf B} \right) = \mbox{\bf v}\cdot {\bf f_r} + \dot{q},
\end{split}
\end{equation}
\begin{equation}\label{eq:mhd_mag}
\frac{\partial{\mbox{\bf B}}}{\partial t} + {\nabla}\cdot \left(\mbox{\bf v} \mbox{\bf B} - \mbox{\bf B} \mbox{\bf v}\right) = \bf 0.
\end{equation}
Here, Equations~(\ref{eq:mhd_mass}),~(\ref{eq:mhd_mom}),~(\ref{eq:mhd_energy}) and~(\ref{eq:mhd_mag}) when omitting the right-hand-side terms represent conservation of mass, momentum, energy and magnetic flux for the plasma, respectively. Equation~(\ref{eq:mhd_mag}) describes the time-evolution of the magnetic field given by Faraday's law. These equations are supplemented by the solenoidality condition for the divergence-free magnetic field, or the Gauss's law of magnetism, given by
\begin{equation}\label{eq:div_B}
{\nabla} \cdot {\mbox{\bf B}} = 0.
\end{equation}
Here, ${\rho}$, ${\mbox{\bf v}} = (v_x, v_y, v_z)$, $e$, ${p}$ and ${\mbox{\bf B}} = (B_x, B_y, B_z)$ are the plasma density, velocity, plasma energy (composed of internal energy, kinetic energy and magnetic energy), plasma thermal pressure and magnetic field, respectively. The $\bf f_r$ source term on the right-hand side of the momentum equation is the radiation force density. In the energy equaton, $\dot{q}$ is the radiative heating and cooling term, which is a function of the plasma density, temperature and the radiation energy. These two terms are evaluated by solving the radiation energy and radiation momentum evolution equations described below. The plasma energy can be written in terms of its components as
\begin{equation}\label{eq:energy_density}
e = \frac{p}{\gamma-1} + \frac{\rho v^2}{2} + \frac{B^2}{2}.
\end{equation}
where $v^2 = v_x^2 + v_y^2 + v_z^2$ and $B^2 = B_x^2 + B_y^2 + B_z^2$. The constant, $\gamma = C_p/C_v$, is the ratio of specific heats or adiabatic index of the gas. The relation between plasma temperature, density and thermal pressure is given by the ideal gas law 
\begin{equation}\label{eq:ideal_gas_law}
p = \frac{k_B T_g}{m_p \mu} \rho \,,
\end{equation}
where $k_B$ is the Boltzmann constant, $m_p$ is the mass of a proton, $\mu$ is the mean molecular weight and $T_g$ is the plasma temperature.

The radiation energy equation is solved in the co-moving frame (CMF) of the fluid. \nmn{The spatial and time derivatives in the radiation energy equation shown below are in the inertial frame, just as in Equations~(\ref{eq:mhd_mass})--(\ref{eq:div_B}) shown above, but radiation-related quantities are as evaluated in the CMF.} This is because it is straightforward to compute opacities in the CMF, rather than the observer's frame. \nmn{For a detailed description of transformation of radiation-related quantities between the inertial lab frame and co-moving frame, the reader is referred to Appendix A.} In this frame, the frequency-integrated radiation energy equation, as specified by \citet{mihalas1984foundations} and \citet{castor2004radiation}, is given by
\begin{equation}\label{eq:mhd_r_e}
\frac{\partial {E}}{\partial {t}} + {\nabla} \cdot \left({E} {\mbox{\bf v}}\right) + {\nabla} \cdot \mbox{\bf F} + \mbox{\bf P} : {\nabla} \mbox{\bf v}  = -\dot{q},
\end{equation}
%
%
%
where $E$, $\mbox{\bf F}$ and $\mbox{\bf P}$ are the frequency-integrated radiation energy density, flux vector, and pressure tensor\nmn{, as evaluated in the CMF,} respectively. The dyadic product $\mbox{\bf P} : {\nabla} \mbox{\bf v}$ is the radiation work term also known as `photon tiring'. This radiation equation is coupled with Equations~(\ref{eq:mhd_mass})--(\ref{eq:mhd_mag}) through the heating and cooling term $\dot{q}$ and the radiation force density term $\bf f_r$. These terms are given by
\begin{equation}\label{eq:heating}
\dot{q} = c \kappa_E \rho E - 4 \kappa_P \rho \sigma T_g^4\,,
\end{equation}
and
\begin{equation}\label{eq:rad_force}
{\bf f_r} = \frac{\rho \kappa_F \mbox{\bf F}}{c}\,,
\end{equation}
where $\sigma$ is the Stefan-Boltzmann constant, \nmn{$c$ is the speed of light,} and $\kappa_P$, $\kappa_E$ and $\kappa_F$ are the Planck, energy density and flux mean opacities, respectively. For the radiation force vector $\bf f_r$, the flux mean opacity $\kappa_F$ is, in general, dependent on the direction. However, in this paper, for the sake of simplicity and reproducability, we ignore these complexities and we assume all of the above opacities to be equal i.e., $\kappa_P = \kappa_E = \kappa_F = \kappa$. In actual astrophysical applications, one typically uses precomputed opacity tables (valid under certain assumptions, such as e.g. adopting static media), and it is important to note that our implementation allows for such treatments. However, as we here want to validate and introduce standard benchmark cases for the RMHD equations, we will use constant values for the opacity, thereby deliberately eliminating many relevant routes for interesting opacity-driven radiation-matter instabilities. The radiation energy density can be defined in terms of a radiation temperature as $
E = a_r T_r^4$
where $T_r$ is the radiation temperature and $a_r = 4\sigma/c$ is the radiation constant. Therefore, under the current assumptions of equal opacities $\kappa$, the heating and cooling term can be reformulated as
\begin{equation}\label{eq:qdot}
\dot{q} = 4 \rho \kappa \sigma (T_r^4 - T_g^4) = c\rho \kappa a_r (T_r^4 - T_g^4)\,.
\end{equation}
This term is non-zero when the radiation and plasma temperatures are different, i.e. in a state of radiative non-equilibrium. This is therefore a so-called two-temperature, non-equilibrium FLD approach (e.g. \cite{mihalas1984foundations}).

Apart from $\dot{q}$ and $\bf f_r$, we also need relations for the radiation flux vector $\mbox{\bf F}$ and pressure tensor $\mbox{\bf P}$ in order to close the set of RMHD equations. We now close the system of RMHD equations using the FLD-approximation \citep{levermore1981flux}. In this method, the radiation flux $\mbox{\bf F}$ is written as a diffusive flux in line with Fick's diffusion law:
\begin{equation}\label{eq:fld_F}
\mbox{\bf F} = -D \nabla E = -\frac{c \lambda}{\rho \kappa} \nabla E\,,
\end{equation}
where $D = (c \lambda)/(\rho \kappa)$ is the diffusion coefficient, with $\lambda$ being the so-called flux limiter. We use the flux limiter suggested by \citet{levermore1981flux}, given by
\begin{equation}\label{eq:fld_lambda}
\lambda = \frac{2 + R}{6 + 3R +R^2}\,,
\end{equation}
where $R$ is the dimensionless gradient of the radiation energy density, given by
\begin{equation}\label{eq:fld_R}
R = \frac{|\nabla E|}{\rho \kappa E}.
\end{equation}
$R$ can also be interpreted as the ratio of the photon mean free path, $\mathcal{l} = 1/(\rho \kappa)$, to the radiation energy density scale height, $ H_R = E/|\nabla E|$. With this formulation, we can see that in the optically thick limit, $R \rightarrow 0$, and therefore $\lambda \rightarrow 1/3$. This gives the proper value of the radiation flux in the diffusive regime, $\mbox{\bf F} = -c \nabla E/(3\rho \kappa)$. In the optically thin limit, $\lambda \rightarrow 1/R$, and the radiation flux also recovers its free-streaming limit value, $|\mbox{\bf F}| \rightarrow cE$. There are several other formulations available for the flux-limiter $\lambda$ that retain these properties. For example, the Minerbo flux-limiter \citep{minerbo1978maximum} has also been implemented and is readily available for use. Lastly, we need a closure relation for the radiation pressure tensor, $\mbox{\bf P}$. In the current FLD approach, $\mbox{\bf P}$ is expressed in terms of the radiation energy density as
\begin{equation}\label{eq:fld_P}
\mbox{\bf P} = \mbox{\bf f} E,
\end{equation}
where $\mbox{\bf f}$ is the Eddington tensor given by
\begin{equation}\label{eq:fld_Eddington_tensor}
\mbox{\bf f} = \frac{1}{2}(1-f)\mbox{\bf I} + \frac{1}{2}(3f-1){\hat{\bf n}} {\hat{\bf n}}.
\end{equation}
Here, ${\hat{\bf n}} = \nabla E/|\nabla E|$ is the unit vector in direction of the gradient of radiation energy, $\mbox{\bf I}$ is the unit tensor, and $f$ is the scalar Eddington factor given by
\begin{equation}\label{eq:fld_f}
f = \lambda + \lambda^2 R^2.
\end{equation}
In the optically thick limit, $f \rightarrow 1/3$, whereas in the optically thin free-streaming limit, $f \rightarrow 1$. These radiation energy and momentum equations and the heating and FLD closure terms have been described in detail by \citet{moens2022radiation}.

\section{Numerical implementation}\label{sec:num_imp}

The RMHD system is described by Equations~(\ref{eq:mhd_mass})--(\ref{eq:div_B}) and Equation~(\ref{eq:mhd_r_e}). The left hand sides of Equations~(\ref{eq:mhd_mass})--(\ref{eq:mhd_mag}) and the advection term in Equation~(\ref{eq:mhd_r_e}) are hyperbolic and operate on the gas-dynamical timescale. They are evaluated using the higher-order, finite-volume, shock-capturing approximate Riemann solvers already available in \texttt{MPI-AMRVAC} \citep{keppens2023mpi}. The HLL approximate Riemann solver \citep{einfeldt1988godunov,linde2002practical} and the third order weighted non-oscillating (weno3) reconstruction scheme \citep{liu1994weighted} were used for this work. This reconstruction is operative when evaluating fluxes at cell interfaces, while all quantities are stored cell-centered.

The other terms, including the divergence of the radiation flux vector $\nabla \cdot \mbox{\bf F}$, the radiative force density $\bf f_r$, its work term ${\bf v} \cdot {\bf f_r}$, the heating and cooling term $\dot{q}$, and the photon tiring term $\mbox{\bf P} : {\nabla} \mbox{\bf v}$
are typically non-hyperbolic and effectively introduce stiff source terms. The radiative force density ${\bf f_r}=-\lambda \nabla E$, the radiation flux vector $\mbox{\bf F}=-D\nabla E$, and the flux-limiter $\lambda$ that appears in these terms, all depend on the gradient of the radiation energy $\nabla E$. This gradient is numerically calculated using a fourth order central difference scheme with a five-point stencil \citep{fornberg1988generation}. The photon-tiring term $\mbox{\bf P} : {\nabla} \mbox{\bf v}$, requires the gradient of the velocity $\nabla \mbox{\bf v}$, which is calculated similarly using a five-point stencil. The terms $\bf f_r$, ${\bf v} \cdot {\bf f_r}$ and $\mbox{\bf P} : {\nabla} \mbox{\bf v}$ are added explicitly as they operate on a dynamical timescale. In typically interesting stellar applications, the $\dot{q}$ term operates on a timescale much smaller than the dynamical timescale, and is added implicitly for both Equations (\ref{eq:mhd_energy}) and (\ref{eq:mhd_r_e}). The implicit approach used for this term involves solving a fourth-degree polynomial equation which has been described by \citet{moens2022first} and \citet{turner2001module}. The radiation divergence term, $\nabla \cdot \mbox{\bf F}$, is parabolic in nature and is computed using \texttt{MPI-AMRVAC}'s geometric multigrid method library \texttt{octree-mg}, introduced in \citet{teunissen2019geometric}. The incorporation of all these terms, namely the hyperbolic terms that are computed explicitly, and all other non-hyperbolic terms that are computed explicitly as well as implicitly, is done through an operator split implicit-explicit (IMEX) scheme. The various steps of this IMEX scheme, as well as the numerical details of computing each of the above terms have been described by \citet{moens2022radiation} and the various choice of IMEX schemes that \texttt{MPI-AMRVAC} offers are detailed in \citet{keppens2023mpi}. Here, we use the 3-step ARS3 IMEX scheme.

\nmn{It must be noted that, in certain regions such as the transition regions between optically thin and optically thick regimes, the coupling terms $\bf f_r$ and ${\bf v} \cdot {\bf f_r}$ can also be significantly stiff. In such regions, an explicit treatment of these terms, as described above, can be detrimental to accurately resolving radiation-momentum coupling. This issue can also be encountered in cases with large radiation pressures, such as numerical simulations of the inner regions of an accretion disc of a supermassive black hole (e.g., \citet{shakura1973black}; \citet{turner2003}). There are methods to deal with the treatment of such terms, such as the modified Godunov method of \citet{miniati2007}. This method was also incorporated by \citet{jiang2012godunov} into their RMHD algorithm based on the Variable Eddington Tensor. However, we do not employ such a technique here and leave such advancements to future work.}

Lastly, in multi-dimensional setups, the discretization of the solenoidality condition, Equation~(\ref{eq:div_B}), requires special treatment to control the discrete divergence of the magnetic field to be at an acceptable (truncation) level. While choices exist to keep this to machine precision zero in one prechosen discretization, here this is done using the parabolic diffusion (linde) method \citep{keppens2003adaptive,keppens2023mpi}. \nmn{It must be noted that the Linde method introduces truncation errors in the magnetic field, and thus in the magnetic energy density. These errors are more pronounced in regions with complex magnetic field structures, such as those involving chaotic magnetic reconnection. This can lead to errors in the numerical evaluation of plasma thermal energies, in turn leading to artificial heating or cooling in the energy exchange between plasma and radiation. However, in combination with effectively high resolution locally (achieved by AMR), these errors are quite acceptable, and they do not differ substantially when insisting on other means to control monopole errors. \texttt{MPI-AMRVAC} has ten distinct choices of methods for divergence control as described in \citet{keppens2023mpi}. These include the constrained transport (CT) method \citep{olivares2019}, Powell source term method \citep{powell1999}, Janhunen method \citep{janhunen2000}, Generalized Lagrange Multiplier (GLM) method \citep{dedner2002} and the multigrid method \citep{teunissen2019geometric}, among others. A brief comparison between some of these methods in terms of the magnitudes of the truncation errors produced, when used in the resistive MHD tilt instability simulation, can be found in \citet{keppens2023mpi}.} AMR, where applicable, is driven using the L{\"o}hner's criterion \citep{lohner1987adaptive}. It must be noted that all equations are solved using non-dimensional quantities for all variables, but the full dimensionalization of quantities is important to mention further on, as opacities will e.g. be given in dimensional units.

%

\section{Test cases}\label{sec:tests}
This section describes various benchmark tests. Section~\ref{sec:shocks} first presents the radiation-modified Rankine-Hugoniot jump conditions for MHD shocks. We then validate the code with simulations of various categories of stationary radiation-dominated MHD shock solutions obtained from these jump conditions. \nmn{Section~\ref{sec:heatcool} studies the energy exchange between plasma and radiation initialized out of radiative equilibrium, through heating and cooling processes. The code is validated by matching the plasma energy evolution rates with theoretical predictions.} Section~\ref{sec:waves} derives the dispersion relation governing all linear modes in a stagnant and uniform radiation-plasma background, to quantify the effect of radiation on slow and fast magnetosonic and Alfv\'en modes. We then look at simulations of linear MHD waves in a radiative equilibrium background medium and compare with analytical damping rates obtained from the solution of the dispersion relation. \nmn{This is done for a weakly radiative as well as a strongly radiative background plasma.} Section~\ref{sec:riemann} adds radiation to standard MHD Riemann shock tube problems and makes observations on the several waves thus formed. As a final multi-dimensional application of the code, Section~\ref{sec:convection} considers radiation-modified magnetoconvection.

\subsection{Radiation-modified steady shocks in the diffusion limit}\label{sec:shocks}

\begin{figure*} 
   \centering
   \FIG{
   \includegraphics[width=6cm,clip]{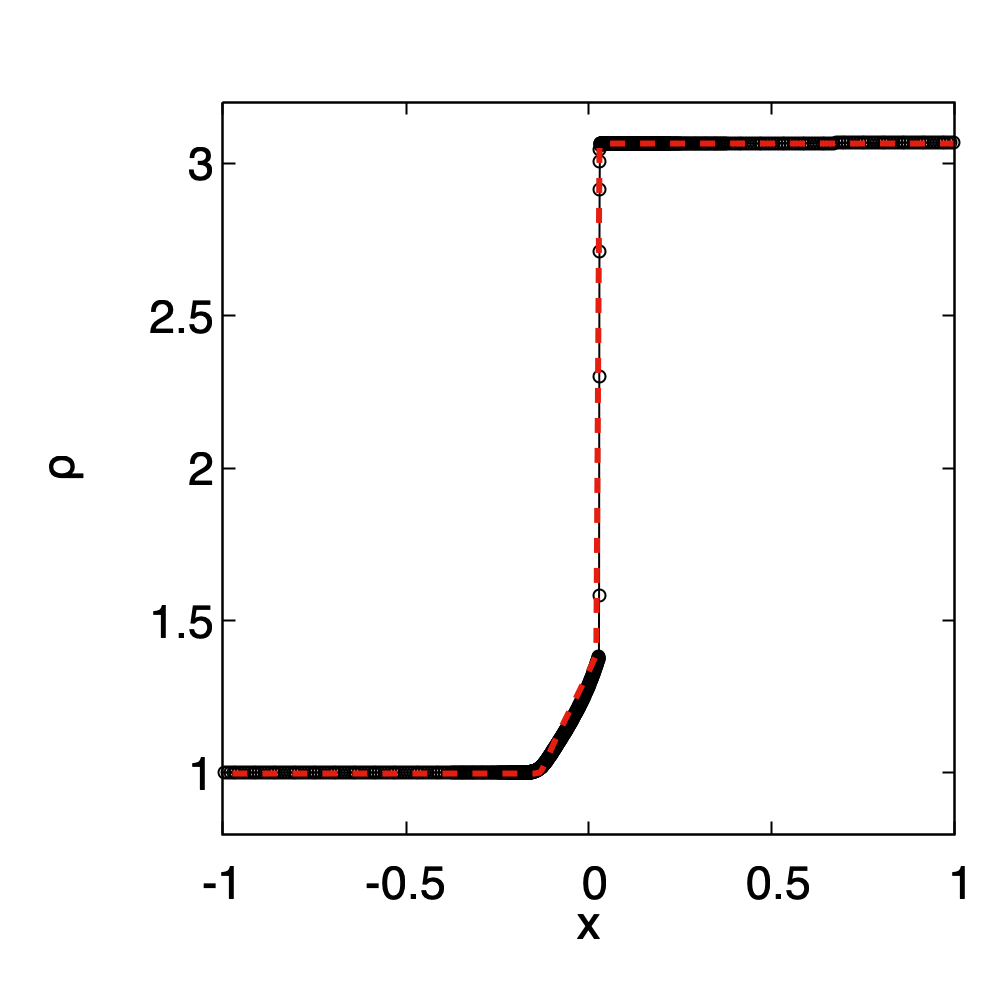}
   \includegraphics[width=6cm,clip]{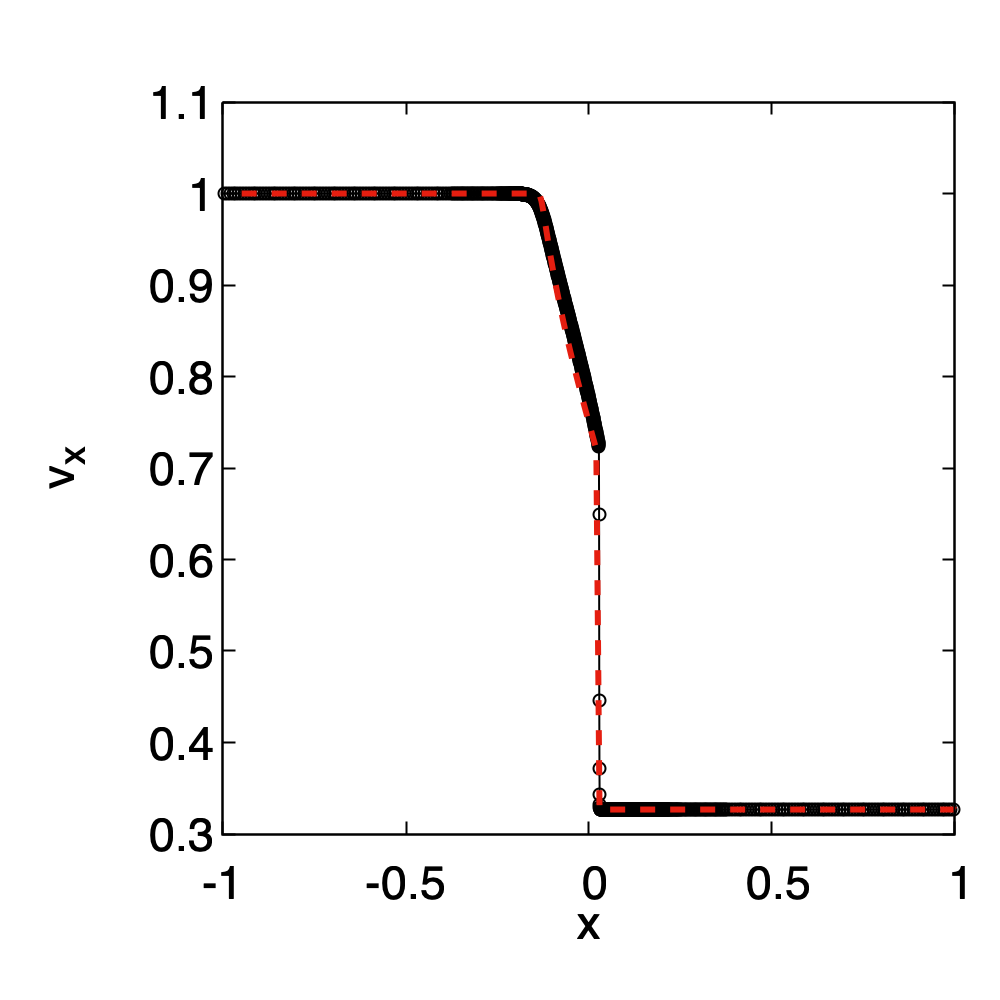}
   \includegraphics[width=6cm,clip]{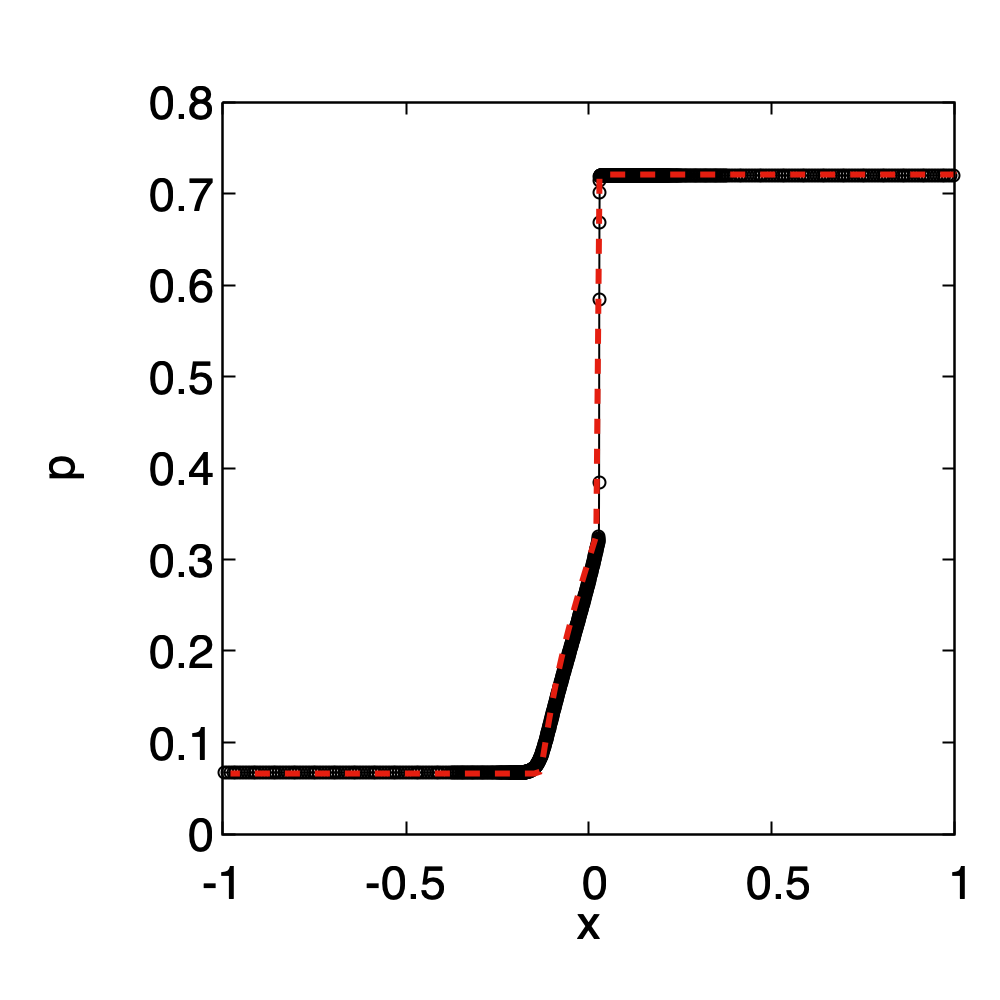}
   \includegraphics[width=6cm,clip]{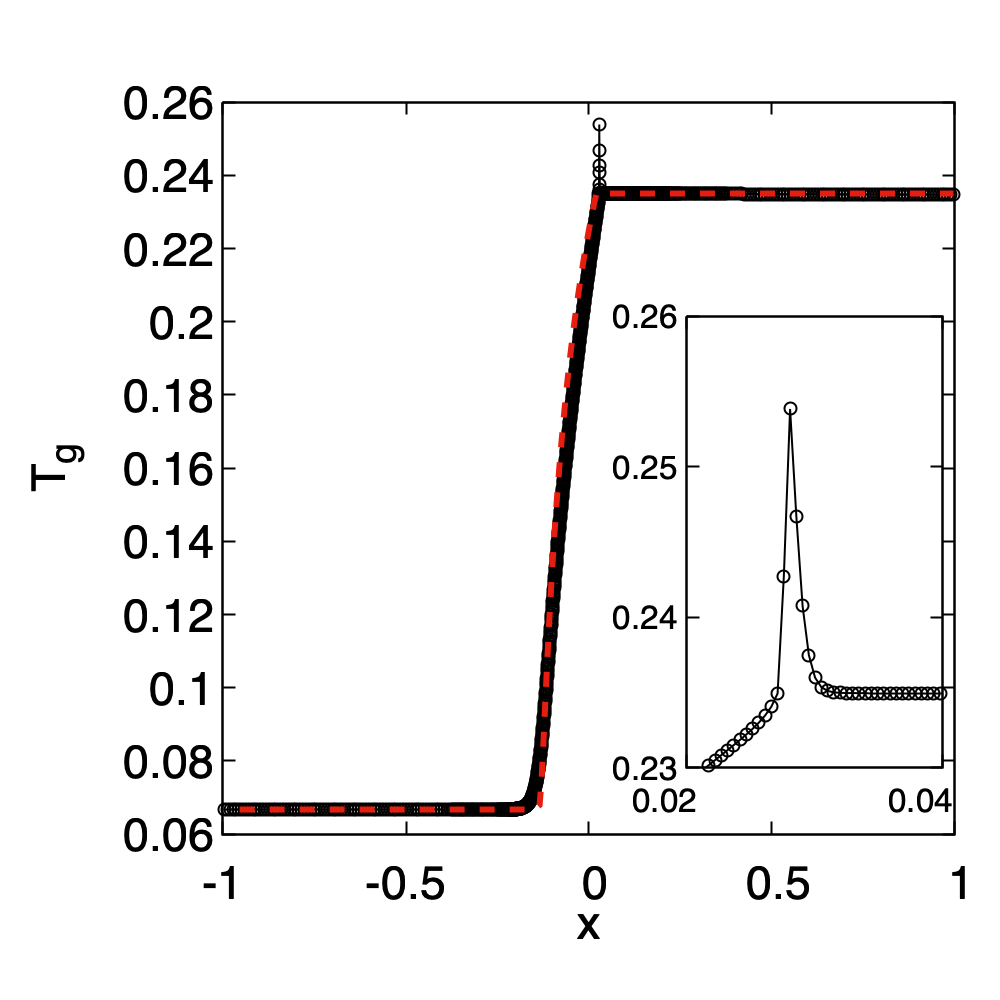}
   \includegraphics[width=6cm,clip]{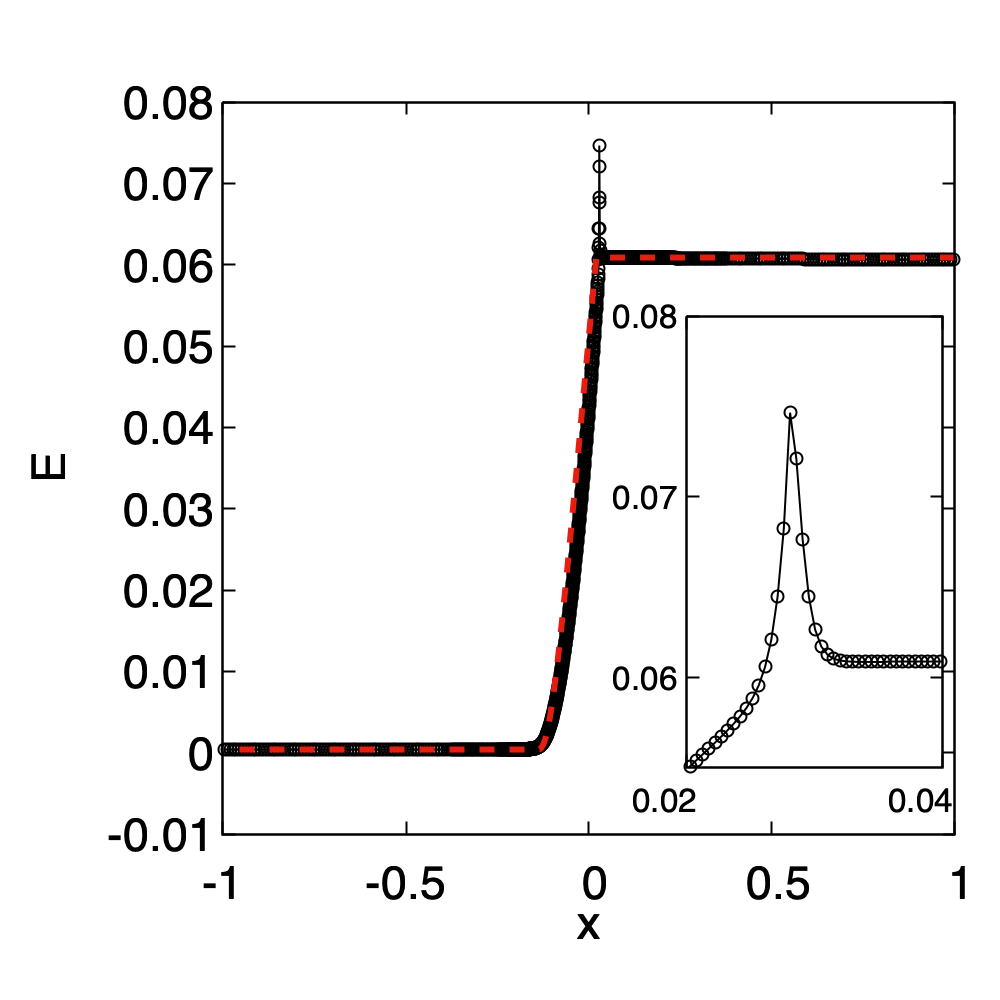}
   }
   \caption{Normalized density, $x$-velocity, plasma pressure, plasma temperature and radiation energy profiles computed for the relaxed state of the hydrodynamic shock. \nmn{The semi-analytic solution of \citet{lowrie2007radiative} is also shown as a dashed red line.}}
\label{fig:shock_Jiang}%
\end{figure*}

As a first test case, we describe several steady-state, stationary, radiation-modified magnetohydrodynamic (RMHD) shocks. The radiation-modified Rankine-Hugoniot jump conditions for hydrodynamical shocks in the optically thick diffusion limit have been discussed by several authors such as \citet{mihalas1984foundations}, \citet{coggeshall1986lie}, \citet{bouquet2000analytical}, \citet{lowrie2007radiative} and \citet{lowrie2008radiative}. For a plane shock contained in the $yz-$plane, the RMHD generalizations of these shock frame relations are given by
\begin{equation}\label{eq:RH_mass}
(\rho v_x)_{\rm left} = (\rho v_x)_{\rm right},
\end{equation}
\begin{equation}\label{eq:RH_momx}
(\rho v_x^2 + p^\ast - B_x^2)_{\rm left} = (\rho v_x^2 + p^\ast - B_x^2)_{\rm right},
\end{equation}
\begin{equation}\label{eq:RH_momy}
(\rho v_x v_y - B_x B_y)_{\rm left} = (\rho v_x v_y - B_x B_y)_{\rm right},
\end{equation}
\begin{equation}\label{eq:RH_momz}
(\rho v_x v_z - B_x B_z)_{\rm left} = (\rho v_x v_z - B_x B_z)_{\rm right},
\end{equation}
\begin{equation}\label{eq:RH_Bx}
(B_x)_{\rm left} = (B_x)_{\rm right},
\end{equation}
\begin{equation}\label{eq:RH_By}
(B_y v_x - B_x v_y)_{\rm left} = (B_y v_x - B_x v_y)_{\rm right},
\end{equation}
\begin{equation}\label{eq:RH_Bz}
(B_z v_x - B_x v_z)_{\rm left} = (B_z v_x - B_x v_z)_{\rm right},
\end{equation}
\begin{equation}\label{eq:RH_energy}
((e^\ast + p^\ast) v_x - B_x(\mbox{\bf v} \cdot \mbox{\bf B}))_{\rm left} = ((e^\ast + p^\ast) v_x - B_x(\mbox{\bf v} \cdot \mbox{\bf B}))_{\rm right}.
\end{equation}
Here, $p^\ast$ is the total pressure, comprising the plasma thermal pressure, magnetic pressure and the radiation pressure:
\begin{equation}\label{eq:p_star}
p^\ast = p + \frac{B^2}{2} + \frac{E}{3},
\end{equation}
where $E/3$ is the radiation pressure in the diffusion limit. Similarly, $e^\ast$ is the total energy density, comprising the plasma internal energy, kinetic energy, magnetic energy and radiation energy:
\begin{equation}\label{eq:e_star}
e^\ast = \frac{p}{\gamma-1} + \frac{\rho v^2}{2} + \frac{B^2}{2} + E.
\end{equation}
Also, radiative equilibrium on either side of the shock requires the radiation and plasma temperatures to be equal:
\begin{equation}\label{eq:rad_equi_l}
E_{\rm left} = a_r T_{g,\rm left}^4,
\end{equation}
\begin{equation}\label{eq:rad_equi_r}
E_{\rm right} = a_r T_{g,\rm right}^4.
\end{equation}
It is possible to manipulate these equations further analytically, e.g. using the constancy of normal mass flux and magnetic field component as expressed by Equations~(\ref{eq:RH_mass})-(\ref{eq:RH_Bx}). In pure ideal MHD, they can be further manipulated into the convenient de Hoffmann-Teller frame expressions involving a transformation to a tangentially moving frame, as found in various textbooks, e.g. \citet{GoedbloedKeppensPoedts2019}. The added radiation complicates these manipulations significantly, but these shock-frame Rankine-Hugoniot equations can be solved numerically to obtain the right state for a given left state. In a pure MHD case, the steady shock solution would be a discontinuous single jump from left to right state obeying Rankine-Hugoniot relations. Here, we expect these radiation-augmented relations to connect left and right states with non-trivial variations in between them, which must be computed numerically. Therefore a combination of left and right states obeying our augmented Rankine-Hugoniot relations can be used as an initial condition and then allowed to interact with the radiation and relax to a steady state solution. These will be good tests for the conservation properties and robustness of the numerical schemes used. Five such solution states that satisfy Equations~(\ref{eq:RH_mass})--(\ref{eq:rad_equi_r}) are shown in Table~(\ref{table:RMHD_shocks}). In this table, the properties in the left solution state that can be decided are the density $\rho$, plasma temperature $T_g$, velocity $\bf v$, and magnetic field $\bf B$. The radiation temperature $T_r$ equals the plasma temperature, and the pressure must satisfy Equation (\ref{eq:ideal_gas_law}). The ratio of the thermal to the magnetic pressure is given by $\beta$. The dimensionless numbers $\mathcal{M}_s$, $\mathcal{M}_a$ and $\mathcal{M}_f$ are the slow magnetosonic, $x$-Alfv\'en and fast magnetosonic Mach numbers, respectively. The Mach numbers that jump from higher than 1 to lower than 1 across the shock are marked in bold for each case. In this table, the physical length of the computational domain, $L$, is also given for each case. The value of $L$ must be chosen such that it is sufficient to capture the thickness of the shock for the given left and right states and opacity $\kappa$. These 5 shocks are simulated in the optically thick diffusion limit for this case. 
For all these shocks, the computational domain was made up of 256 cells on the initial mesh, and 5 AMR levels were used, leading to an effective resolution of 4096 cells.
Zero-gradient boundary conditions are applied at the left and right boundaries. A mean molecular weight of $\mu = 0.5$ and a constant background opacity of $\kappa = 0.4$ cm$^2$/g is used for all these shock solutions, unless otherwise specified.

Note that RHD shocks are special cases of the above relations with vanishing magnetic field throughout. As we are not aware of steady RMHD shock tests in the literature that we can readily reproduce, the first
test here is a Mach 3 radiation-dominated, purely hydrodynamic 1D shock without any magnetic field, as described by \citet{jiang2014radiation}. The relaxed shock, obtained after several passing times is shown in Figure~(\ref{fig:shock_Jiang}). \nmn{For comparison, Figure~(\ref{fig:shock_Jiang}) also shows the diffusion limit semi-analytic solution for radiative shocks, as described in the work of \citet{lowrie2007radiative}. This shows a very strong match between the semi-analytic and numerical solutions.} When the shape of the shock stopped changing visually and reached a clear steady state, the simulation was stopped. Unlike ideal hydrodynamic and magnetohydrodynamic shocks which are simple discontinuities, radiation-dominated shocks have a much more complicated structure. Radiation diffuses upstream of the shock, causing the plasma immediately upstream of the shock to heat up and form a precursor region in plasma and radiation temperatures. The other plasma properties such as density, velocity and pressure, vary smoothly and monotonously in this precursor region, before undergoing a discontinuous jump at the shock. We notice that the plasma and radiation temperatures exceed their respective downstream value at the end of the precursor region. This phenomenon is called a Zel'dovich spike~\citep{zeldovich1967,mihalas1984foundations}, and is followed by a relaxation region where they cool down to the downstream values. This Zel'dovich spike is shown in a magnified view in the insets of the plasma temperature and radiation energy solutions. \nmn{Although the semi-analytic solution of \citet{lowrie2007radiative} does not include a Zel'dovich spike, they provide an approximate model to calculate the plasma temperature of the spike. Using this model, the non-dimensional, analytical, plasma temperature we obtain for the spike is 0.26961, which is somewhat higher than that found in the numerical solution of Figure~(\ref{fig:shock_Jiang}).} This relaxation region is much smaller than the precursor region. If the downstream plasma temperature is larger than the plasma temperature immediately upstream of the shock at the end of the precursor region, the shock is called a subcritical shock. This shock is one such subcritical shock. 

\begin{figure*} 
   \centering
   \FIG{\includegraphics[width=6cm,clip]{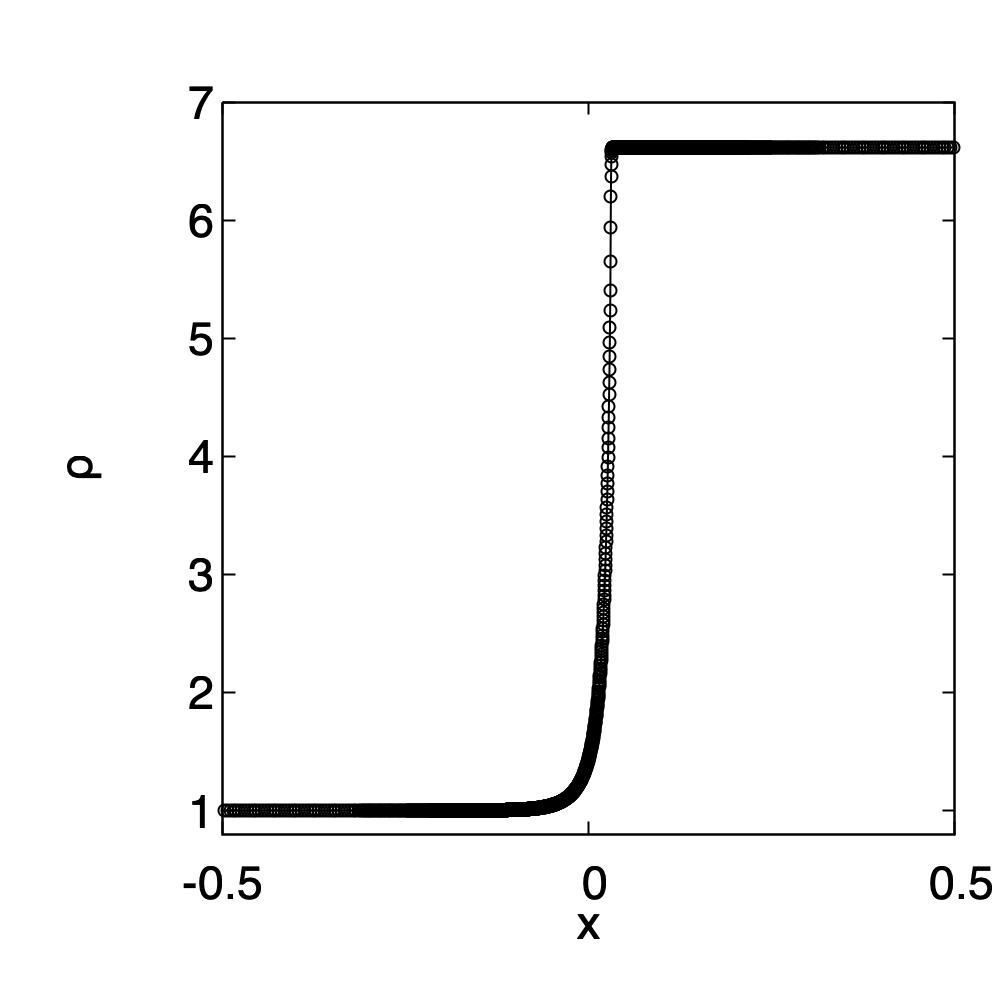}
   \includegraphics[width=6cm,clip]{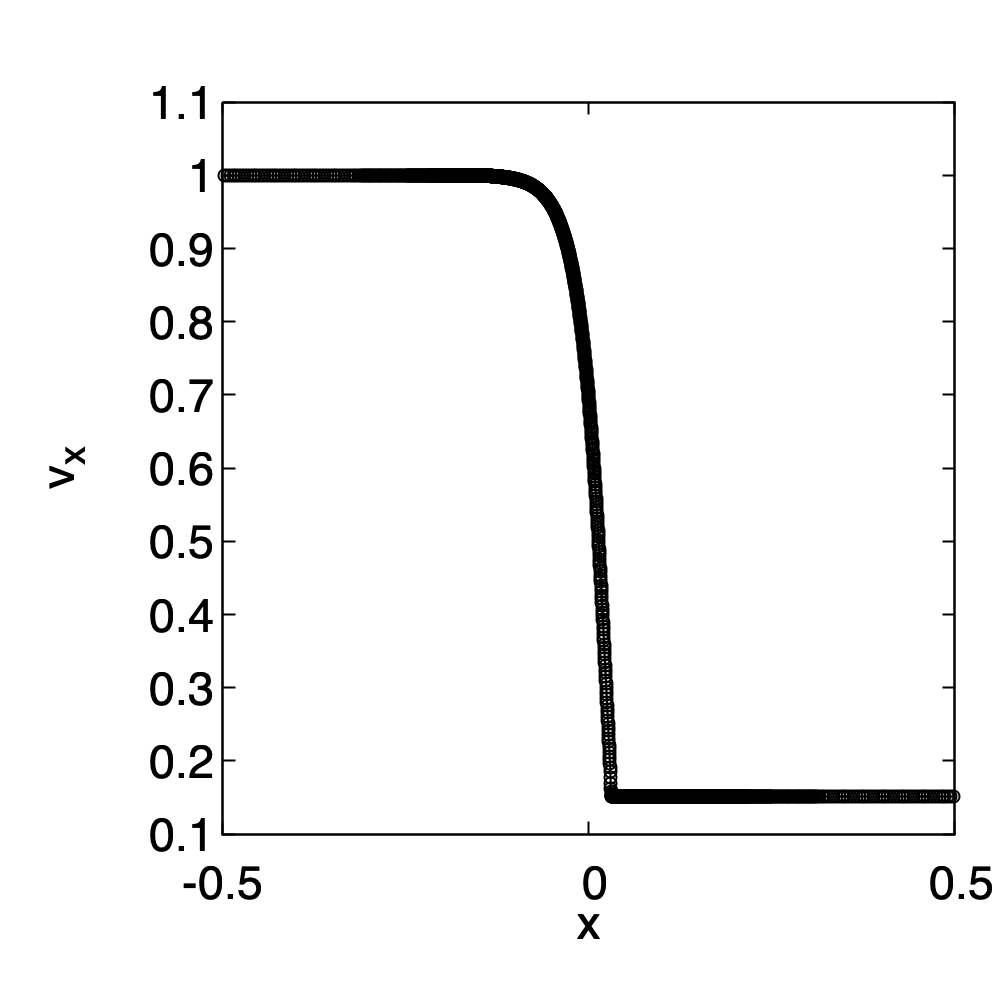}
   \includegraphics[width=6cm,clip]{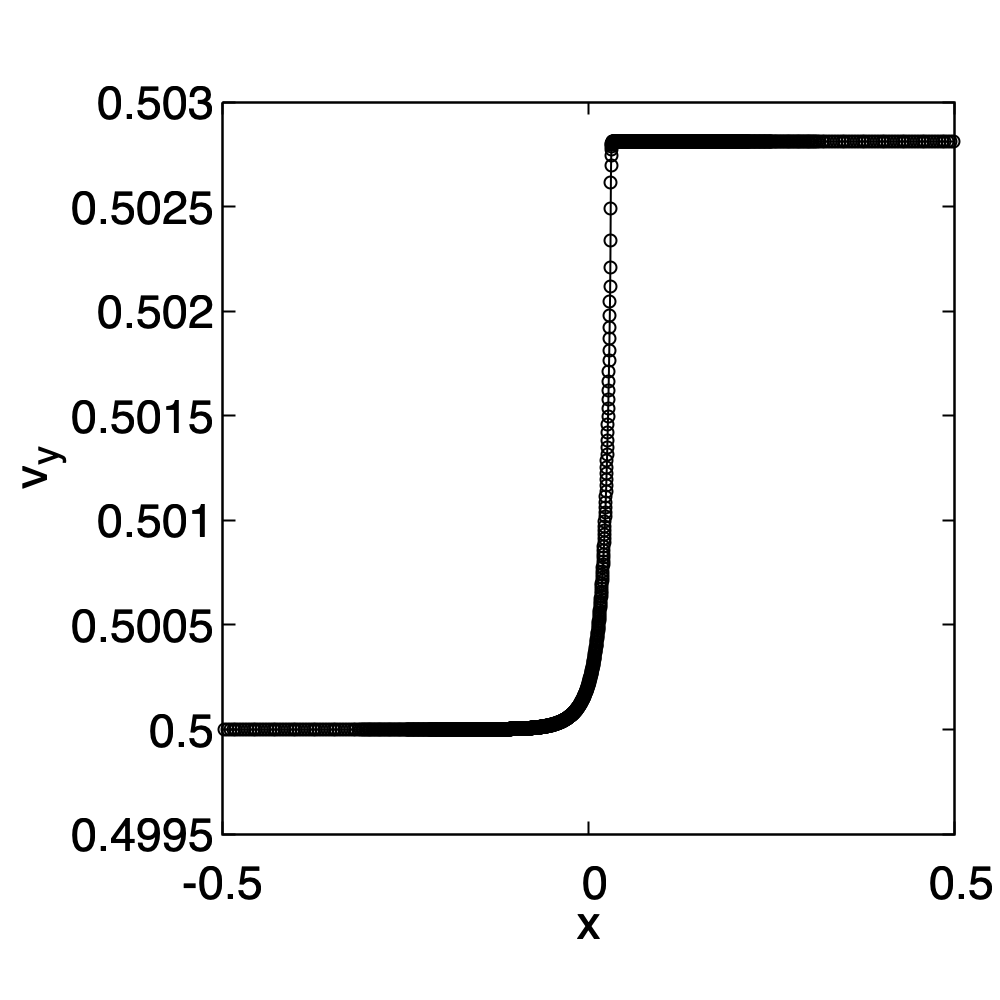}
   \includegraphics[width=6cm,clip]{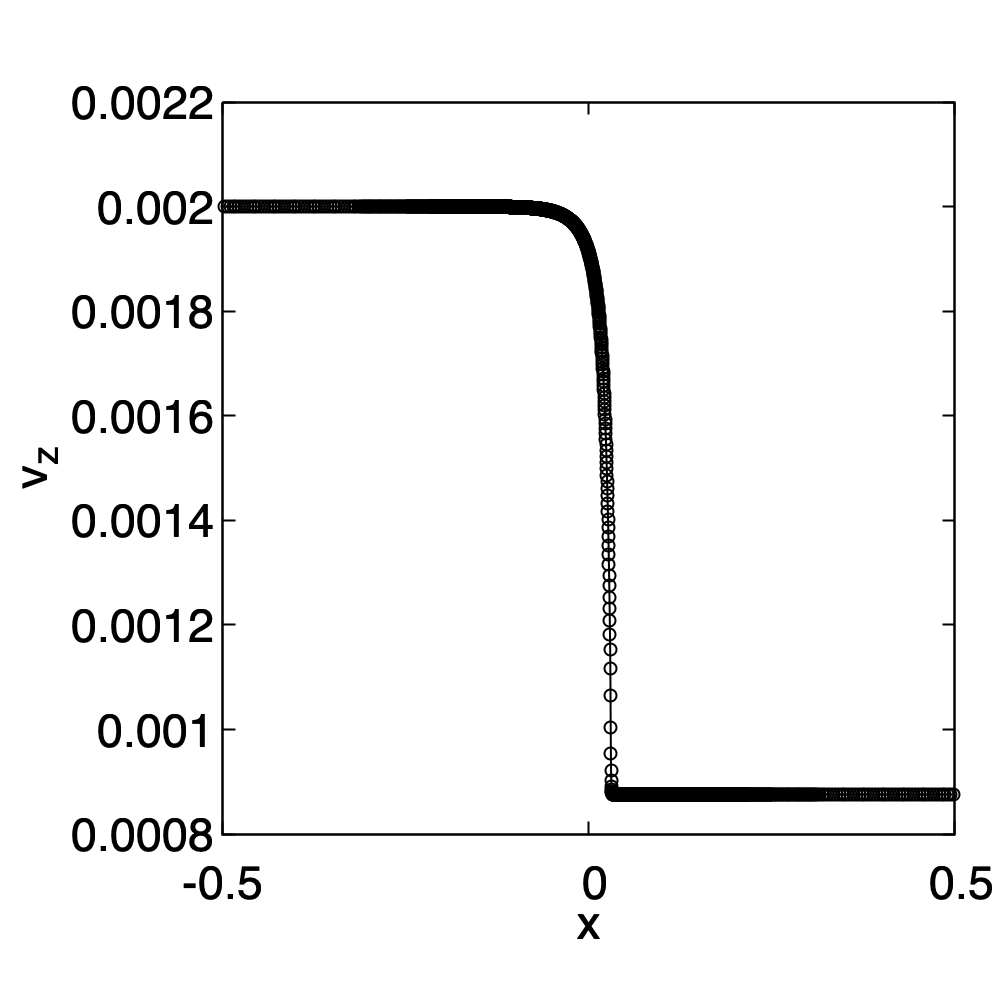}
   \includegraphics[width=6cm,clip]{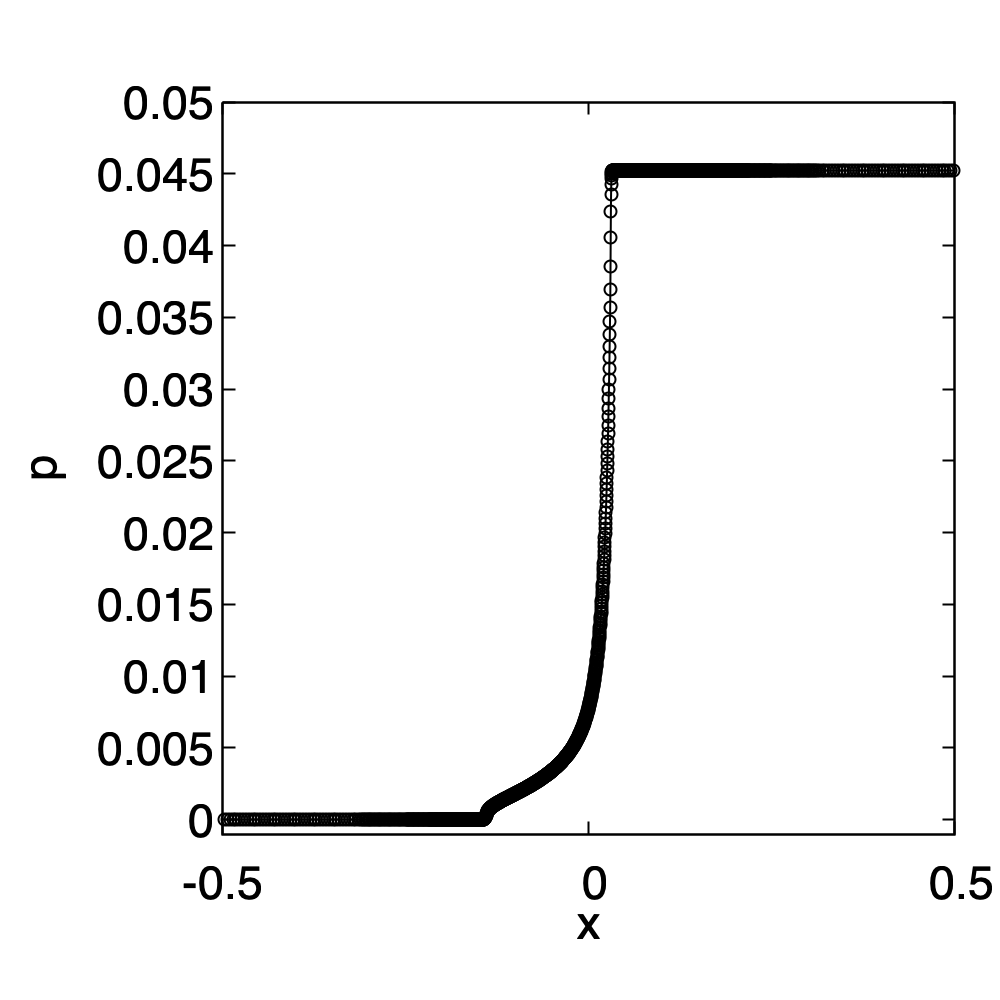}
   \includegraphics[width=6cm,clip]{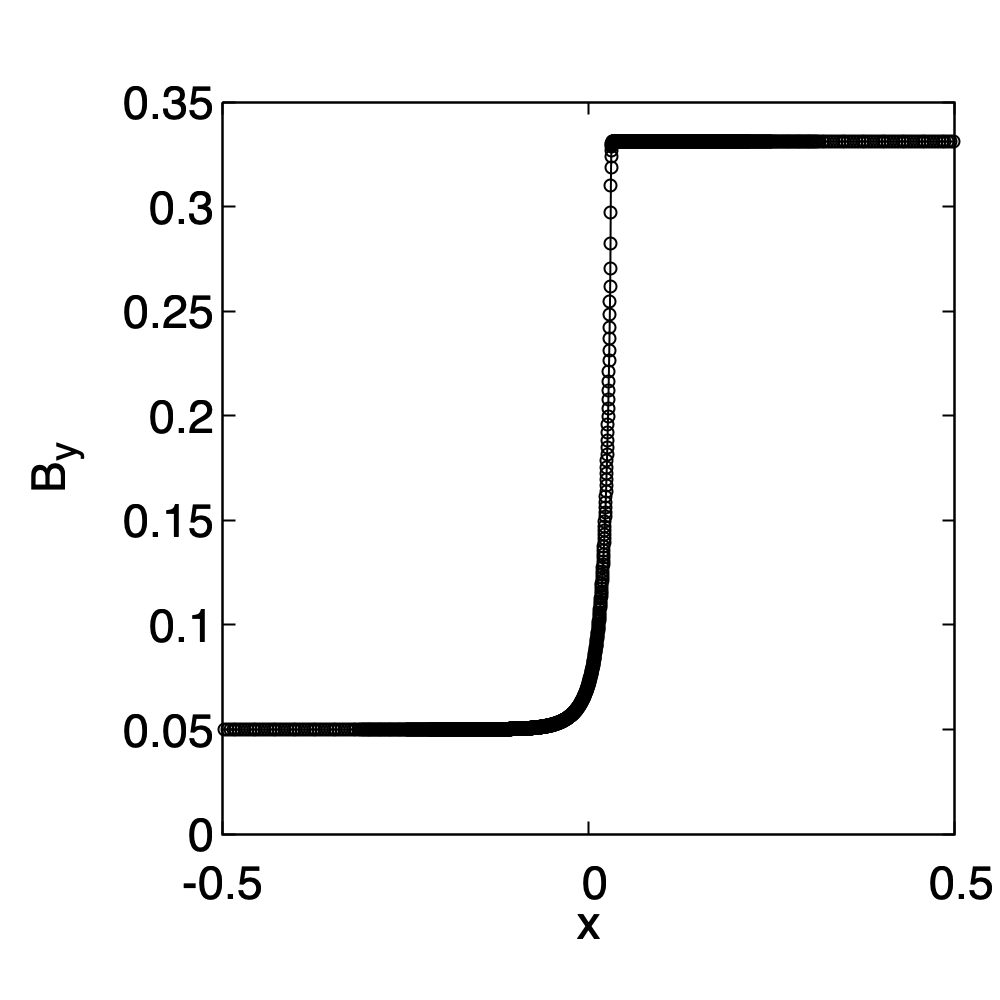}
   \includegraphics[width=6cm,clip]{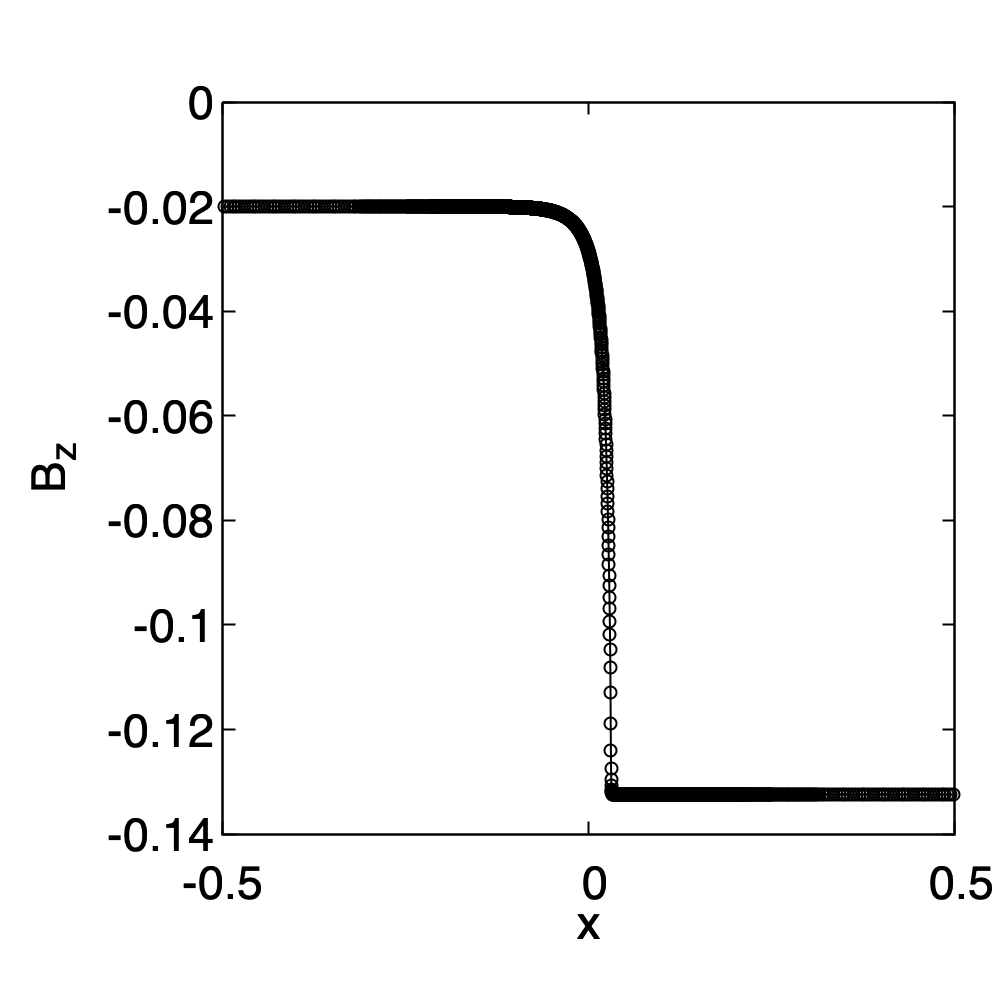}
   \includegraphics[width=6cm,clip]{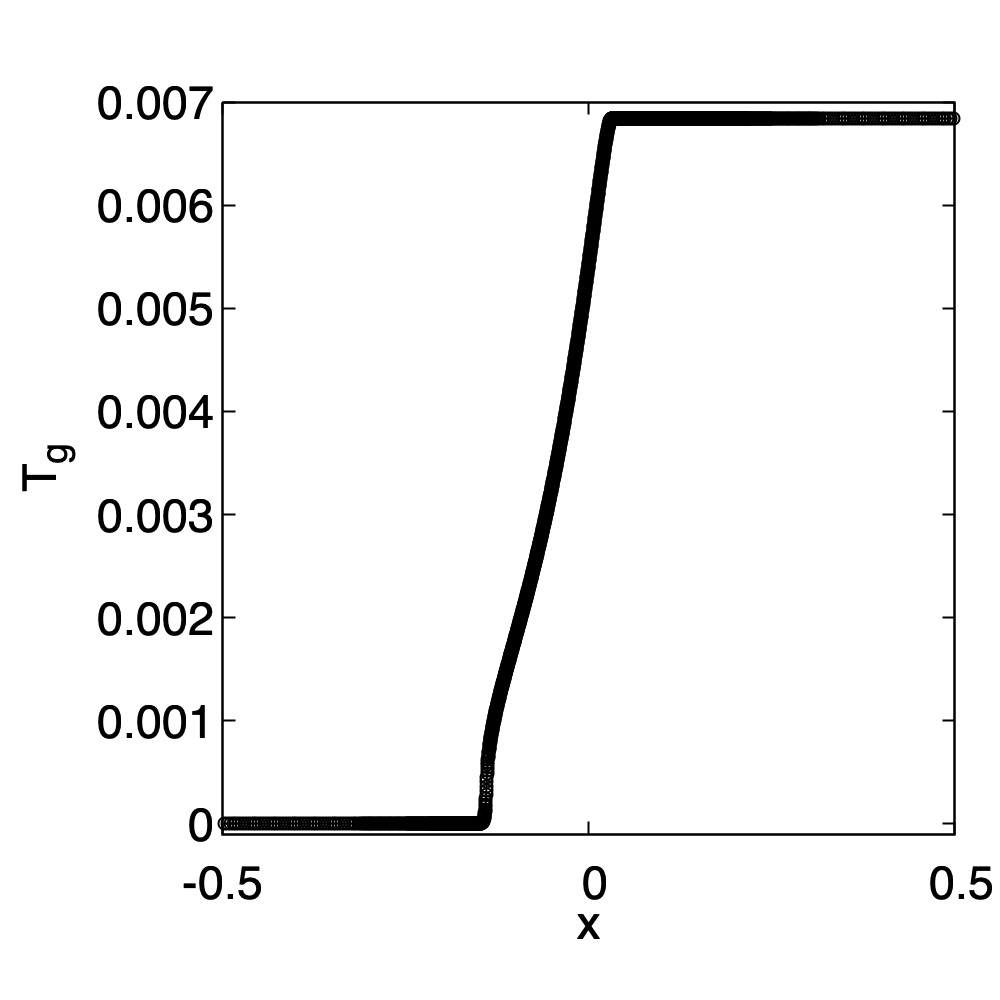}
   \includegraphics[width=6cm,clip]{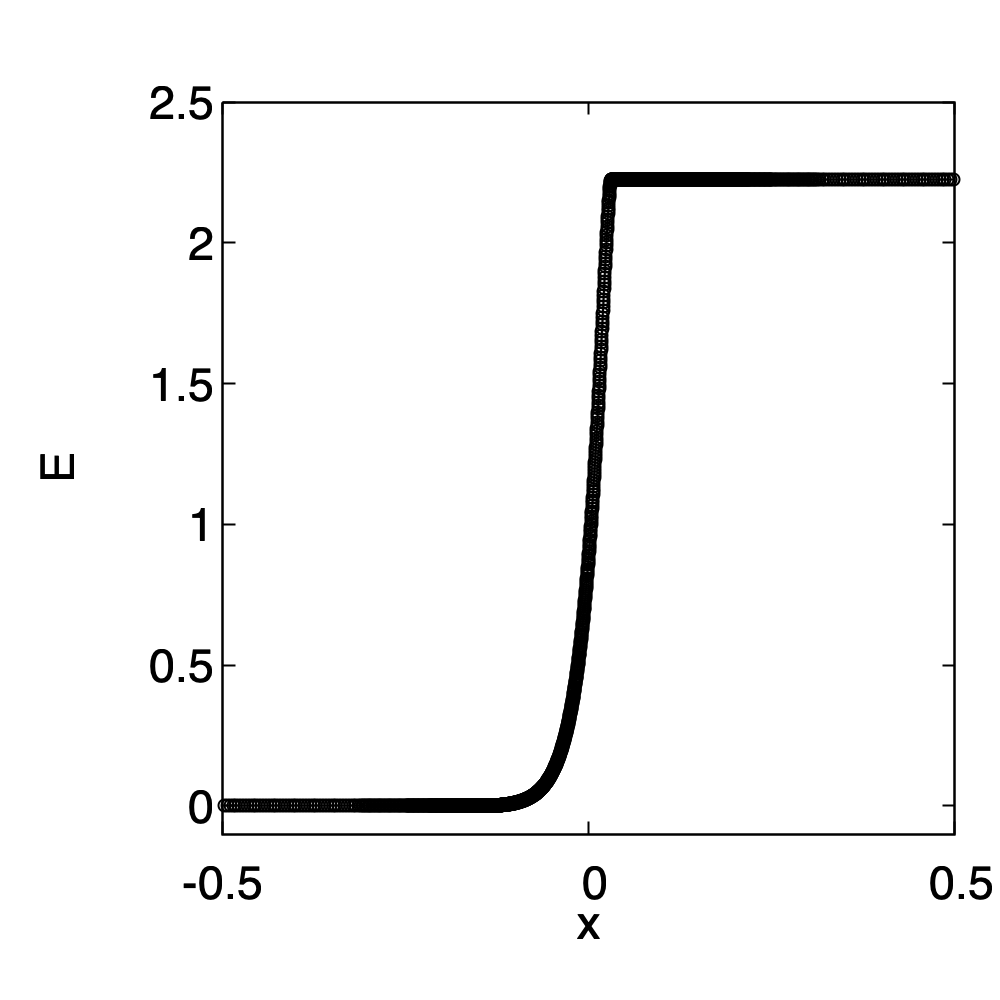}
   }

   \caption{Normalized density, $x$-velocity, $y$-velocity, $z$-velocity, plasma pressure, $y$-magnetic field, $z$-magnetic field, plasma temperature and radiation energy \nmn{density} profiles computed for the relaxed state of the fast magnetosonic shock. }
\label{fig:shock_fast}%
\end{figure*}

We now introduce various actual RMHD steady shock solutions, chosen to illustrate the variety of shock types familiar from MHD only.
Our first RMHD shock consists of a fast magnetosonic shock, where the plasma goes from an upstream superfast ($\mathcal{M}_f>1$) state to a downstream subfast, but still super-Alfv\'enic ($\mathcal{M}_f<1$, $\mathcal{M}_a>1$) state. The numerically settled solution for this fast shock case, is shown in Figure~(\ref{fig:shock_fast}). The upstream flow is a superfast magnetically dominated flow, thus $\beta<1$. The plasma temperature undergoes a monotonous increase in the precursor region, without overshooting the downstream value. 
All other plasma properties undergo a smooth transition with an ever-increasing slope in the precursor region, eventually approaching their respective downstream values. In the upstream region, the radiation pressure is negligible as compared to the plasma thermal pressure. In the downstream region, however, the radiation pressure dominates the plasma thermal pressure by an order of magnitude. A defining characteristic of a fast magnetosonic shock is the increase in magnitude of the tangential magnetic field components in the downstream region. This can be clearly observed in the $B_y$ and $B_z$ profiles in Figure~(\ref{fig:shock_fast}), as well as in Table~(\ref{table:RMHD_shocks}). This increase in the tangential magnetic field causes the magnetic field vector to bend away from the shock normal in the downstream region, as can be seen in Figure (\ref{fig:shock_fast}).

\begin{figure*} 
   \centering
   \FIG{
   \includegraphics[width=6cm,clip]{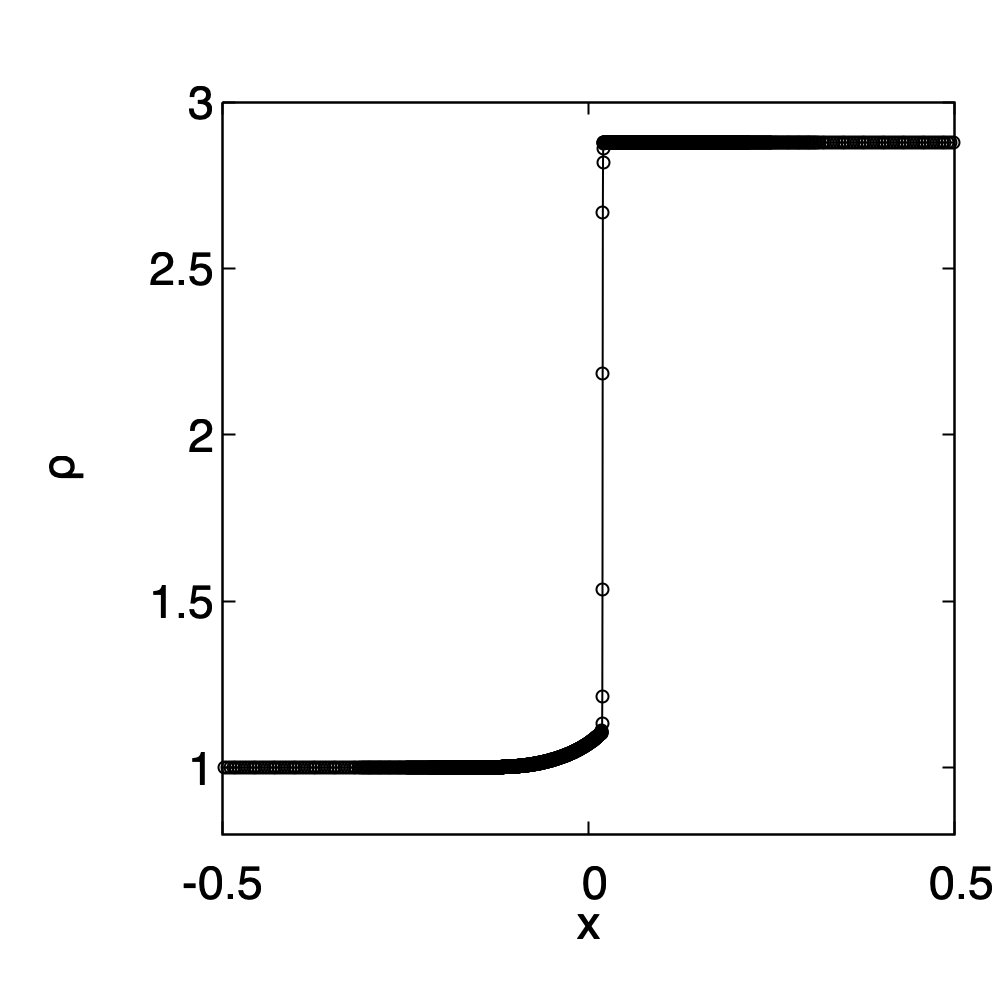}
   \includegraphics[width=6cm,clip]{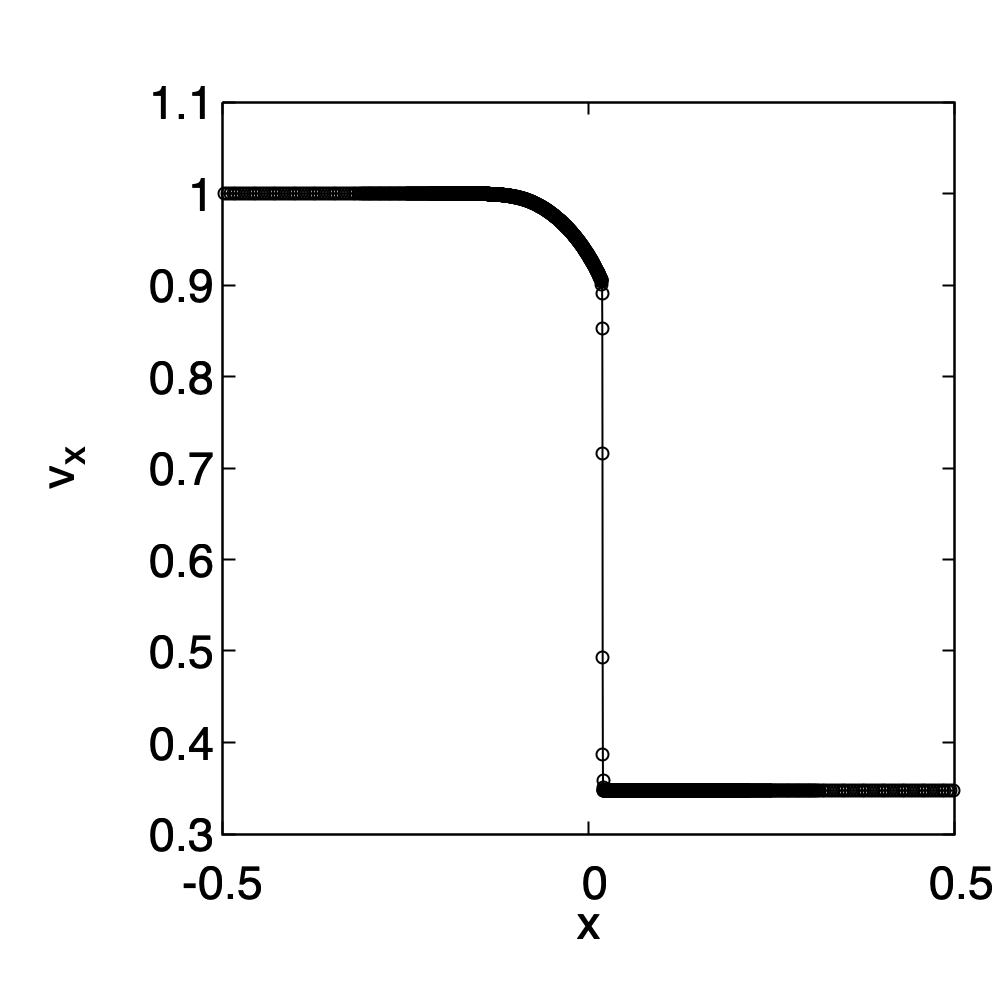}
   \includegraphics[width=6cm,clip]{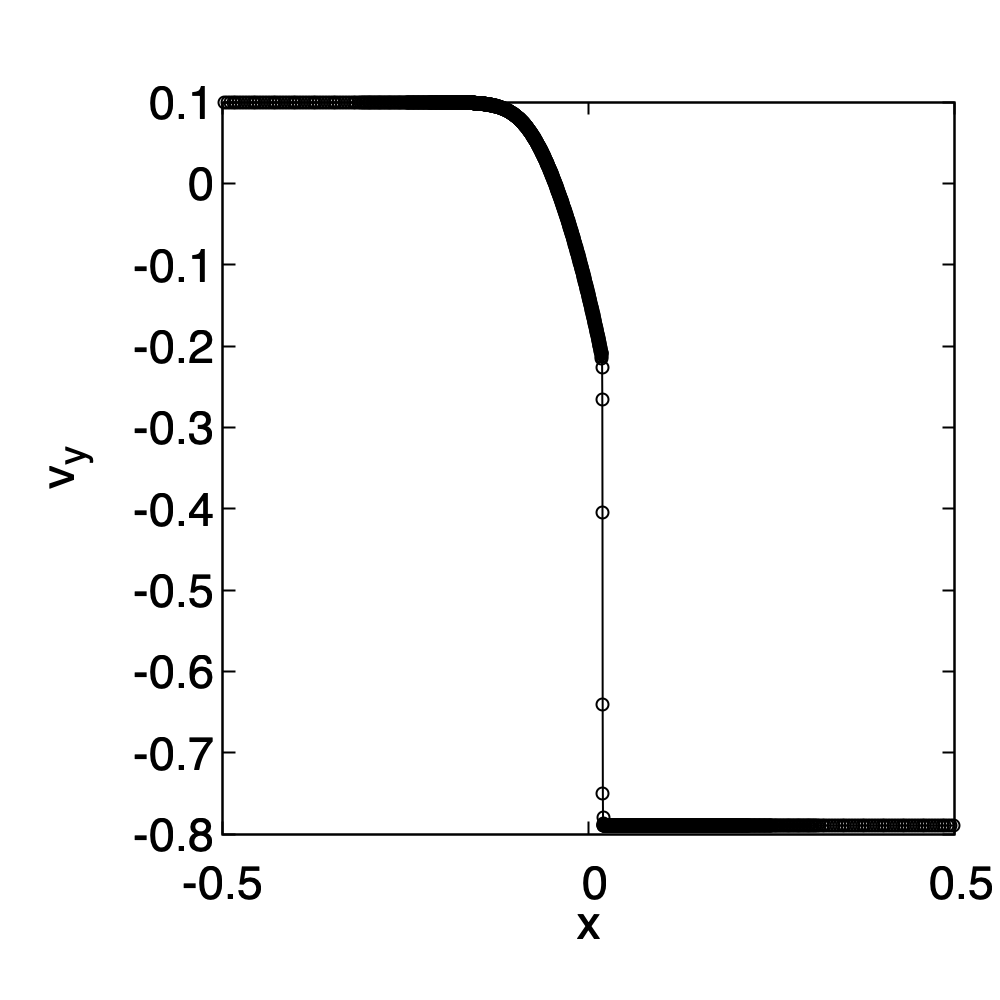}
   \includegraphics[width=6cm,clip]{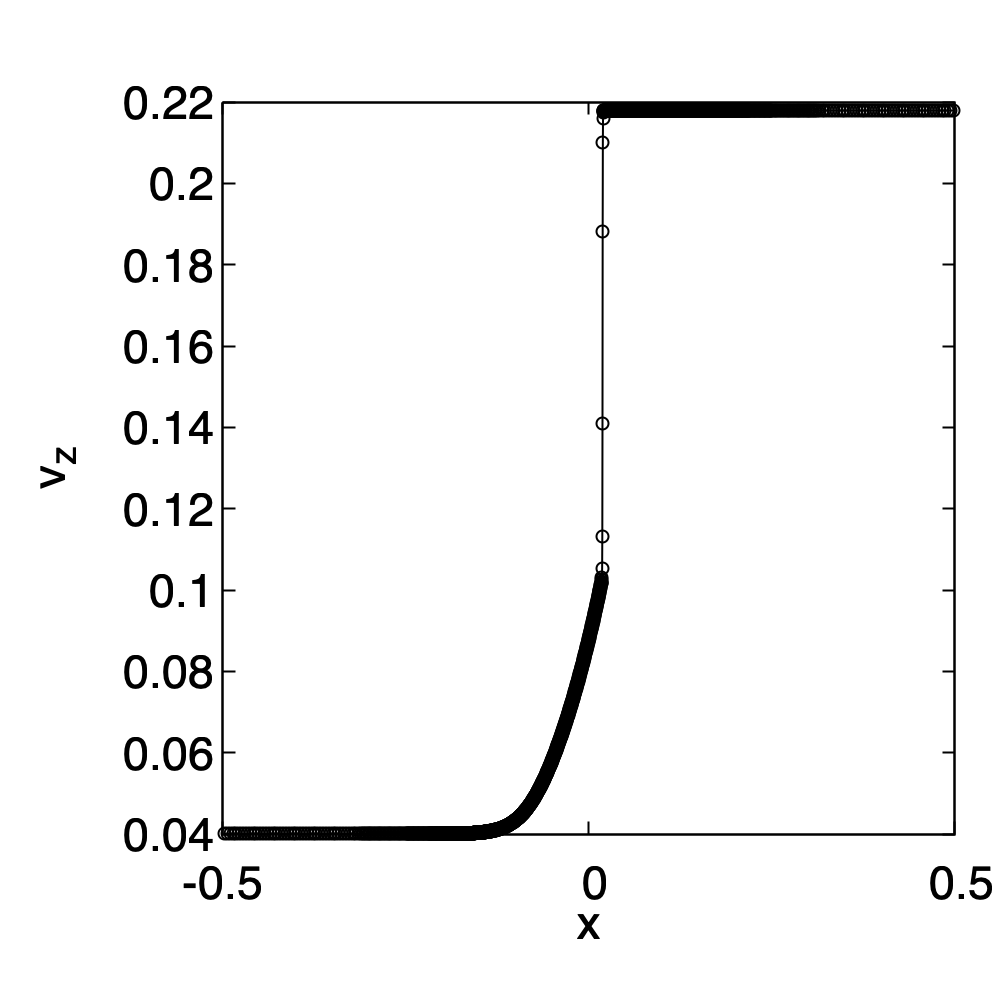}
   \includegraphics[width=6cm,clip]{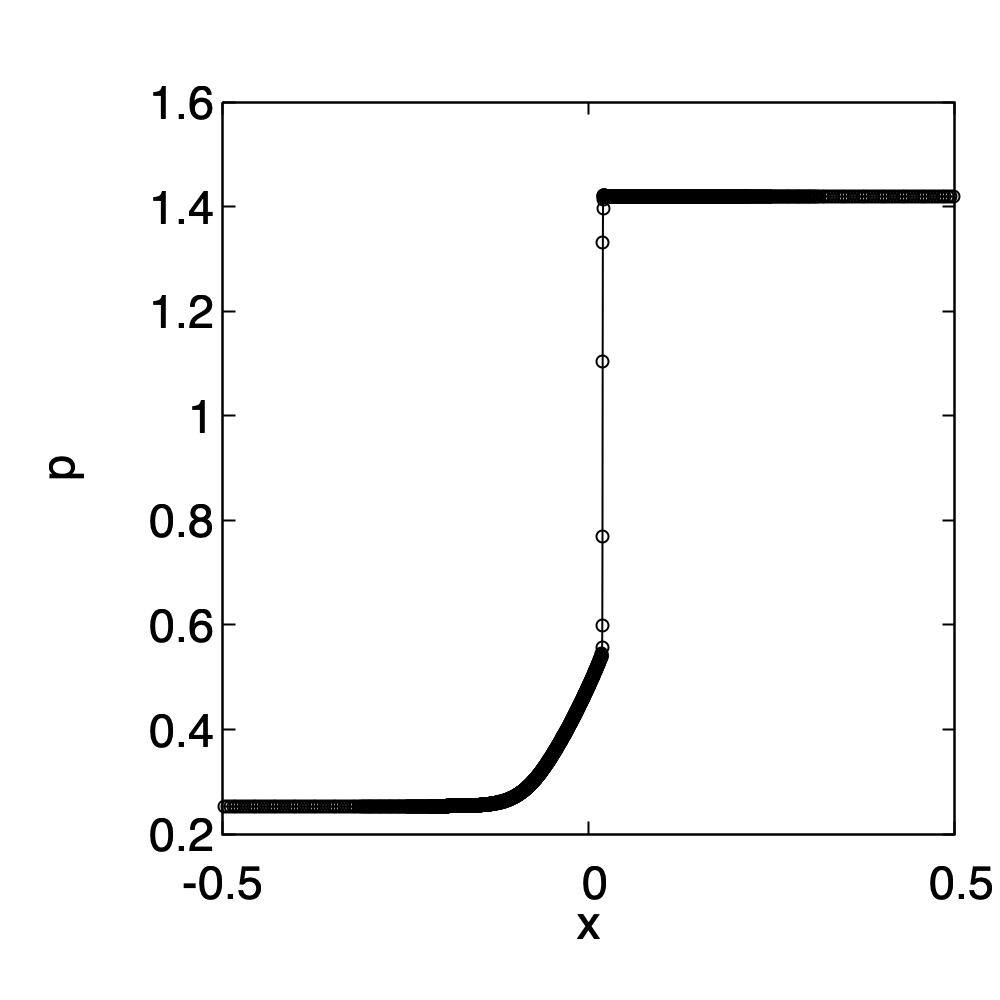}
   \includegraphics[width=6cm,clip]{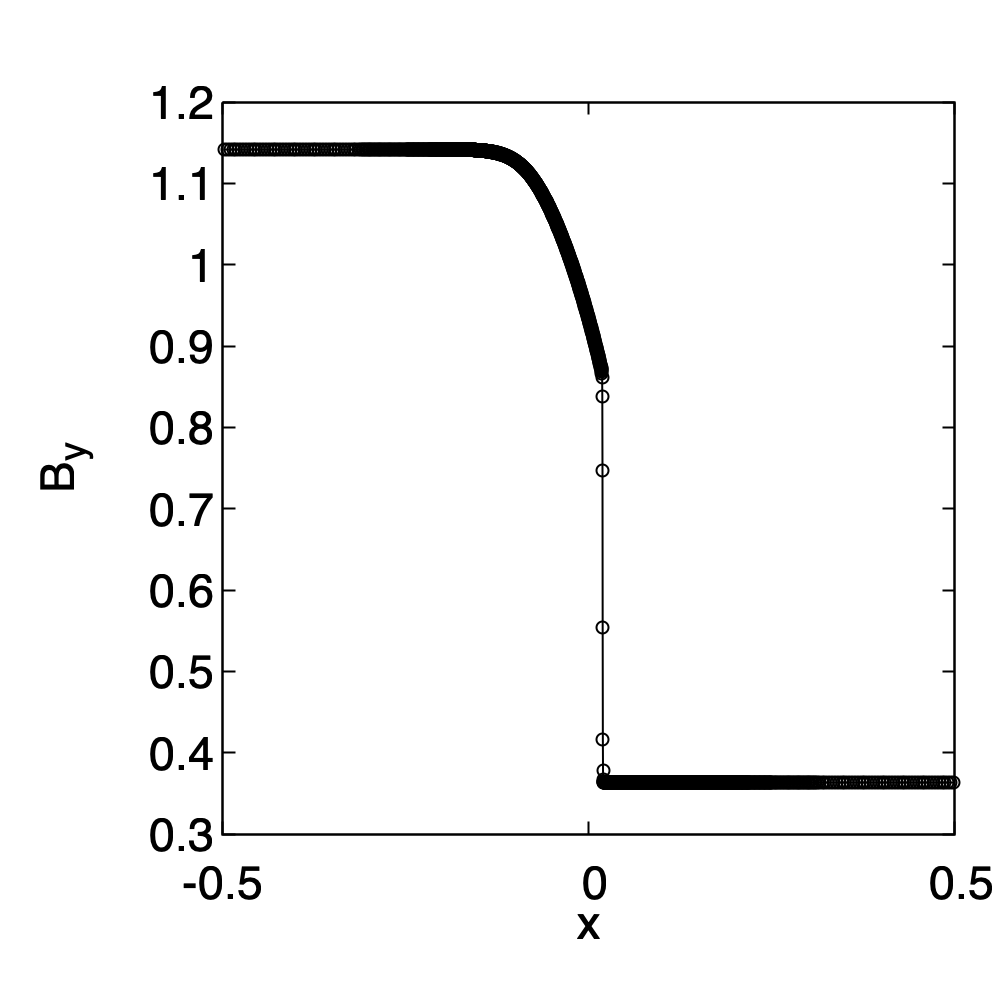}
   \includegraphics[width=6cm,clip]{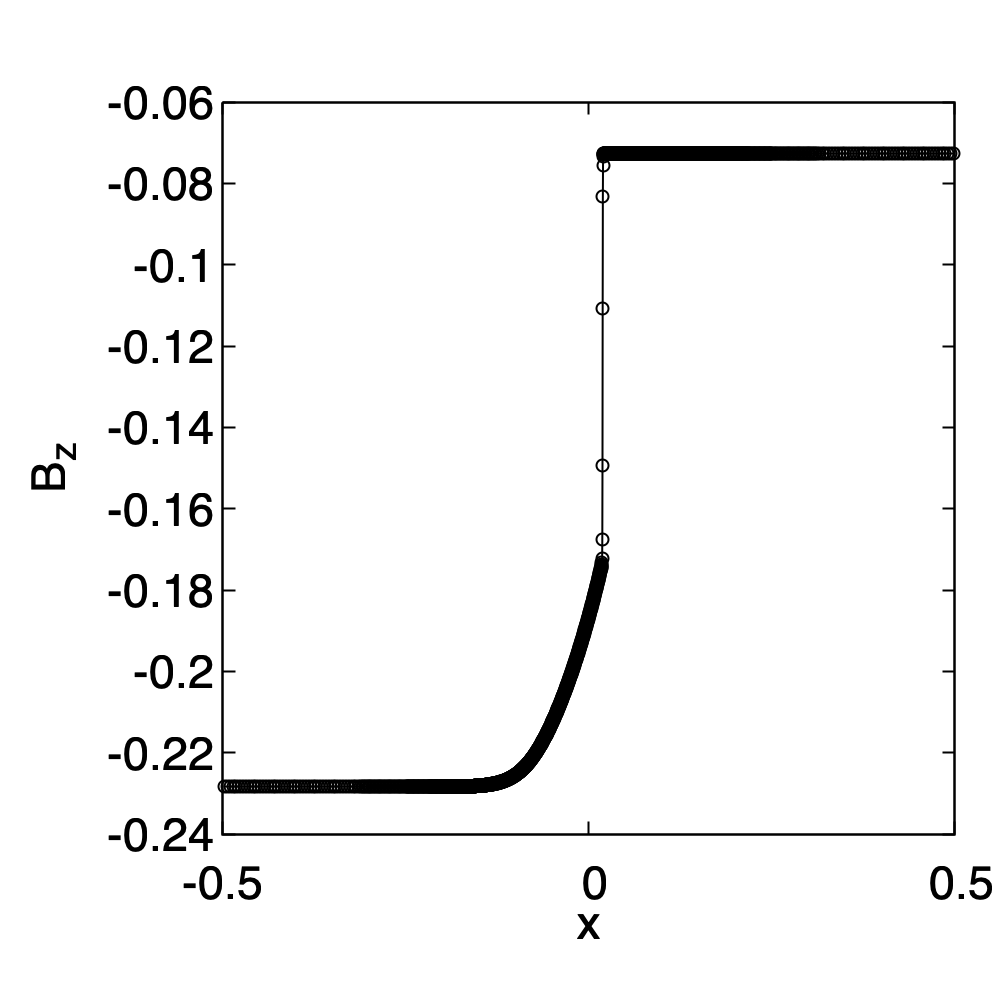}
   \includegraphics[width=6cm,clip]{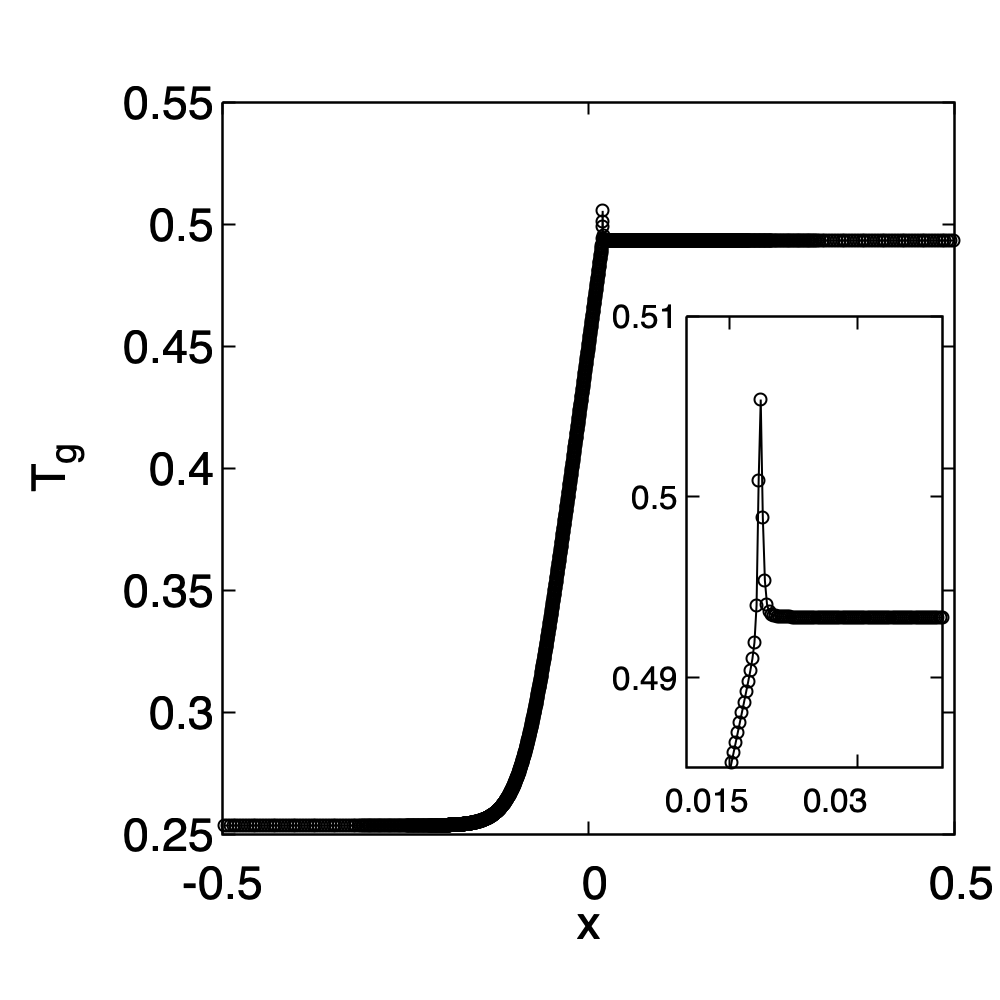}
   \includegraphics[width=6cm,clip]{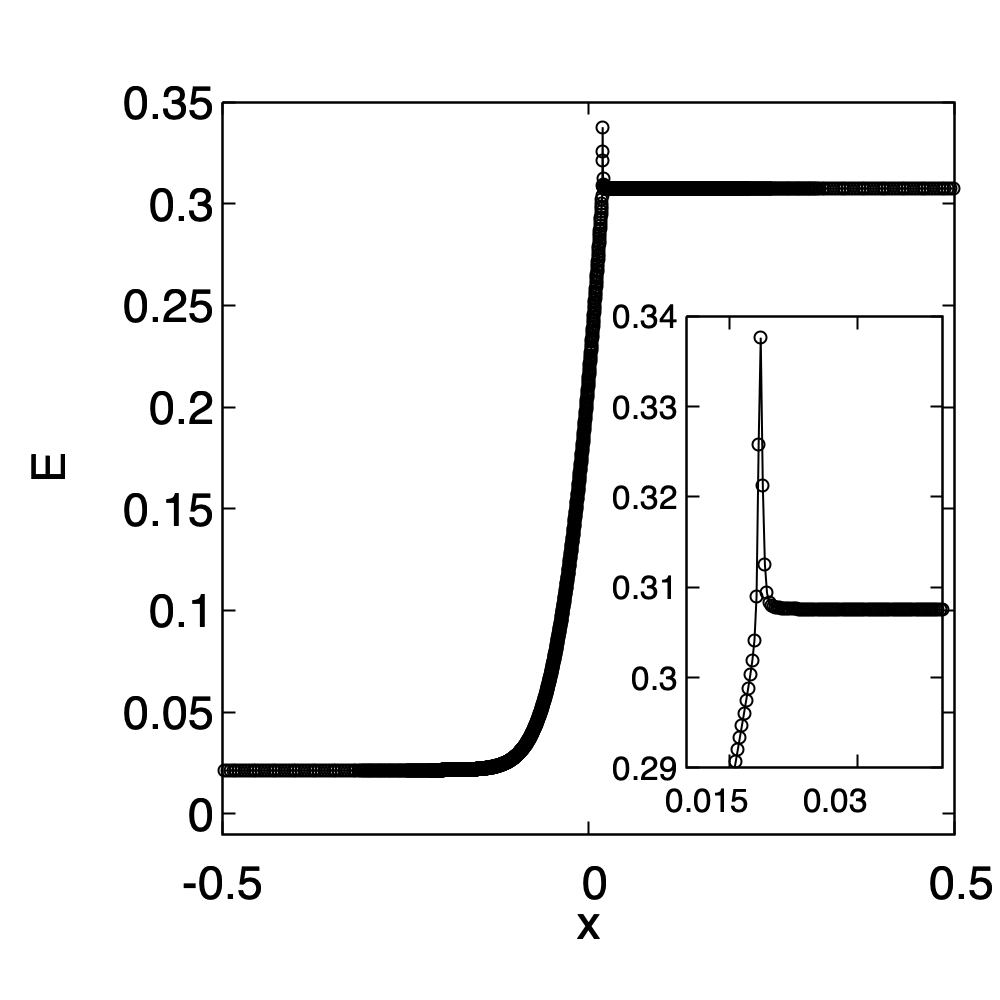}
   }
   \caption{Normalized density, $x$-velocity, $y$-velocity, $z$-velocity, plasma pressure, $y$-magnetic field, $z$-magnetic field, plasma temperature and radiation energy \nmn{density} profiles computed for the relaxed state of the slow magnetosonic shock.}
\label{fig:shock_slow}%
\end{figure*}

Our second RMHD test consists of a slow magnetosonic shock, where the plasma goes from an upstream superslow, sub-Alfv\'enic state ($\mathcal{M}_a<1, \mathcal{M}_s>1$) to a downstream subslow state ($\mathcal{M}_s<1$). The shock solution for this subcase is shown in Figure~(\ref{fig:shock_slow}). Similar to the earlier hydrodynamic shock, we notice a Zel'dovich spike in the plasma and radiation temperatures, as shown in the insets of the plasma temperature and radiation energy solutions, whereas all other properties vary smoothly and monotonously in the precursor region before undergoing a discontinuous jump. From the temperature profile, we infer that this shock is subcritical. An interesting feature of this shock is that it connects a magnetic pressure-dominated ($\beta<1$) upstream state to a plasma pressure-dominated ($\beta>1$) downstream state. As is expected in a slow magnetosonic shock, the tangential magnetic field components decrease in magnitude in the downstream region. This also causes the magnetic field vector to bend towards the shock normal, as can be seen in Figure (\ref{fig:shock_slow}).

\begin{figure*} 
   \centering
   \FIG{
   \includegraphics[width=6cm,clip]{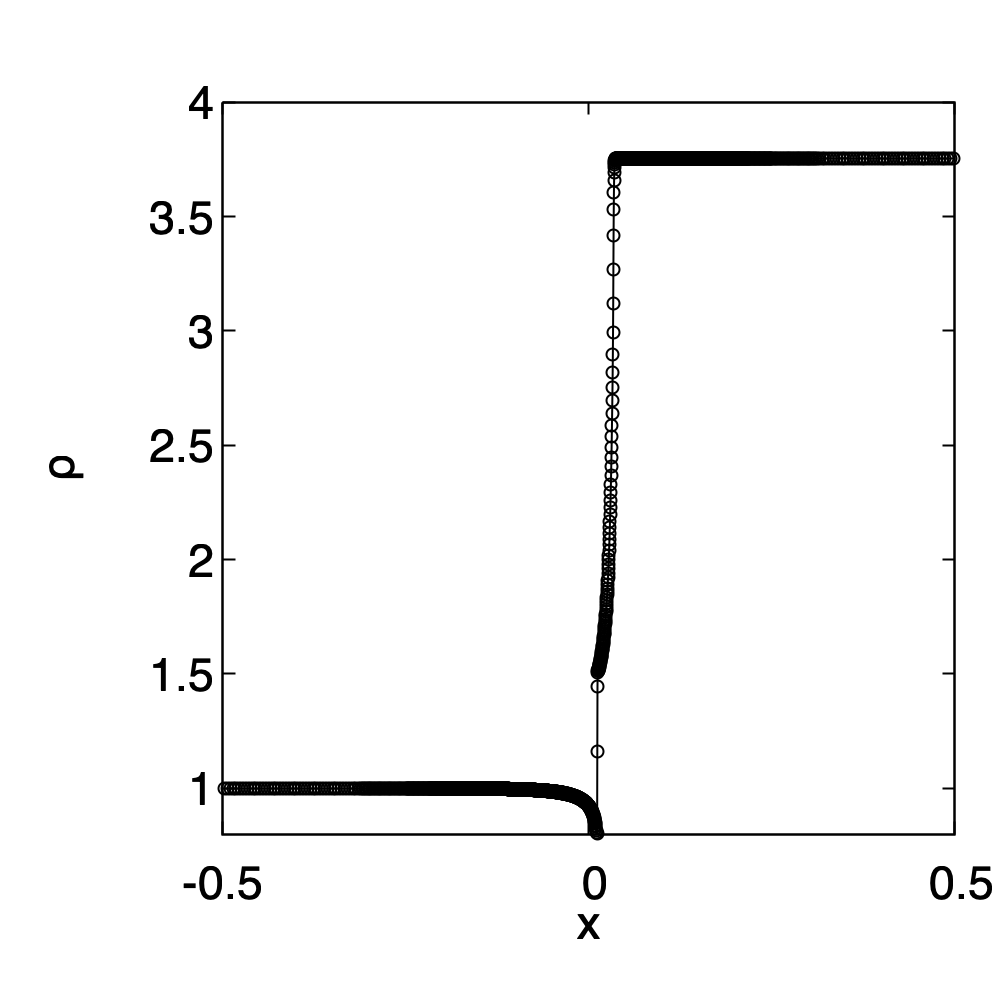}
   \includegraphics[width=6cm,clip]{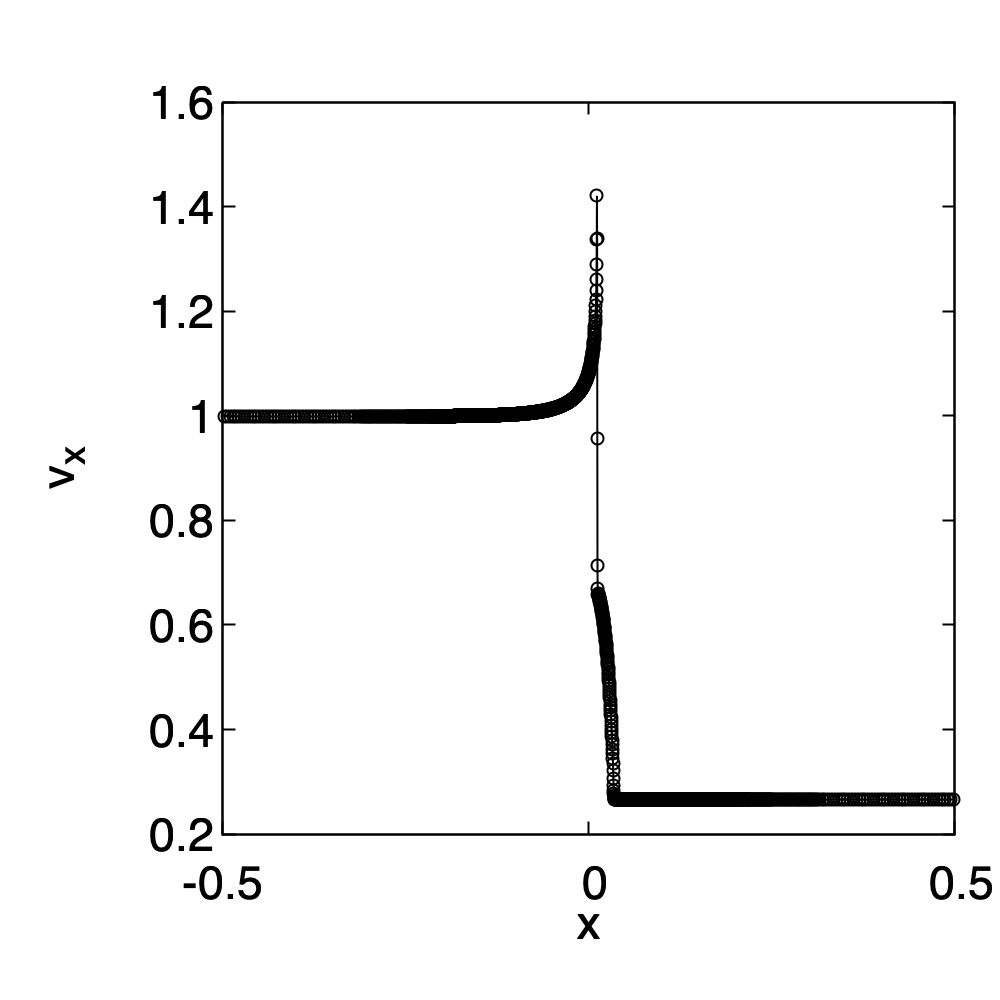}
   \includegraphics[width=6cm,clip]{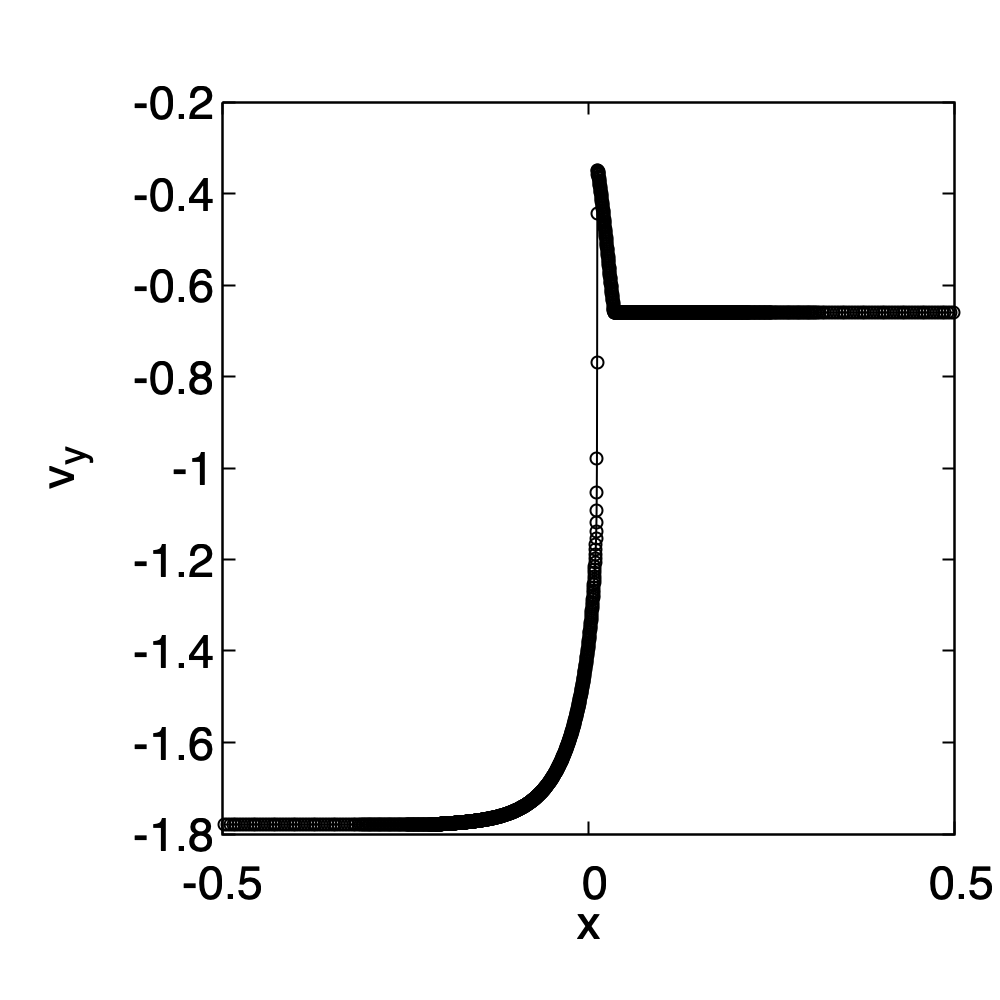}
   \includegraphics[width=6cm,clip]{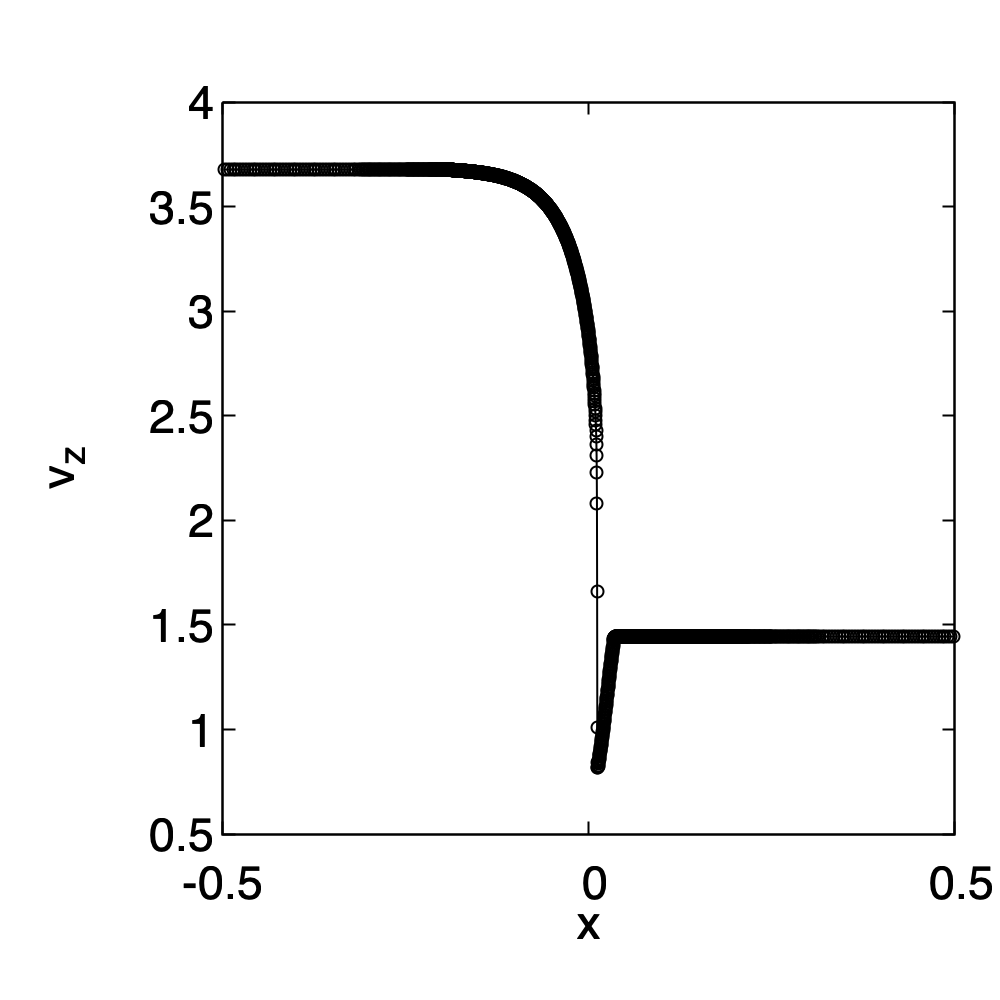}
   \includegraphics[width=6cm,clip]{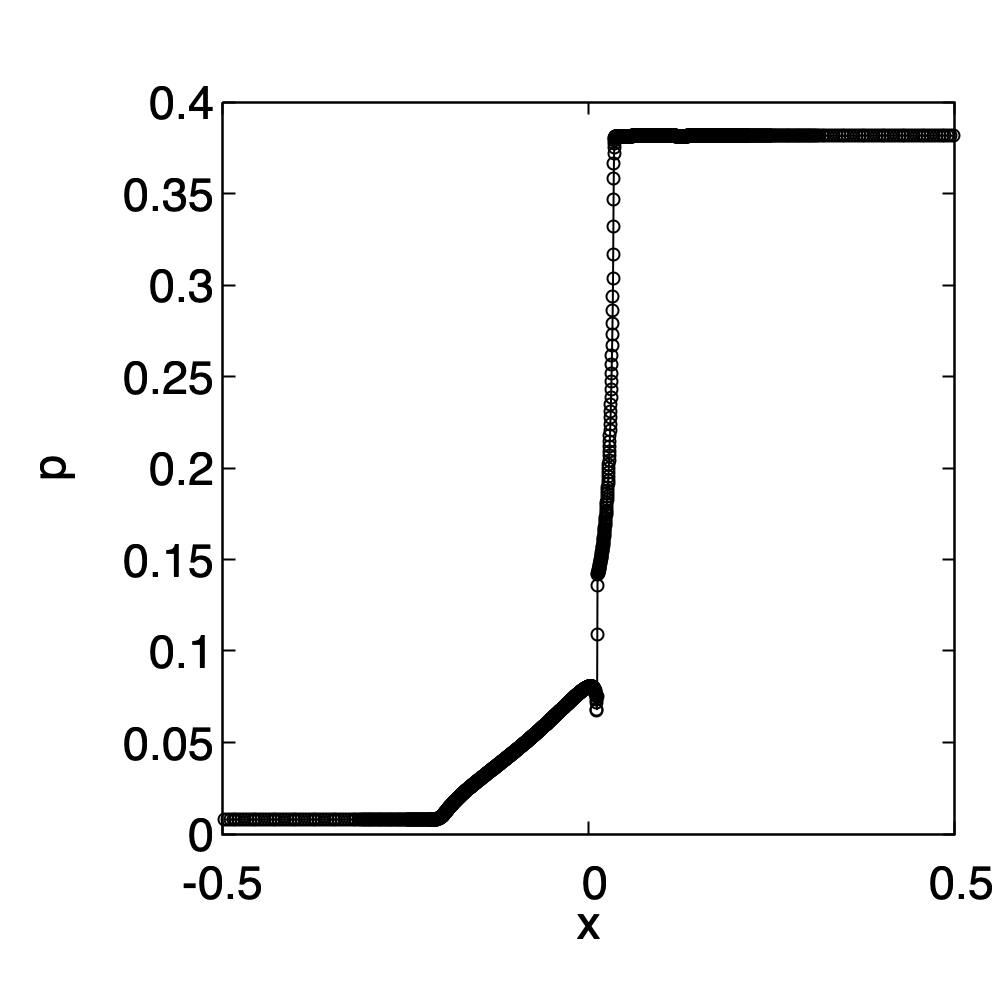}
   \includegraphics[width=6cm,clip]{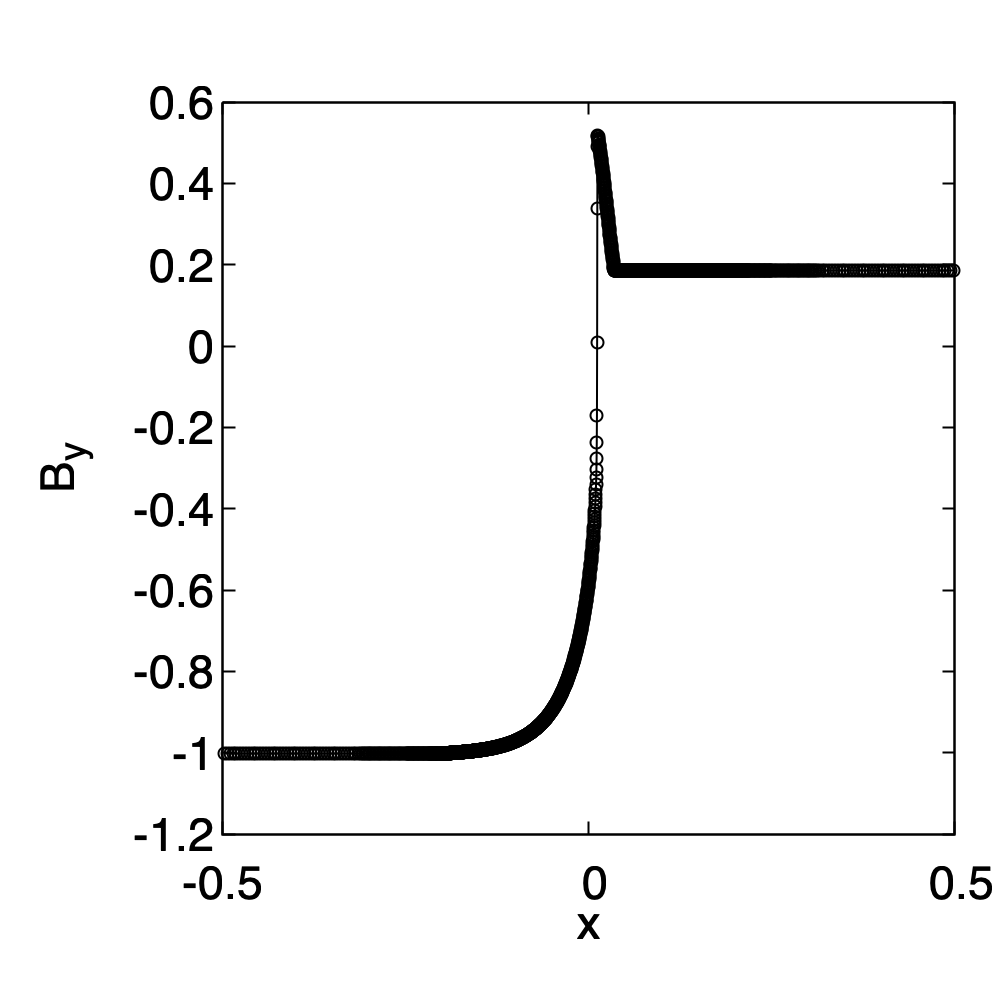}
   \includegraphics[width=6cm,clip]{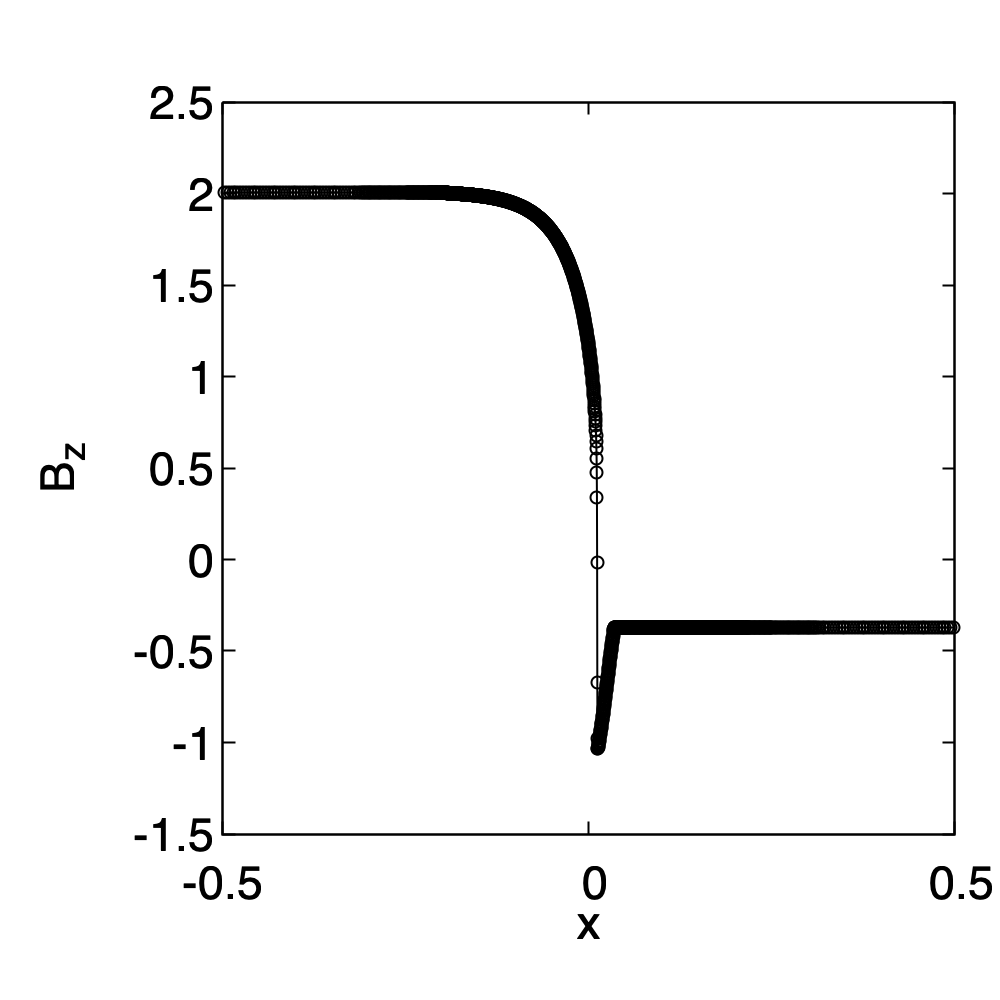}
   \includegraphics[width=6cm,clip]{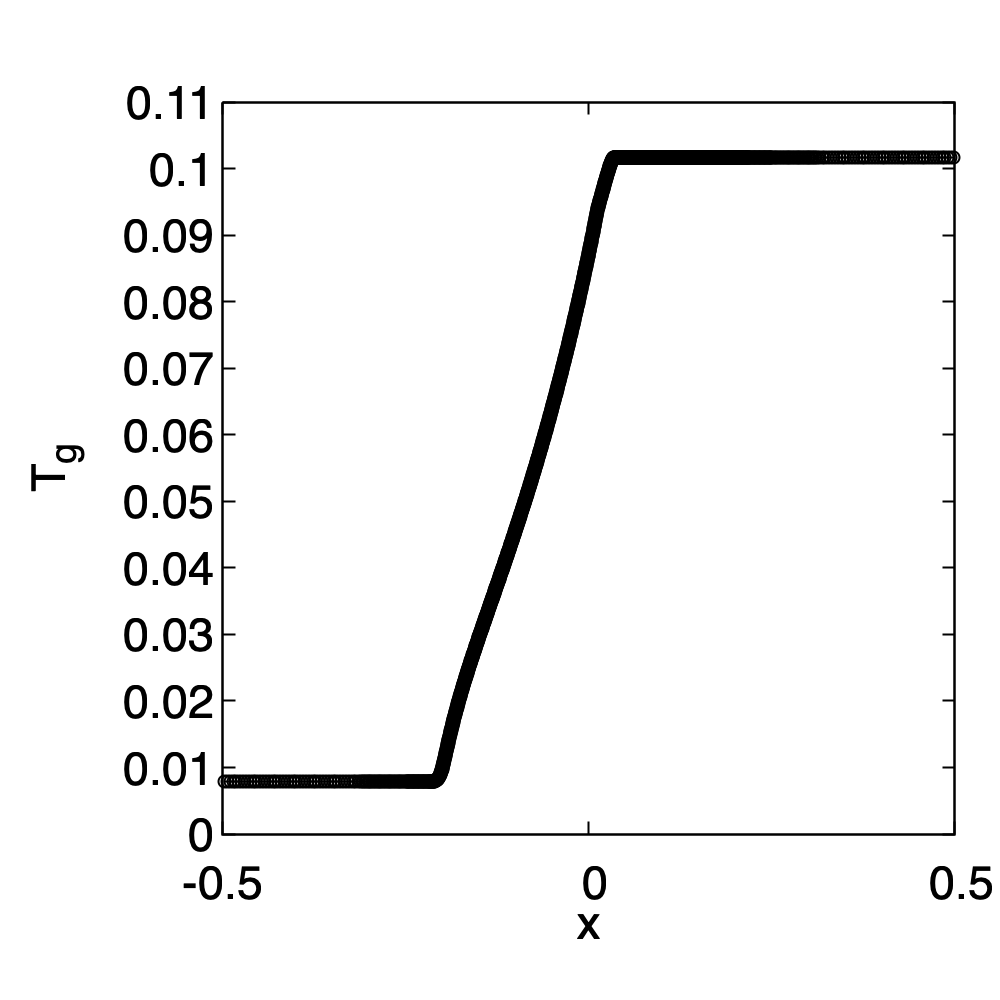}
   \includegraphics[width=6cm,clip]{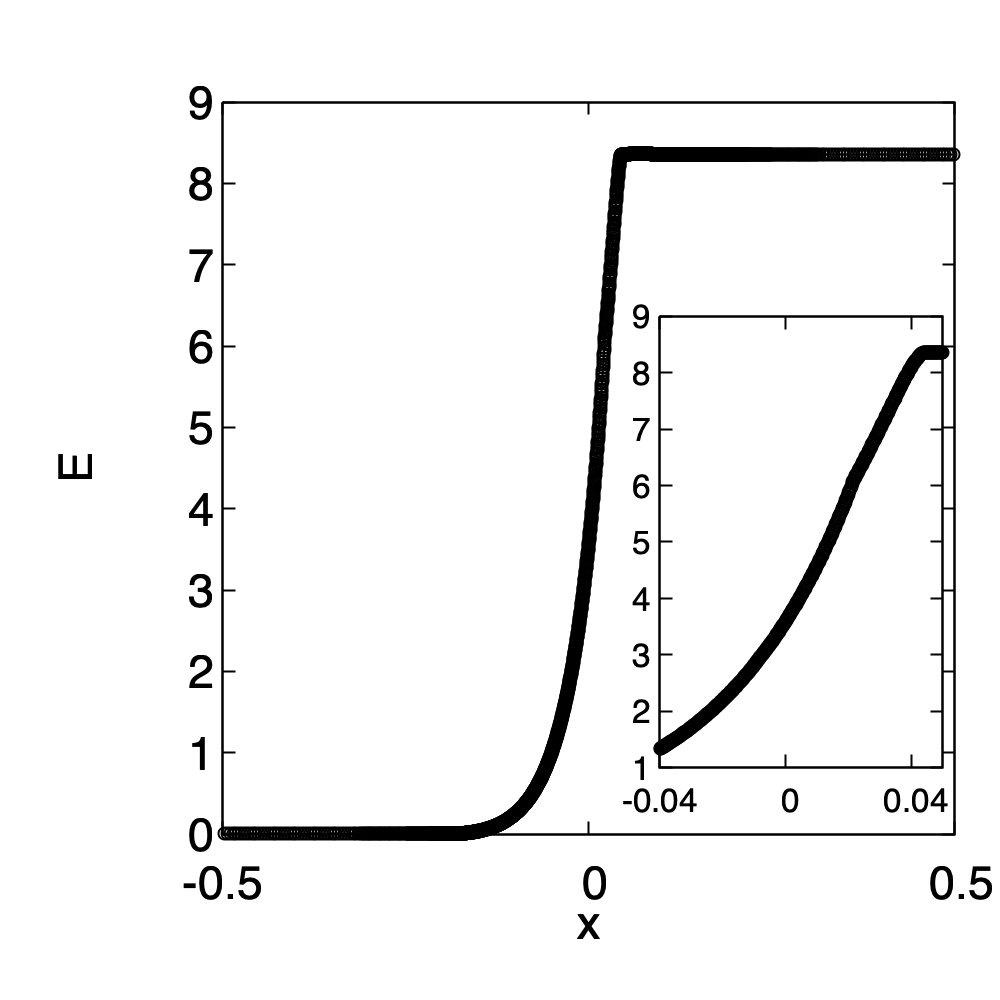}
   }
   \caption{Normalized density, $x$-velocity, $y$-velocity, $z$-velocity, plasma pressure, $y$-magnetic field, $z$-magnetic field, plasma temperature and radiation energy \nmn{density} profiles computed for the relaxed state of the intermediate magnetosonic shock.}
\label{fig:shock_alfslow}%
\end{figure*}

Next, we turn attention to an intermediate magnetosonic shock, where the plasma goes from an upstream, low-$\beta$, super-Alfv\'enic ($\mathcal{M}_f<1$, $\mathcal{M}_a>1$) state to a downstream subslow state ($\mathcal{M}_s<1$). A constant background opacity of $\kappa = 200$ cm$^2$/g is used for this case. The shock solution obtained is shown in Figure~(\ref{fig:shock_alfslow}). In such an intermediate shock, the tangential magnetic field flips its direction across the shock, as seen by the sign changes of $B_y$ and $B_z$ in Figure~(\ref{fig:shock_alfslow}) and in Table~(\ref{table:RMHD_shocks}). This causes the magnetic field vector to flip across the shock normal in the downstream region. This case involves a precursor region and a short but observable relaxation region. The plasma accelerates to a higher velocity in the precursor region, before undergoing sudden deceleration through the shock. The density, thermal pressure and normal velocity undergo relatively smaller jumps through the shock, and most of the transition to their far downstream values occurs in the relaxation region. After going through the shock, the tangential velocity and magnetic field components overshoot their far downstream values, before monotonically approaching their downstream values in the relaxation region. The plasma temperature undergoes a monotonous transition between the upstream and downstream values. The radiation energy increases with an increasing slope in the precursor region, but shows a linear increase in the relaxation region. This can be observed in the magnified view in the inset of the radiation energy solution shown in Figure~(\ref{fig:shock_alfslow}). The thermal pressure dominates the radiation pressure on the upstream side, whereas the radiation pressure dominates by an order of magnitude on the downstream side.

\begin{figure*} 
   \centering
   \FIG{
   \includegraphics[width=6cm,clip]{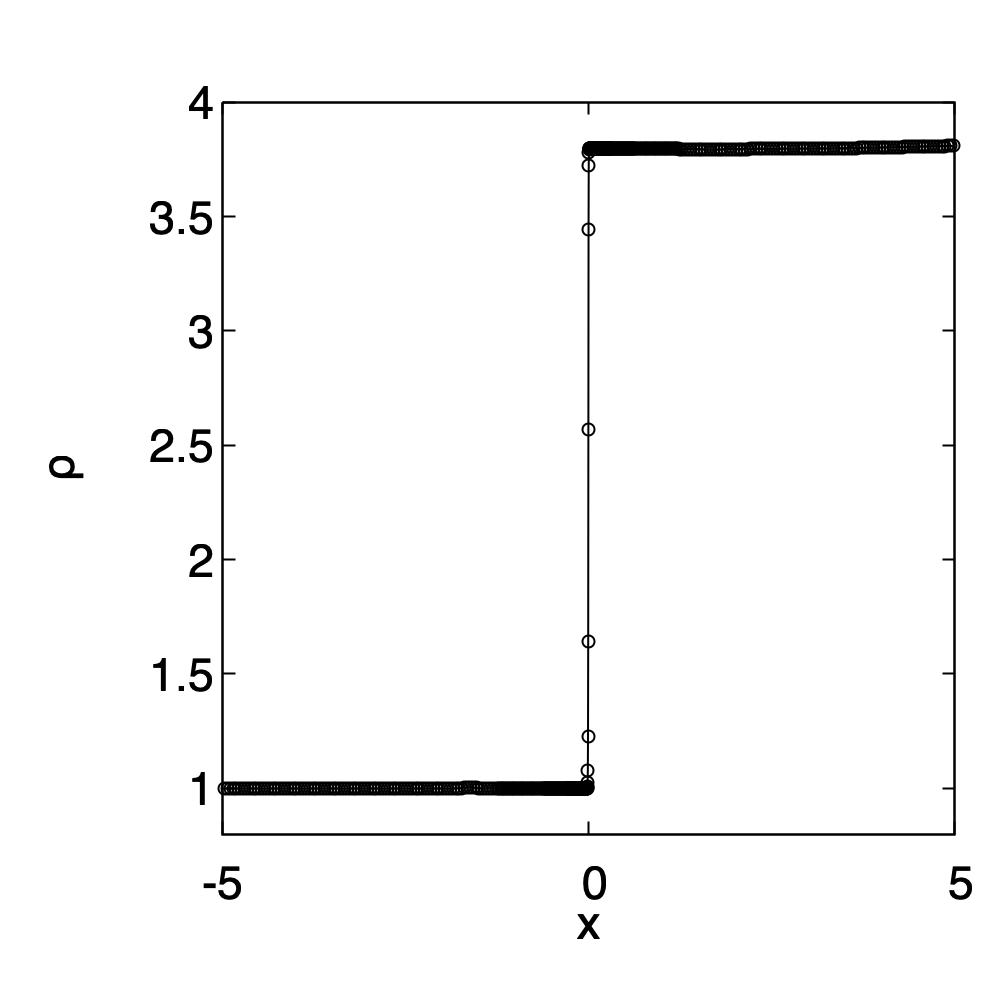}
   \includegraphics[width=6cm,clip]{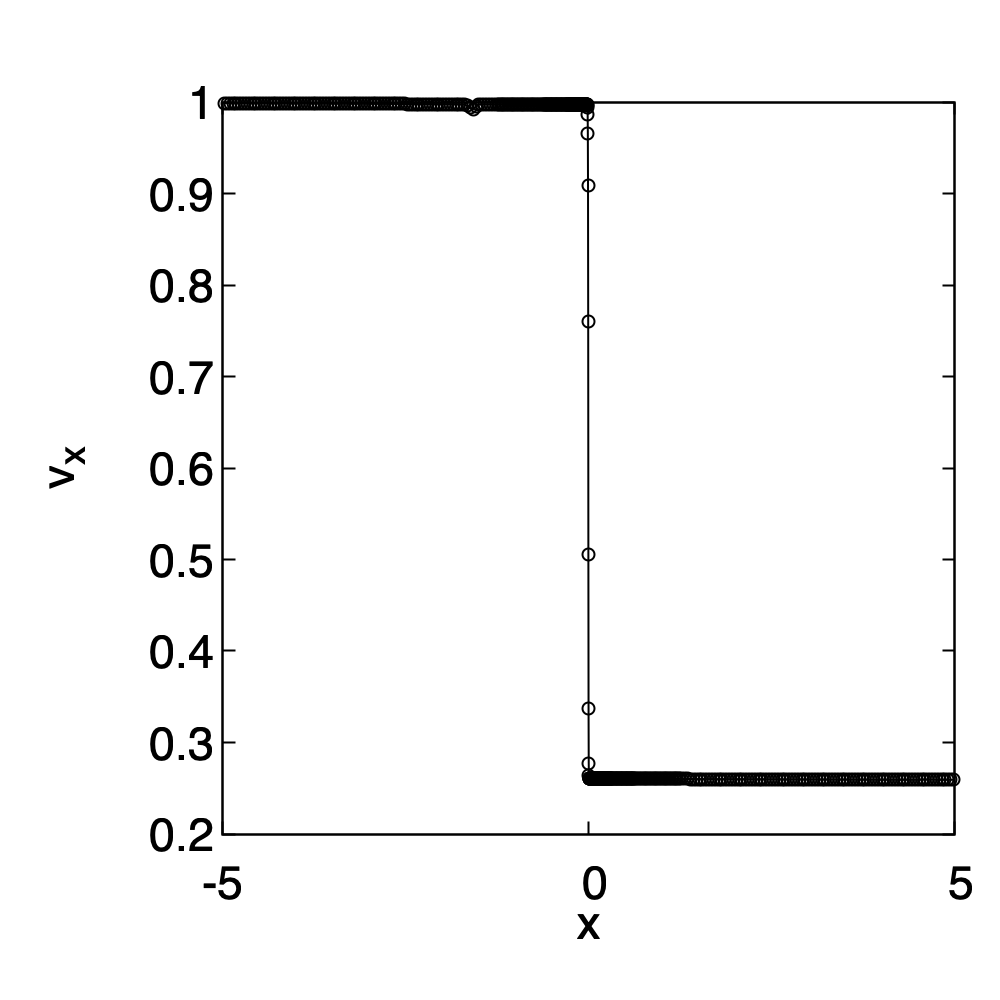}
   \includegraphics[width=6cm,clip]{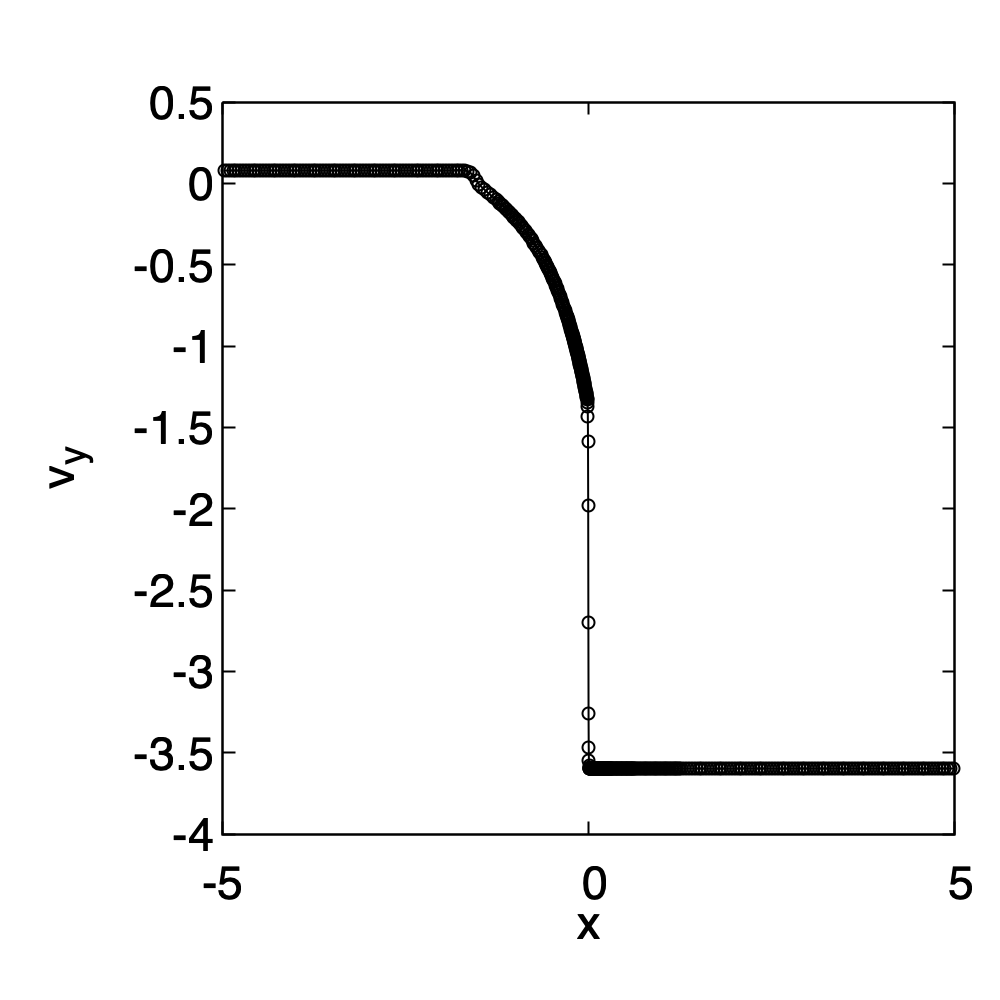}
   \includegraphics[width=6cm,clip]{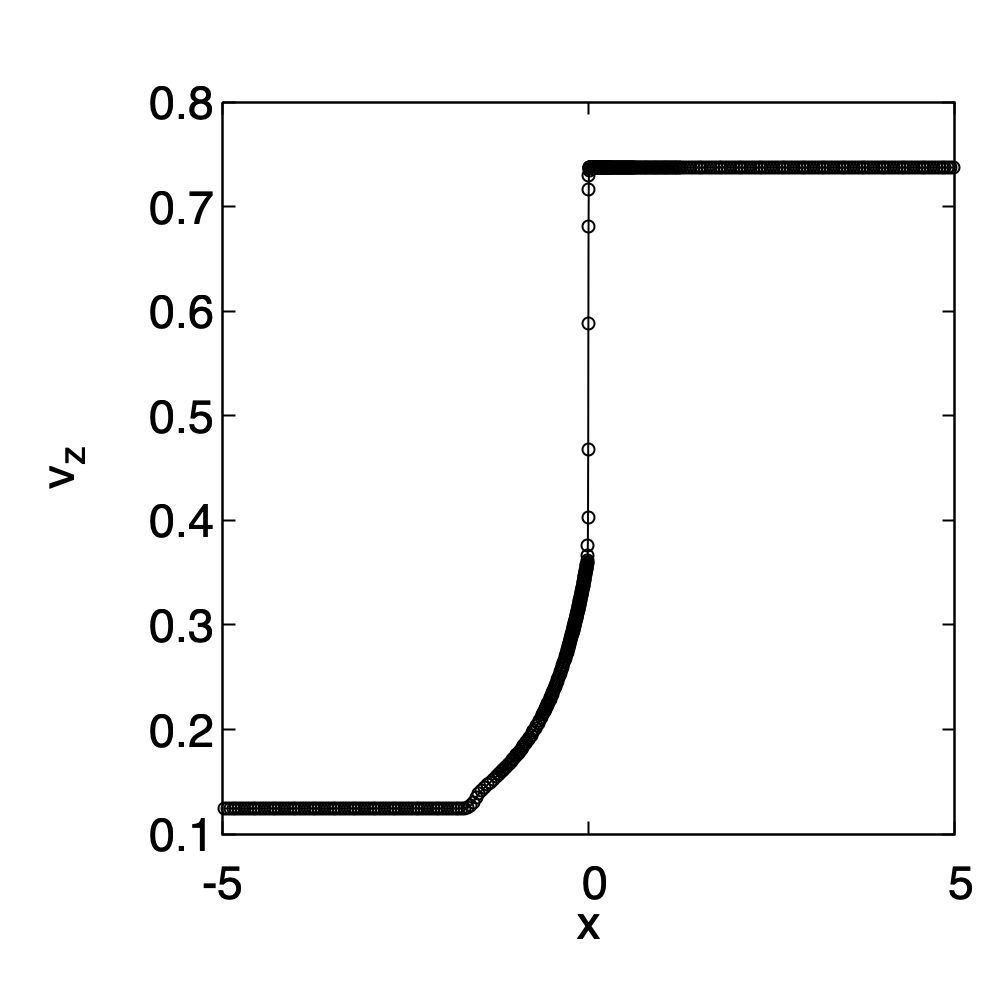}
   \includegraphics[width=6cm,clip]{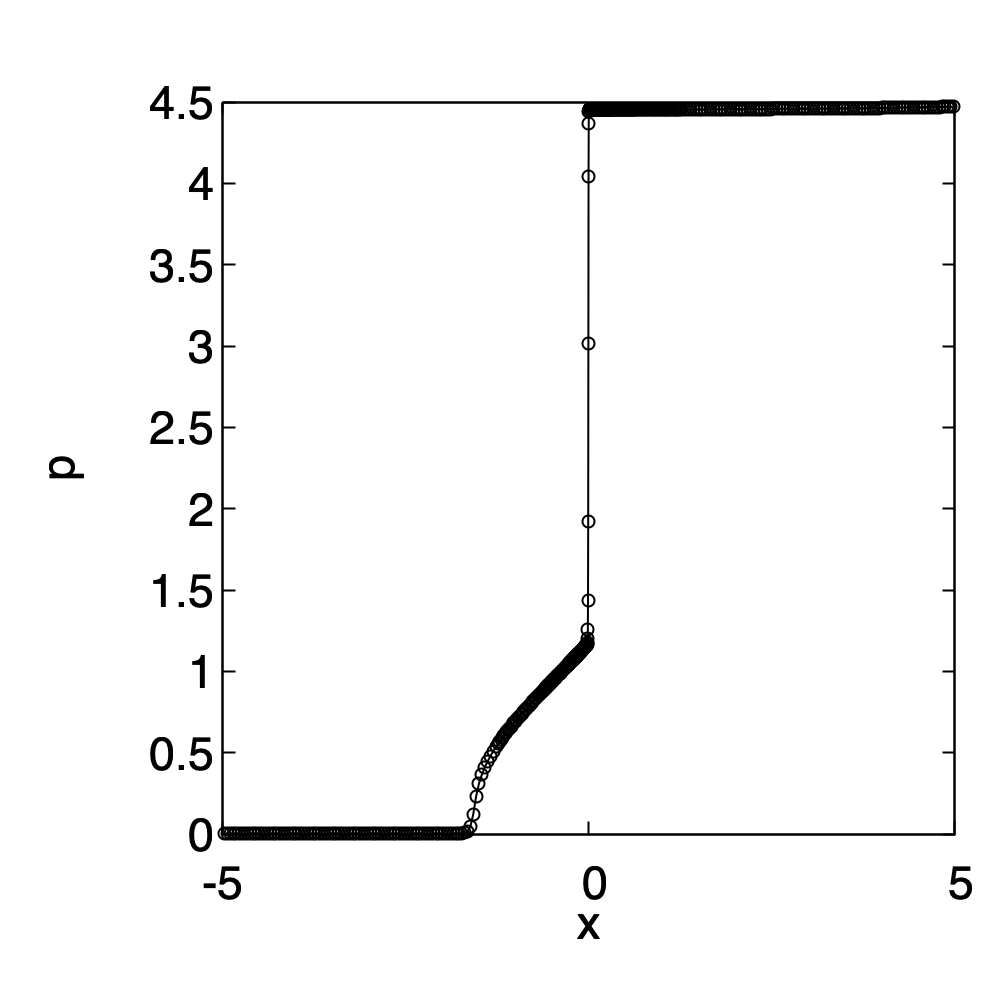}
   \includegraphics[width=6cm,clip]{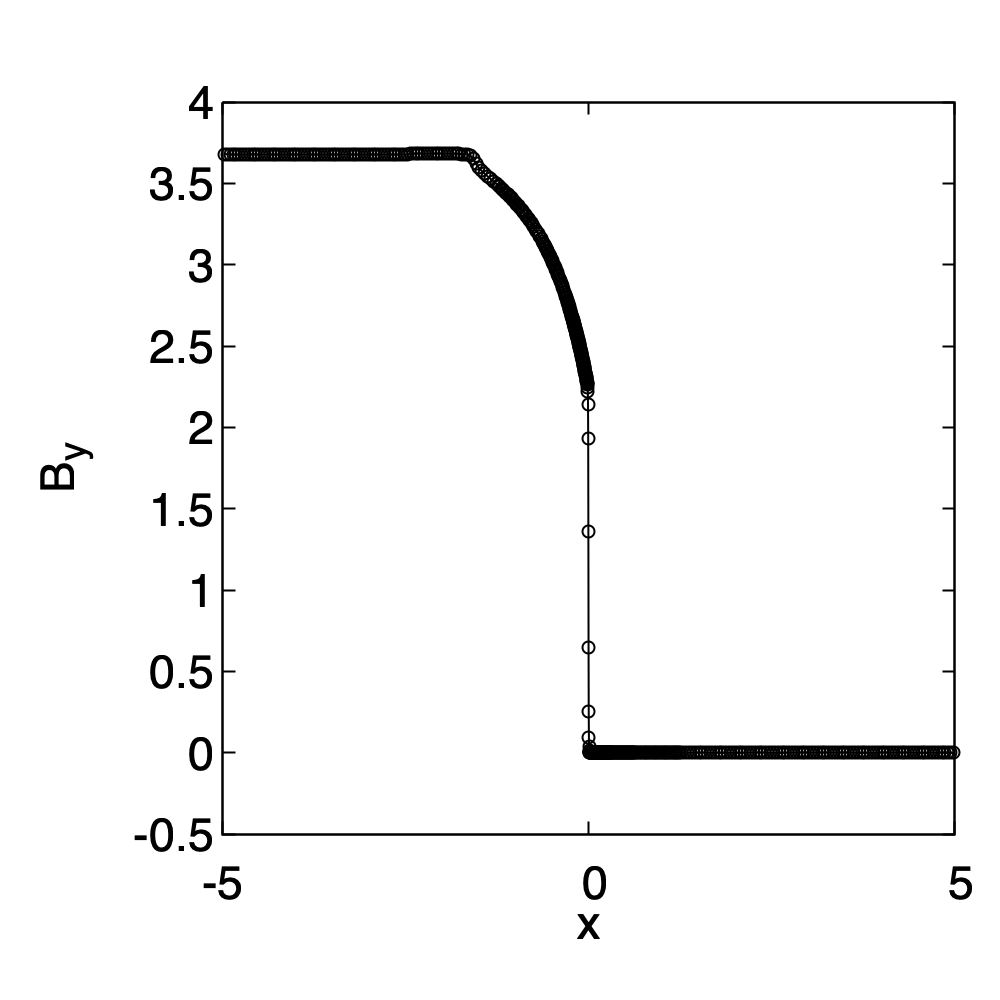}
   \includegraphics[width=6cm,clip]{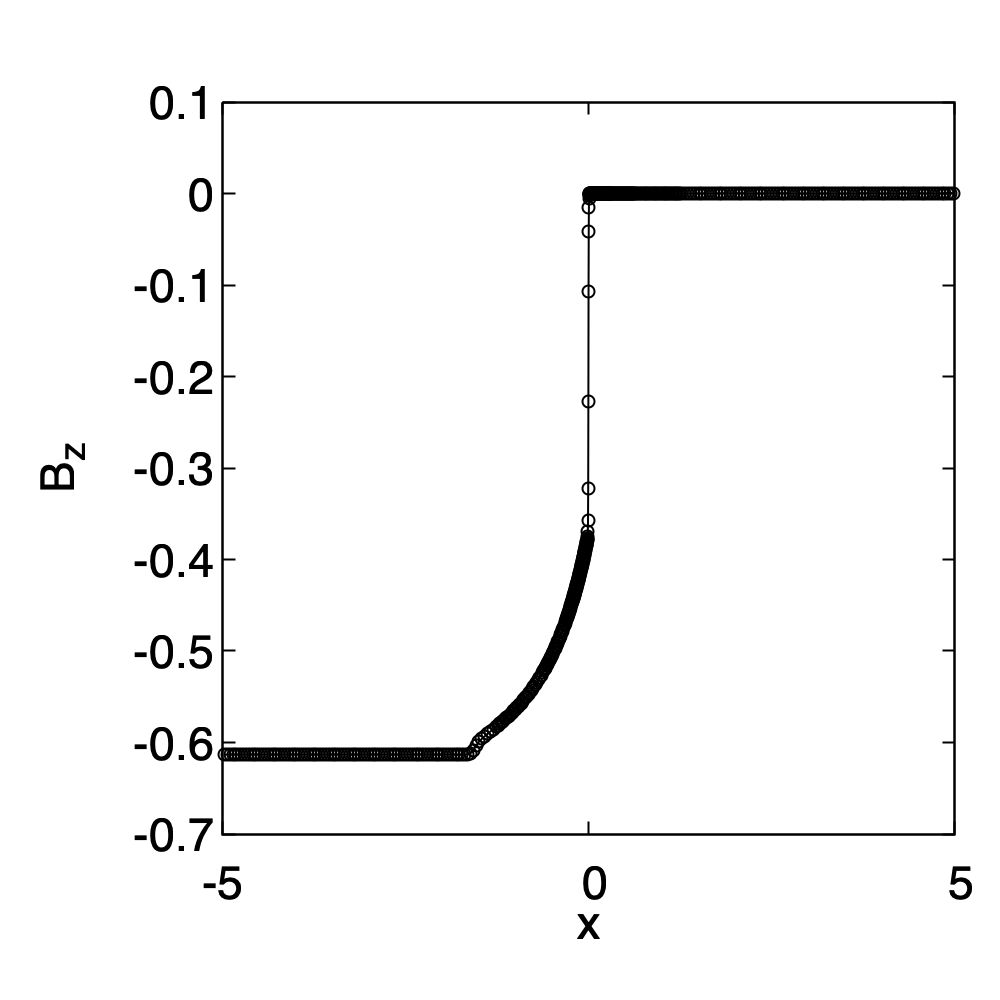}
   \includegraphics[width=6cm,clip]{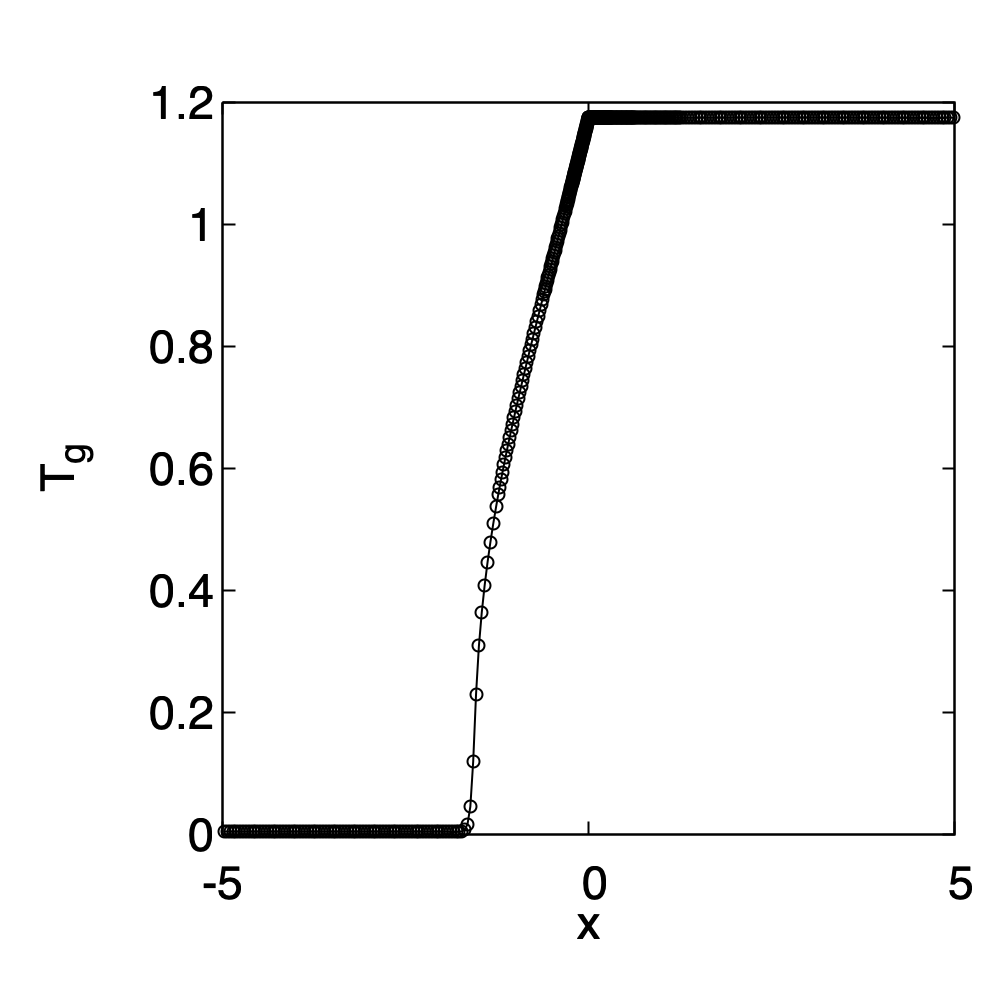}
   \includegraphics[width=6cm,clip]{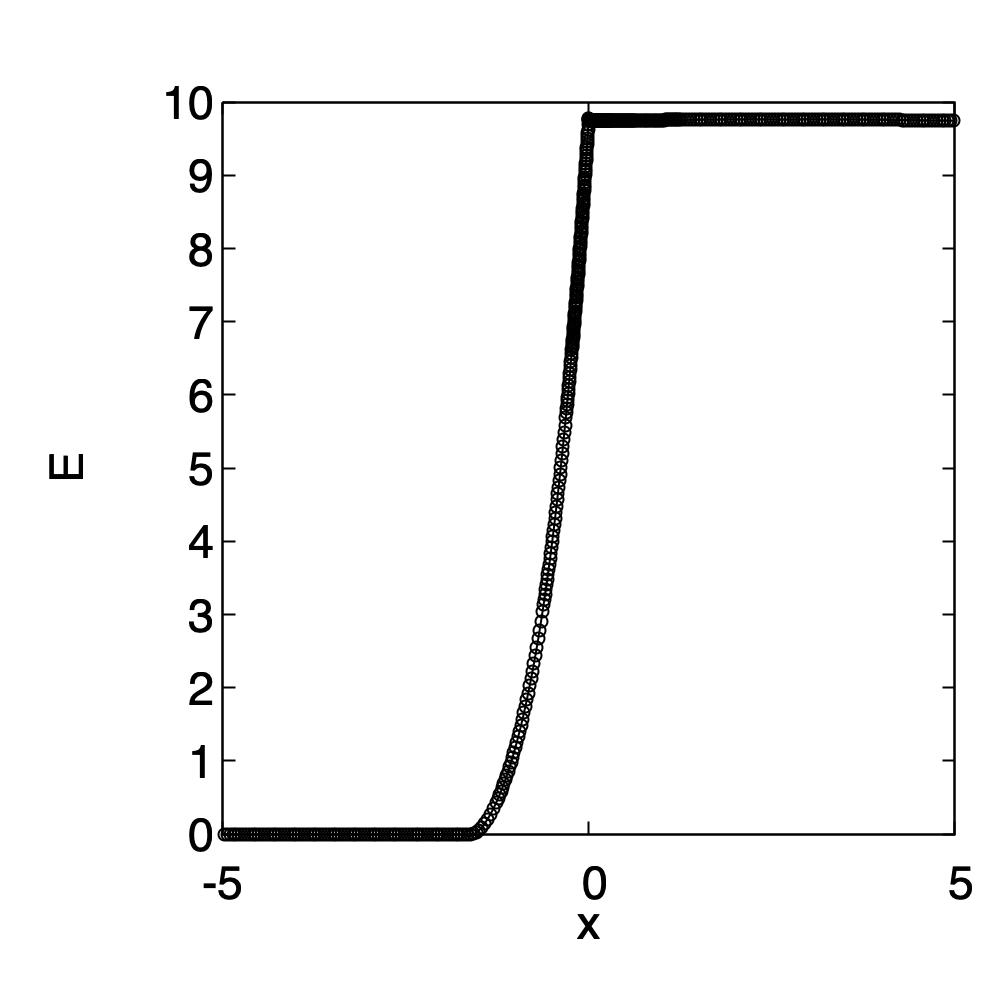}
   }
   \caption{Normalized density, $x$-velocity, $y$-velocity, $z$-velocity, plasma pressure, $y$-magnetic field, $z$-magnetic field, plasma temperature and radiation energy \nmn{density} profiles computed for the relaxed state of the slow switch-off magnetosonic shock.}
\label{fig:shock_switchoff}%
\end{figure*}

To demonstrate we can handle special RMHD shock solutions as well, we simulate a switch-off slow shock formed by a high-$\beta$, superslow, Alfv\'enic ($\mathcal{M}_s>1$, $\mathcal{M}_a=1$) upstream plasma decelerating through a shock to a subslow downstream state ($\mathcal{M}_s<1$). The shock solution for this case is shown in Figure (\ref{fig:shock_switchoff}). As the name suggests, in a switch-off shock, the tangential magnetic field components become zero after going through the shock, causing the magnetic field vector to align with the shock normal in the downstream region. The plasma transitions from a highly magnetically dominated ($\beta<<1$) upstream state to a plasma pressure-dominated downstream state ($\beta>1$). It must also be noted that the slow magnetosonic speed equals the $x$-Alfv\'en speed in the downstream region, since the tangential magnetic field vanishes.

In all the \nmn{stationary} shock simulations shown here, the diffusion limit i.e. $\lambda = 1/3$ was assumed. However, we also ran all these with full FLD, using the flux limiter suggested by \citet{levermore1981flux}. The shock structure stayed the same and well within the diffusion limit, with minor changes in the value of $\lambda$. The maximum change in $\lambda$ observed for the hydrodynamic, fast magnetosonic, slow magnetosonic, intermediate and slow switch-off shock was 1.6\%, 28\%, 0.011\%, 0.13\% and 3.4\%, respectively.

\begin{figure*} 
   \centering
   \FIG{
   \includegraphics[width=12.7cm,clip]{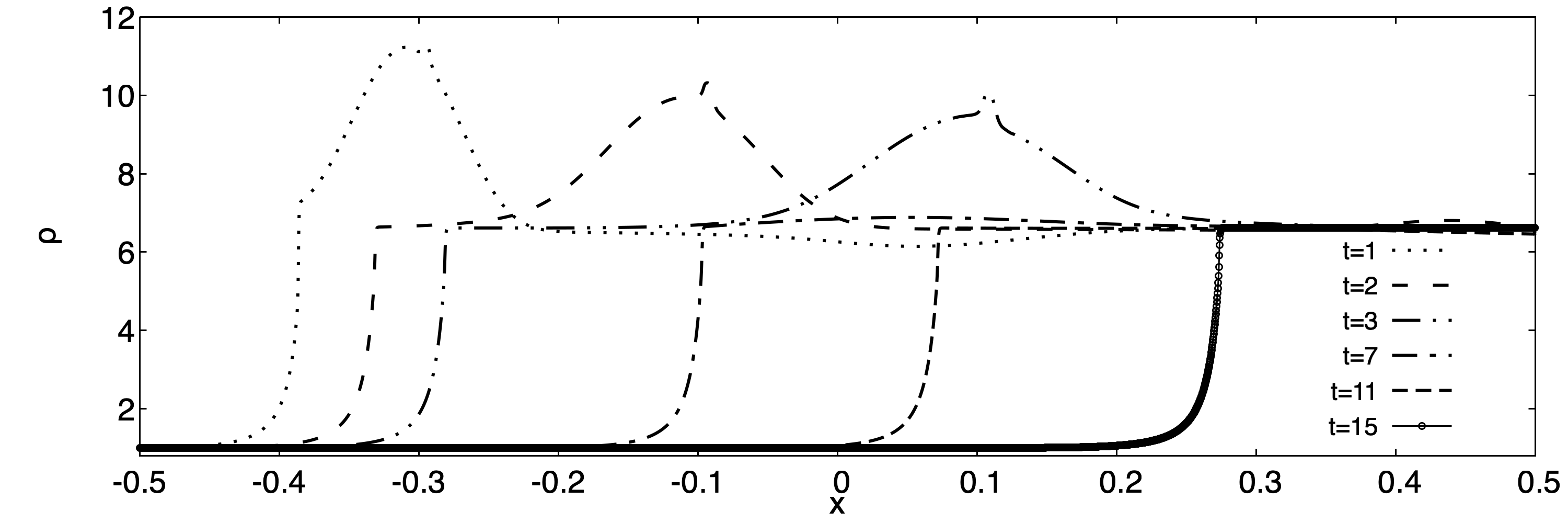}
   \includegraphics[width=12.7cm,clip]{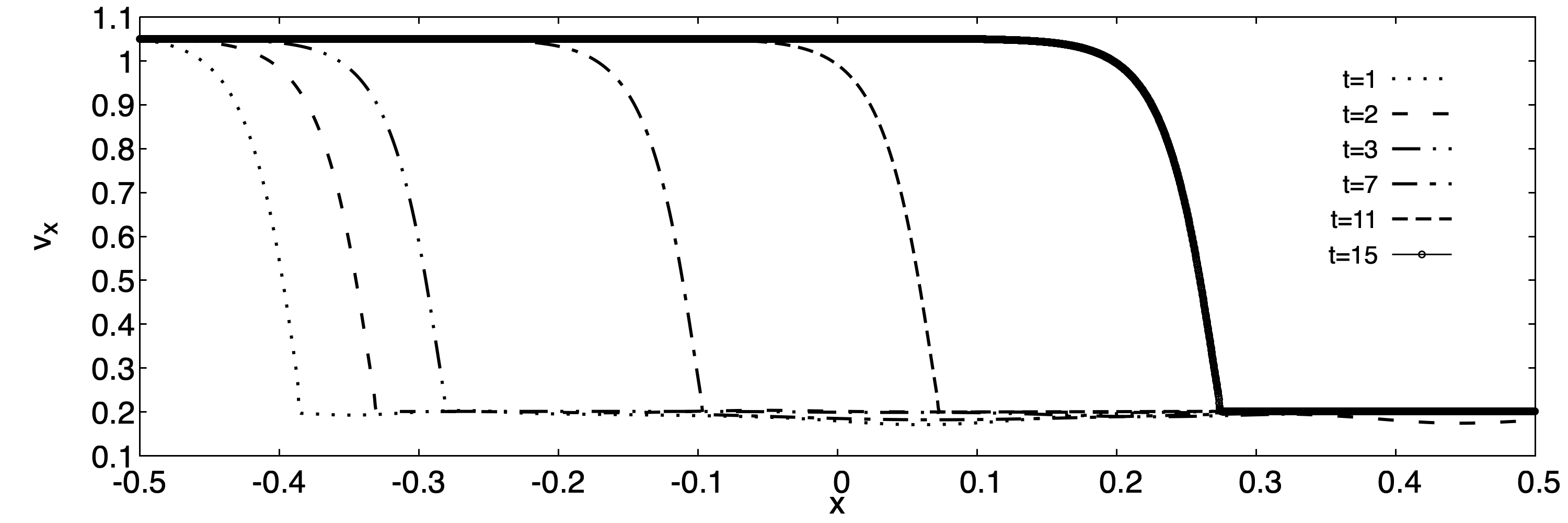}
   \includegraphics[width=12.7cm,clip]{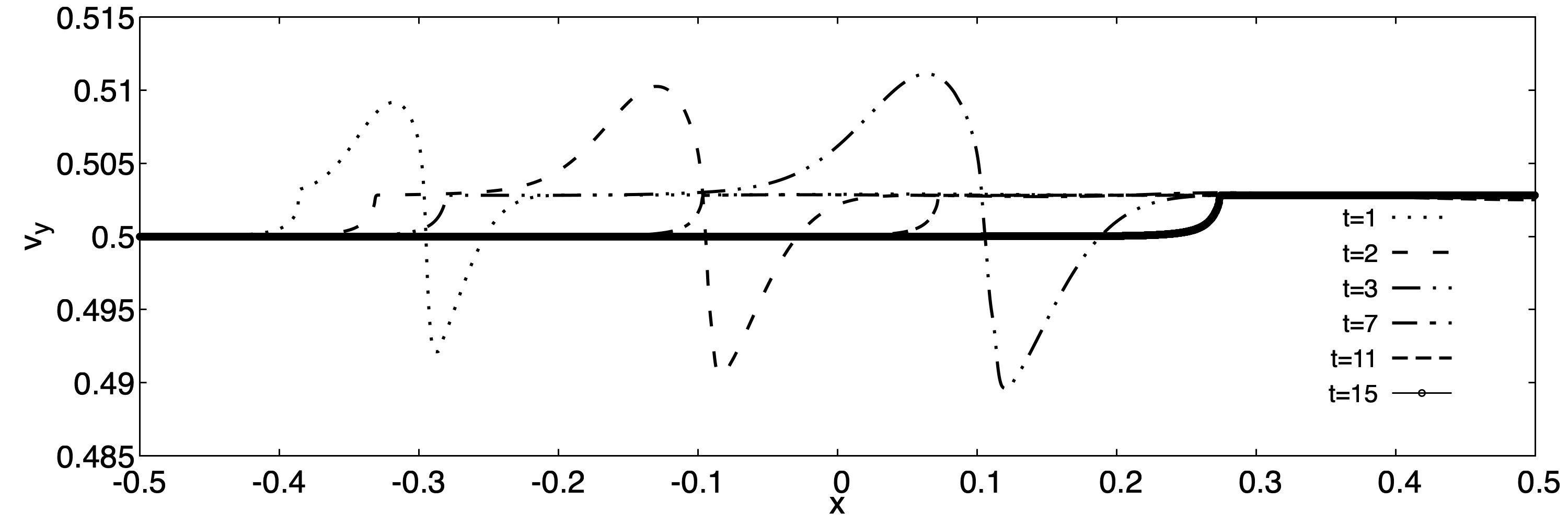}
   \includegraphics[width=12.7cm,clip]{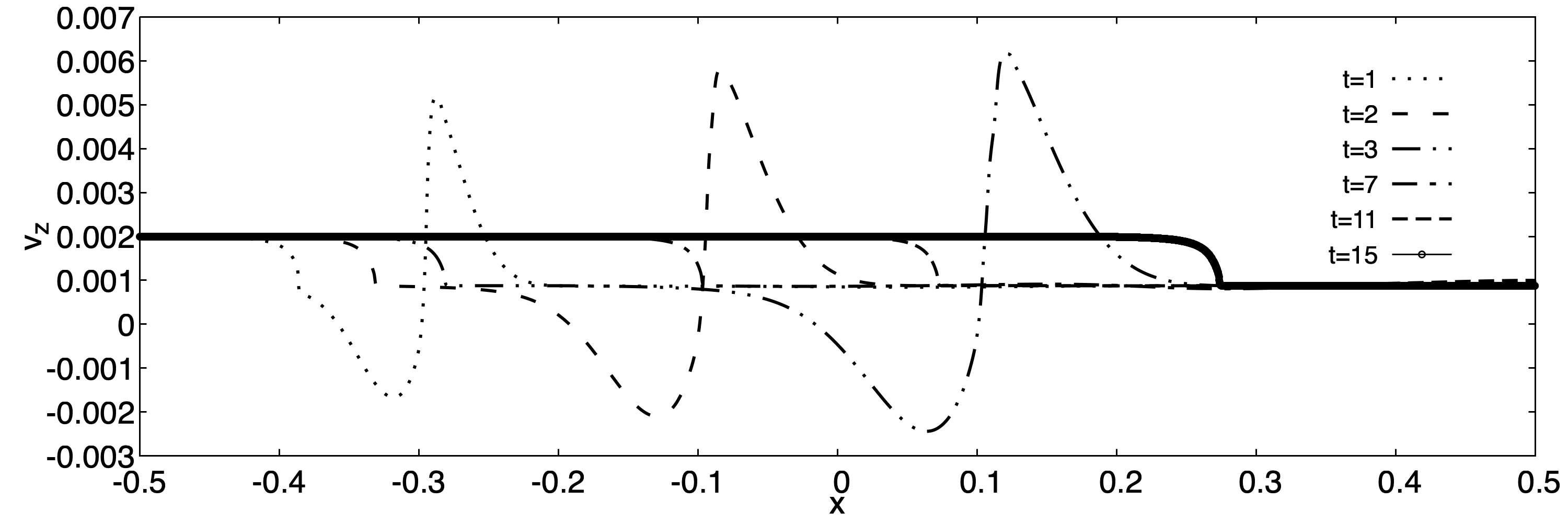}
   \includegraphics[width=12.7cm,clip]{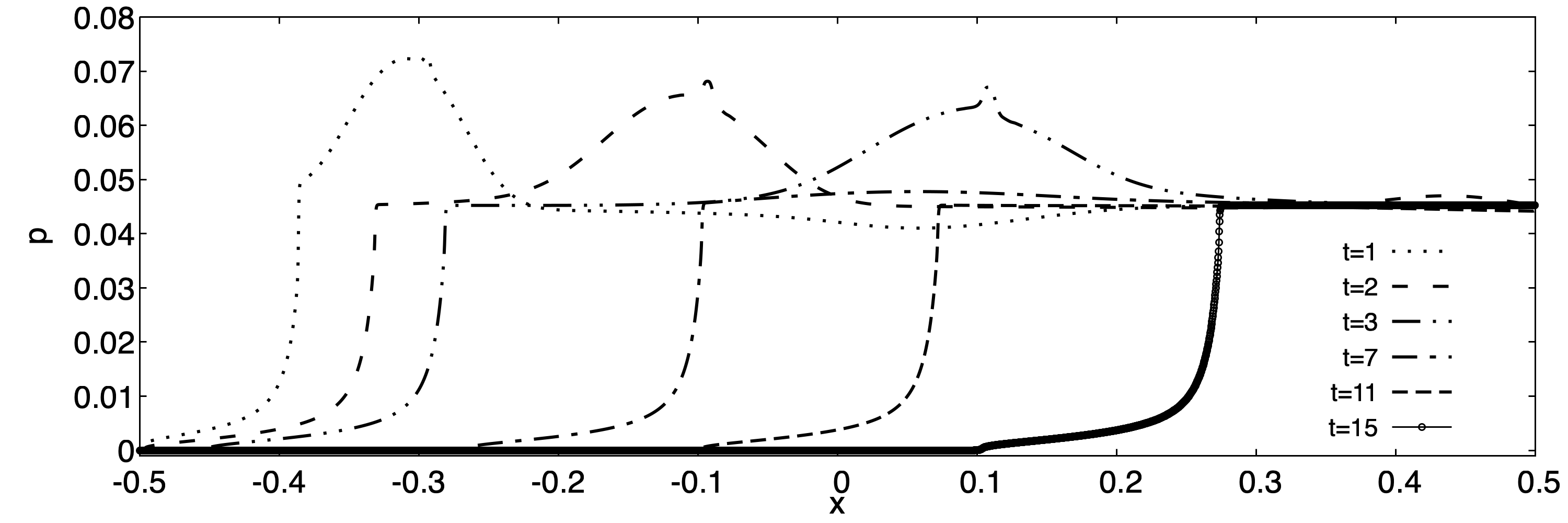}
   }

   \caption{\nmn{Normalized density, $x$-velocity, $y$-velocity, $z$-velocity and plasma pressure profiles computed for the moving fast magnetosonic shock at times $t = 1.0$, $t = 2.0$, $t = 3.0$, $t = 7.0$, $t = 11.0$ and $t = 15.0$.}}
\label{fig:shock_fast_moving1}%
\end{figure*}

\begin{figure*} 
   \centering
   \FIG{
   \includegraphics[width=12.7cm,clip]{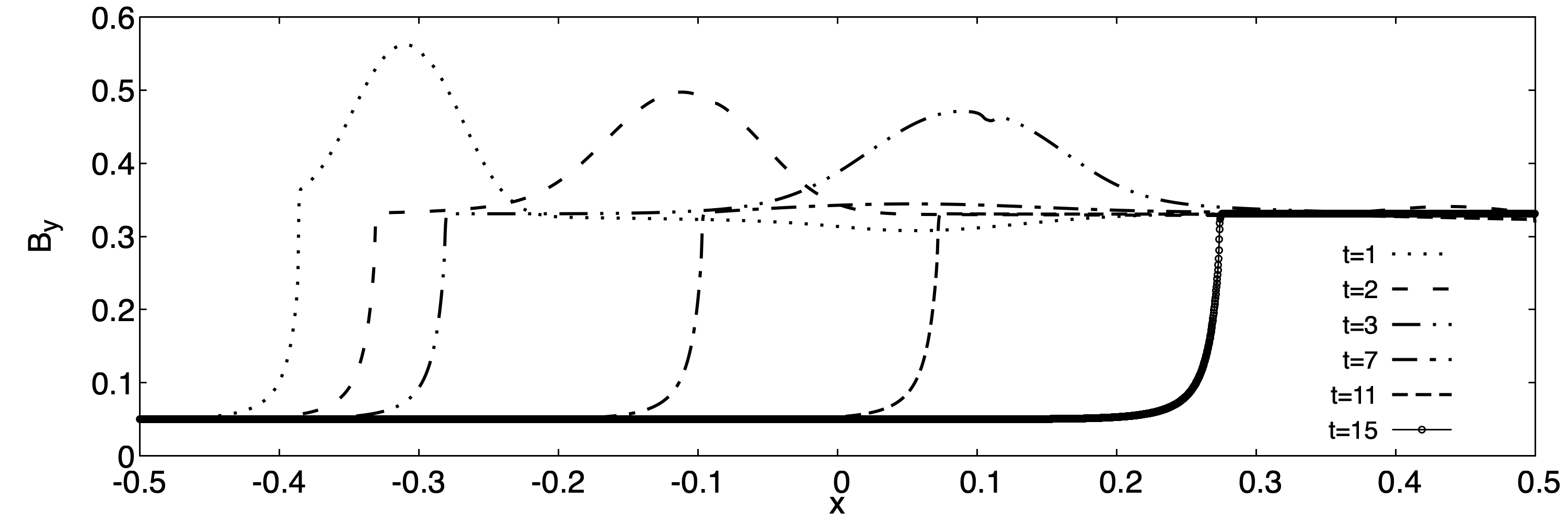}
   \includegraphics[width=12.7cm,clip]{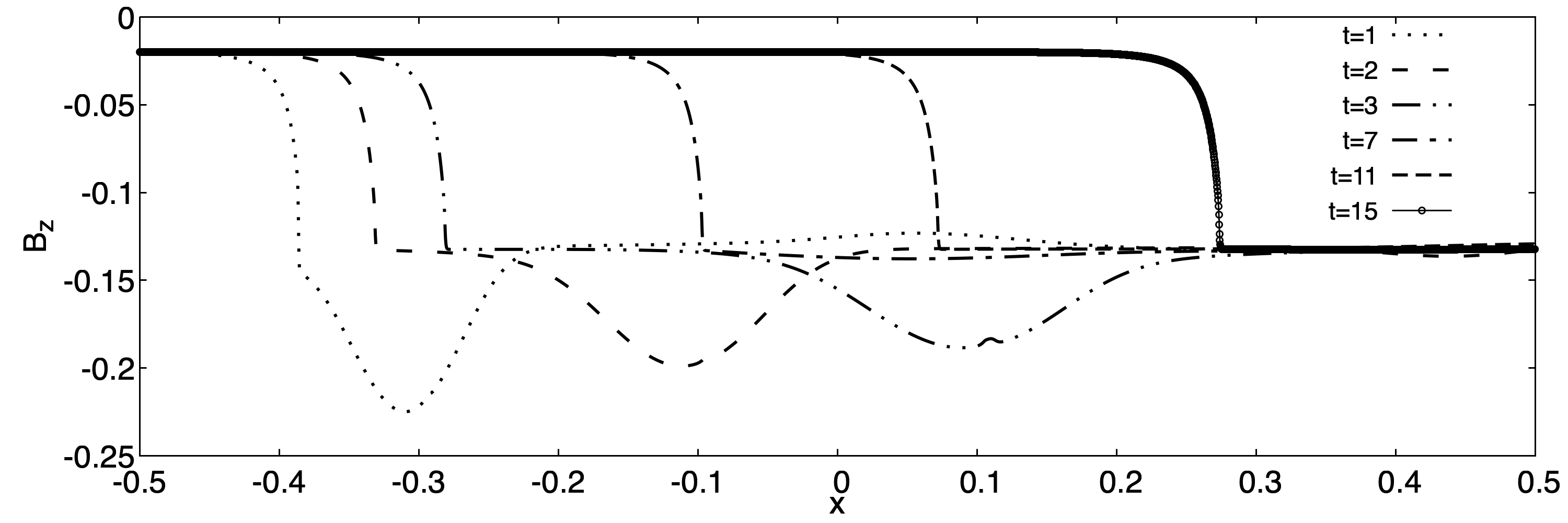}
   \includegraphics[width=12.7cm,clip]{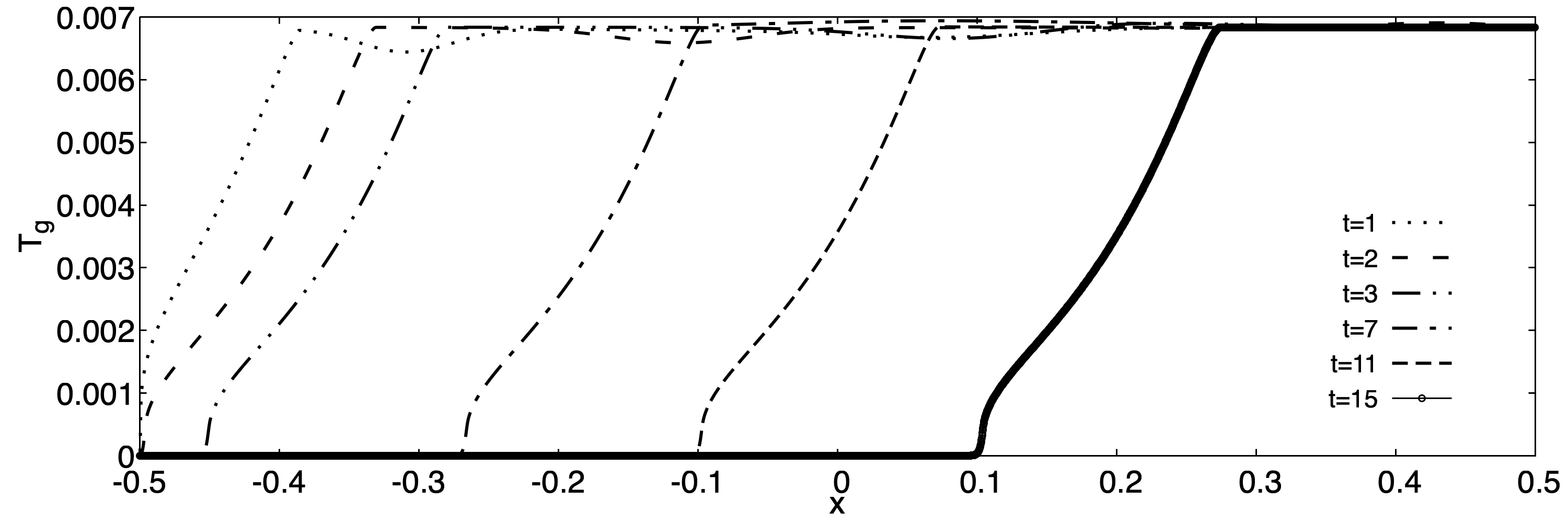}
   \includegraphics[width=12.7cm,clip]{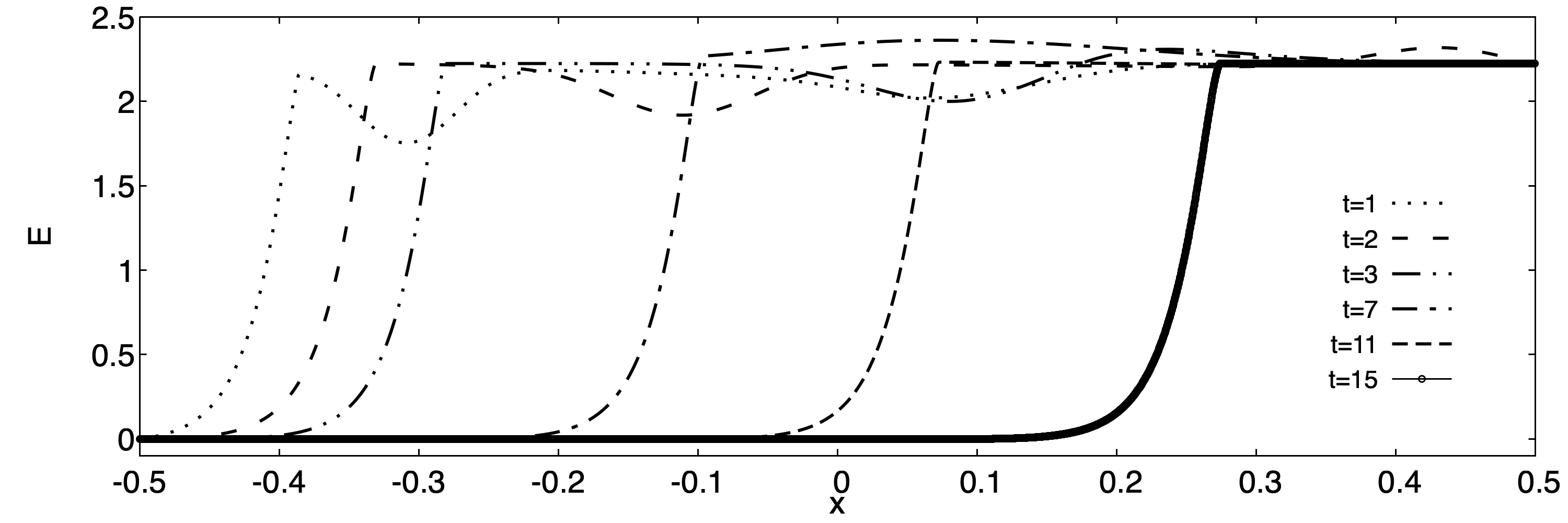}
   }

   \caption{\nmn{Normalized $y$-magnetic field, $z$-magnetic field, plasma temperature and radiation energy density profiles computed for the moving fast magnetosonic shock at times $t = 1.0$, $t = 2.0$, $t = 3.0$, $t = 7.0$, $t = 11.0$ and $t = 15.0$.}}
\label{fig:shock_fast_moving2}%
\end{figure*}

\nmn{In order to demonstrate the capability of the code to handle moving shocks, the fast magnetosonic shock described above is also simulated on an inertial frame moving at a constant velocity. Here, the frame moves at a constant velocity along the leftward direction perpendicular to the shock front. The shock therefore proceeds to the right. The right state, shown in Table~(\ref{table:RMHD_shocks}), is used as the initial condition throughout the domain. A zero-gradient boundary condition is imposed at the right boundary. At the left boundary, the left state is used as a fixed boundary condition to generate and drive the shock. The frame is assumed to move at a velocity of $5\times10^7$ cm/s to the left, with respect to the shock frame, and this velocity must be added to the stationary shock velocities shown in Table~(\ref{table:RMHD_shocks}). With this added velocity, the left and right state plasma velocities are now $1.05\times10^9$ cm/s and $2.01157\times10^8$ cm/s, respectively. All other properties remain the same as in Table~(\ref{table:RMHD_shocks}). The shock is created from the initial discontinuity located at the left boundary, and is expected to relax into the non-trivial structure seen in Figure~(\ref{fig:shock_fast}) as it moves to the right. Figure~(\ref{fig:shock_fast_moving1}) shows snapshots of plasma density, $x$-velocity, $y$-velocity, $z$-velocity and plasma pressure solutions at non-dimensional times $t = 1.0$, $t = 2.0$, $t = 3.0$, $t = 7.0$, $t = 11.0$ and $t = 15.0$. The corresponding snapshots for $y$-magnetic field, $z$-magnetic field, plasma temperature and radiation energy density solutions are shown in Figure~(\ref{fig:shock_fast_moving2}). At $t = 15.0$, the shock has reached its terminal, relaxed state at $x = 0.25$. This corresponds to $x = 2.5\times10^4$ cm and $t = 1.5$ ms in dimensional form.}

\begin{table*}
\caption{Initial states for radiation-modified shocks that satisfy the Rankine-Hugoniot relations for RMHD, and length of computational domain for each case. The Mach numbers marked in bold are those that jump from higher than 1 to lower than 1 across the shock.} 
\label{table:RMHD_shocks}      
\centering   
\begin{tabular}{c | c c | c c} 
\hline\hline 
&\multicolumn{2}{c|}{Hydrodynamic shock ($\gamma = 5/3$)}\\
\hline
Variable & Left state & Right state \\ 
\hline  
   $\rho$ (g/cm$^3$) & $1.0\times10^{-2}$ & $3.06495\times 10^{-2}$ \\ 
   T$_g$, T$_r$ (K)  & $1.08899\times10^{6}$     & $3.83849\times10^{6}$ \\
   v$_x$ (cm/s)      & $5.19271\times10^{7}$    & $1.69422\times10^{7}$ \\
   v$_y$ (cm/s)      & 0.0    & 0.0 \\
   v$_z$ (cm/s)      & 0.0    & 0.0 \\
   B$_x$ (Gauss)     & 0.0    & 0.0 \\
   B$_y$ (Gauss)     & 0.0    & 0.0 \\
   B$_z$ (Gauss)     & 0.0    & 0.0 \\
   p (erg/cm$^3$)    & $1.79762\times10^{12}$ & $1.94203\times10^{13}$ \\
   E (erg/cm$^3$)    & $1.06401\times10^{10}$ & $1.64244\times10^{12}$ \\
   $\beta$           & undefined    & undefined \\
   $\mathcal{M}_s$ & undefined    & undefined \\
   $\mathcal{M}_a$ & undefined    & undefined \\
   $\mathcal{M}_f$ & $\bf{3.0}$   & $\bf{0.52135}$ \\
   \hline
   $L$ (cm) &\multicolumn{2}{c|}{$2\times10^5$}\\
\hline\hline 
&\multicolumn{2}{c|}{Fast magnetosonic shock ($\gamma = 1.4$)} & \multicolumn{2}{c|}{Slow magnetosonic shock ($\gamma = 1.4$)}\\
\hline
Variable & Left state & Right state & Left state & Right state \\ 
\hline  
   $\rho$ (g/cm$^3$) & $1.0\times10^{-2}$ & $6.60865\times 10^{-2}$ & $3.06495\times10^{-2}$ & $8.82626\times 10^{-2}$  \\ 
   T$_g$, T$_r$ (K)  & $1.0\times10^{4}$     & $4.14098\times10^{7}$ & $3.83849\times10^{6}$     & $7.47109\times10^{6}$ \\
   v$_x$ (cm/s)      & $1.0\times10^{9}$    & $1.51157\times10^{8}$ & $5.0\times10^{7}$    & $1.73627\times10^{7}$ \\
   v$_y$ (cm/s)      & $5.0\times10^{8}$    & $5.02805\times10^{8}$ & $5.0\times10^{6}$    & $-3.94698\times10^{7}$ \\
   v$_z$ (cm/s)      & $2.0\times10^{6}$    & $8.77613\times10^{5}$ & $2.0\times10^{6}$    & $1.08940\times10^{7}$ \\
   B$_x$ (Gauss)     & $3.54491\times10^{6}$    & $3.54491\times10^{6}$ & $3.54491\times10^{7}$    & $3.54491\times10^{7}$ \\
   B$_y$ (Gauss)     & $1.77245\times10^{7}$  & $1.17215\times10^{8}$ & $3.54491\times10^{7}$  & $1.12909\times10^{7}$ \\
   B$_z$ (Gauss)     & $-7.08982\times10^{6}$  & $-4.68853\times10^{7}$ &  $-7.08982\times10^{6}$  & $-2.25819\times10^{6}$ \\
   p (erg/cm$^3$)    & $1.65088\times10^{10}$ & $4.52107\times10^{14}$ & $1.94203\times10^{13}$ & $1.08851\times10^{14}$ \\
   E (erg/cm$^3$)    & $7.56570\times10^{1}$ & $2.22465\times10^{16}$ & $1.64244\times10^{12}$ & $2.35713\times10^{13}$ \\
   $\beta$             & $1.10059\times10^{-3}$    & $7.11318\times10^{-1}$ & $1.90414\times10^{-1}$    & $1.96925\times10^{0}$ \\
   $\mathcal{M}_s$ & $3605.09235$ & $67.36428$ & ${\bf 2.48298}$ & ${\bf 0.55544}$ \\
   $\mathcal{M}_a$ & $100$     & $38.85895$ & $0.87535$  & $0.51583$ \\
   $\mathcal{M}_f$ & $\bf{18.25063}$ & $\bf{0.89141}$ & $ 0.59183$ & $0.38805$ \\
   \hline
   $L$ (cm) &\multicolumn{2}{c|}{$10^5$}&\multicolumn{2}{c|}{$10^5$}\\
\hline\hline 
&\multicolumn{2}{c|}{Intermediate magnetosonic shock ($\gamma = 1.4$)} & \multicolumn{2}{c|}{Slow switch-off shock ($\gamma = 1.4$)}\\
\hline
Variable & Left state & Right state & Left state & Right state \\ 
\hline  
   $\rho$ (g/cm$^3$) & $1.16963\times10^{-4}$ & $4.39947\times 10^{-4}$ & $7.96933\times10^{-3}$ & $3.06381\times 10^{-2}$  \\ 
   T$_g$, T$_r$ (K)  & $3.46709\times10^{5}$     & $5.94400\times10^{6}$ & $6.14135\times10^{3}$     & $1.12616\times10^{7}$ \\
   v$_x$ (cm/s)      & $9.82668\times10^{7}$    & $2.61249\times10^{7}$ & $3.98682\times10^{7}$    & $1.03724\times10^{7}$ \\
   v$_y$ (cm/s)      & $-1.74756\times10^{8}$    & $-6.48489\times10^{7}$ & $3.22643\times10^{6}$    & $-1.43474\times10^{8}$ \\
   v$_z$ (cm/s)      & $3.61511\times10^{8}$    & $1.41697\times10^{8}$ & $4.95596\times10^{6}$    & $2.94061\times10^{7}$ \\
   B$_x$ (Gauss)     & $ 3.54491\times10^{6}$    & $3.54491\times10^{6}$ & $1.26166\times10^{7}$    & $1.26166\times10^{7}$ \\
   B$_y$ (Gauss)     & $-3.77898\times10^{6}$  & $6.99041\times10^{5}$ & $4.64246\times10^{7}$  & $0.0$ \\
   B$_z$ (Gauss)     & $7.55796\times10^{6}$  & $-1.39808\times10^{6}$ &  $-7.73743\times10^{6}$  & $0.0$ \\
   p (erg/cm$^3$)    & $6.69468\times10^{9}$ & $4.31669\times10^{11}$ & $8.07981\times10^{9}$ & $5.69535\times10^{13}$ \\
   E (erg/cm$^3$)    & $1.09323\times10^{8}$ & $9.44419\times10^{12}$ & $1.07623\times10^{1}$ & $1.25792\times10^{14}$ \\
   $\beta$             & $2.00356\times10^{-3}$    & $7.22803\times10^{-1}$ & $8.55309\times10^{-5}$    & $8.99233\times10^{0}$ \\
   $\mathcal{M}_s$ & ${\bf28.39485}$ & ${\bf0.54797}$ & ${\bf129.24135}$  & ${\bf0.51006}$ \\
   $\mathcal{M}_a$ & ${\bf1.06275}$ & ${\bf0.50139}$ & ${\bf1}$ & ${\bf0.51006}$ \\
   $\mathcal{M}_f$ & $0.41088$ & $0.47132$ & $0.25892$ & $0.20330$ \\
   \hline
   $L$ (cm) &\multicolumn{2}{c|}{$10^5$}&\multicolumn{2}{c|}{$10^6$}\\
\hline\hline 
\hline                                   
\end{tabular}
\end{table*}

\subsection{Heating and cooling}\label{sec:heatcool}

\begin{figure}
   \centering
  \FIG{ \includegraphics[width=\hsize]{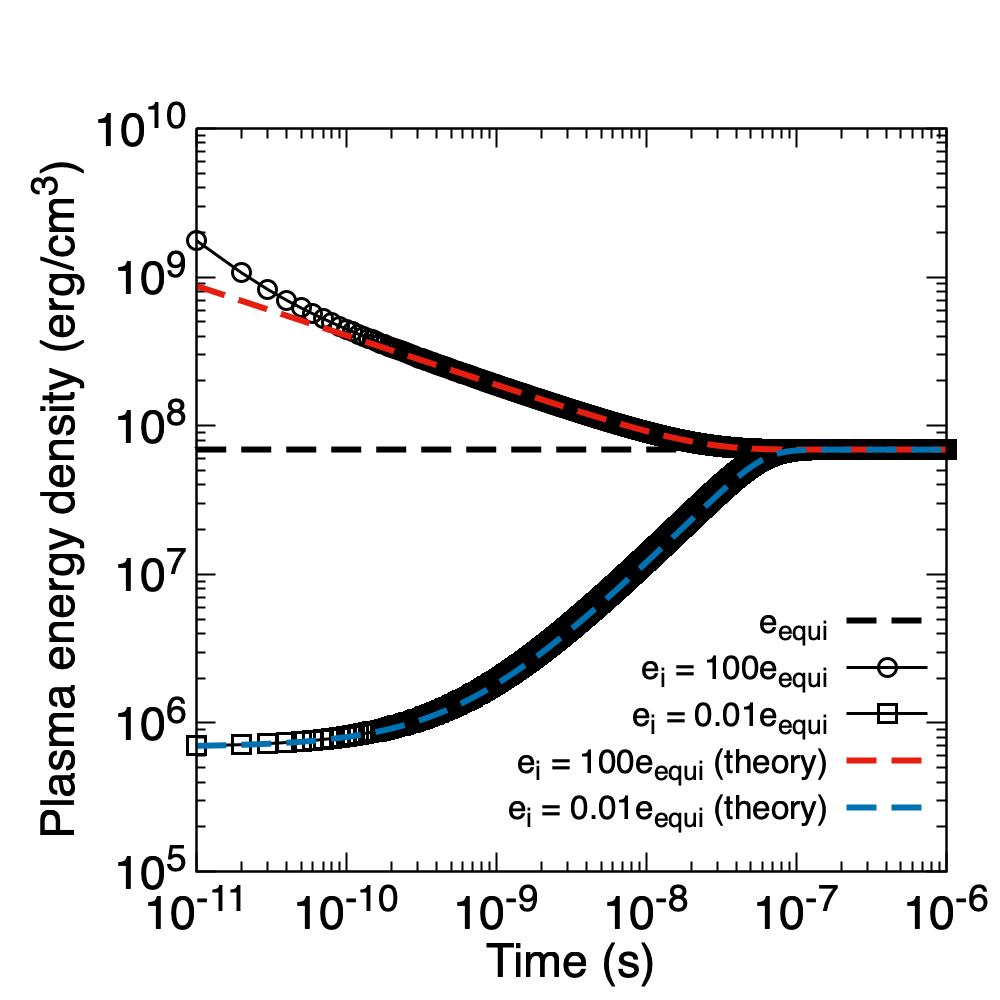}}
      \caption{\nmn{Time evolution of plasma energy density for a plasma initialized with $e_i = 10^{2} e_{equi}$ (cooling) and $e_i = 10^{-2} e_{equi}$ (heating). The red and blue dashed lines are the semi-analytical solutions for the cooling and heating cases, respectively. The black dashed line shows the equilibrium plasma thermal energy. Due to the logarithmic scaling of the time axis, the initial conditions at $t = 0$ are not seen here.}}
         \label{fig:heatcool}
\end{figure}

\nmn{We now test the process of energy exchange between plasma and radiation through the heating and cooling term $\dot{q}$. A 2D stationary, non-magnetized plasma is considered with uniform properties and zero velocity. The density of the gas is taken to be $\rho_0$ = 10$^{-7}$ g/cm$^3$. The adiabatic index is $\gamma = 5/3$, the opacity is $\kappa = 0.4$ cm$^2$/g and the mean molecular weight is $\mu = 0.6$. The computational domain consists of 10 cells per direction. Initially, the plasma and radiation energies are out of radiative equilibrium i.e. the plasma and radiation temperatures are different. The total energy, i.e. the sum of the plasma thermal energy and radiation energy is initially taken to be $(e_0+E_0)$ = 10$^{12}$ erg/cm$^3$. Given the total energy, for a given density, it is possible to find the equilibrium temperature, $T_{eq}$, by equating the total energy to the sum of $e = \rho k_B T_{eq}/(m_p \mu)$ and $E = a_r T_{eq}^4$. The corresponding plasma thermal energy at equilibrium is found out to be $e_{equi}$ = $6.89706\times10^7$ erg/cm$^3$. The thermal energy of the stationary plasma considered here would eventually reach this value, after exchanging energy with radiation through heating or cooling processes.

We consider two different scenarios in terms of the initial value of the plasma thermal energy density. In the first scenario, the initial plasma thermal energy is taken to be $e_i = 10^{-2} e_{equi}$, i.e. the plasma is cooler than the radiation and is expected to absorb energy from radiation through heating until radiative equilibrium is achieved. In the second scenario, we have $e_i = 10^{2} e_{equi}$, and therefore the initially hotter plasma is expected to expel the excess energy through cooling and reach equilibrium. For both these conditions, the initial radiation energy density can be calculated by subtracting the initial plasma thermal energy density from the total energy density i.e. $E_i = e_0 + E_0 - e_i$. We initialize the system with these values and let it relax until equilibrium is achieved. For the heating case, $e_i$ increases while $E_i$ decreases with time, with both reaching their respective equilibrium values after some time. For the cooling case, the trend is opposite. Since for both scenarios $e_i \ll 10^{12}$ erg/cm$^3$, $E_i$ stays $\approx 10^{12}$ erg/cm$^3$ throughout the entire simulation. In this case, the radiation field is effectively a large reservoir, that plays the role of a heat source or sink. The time evolution of the plasma thermal energy density for both scenarios is shown on a logarithmic scale in Figure (\ref{fig:heatcool}), along with expected, semi-analytical evolution profiles. The case with $e_i = 10^{2} e_{equi}$ seems to lag behind the semi-analytical solution in the beginning i.e. the plasma cools slower than expected. However, the observed solution eventually catches up with the semi-analytical solution and asymptotes to equilibrium at the expected rate. On the other hand, the case where $e_i = 10^{-2} e_{equi}$ seems to perfectly align with the semi-analytical solution. This test was also performed with a background uniform magnetic field, and it was verified that the presence of a uniform magnetic field makes no difference to the results.}

\subsection{Linear RMHD waves}\label{sec:waves}
It is well known that acoustic and, in general, magnetoacoustic waves may undergo damping when propagating in a radiative background field. \citet{mihalas1984foundations} performed a linear perturbation analysis of the RHD equations in the optically thick diffusion limit. They derived a dispersion relation to quantify the damping rates of acoustic waves in a radiative medium. In this section, we \nmn{
consider} the magnetohydrodynamic generalization of such a dispersion relation for magnetoacoustic and Alfv\'en waves in a radiative medium. Travelling waves are then set up in a magnetized, radiative, stagnant background plasma and their various damping rates are observed. We then validate our simulation results by comparing the observed damping rates with those obtained analytically from the dispersion relation. Note that we will here linearize the governing RMHD set about a static uniform medium, while more general linearizations about a 1D gravitationally stratified medium exist, as presented by \citet{blaes2003local}. Our results connect directly with linear ideal MHD dispersion relations found in textbooks \citep{GoedbloedKeppensPoedts2019}, and may clarify the relation between the thermal instability resulting from optically thin radiative loss prescriptions \citep{claes2019thermal} with the various unstable modes already identified in radiative media. \nmn{For a complete derivation of the dispersion relation, the reader is referred to Appendix B.}

\nmn{This dispersion relation, as shown in Equation (\ref{eq:dispersion_relation_simplified}),} is a polynomial of order 8, as expected. For a uniform background medium and constant opacity $\kappa$, Equations (49) and (53) from \citet{blaes2003local} are equivalent to the above dispersion relation for the magnetohydrodynamic and hydrodynamic cases, respectively. We can clearly see that the dispersion relation factorizes into quadratic and hexic polynomials of $\omega$. The quadratic polynomial factor gives the familiar Alfv\'en mode solution
\begin{equation}\label{eq:alfvenx_speed_solution}
\frac{\omega}{k} = \pm v_{A,x,0}.
\end{equation}
This means that the Alfv\'en mode has no imaginary component of $\omega$ and is therefore a simple forward and backward travelling wave without any exponential growth or decay. Therefore, in the current diffusion approximation, this mode is not influenced by the radiative terms (the reader is referred to the work by \citet{2020MNRAS.499.4282D} demonstrating radiation-driven damping of Alfv\'en modes). The remaining 6 modes in Equation~(\ref{eq:dispersion_relation_simplified}) reduce to the standard MHD dispersion relation covering the (marginal) entropy mode, and the slow and fast pairs, augmented with a marginal frequency $\omega=0$ (the remnant of the radiative diffusion mode) if all radiative terms vanish (i.e. all derivatives of $\dot{q}$ vanish). In future work, it will be of interest to study how the thermal instability (related to the entropy mode, which can cause solar coronal condensations like prominences), and the added radiative diffusion mode relate in more general settings.

\begin{figure}
   \centering
  \FIG{ \includegraphics[width=\hsize]{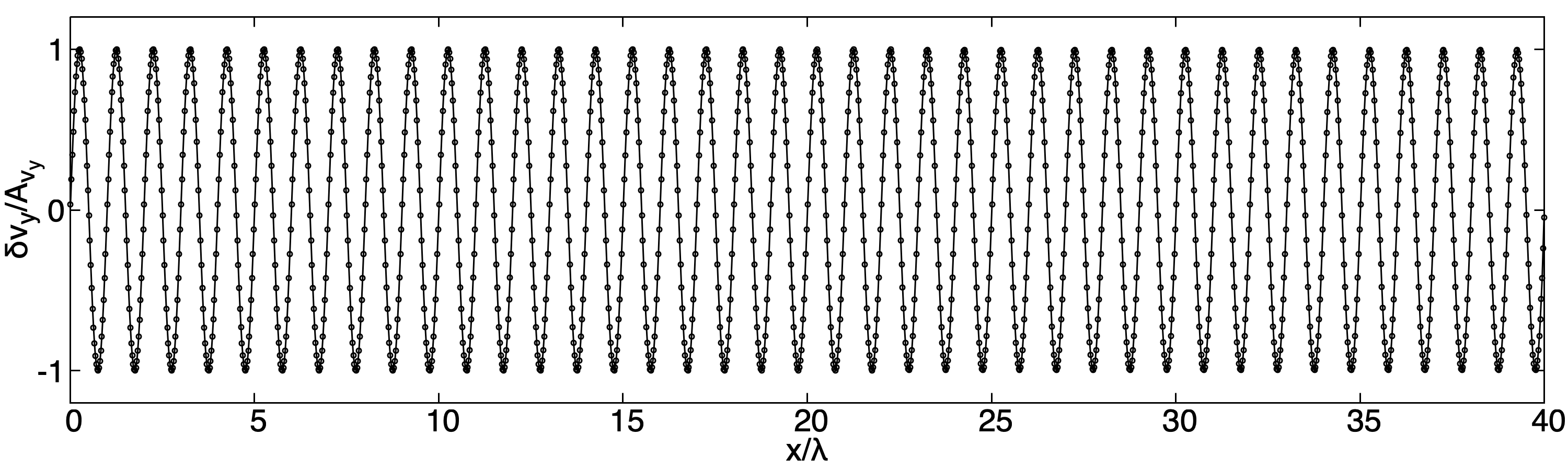}
    \includegraphics[width=\hsize]{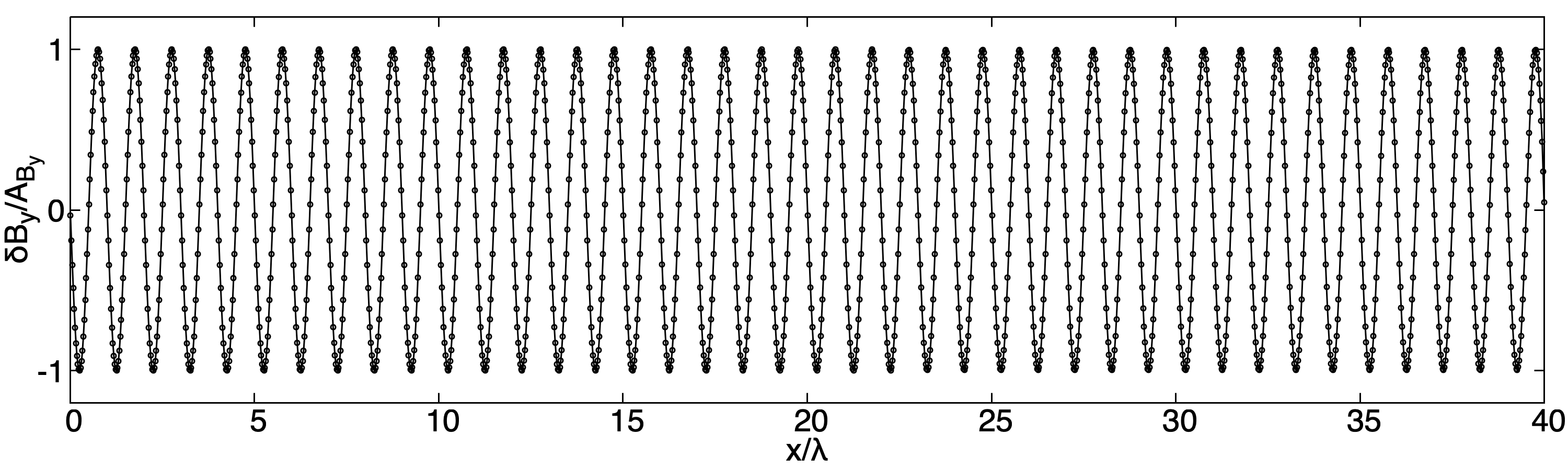}}
      \caption{$y$-velocity (top) and $y$-magnetic field (bottom) perturbation components of the Alfv\'en wave.}
         \label{fig:alfven_propagation}
\end{figure}

\begin{figure}
   \centering
   \FIG{\includegraphics[width=\hsize]{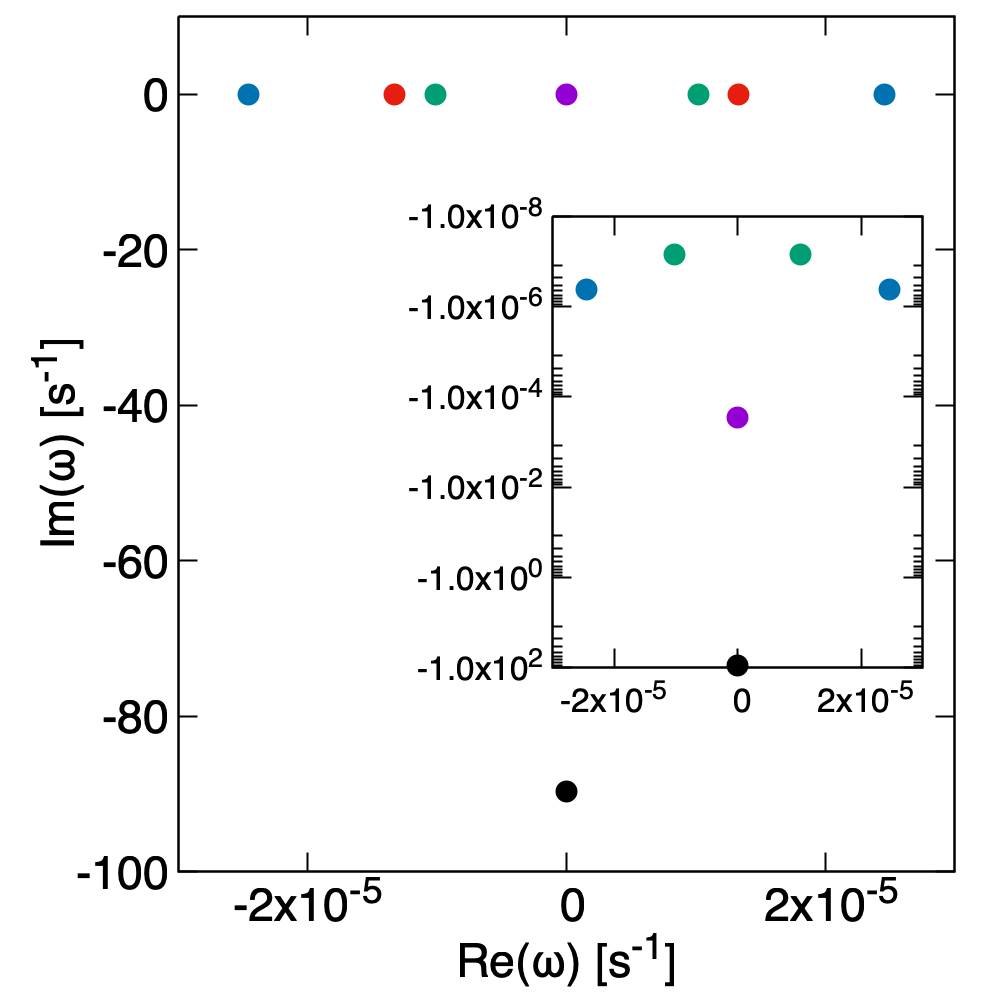}
   \includegraphics[width=\hsize]{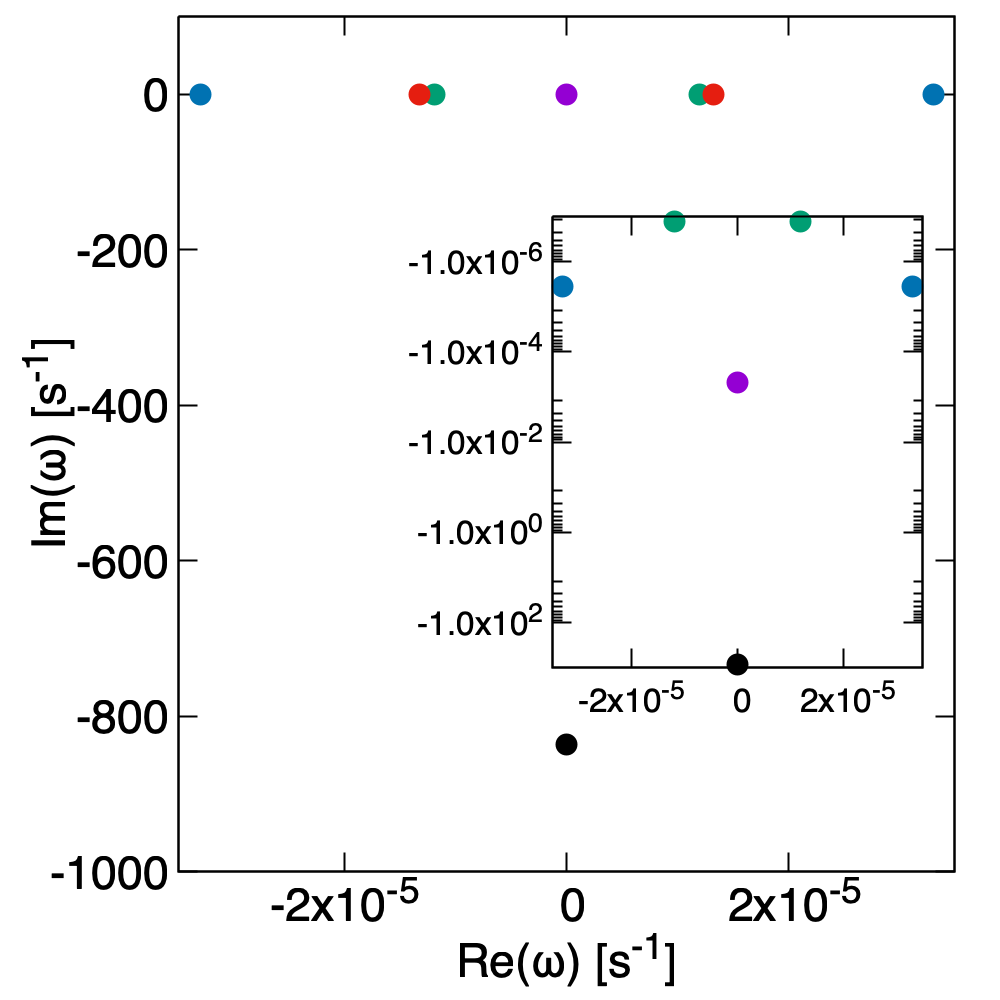}}
      \caption{Complex eigenfrequency plane showing the solutions (solid dots) to the dispersion relation in Equation (\ref{eq:dispersion_relation_simplified}), for the given background plasma properties and wavenumber. \nmn{The top and bottom panels show the solutions for the weakly and strongly radiative cases, respectively.} Green, red, blue, purple and black show the slow, Alfv\'en, fast, thermal and radiative diffusion modes, respectively. The inset plot shows all modes except the Alfv\'en mode using a logarithmic scale to compare the relative magnitudes of the imaginary components of their eigenfrequencies.}
         \label{fig:eigen_solns}
\end{figure}

\begin{table}
\caption{Eigenfrequency solutions (s$^{-1}$) to the dispersion relation in Equation (\ref{eq:dispersion_relation_simplified}), for the given background plasma properties and wavenumber, \nmn{for the weakly radiative case.}}
\label{table:eigen_solns_weak}      
\centering   
\begin{tabular}{c | c} 
\hline
Mode & Value (s$^{-1}$) \\ 
\hline  
   Fast magnetosonic   &$-2.45528\times10^{-5} - 4.10666\times10^{-7}i$\\ 
   Alfv\'en            &$-1.32737\times10^{-5}$\\
   Slow magnetosonic   &$-1.01617\times10^{-5} - 7.05868\times10^{-8}i$\\
   Thermal             &$0.0 - 2.88146\times10^{-4}i$\\
   Radiative diffusion &$0.0 - 8.96558\times10^{1}i$\\
   Slow magnetosonic   &$1.01617\times10^{-5} - 7.05868\times10^{-8}i$\\
   Alfv\'en            &$1.32737\times10^{-5}$\\
   Fast magnetosonic   &$2.45528\times10^{-5} - 4.10666\times10^{-7}i$\\
\hline 
\end{tabular}
\end{table}

\begin{table}
\caption{\nmn{Eigenfrequency solutions (s$^{-1}$) to the dispersion relation in Equation (\ref{eq:dispersion_relation_simplified}), for the given background plasma properties and wavenumber, for the strongly radiative case.}}
\label{table:eigen_solns_strong}
\centering   
\begin{tabular}{c | c} 
\hline
Mode & Value (s$^{-1}$) \\ 
\hline  
   Fast magnetosonic   &$-3.30517\times10^{-5} - 3.50913\times10^{-6}i$\\ 
   Alfv\'en            &$-1.32737\times10^{-5}$\\
   Slow magnetosonic   &$-1.19404\times10^{-5} - 1.27523\times10^{-7}i$\\
   Thermal             &$0.0 - 4.76828\times10^{-4}i$\\
   Radiative diffusion &$0.0 - 8.36612\times10^{2}i$\\
   Slow magnetosonic   &$1.19404\times10^{-5} - 1.27523\times10^{-7}i$\\
   Alfv\'en            &$1.32737\times10^{-5}$\\
   Fast magnetosonic   &$3.30517\times10^{-5} - 3.50913\times10^{-6}i$\\
\hline 
\end{tabular}
\end{table}

To use this dispersion relation to validate our RMHD implementation, we \nmn{first} consider, a static 1D \nmn{weakly radiative} plasma with plasma settings given by $\rho_0 = 3.216 \times 10^{-9}$ g/cm$^3$, $p_0 = 17.34 \times 10^3$ erg/cm$^3$ and ${\bf B}_0 =(330.14,0,330.14)$ Gauss. The heat capacity ratio is $\gamma = 5/3$, and the mean molecular weight is $\mu = 0.5$. The radiation energy density, $E_0 = 8.60834 \times 10^3$ erg/cm$^3$, can be obtained from the radiative equilibrium condition by equating the radiation temperature to the plasma temperature. The plasma beta is $\beta = 1$. The length of the computational domain is $40\lambda_w$, where $\lambda_w$ is the wavelength of the wave excited at the left boundary. The computational grid is made up of 3200 cells, and AMR was not used for this case. At the right-hand side boundary, a Dirichlet boundary condition is used for the radiation energy density, whereas Neumann boundary conditions are applied on all other properties. Using the code's capacity to handle interior boundaries, we actually overwrite the first wavelength within the domain with the purely ideal MHD time-dependent linear wave solution for various kinds of right-travelling waves, as described below, and their interactions with the background radiation field are observed as they propagate to the right. For each of these waves, the wavenumber, $k = 2\pi/\lambda_w$, is chosen in such a way that for the constant background opacity, $\kappa = 0.4$ cm$^2$/g, the optical depth given by $\tau_{\lambda_w} = \kappa \rho_0 \lambda_w$ is 10$^3$. For the above values, $\lambda_w = 7.77363\times10^{11}$ cm, and therefore, $k = 8.08269\times10^{-12}$ cm$^{-1}$.

As a first test, an Alfv\'en wave is excited. The eigenfunctions for the Alfv\'en wave are obtained by setting $\omega/k$ 
to the background 
$x-$Alfv\'en speed shown in Equation~(\ref{eq:char_speeds}), and neglecting all heating and radiation terms. For this value of $\omega$, and since $B_{y,0} = 0$, we can fix the $\hat{v}_y$ perturbation, while the $y$-magnetic field perturbation can be obtained in terms of $\hat{v}_y$ as $\hat{B}_y = -(B_{x,0}k/\omega)\hat{v}_y$. Using these perturbations, the following wave settings are used at the left interior boundary of the domain:
\begin{equation}\label{eq:alfven_pert_v_y}
v_{y,1} = \hat{v}_y sin(kx-\omega t),
\end{equation}
\begin{equation}\label{eq:alfven_pert_B_y}
B_{y,1} = \hat{B}_y sin(kx-\omega t).
\end{equation}
All other perturbations are zero i.e., $\hat{\rho} = \hat{v}_x = \hat{v}_z = \hat{p}_y = \hat{B}_z = \hat{E} = 0$. We use a non-dimensional $\hat{v}_y$ value of 10$^{-2}$. Figure~(\ref{fig:alfven_propagation}) shows the propagation of the ${v}_y$ and $B_y$ components of the Alfv\'en wave, after it crossed the right-hand side boundary of the computational domain. As discussed earlier, the Alfv\'en solution is completely decoupled from the rest of the eigensystem. It is therefore unaffected by the radiation and heating terms and propagates across the domain without undergoing any damping or instability. 

\begin{figure*} 
   \centering
   \FIG{\includegraphics[width=12.5cm,clip]{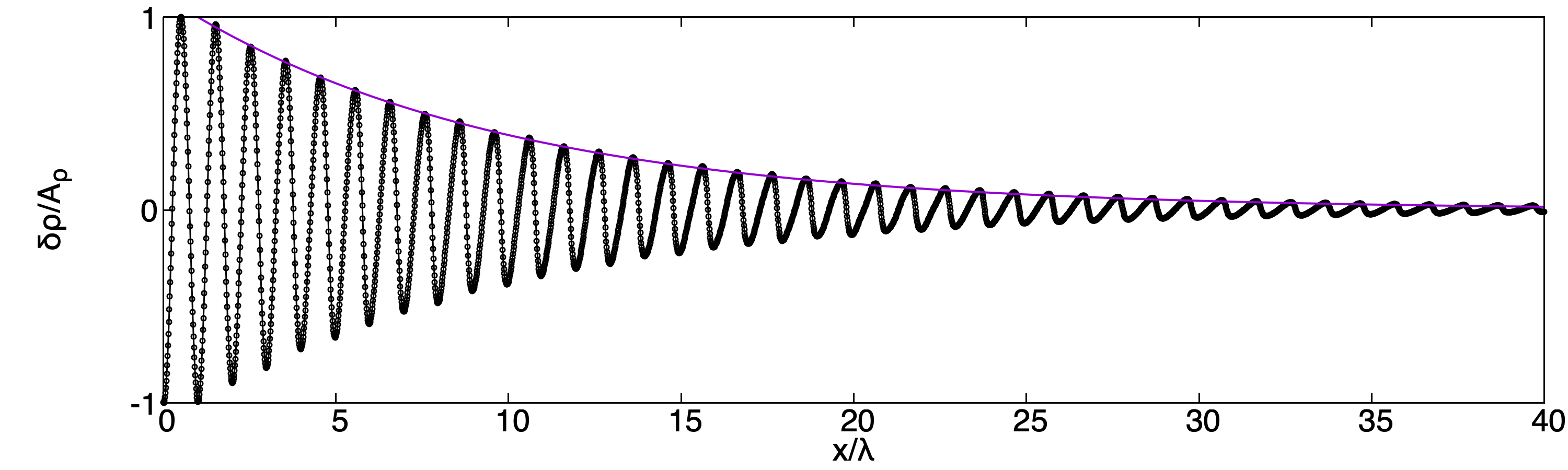}
   \includegraphics[width=12.5cm,clip]{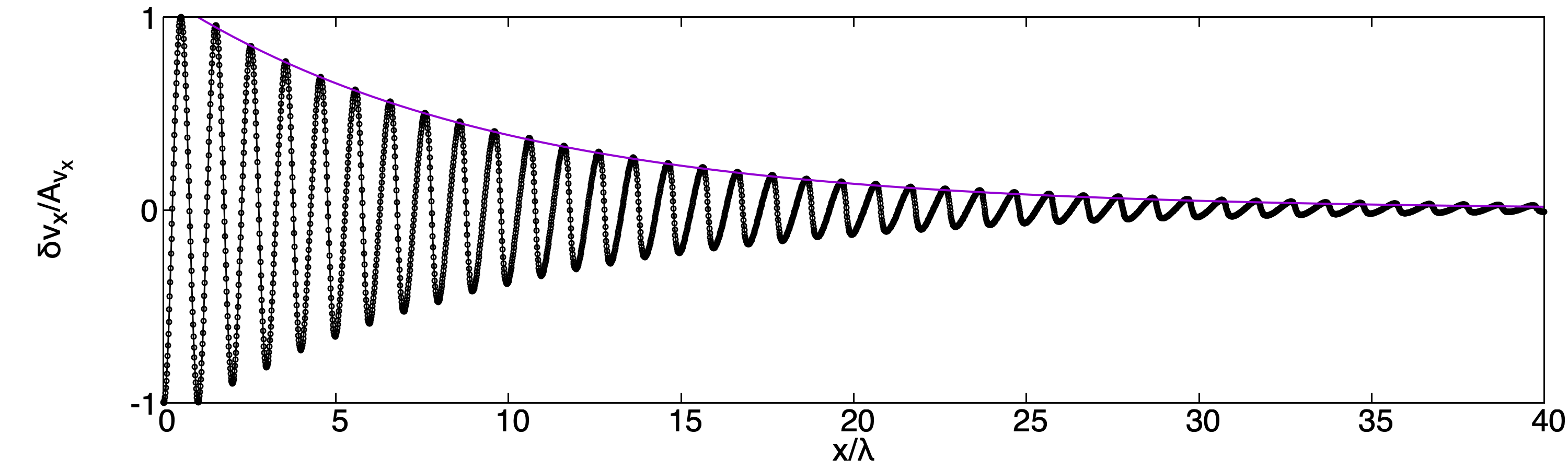}
   \includegraphics[width=12.5cm,clip]{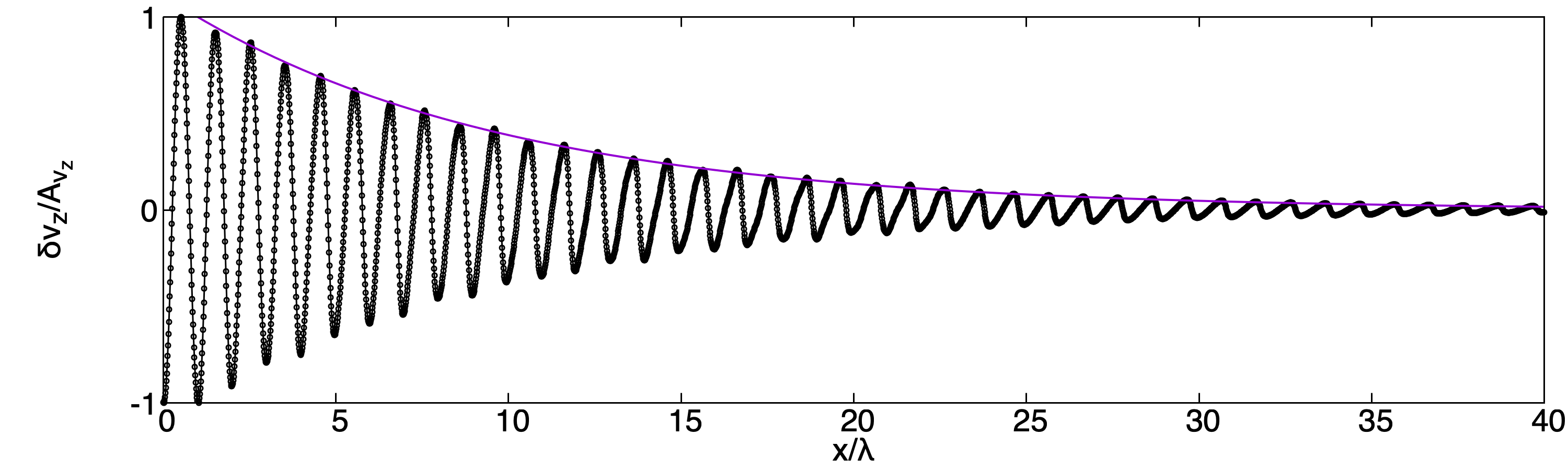}
   \includegraphics[width=12.5cm,clip]{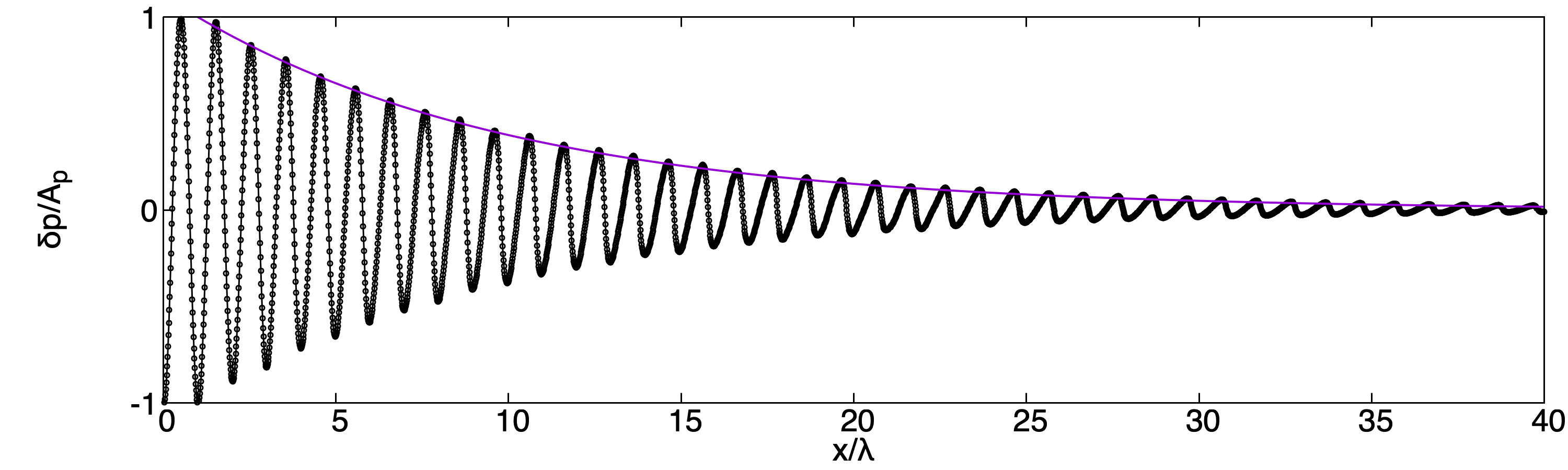}
   \includegraphics[width=12.5cm,clip]{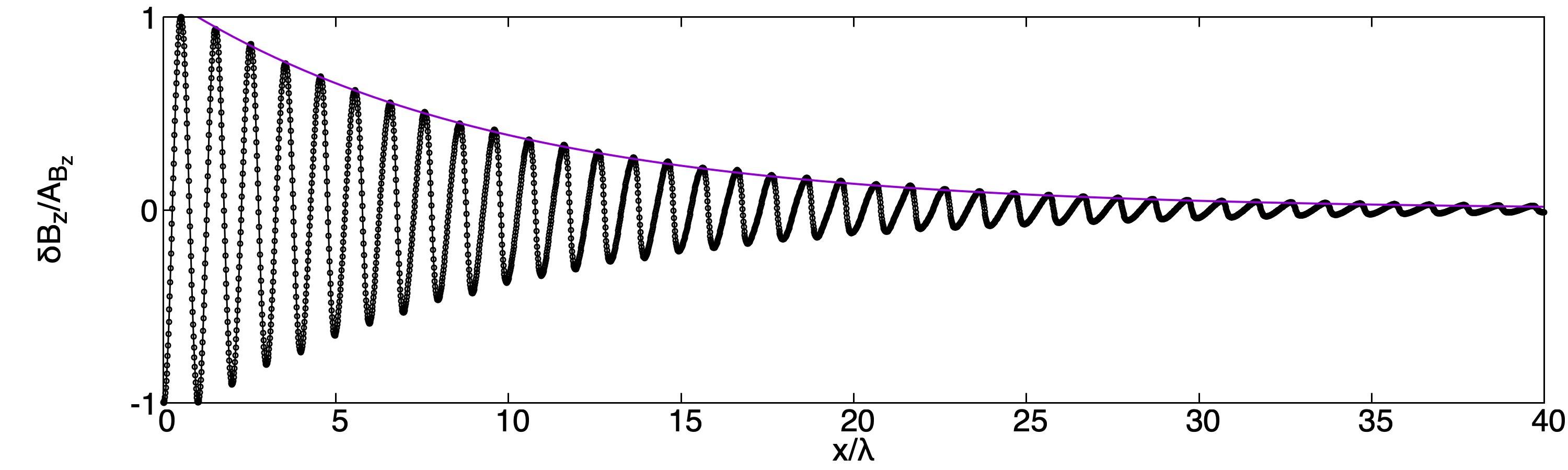}
   \includegraphics[width=12.5cm,clip]{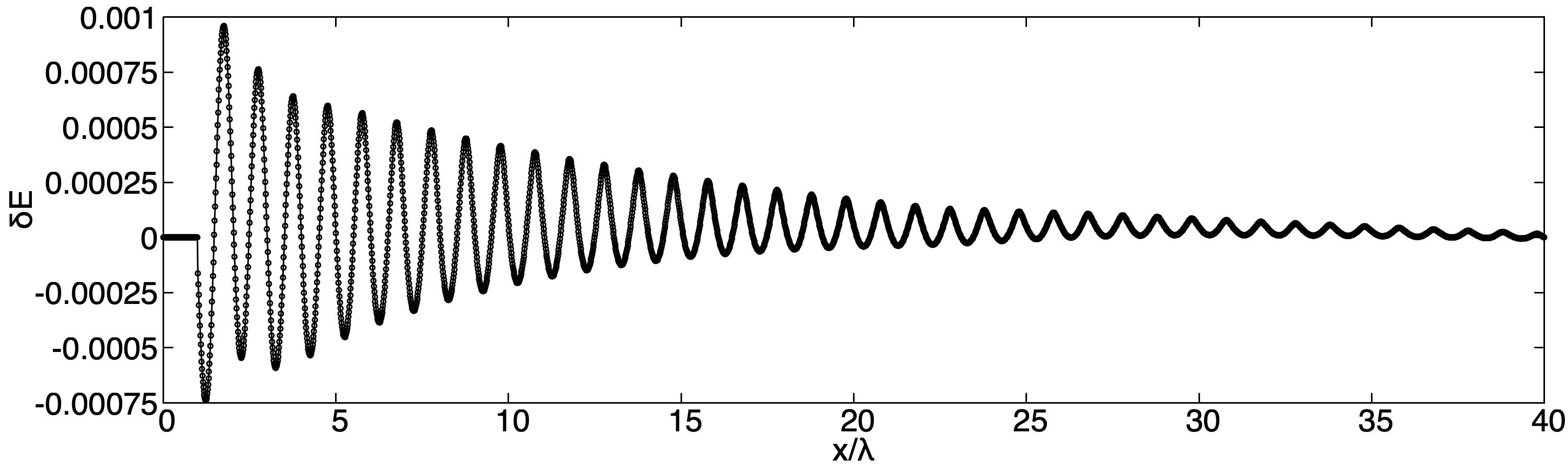}}
   \caption{The density, $x$-velocity, $z$-velocity, plasma pressure, $z$-magnetic field and radiation energy \nmn{density} perturbation components for the fast magnetosonic wave, for \nmn{the weakly radiative background plasma
   }. The analytical, expected damping of the perturbation magnitude is also plotted as a solid purple line, for comparison with simulation results.}
    \label{Fig3}%
    \end{figure*}

\begin{figure*} 
   \centering
   \FIG{\includegraphics[width=12.5cm,clip]{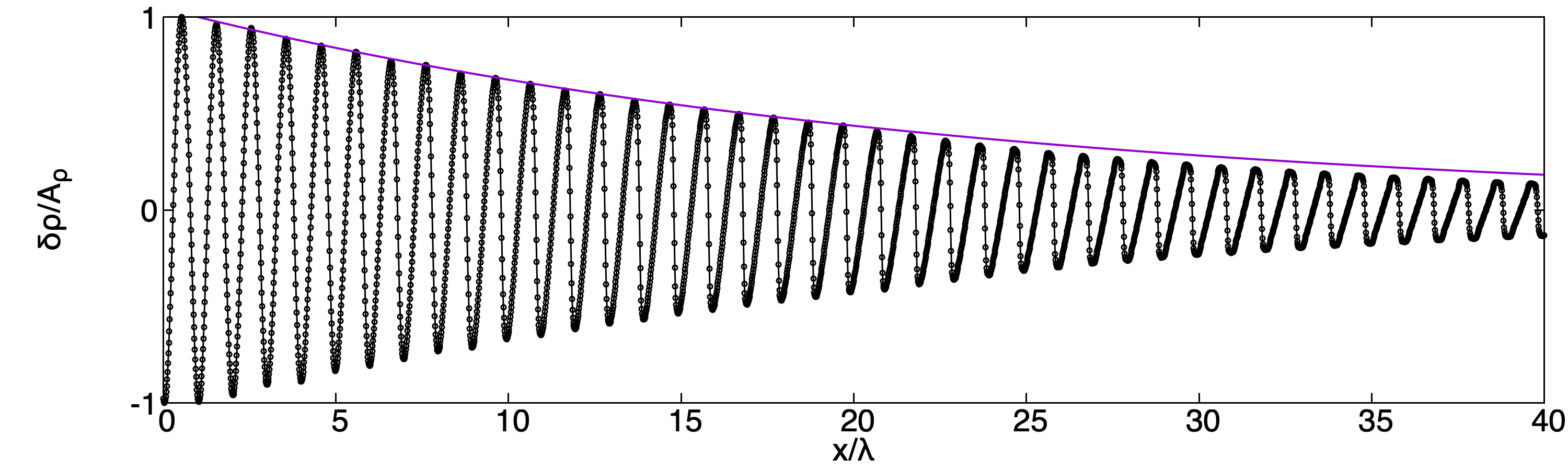}
   \includegraphics[width=12.5cm,clip]{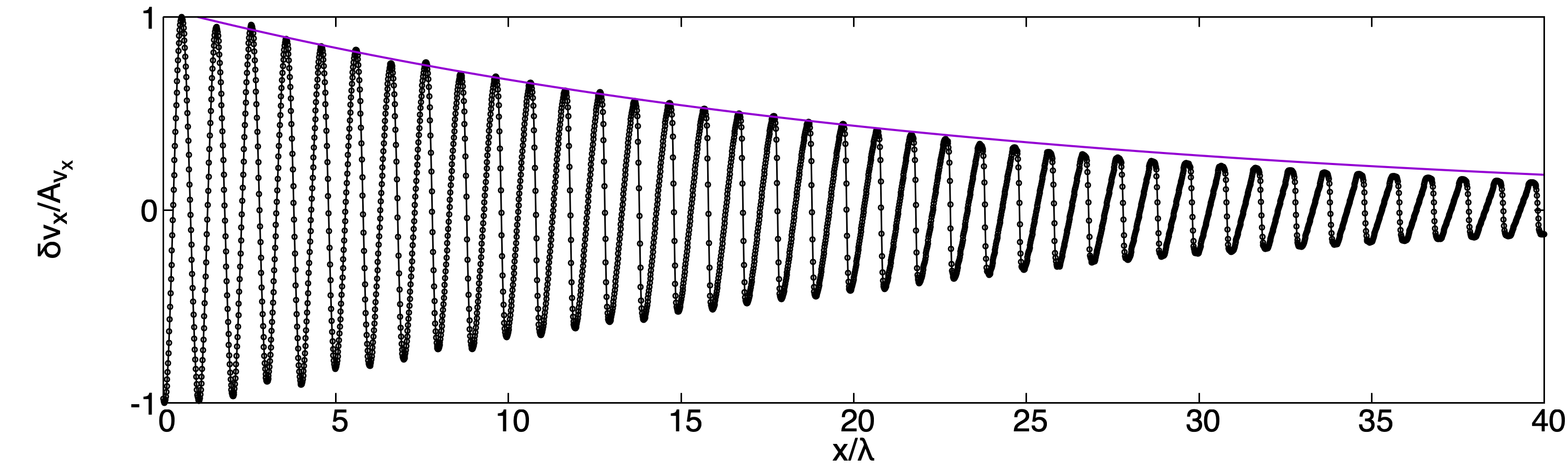}
   \includegraphics[width=12.5cm,clip]{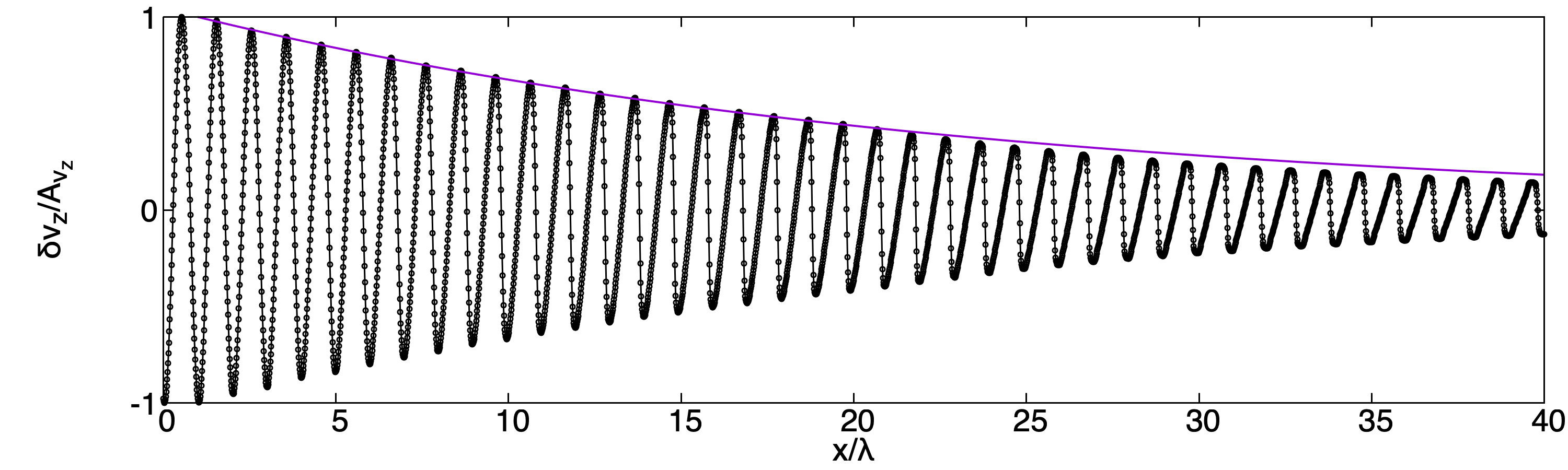}
   \includegraphics[width=12.5cm,clip]{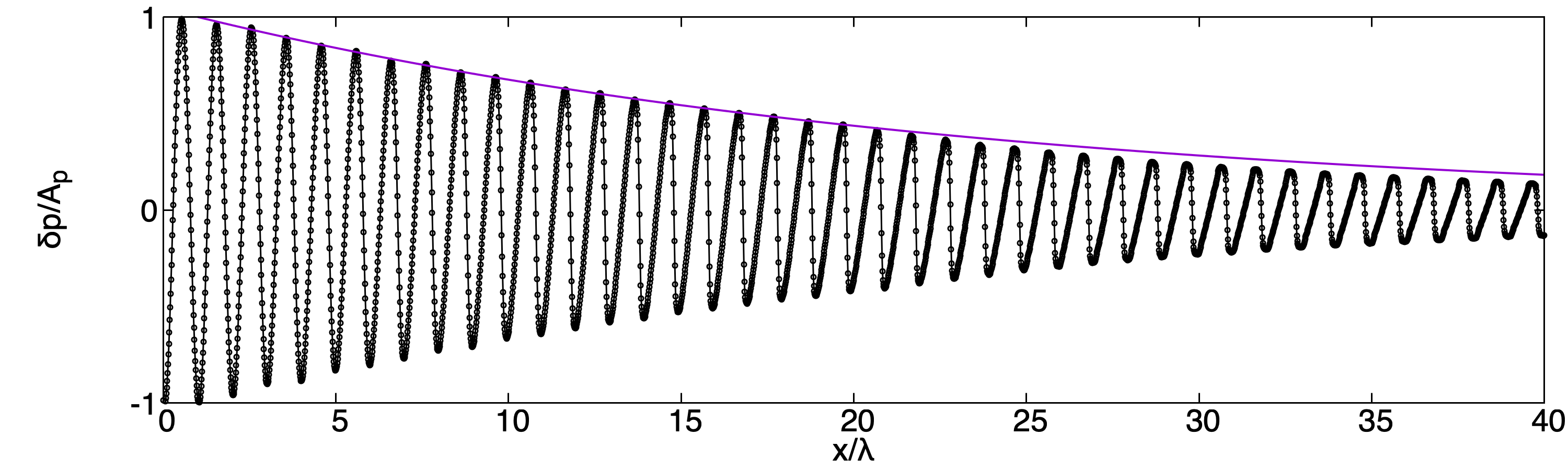}
   \includegraphics[width=12.5cm,clip]{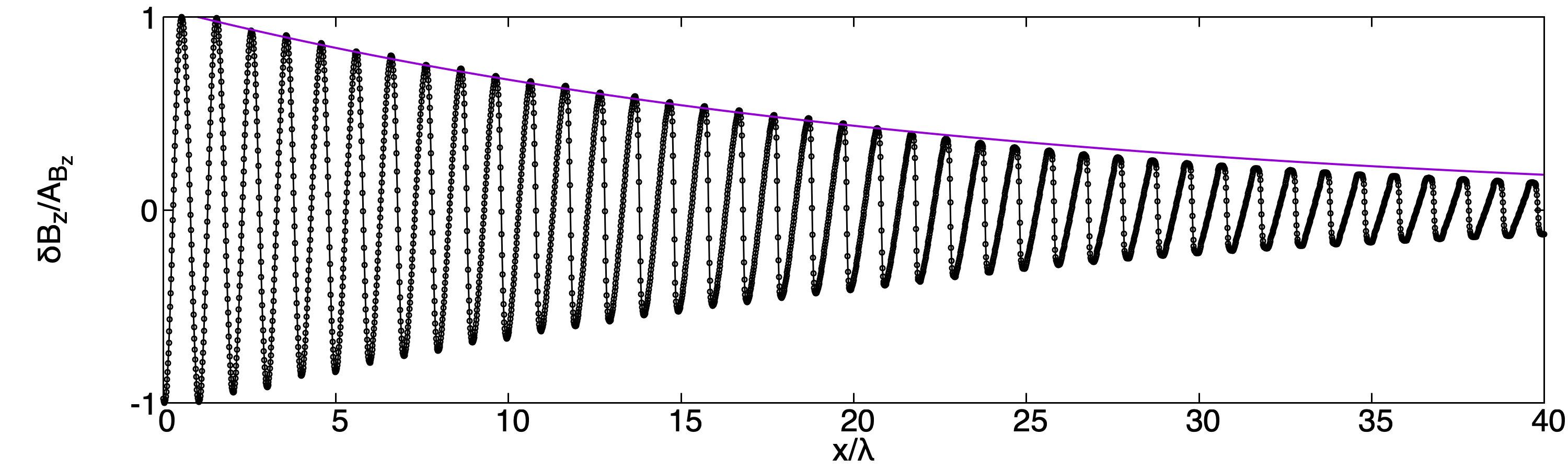}
   \includegraphics[width=12.5cm,clip]{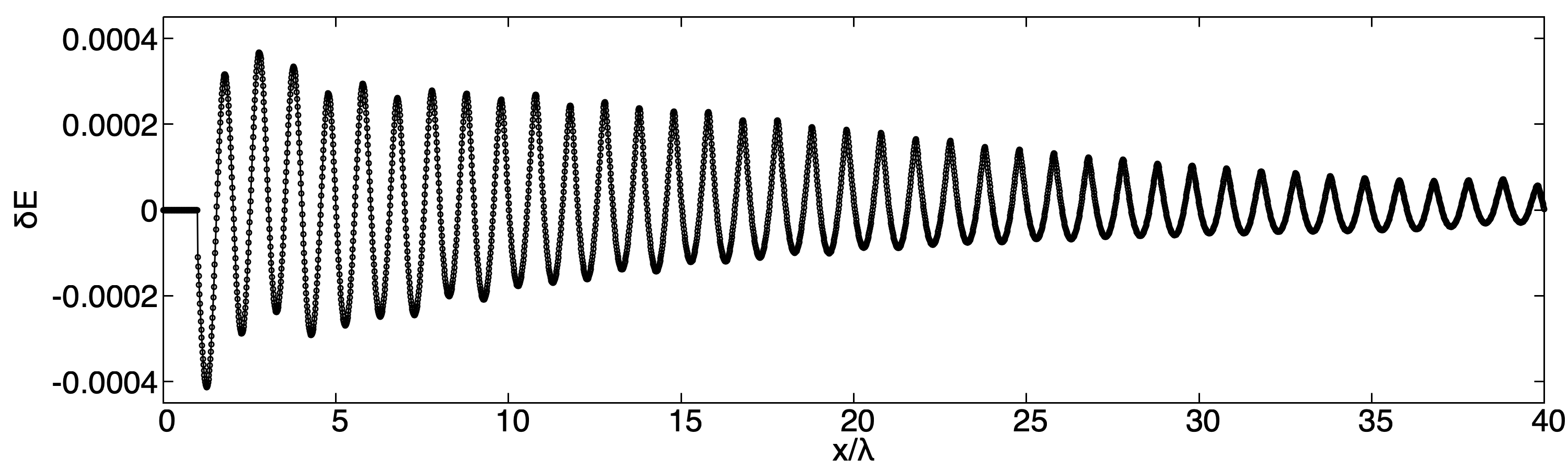}}
   \caption{The density, $x$-velocity, $z$-velocity, plasma pressure, $z$-magnetic field and radiation energy \nmn{density} perturbation components for the slow magnetosonic wave, for \nmn{the weakly radiative background plasma
   }. The analytical, expected damping of the perturbation magnitude is also plotted as a solid purple line, for comparison with simulation results.}
    \label{Fig4}%
    \end{figure*}

Similarly, fast and slow magnetosonic waves, in two separate tests, are excited. The corresponding eigenfunctions are obtained by setting $\omega/k$ 
to the background ideal MHD fast or slow magnetosonic speed, depending on the case. These magnetosonic speeds are given by
\begin{equation}\label{eq:cfs_speed}
c_{f/s,0} = \sqrt{\frac{(c_{g,0}^2 + v_{A,0}^2) \pm \sqrt((c_{g,0}^2 + v_{A,0}^2)^2 - 4 c_{g,0}^2 v_{A,x,0}^2)^2)}{2}},
\end{equation}
where the `+' sign is for the fast wave and the `-' sign is for the slow wave. 
For exciting these waves in a way that avoids gas (and hence radiation) temperature variations, the adiabatic sound speed $c_g$, was replaced with the isothermal sound speed, $c_{\mathrm{iso}}$ in the above equation. Again, the heating and radiation terms are neglected for setting up these eigenfunctions. Fixing with a simple sinusoidal plane wave $\hat{\rho}$ perturbation, all other perturbations can be written in terms of $\hat{\rho}$ as follows:
\begin{equation}\label{eq:cfast_pert_v_x}
\hat{v}_x = \biggl(\frac{\omega}{k}\biggr)\frac{\hat{\rho}}{\rho_0},
\end{equation}
\begin{equation}\label{eq:cfast_pert_p}
\hat{p} = (c_{g,0}^2)\hat{\rho},
\end{equation}
%
%
%
%
\begin{equation}\label{eq:cfast_pert_v_z}
\hat{v}_z = \frac{({B_{z,0}}/{B_{x,0}})({\omega}/{k})}{\biggl(1 - {(\omega/k)^2}/{v_{A,x,0}^2} \biggr)}\frac{\hat{\rho}}{\rho_0},
\end{equation}
%
%
%
%
\begin{equation}\label{eq:cfast_pert_B_z}
\hat{B}_z = \frac{{B_{z,0}}}{\biggl(1 - {v_{A,x,0}^2}/{(\omega/k)^2} \biggr)}\frac{\hat{\rho}}{\rho_0}.
\end{equation}
For $B_{y,0} = 0$, the $\hat{v}_y$ and $\hat{B}_y$ perturbations are both zero i.e., $\hat{v}_y = \hat{B}_y = 0$. We use a non-dimensional $\hat{\rho}$ value of 10$^{-2}$ for both the fast and slow magnetosonic wave tests. For the given values of the background plasma properties, the dispersion relation can be solved and all roots can be quantified.
%
%
These 8 solutions are tabulated in Table~(\ref{table:eigen_solns_weak}) and shown on the complex eigenfrequency plane in \nmn{the top panel of} Figure~(\ref{fig:eigen_solns}). 
The imaginary components of the $\omega$ solutions, corresponding to the slow and fast magnetosonic speeds are $\omega_{Im,slow} = 7.05868\times 10^{-8}$s$^{-1}$ and $\omega_{Im,fast} = 4.10666\times 10^{-7}$s$^{-1}$, respectively. The analytical damping rate per unit length, $\omega_{Im}/(\omega/k)$ is therefore $\omega_{Im,slow}/c_{s,0} = 5.61643\times10^{-14}$cm$^{-1}$ and $\omega_{Im,fast}/c_{f,0} = 1.35347\times10^{-13}$cm$^{-1}$ for the slow and fast waves, respectively. Figures (\ref{Fig3}) and (\ref{Fig4}) show the results for the slow and fast waves, respectively. The expected, theoretical damping of the density perturbation amplitude is also plotted along with the observed results. The observed and theoretical results show an excellent match, for both the slow and fast magnetosonic waves. 


%
%

\begin{figure*} 
   \centering
   \FIG{\includegraphics[width=12.5cm,clip]{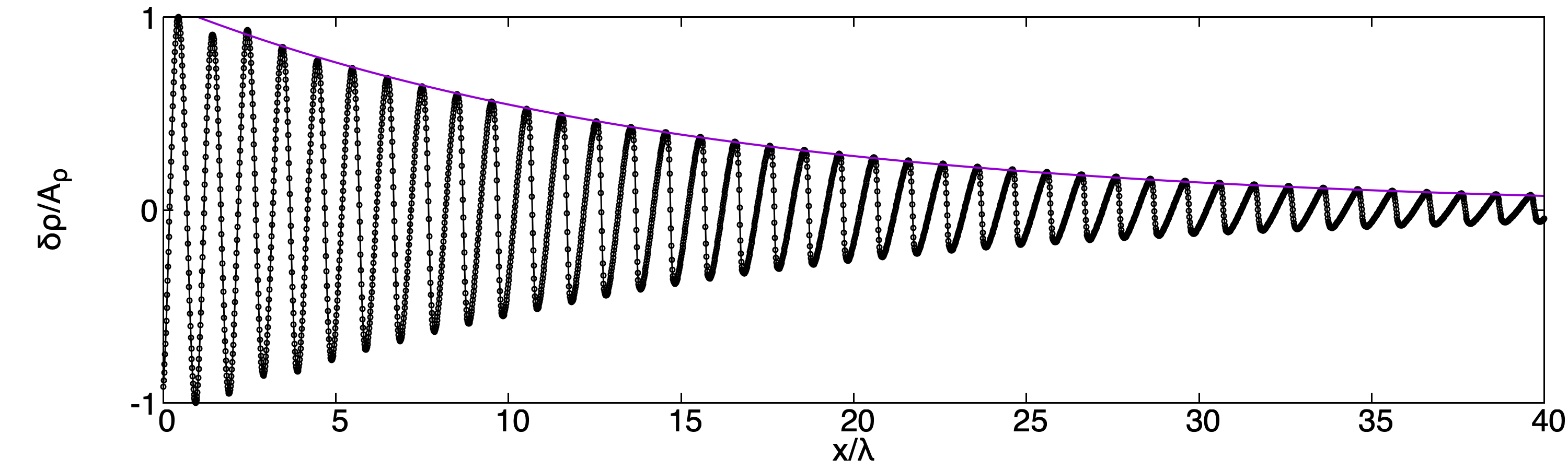}
   \includegraphics[width=12.5cm,clip]{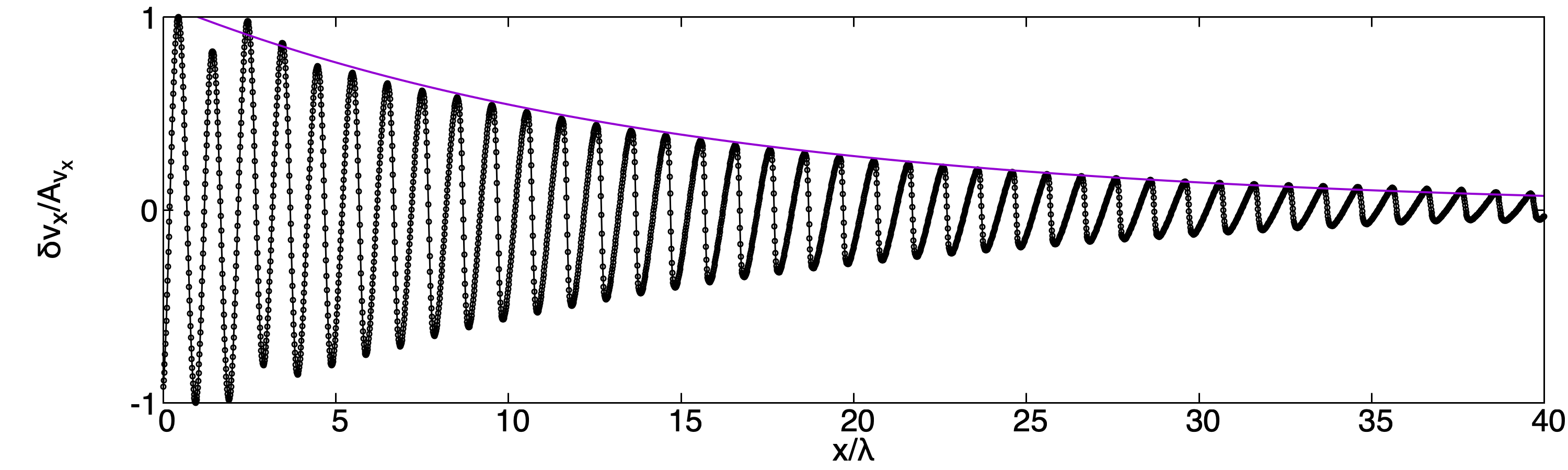}
   \includegraphics[width=12.5cm,clip]{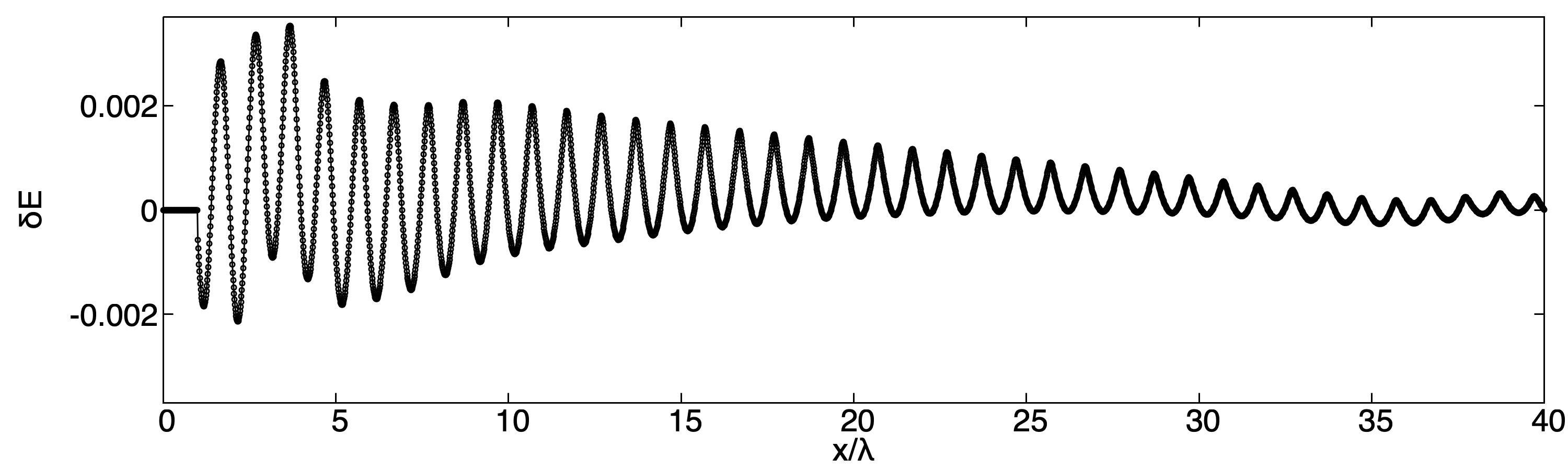}}
   \caption{\nmn{The density, $x$-velocity and radiation energy \nmn{density} perturbation components for the slow magnetosonic wave, for the strongly radiative background plasma. The analytical, expected damping of the perturbation magnitude is also plotted as a solid purple line, for comparison with simulation results.}}
    \label{Fig_slow_raddom}%
    \end{figure*}
\begin{figure*}
   \centering
   \FIG{\includegraphics[width=5.9cm,clip]{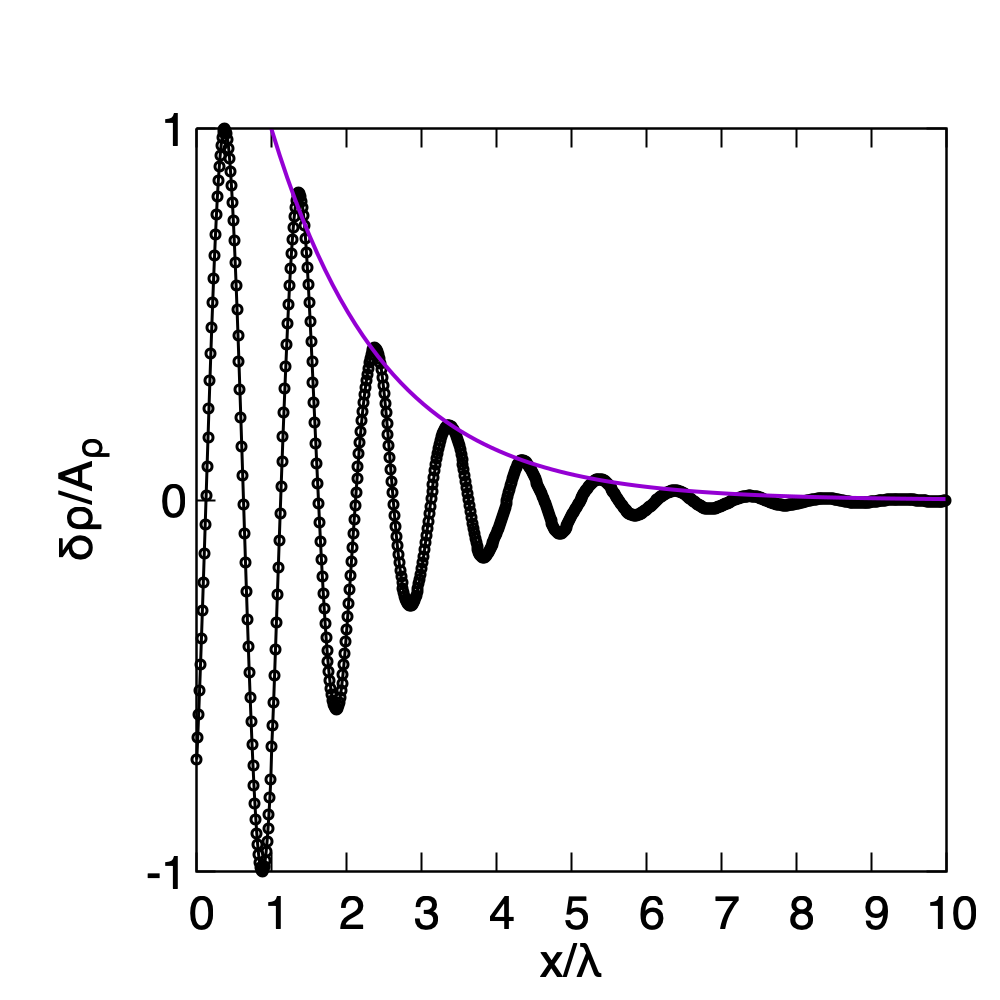}
   \includegraphics[width=5.9cm,clip]{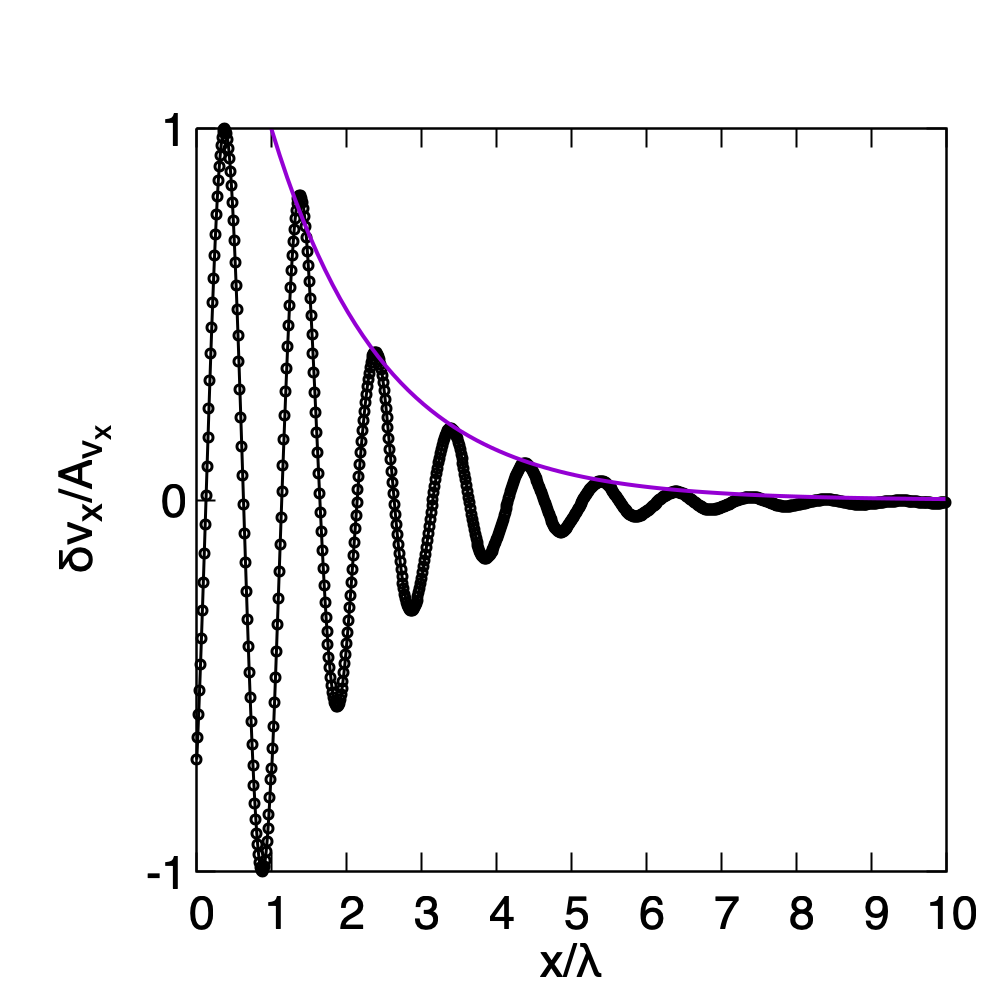}
   \includegraphics[width=5.9cm,clip]{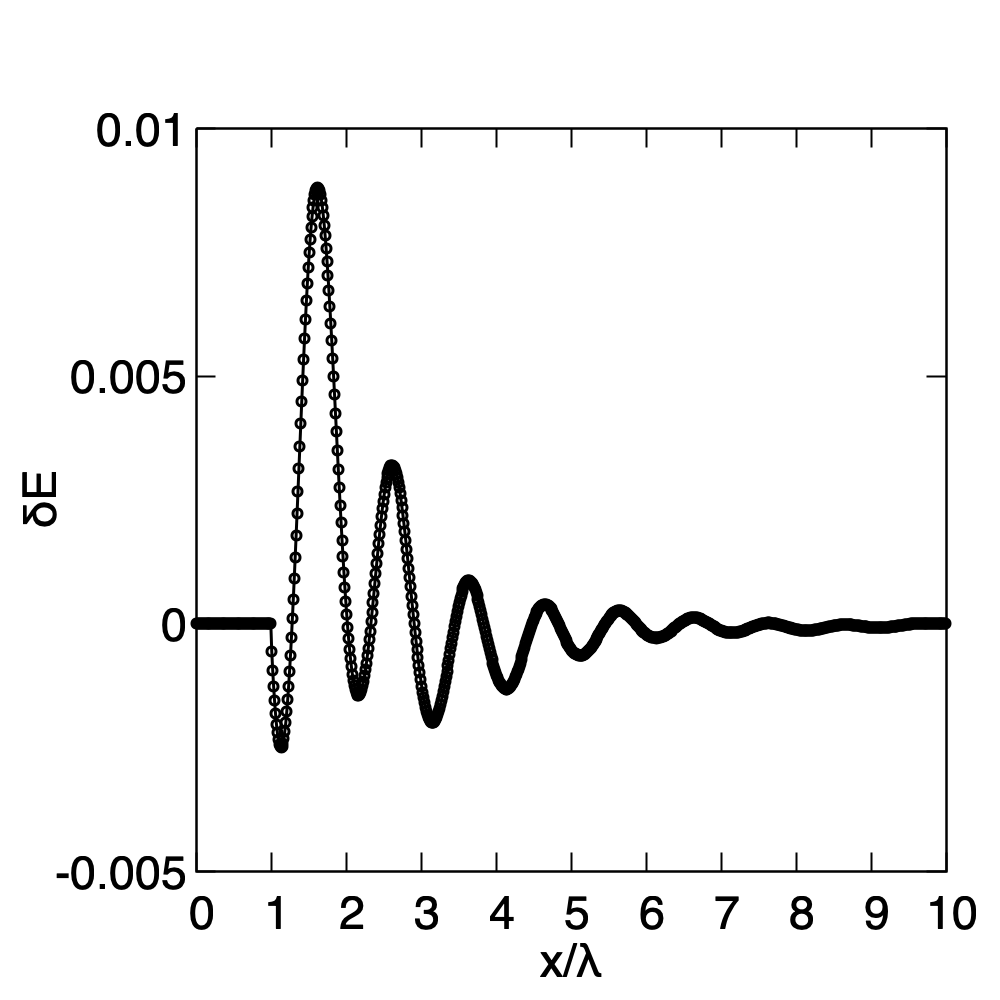}}
      \caption{\nmn{The density, $x$-velocity and radiation energy \nmn{density} perturbation components for the fast magnetosonic wave, for the strongly radiative background plasma. The analytical, expected damping of the perturbation magnitude is also plotted as a solid purple line, for comparison with simulation results.}}
         \label{Fig_fast_raddom}
\end{figure*}
\begin{figure}
   \centering
  \FIG{ \includegraphics[width=\hsize]{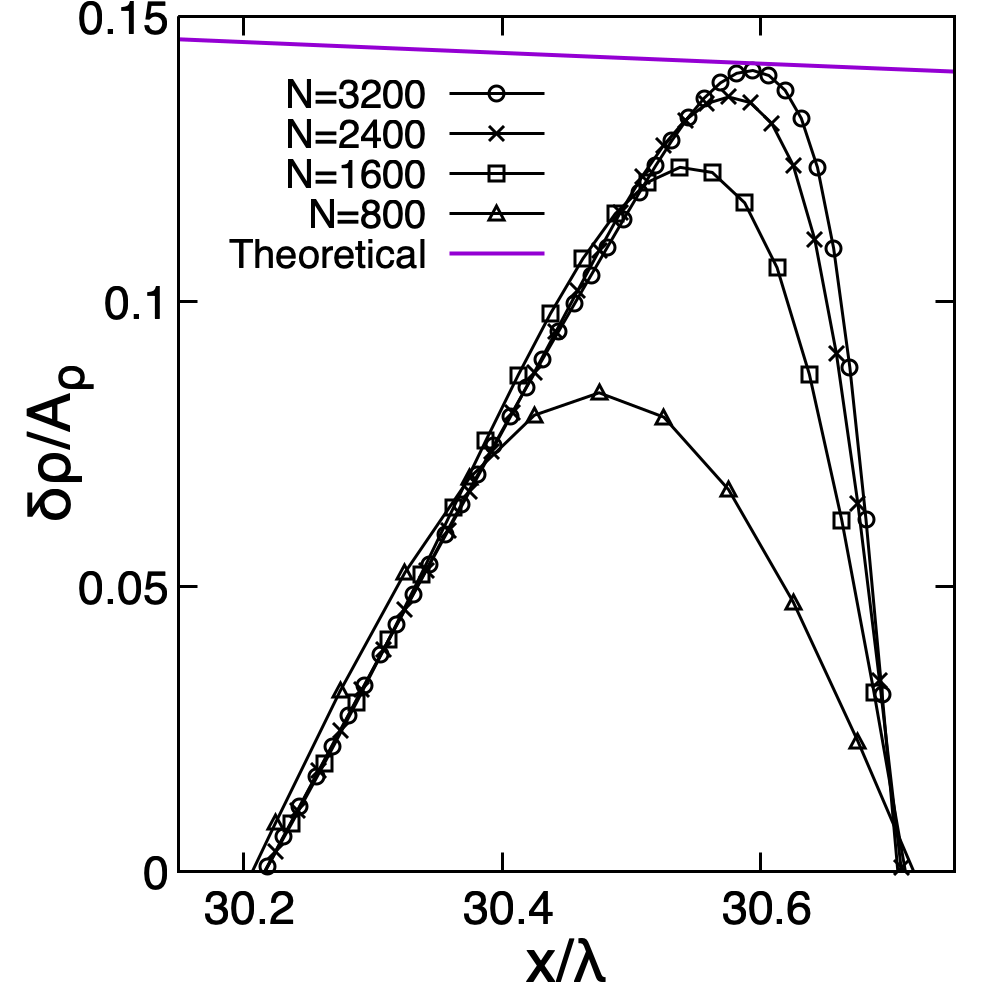}}
      \caption{\nmn{A magnified version of the density solution from Figure (\ref{Fig_slow_raddom}), showing a single half-cycle of the damped slow magnetosonic wave. The analytical, expected damping of the perturbation magnitude is also plotted as a solid purple line. Corresponding results from coarser meshes comprising 800, 1600 and 2400 cells are also shown.}}
         \label{Fig_slow_raddom_crest}
\end{figure}
\begin{figure}
   \centering
  \FIG{ \includegraphics[width=\hsize]{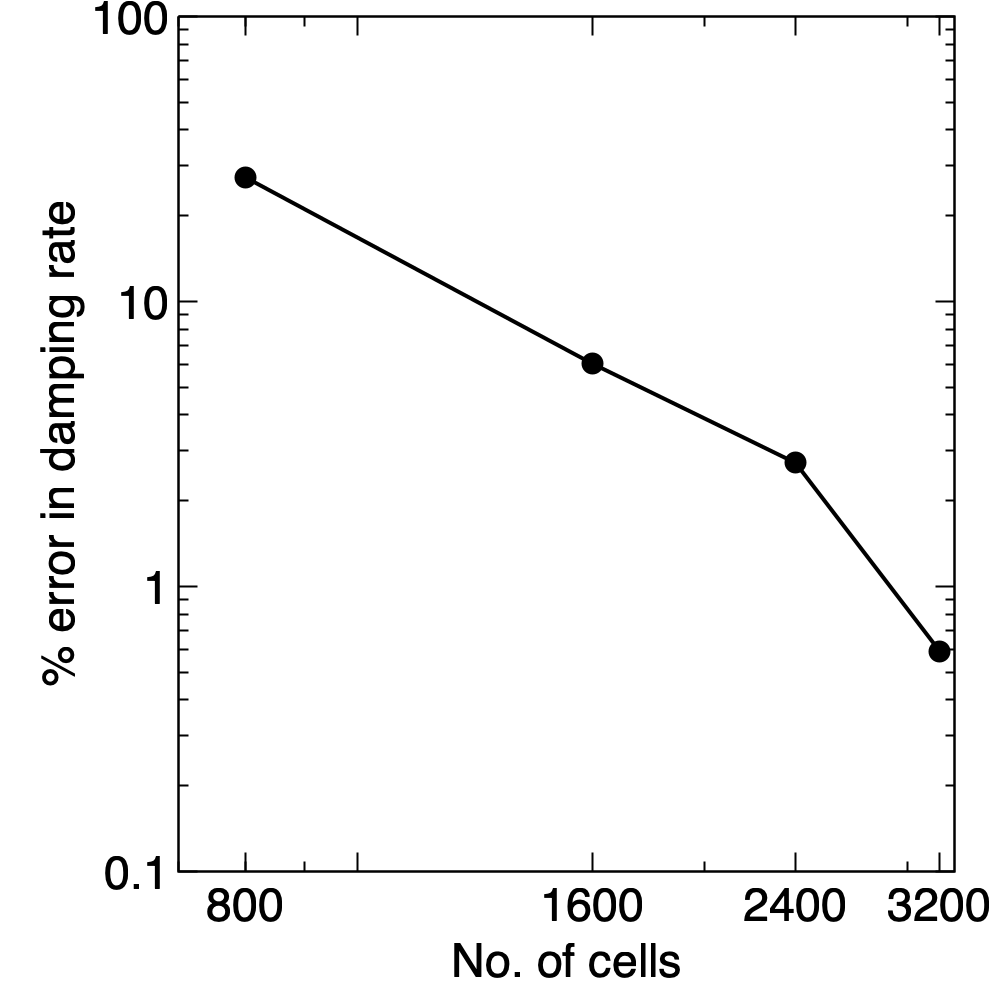}}
      \caption{\nmn{Percent error in the observed damping rates with respect to the theoretical rate, for the slow magnetosonic wave damped by a strongly radiative background plasma.}}
         \label{Fig_slow_raddom_convergence}
\end{figure}

\nmn{We next consider the damping of waves in a radiation-dominant 1D static background plasma. The initial plasma settings are $\rho_0 = 3.216 \times 10^{-9}$ g/cm$^3$, $p_0 = 43.35 \times 10^3$ erg/cm$^3$ and ${\bf B}_0 =(330.14,0,330.14)$ Gauss. With the higher pressure, the initial radiation energy density is now $E_0 = 336.26331 \times 10^3$ erg/cm$^3$, according to the radiative equilibrium condition. All other parameters such as $\mu$ $\gamma$, $k$, $\tau_{\lambda_w}$, $\lambda_w$ and $\kappa$ are the same as in the weakly radiative case described above. The boundary conditions and the prescription for exciting the right-travelling waves are also the same. The computational domain is composed of 3200 cells. The solutions of the resultant dispersion relation, corresponding to these background conditions are tabulated in Table~(\ref{table:eigen_solns_strong}) and shown on the complex eigenfrequency plane in the bottom panel of Figure~(\ref{fig:eigen_solns}). The analytical damping rate per unit length is now $\omega_{Im,slow}/c_{s,0} = 8.64217\times10^{-14}$cm$^{-1}$ for the slow magnetosonic wave and $\omega_{Im,fast}/c_{f,0} = 8.58803\times10^{-13}$cm$^{-1}$ for the fast magnetosonic wave. Since the damping is an order of magnitude higher for the fast wave, the computational domain used for the fast wave damping case is $10\lambda_w$, opposed to $40\lambda_w$ for the slow case damping case. The $\rho$, $v_x$ and $E$ solutions, superimposed with the theoretical damping, for the slow and fast waves are shown in Figures (\ref{Fig_slow_raddom}) and (\ref{Fig_fast_raddom}), respectively. For the slow wave, the case was run with a range of mesh resolutions (800, 1600, 2400 and 3200 cells), to observe the effect of grid sizes on the observed damping rates. A magnified version of the density solution from Figure (\ref{Fig_slow_raddom}) is shown in Figure (\ref{Fig_slow_raddom_crest}), showing a single half-cycle of the damped slow magnetosonic wave, along with the theoretical, expected damping of the wave amplitude. This is also superimposed with the corresponding results obtained using 800, 1600 and 2400 cells. It is observed that the crest of the half-cycle approaches the theoretical damped amplitude with increasing resolution. The $\%$ errors in the observed damping rates with respect to the theoretical rate, for these various grid resolutions, are plotted as a function of grid resolution in Figure (\ref{Fig_slow_raddom_convergence}).}
    
\subsection{Radiation-modified Riemann shock-tube problem}\label{sec:riemann}
We now move over to RMHD Riemann shock tube problems. We add radiation to two standard ideal MHD examples, namely the 1.5D case studied by \citet{brio1988upwind}, and the 1.75D case studied by \citet{torrilhon2003uniqueness}. The initial left and right states for both cases and their non-dimensional equivalent values are given in Table~(\ref{table:RMHD_Riemann}). All velocity components are zero initially for both cases. For the Brio-Wu case, the $z-$magnetic fields are also zero. In this table, the physical length of the computational domain $L$, is also given for each case. A mean molecular weight of $\mu = 0.5$ and a constant background opacity of $\kappa = 400$ cm$^2$/g is used here. The adiabatic index is also listed. The initial computational grid is made up of 1024 cells, and 7 AMR levels are used, leading to an effective resolution of 65536 cells. Zero-gradient boundary conditions are applied at both boundaries for all variables. Without the radiation energy, these cases are identical to the ideal MHD cases by \citet{brio1988upwind} and \citet{torrilhon2003uniqueness}, respectively. This can be clearly observed from the non-dimensional values in Table~(\ref{table:RMHD_Riemann}).

\begin{figure*} 
   \centering
   \FIG{\includegraphics[width=5.9cm,clip]{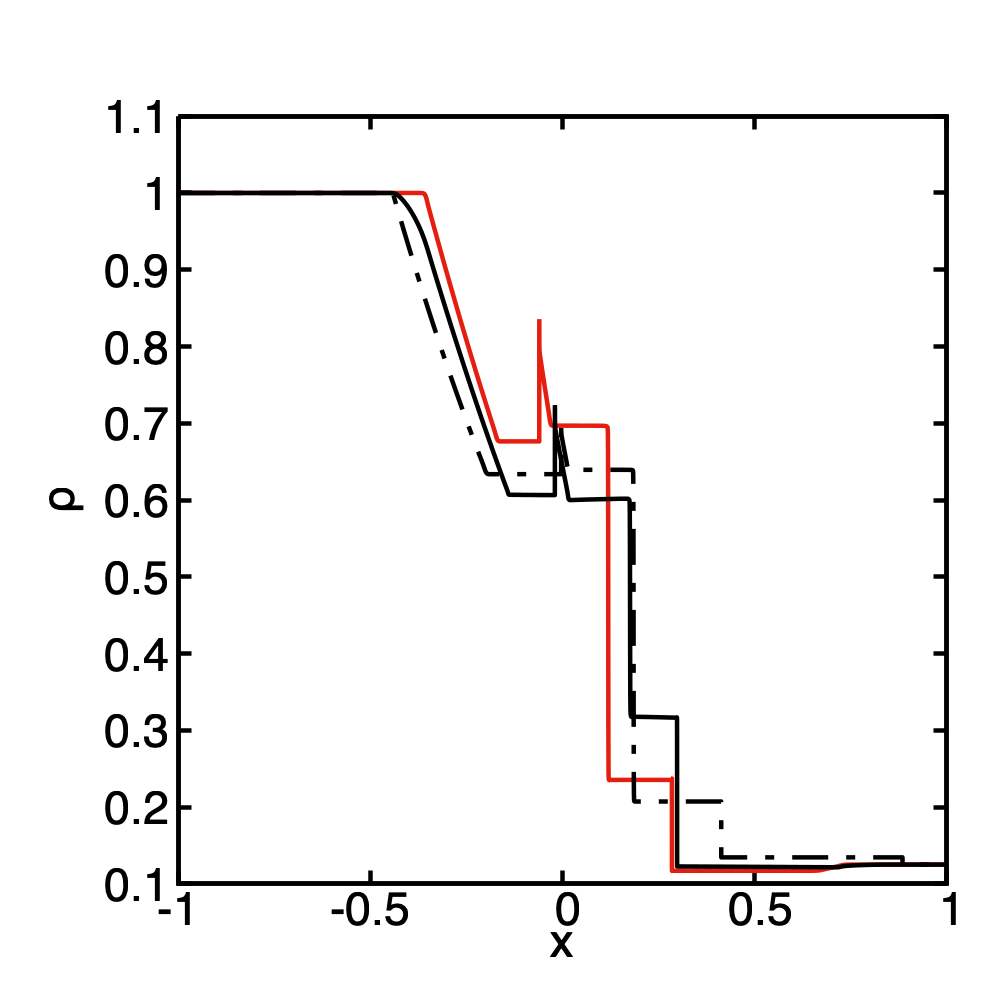}
   \includegraphics[width=5.9cm,clip]{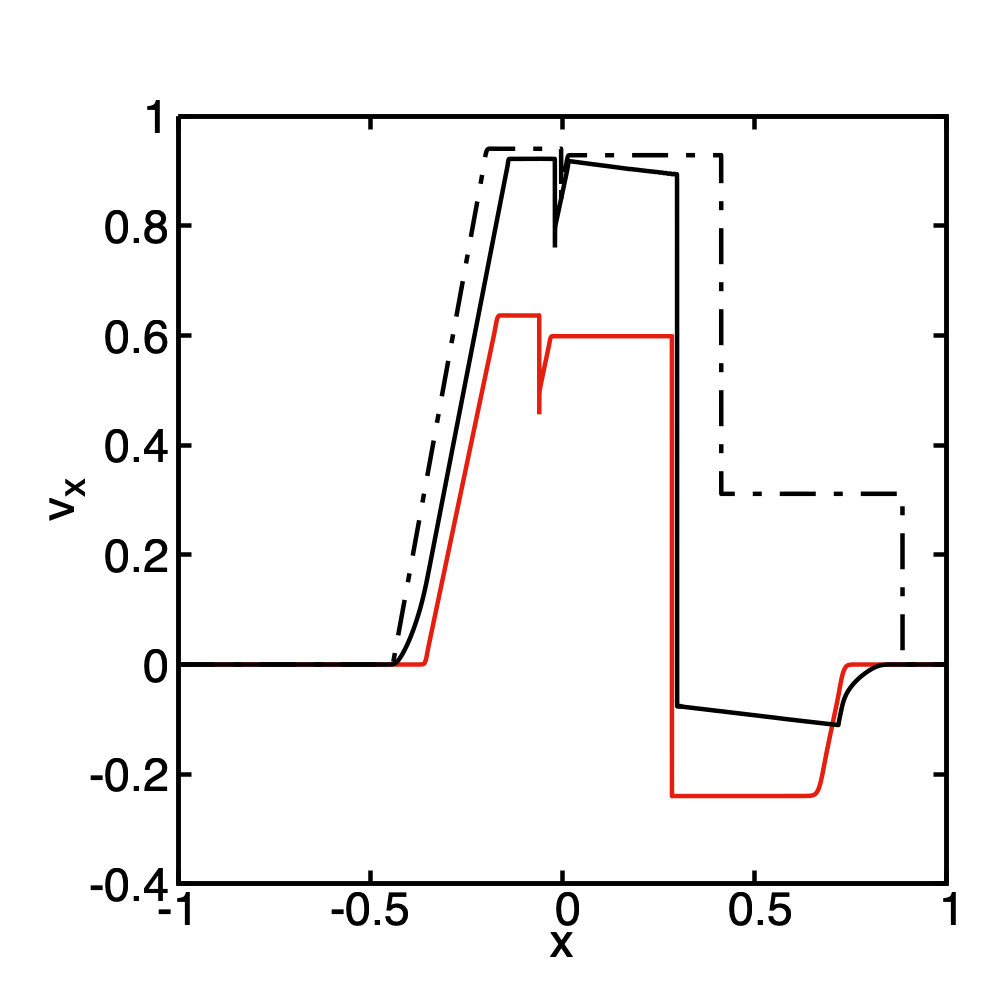}
   \includegraphics[width=5.9cm,clip]{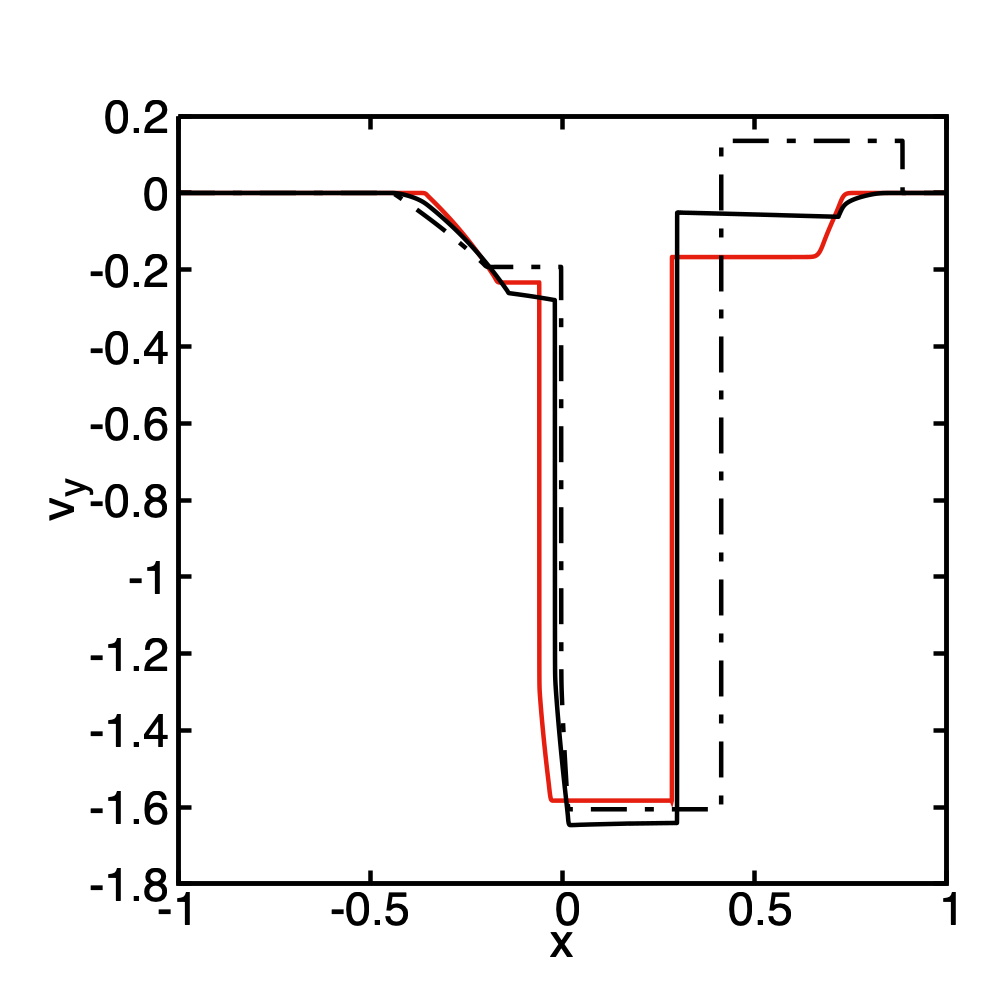}
   \includegraphics[width=5.9cm,clip]{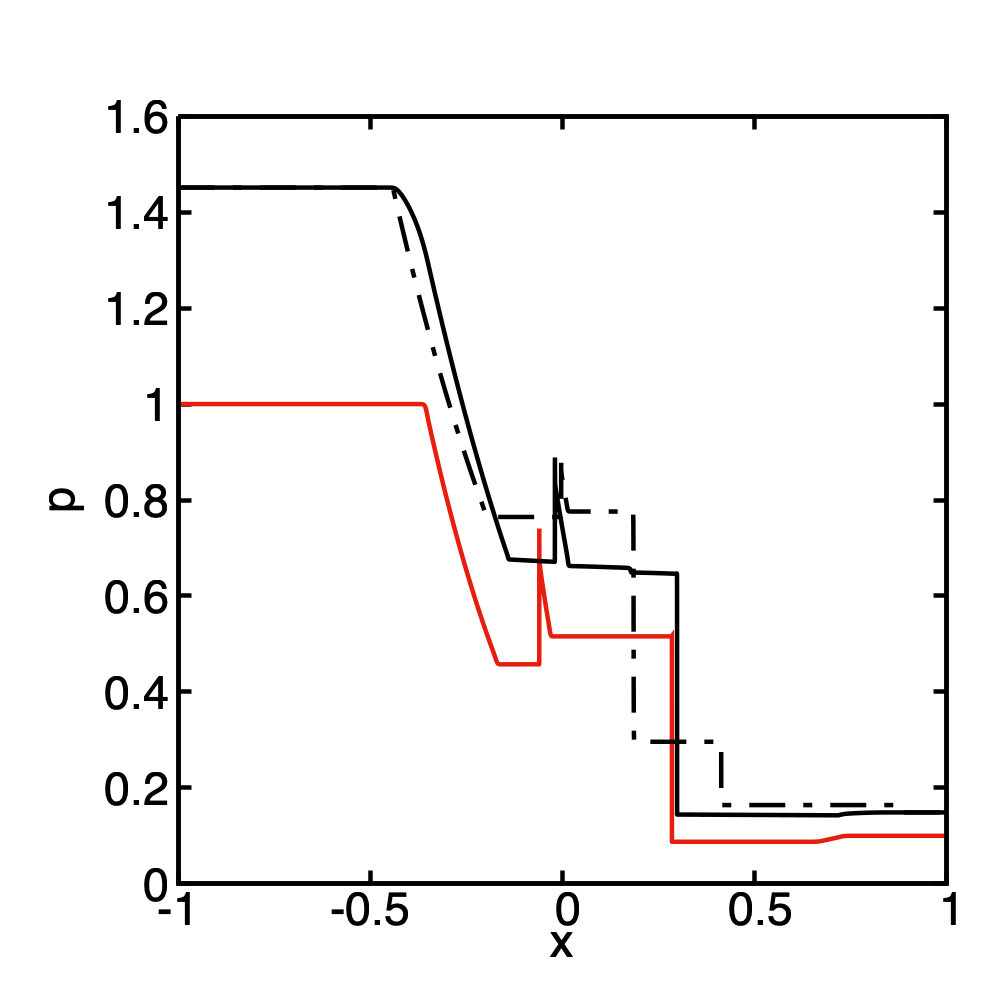}
   \includegraphics[width=5.9cm,clip]{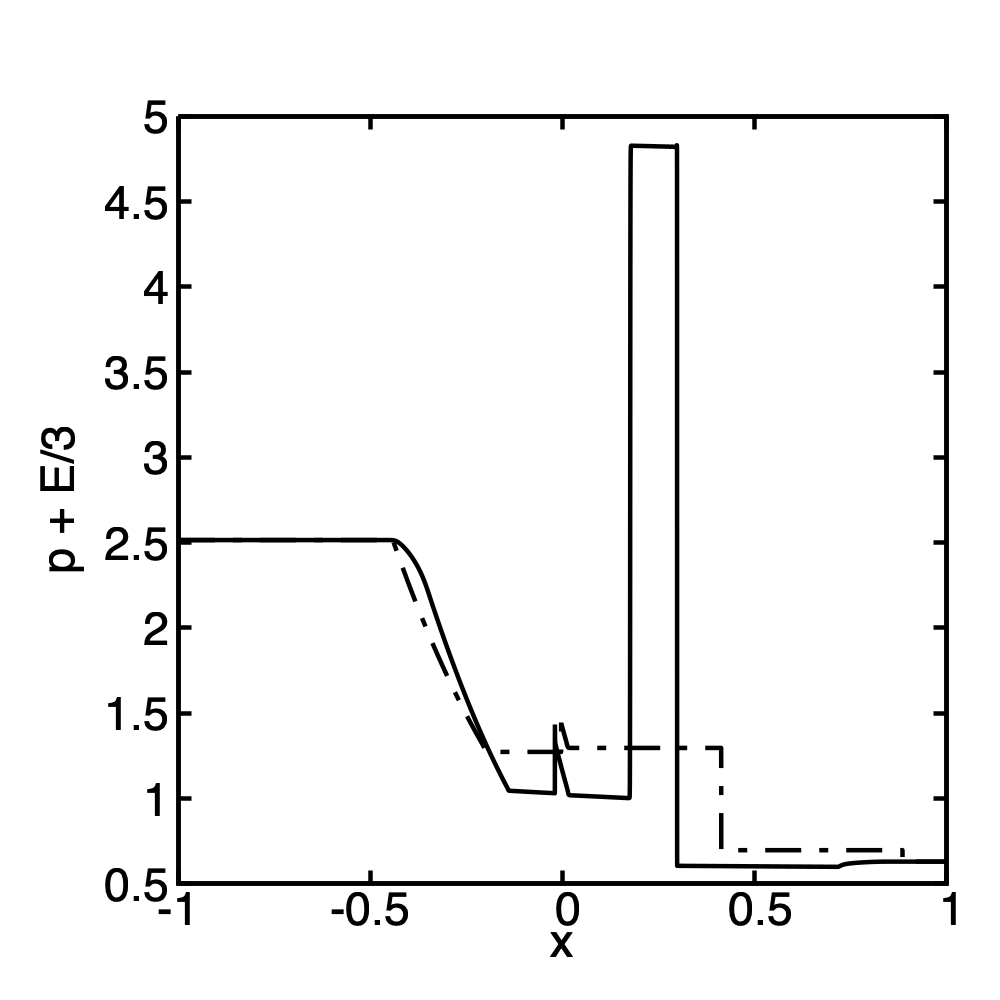}
   \includegraphics[width=5.9cm,clip]{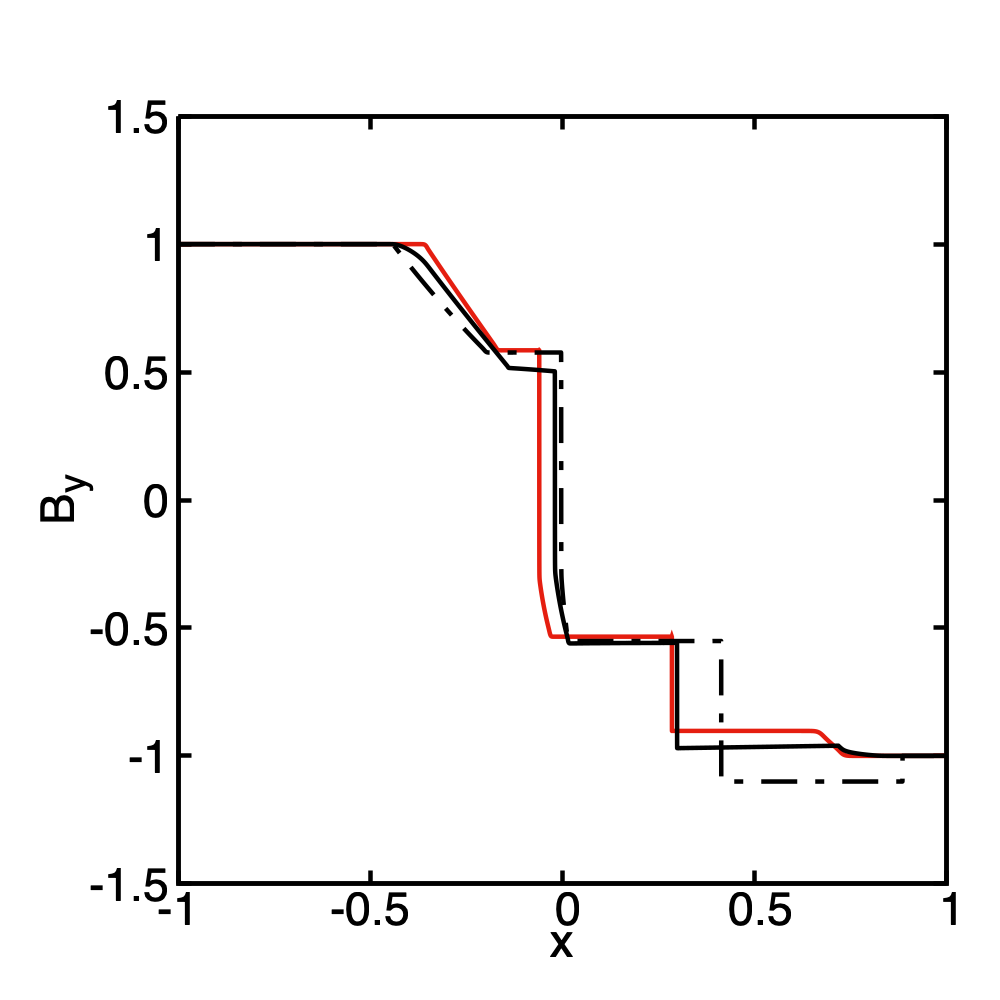}
   \includegraphics[width=5.9cm,clip]{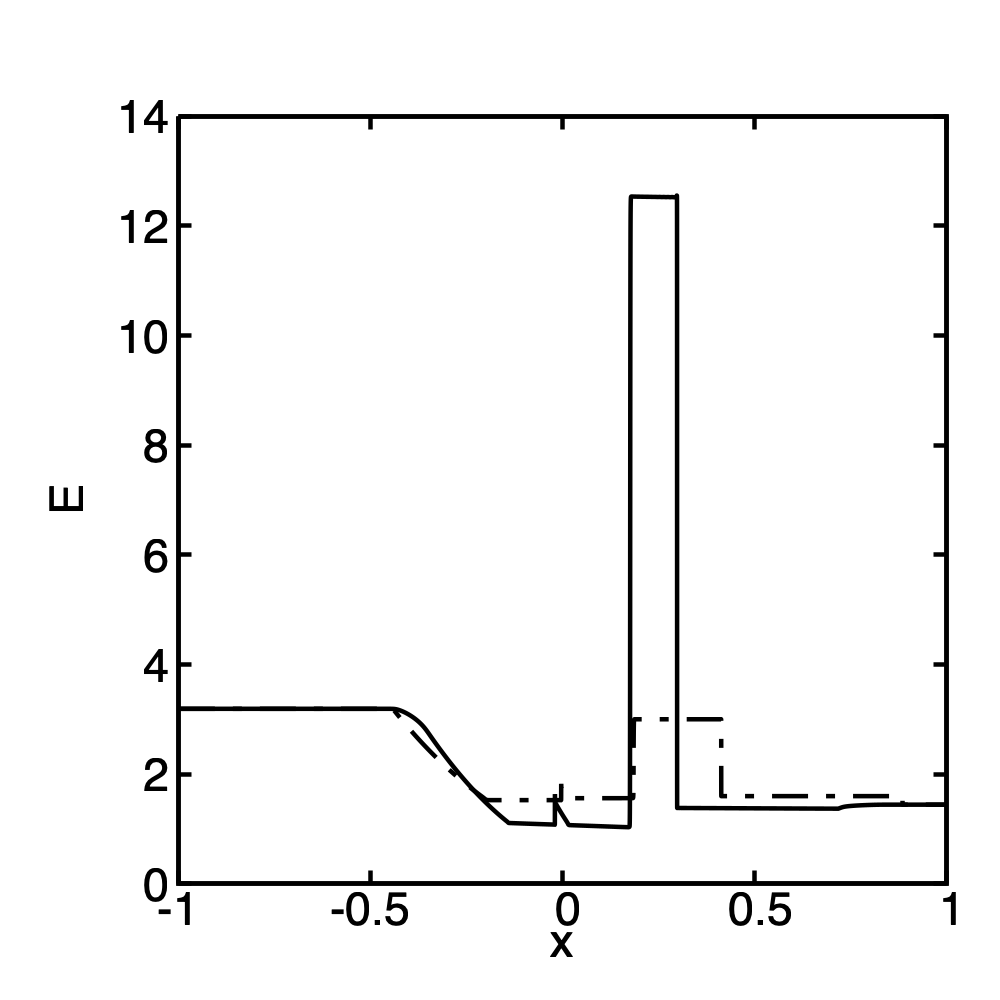}
   \includegraphics[width=5.9cm,clip]{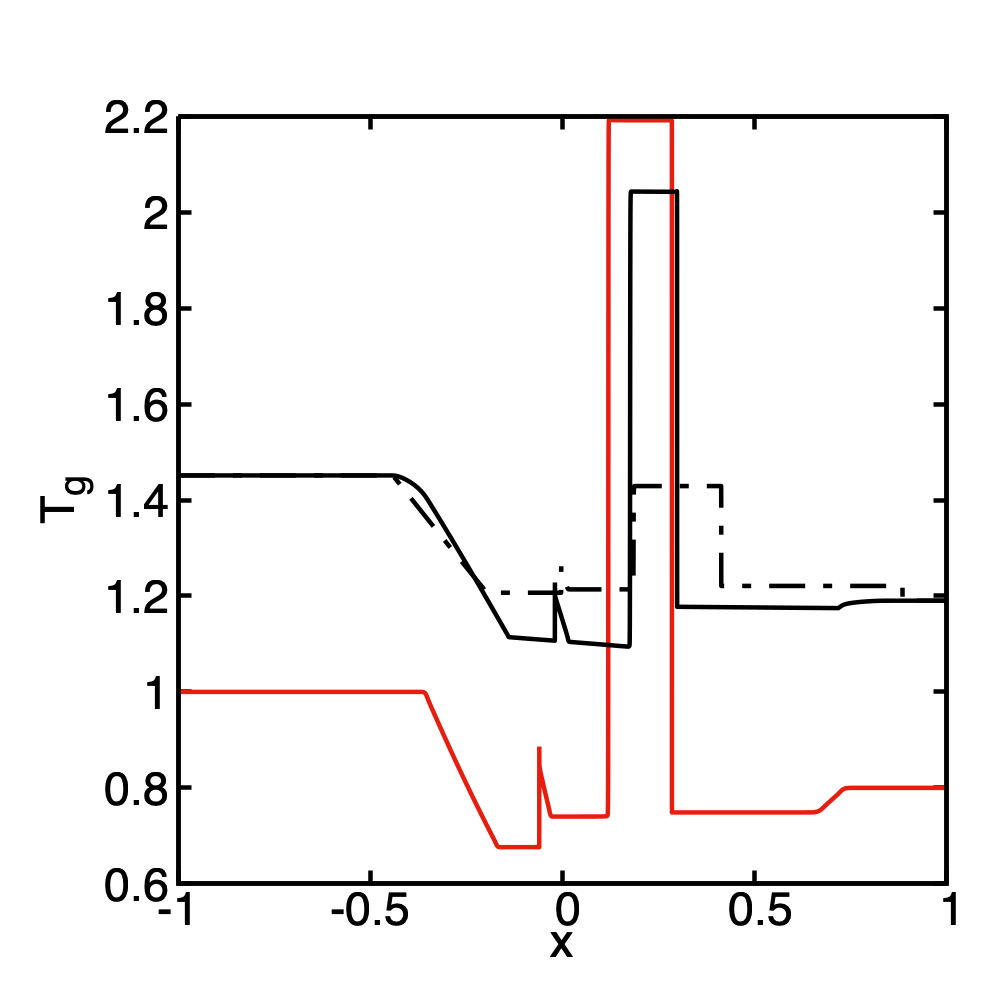}}
   \vspace{-3mm} 
   \caption{The normalized density, $x$-velocity, $y$-velocity, plasma pressure, (plasma + radiation) pressure, $y$-magnetic field, radiation energy \nmn{density and plasma temperature} solutions, at $t = 0.2$ for \citet{brio1988upwind}'s 1.5D Riemann shock-tube problem. Wherever applicable, the corresponding solution for the non-radiative case by \citet{brio1988upwind} is also shown as a solid red line. The solutions we show in black are either full FLD (solid) or in diffusion limit (dot-dashed).}
\label{fig:briowu}%
\end{figure*}

\begin{figure*} 
   \centering
   \FIG{\includegraphics[width=5.9cm,clip]{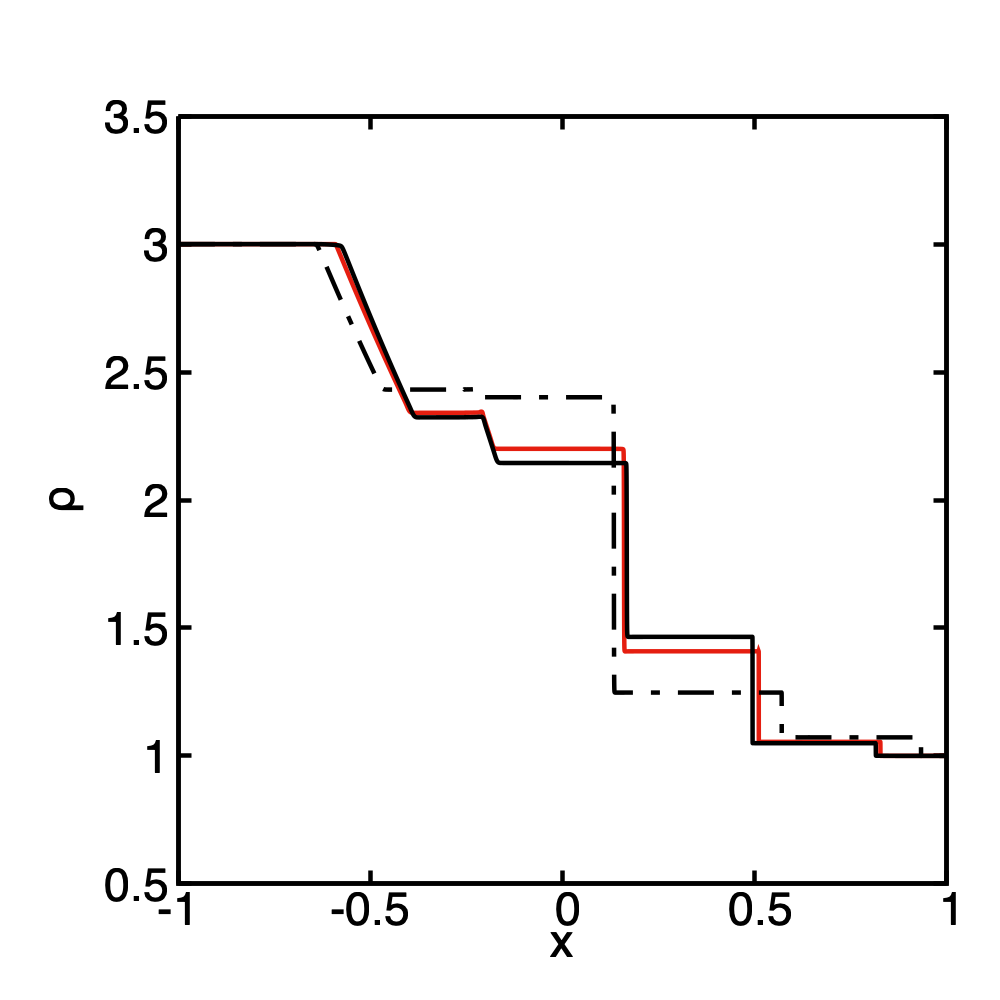}
   \includegraphics[width=5.9cm,clip]{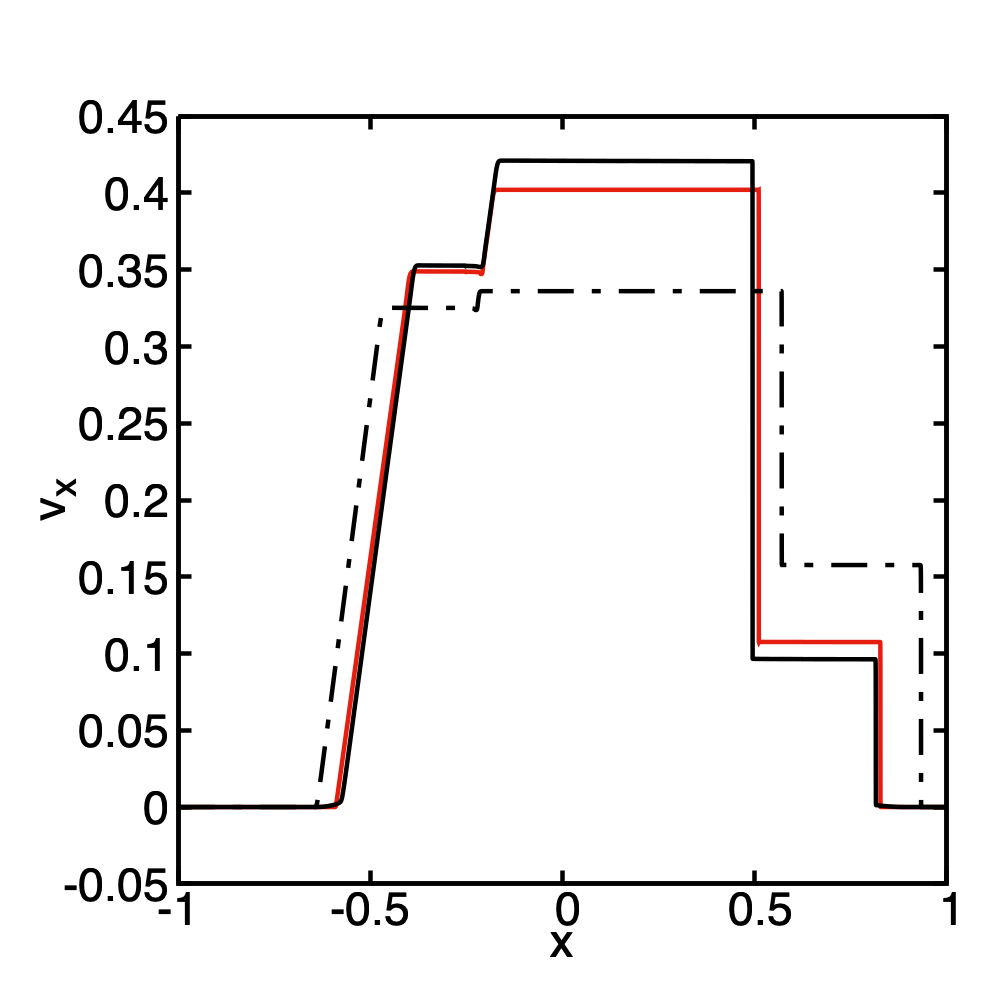}
   \includegraphics[width=5.9cm,clip]{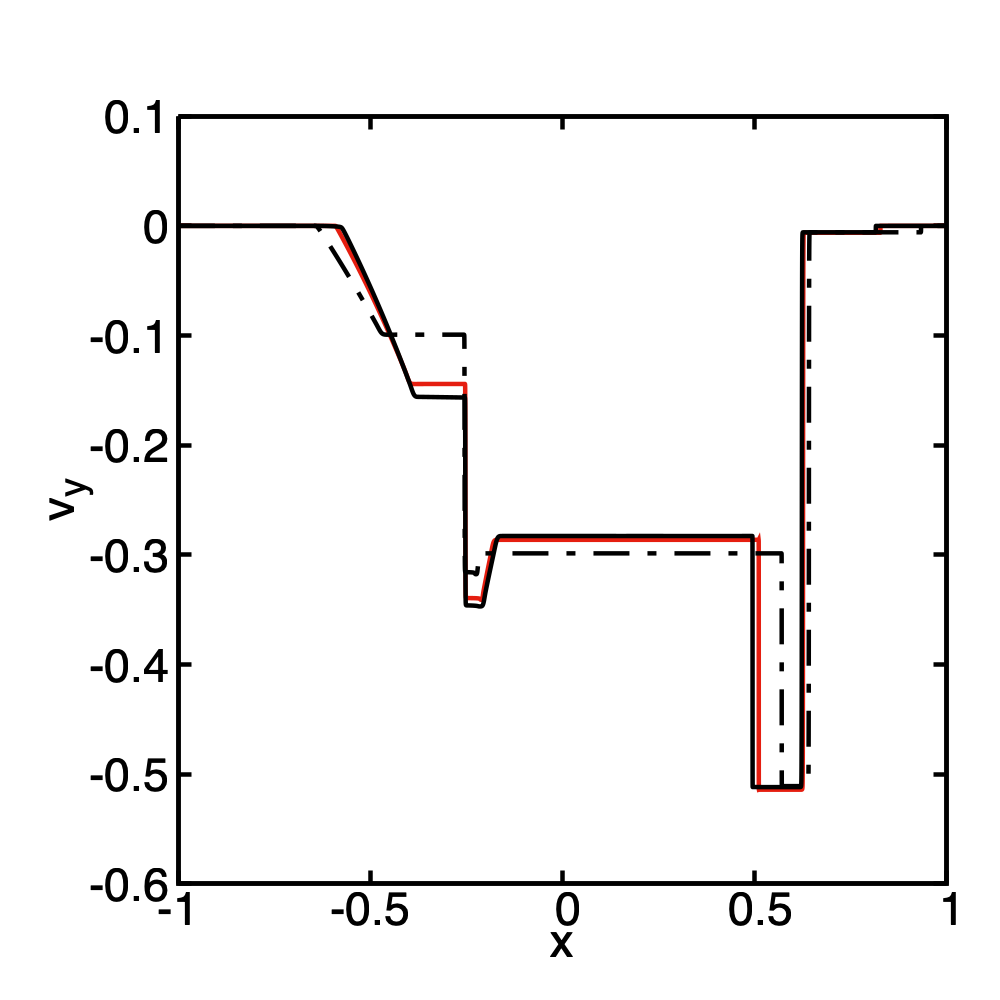}
   \includegraphics[width=5.9cm,clip]{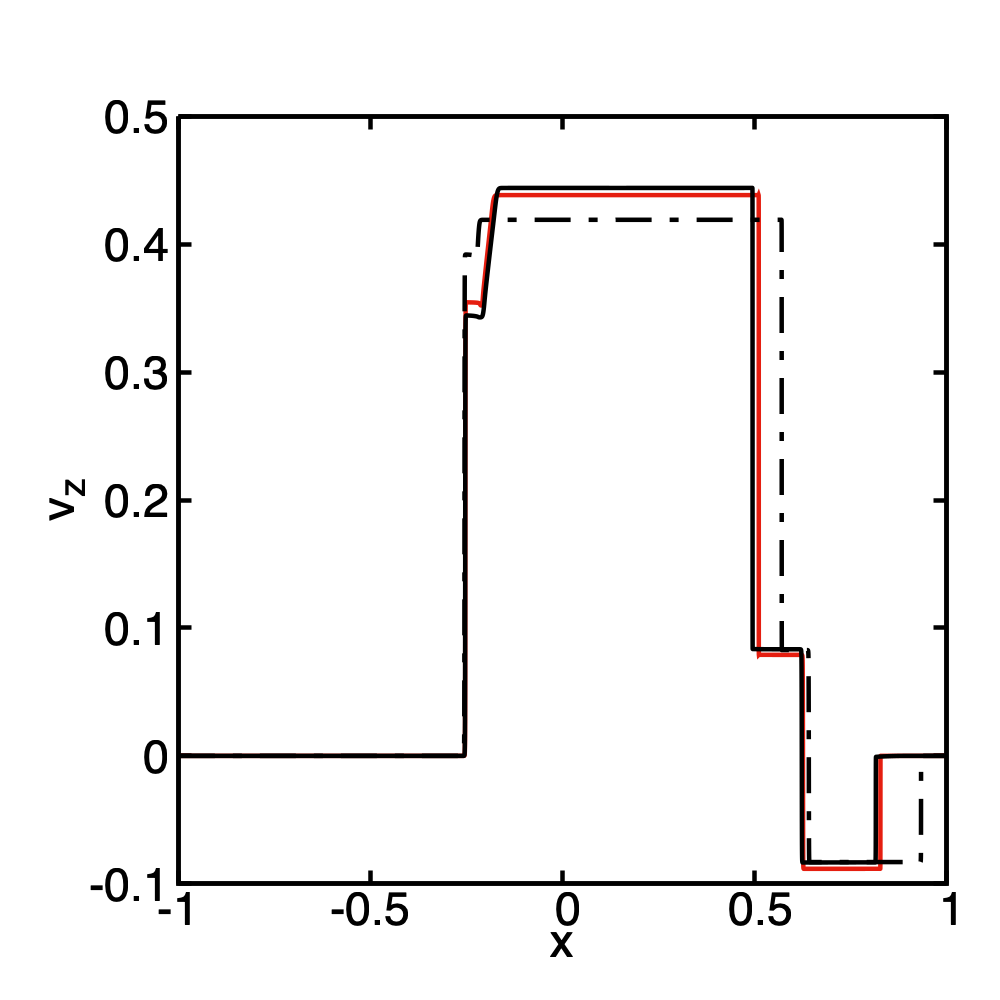}
   \includegraphics[width=5.9cm,clip]{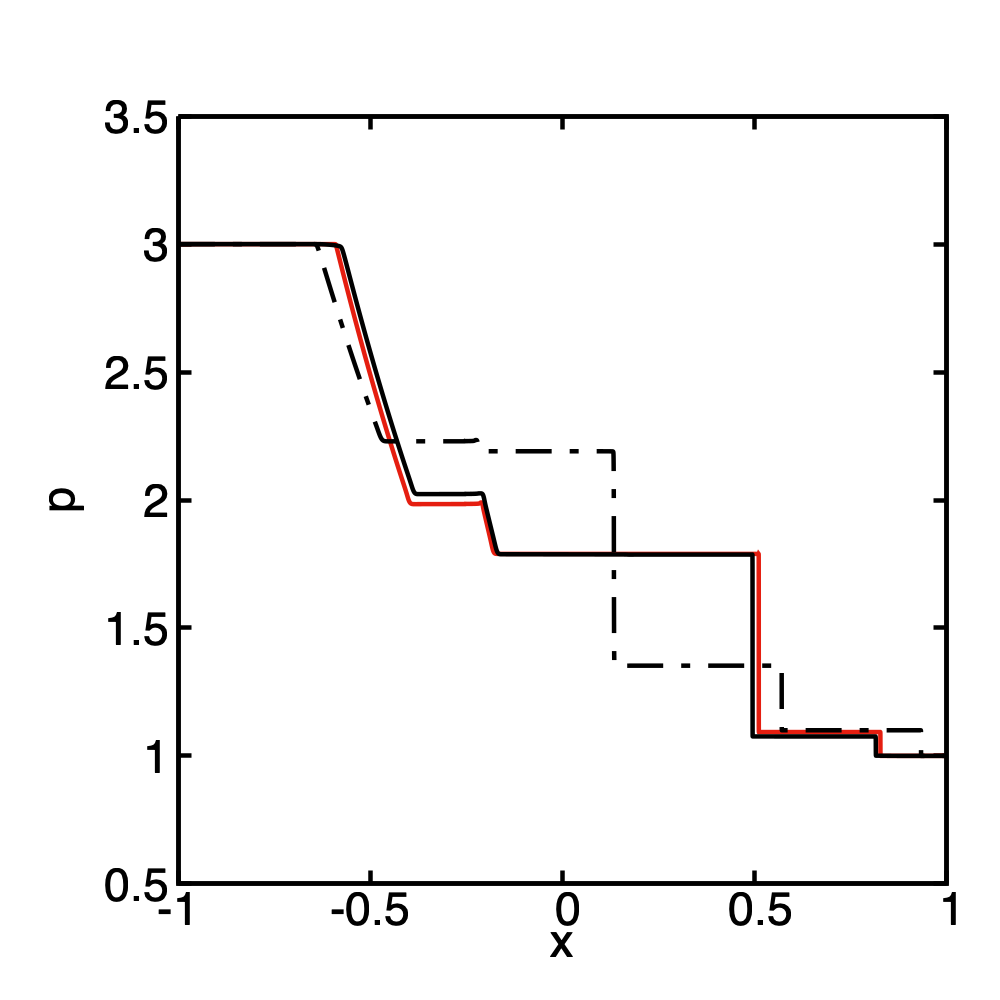}
   \includegraphics[width=5.9cm,clip]{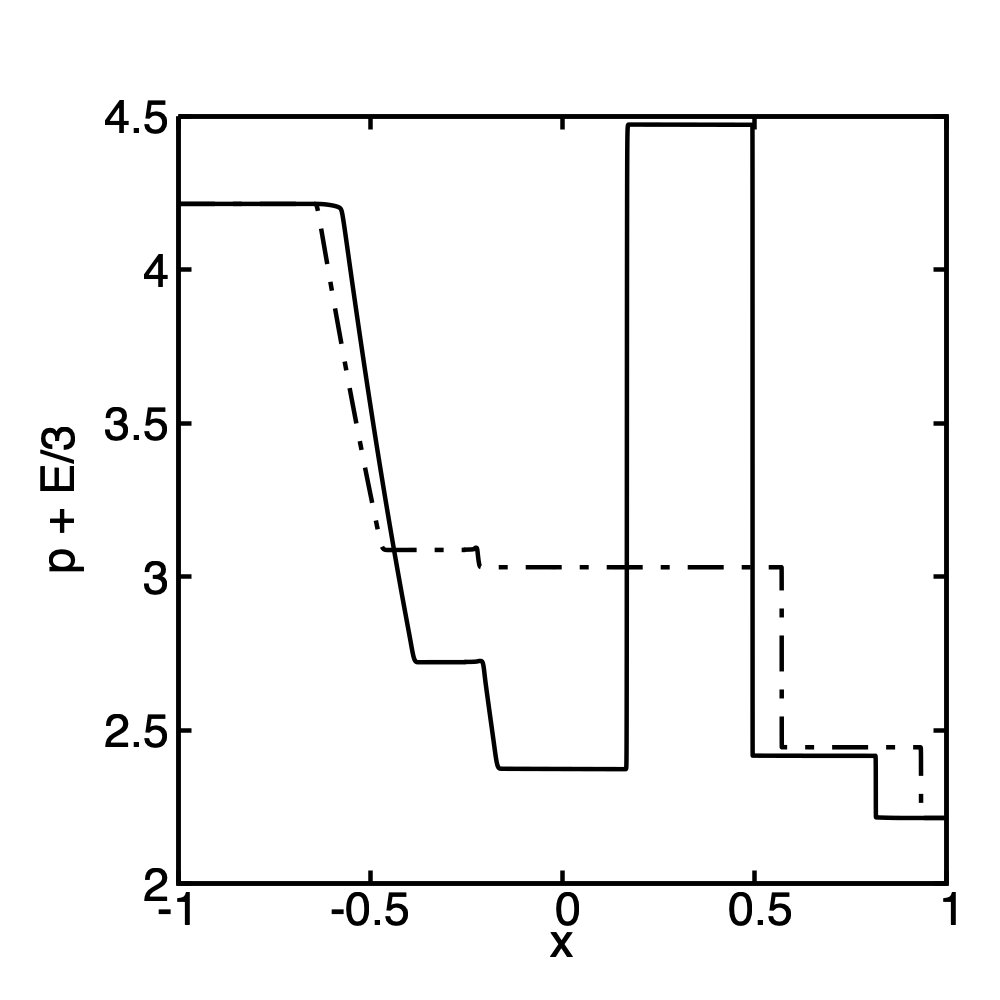}
   \includegraphics[width=5.9cm,clip]{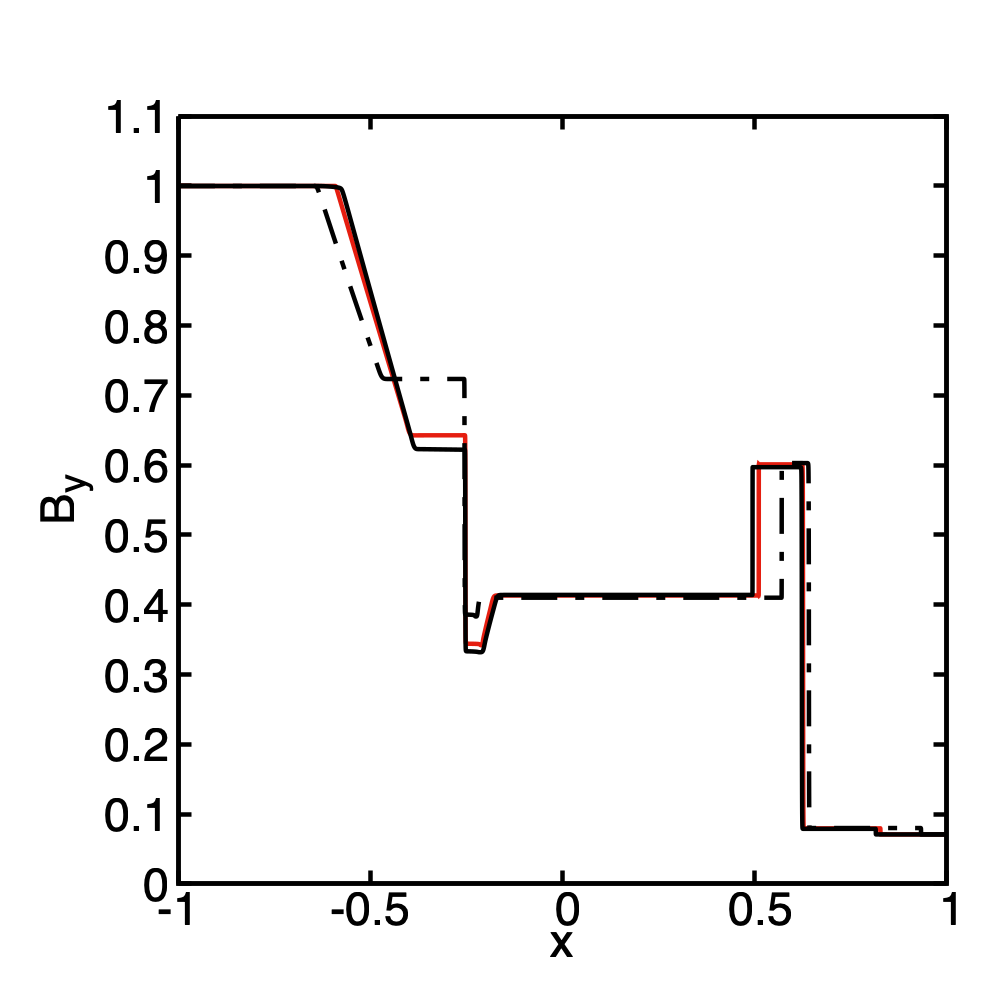}
   \includegraphics[width=5.9cm,clip]{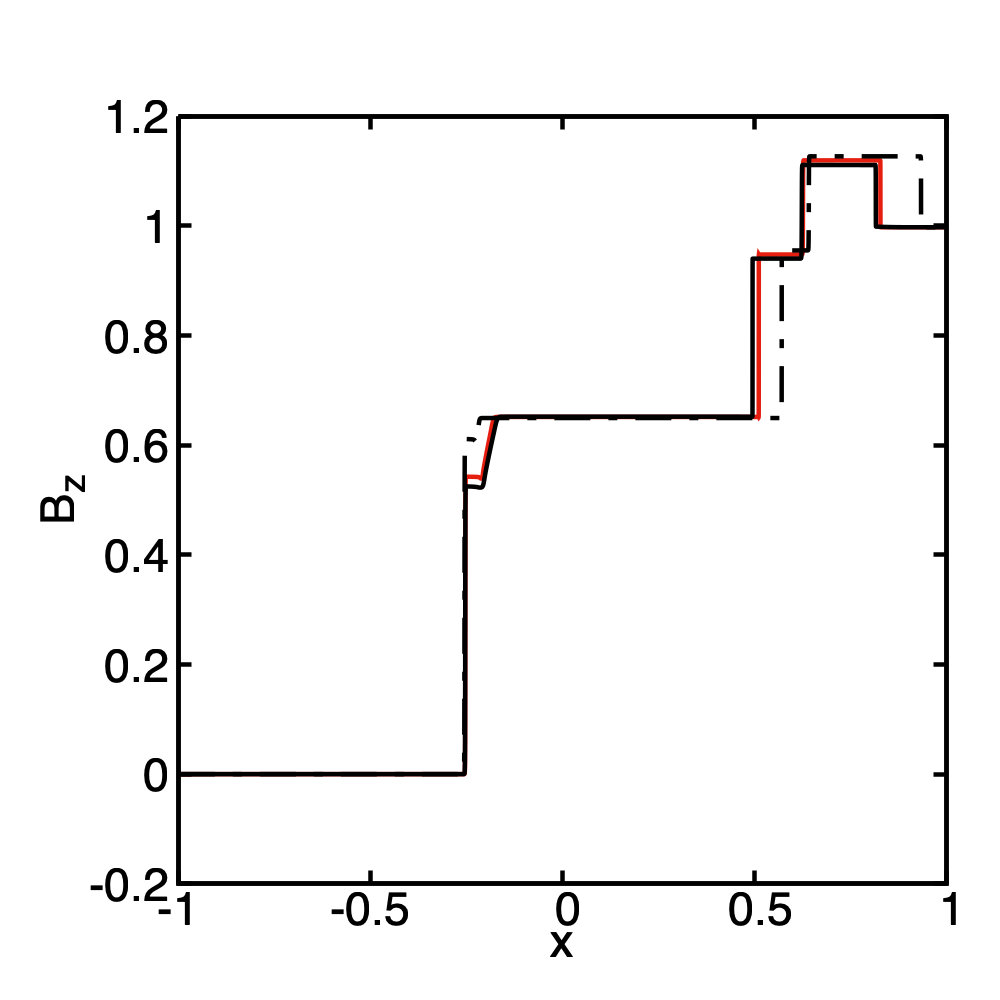}
   \includegraphics[width=5.9cm,clip]{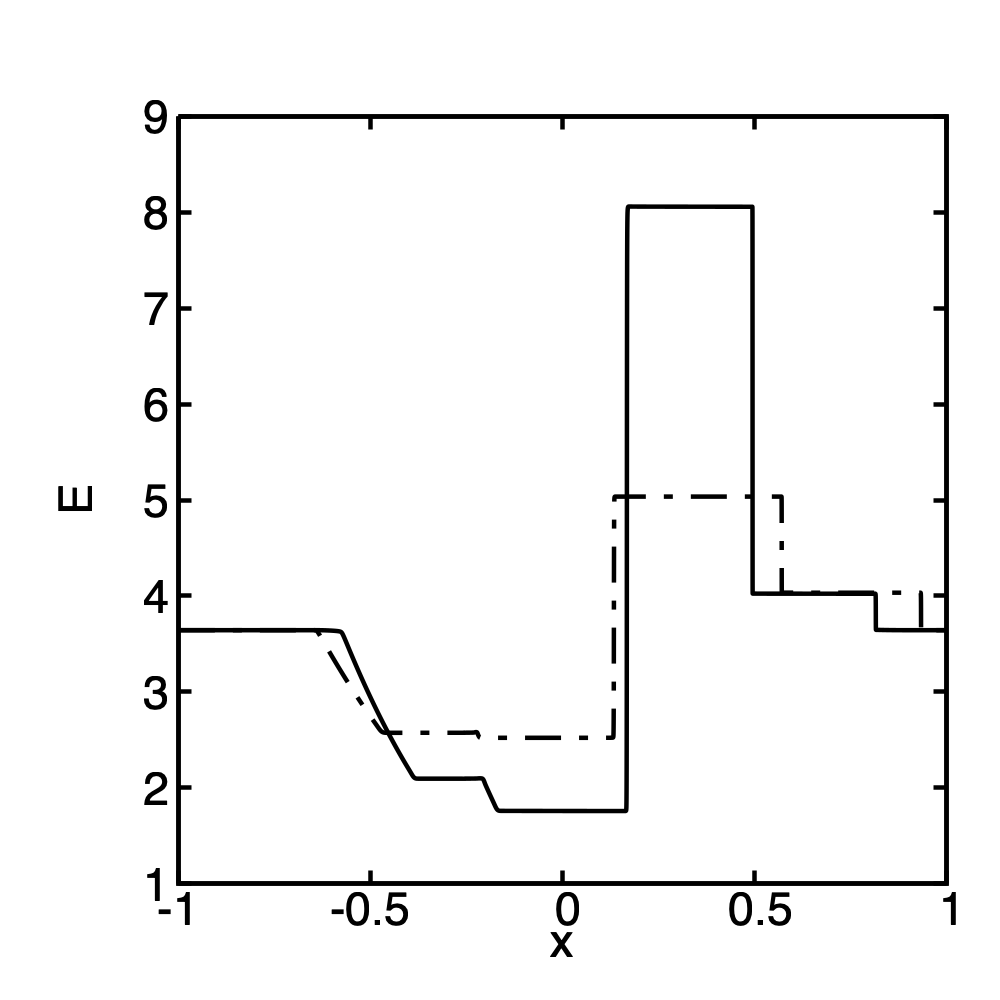}
   \includegraphics[width=5.9cm,clip]{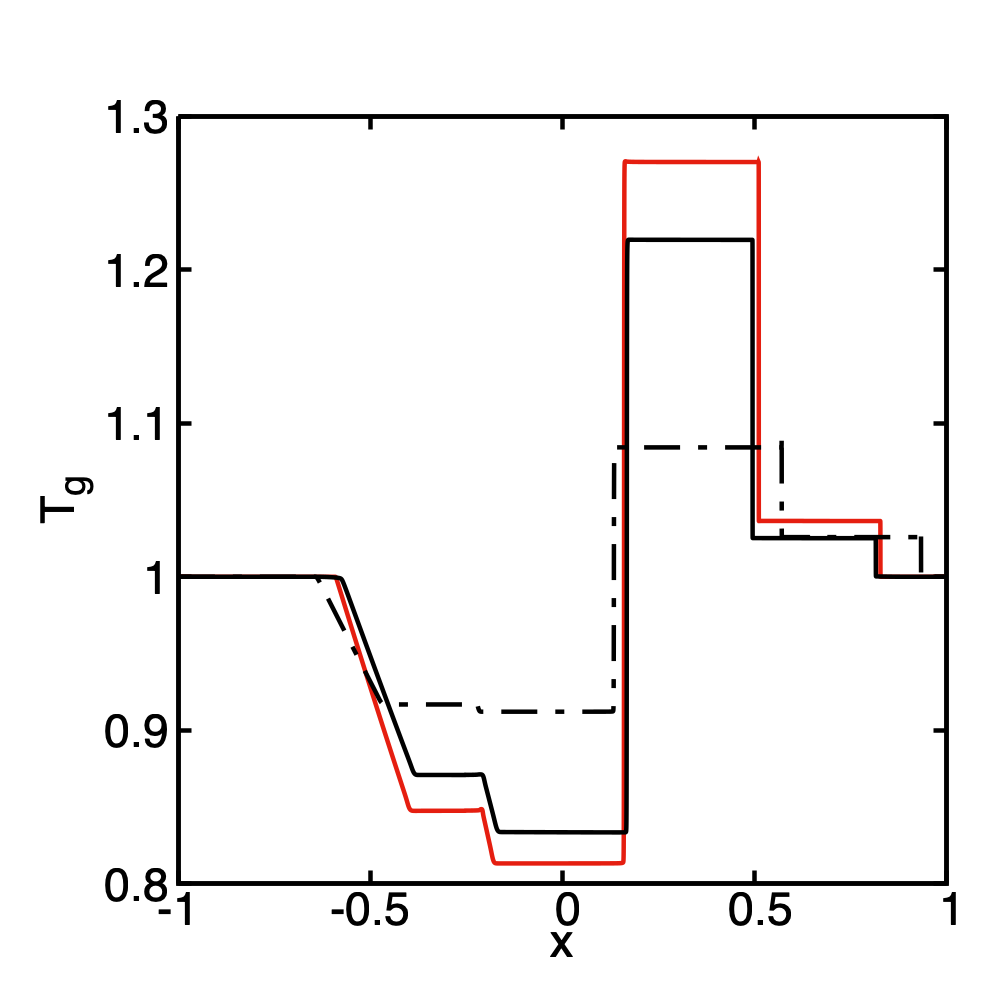}
   \includegraphics[width=5.9cm,clip]{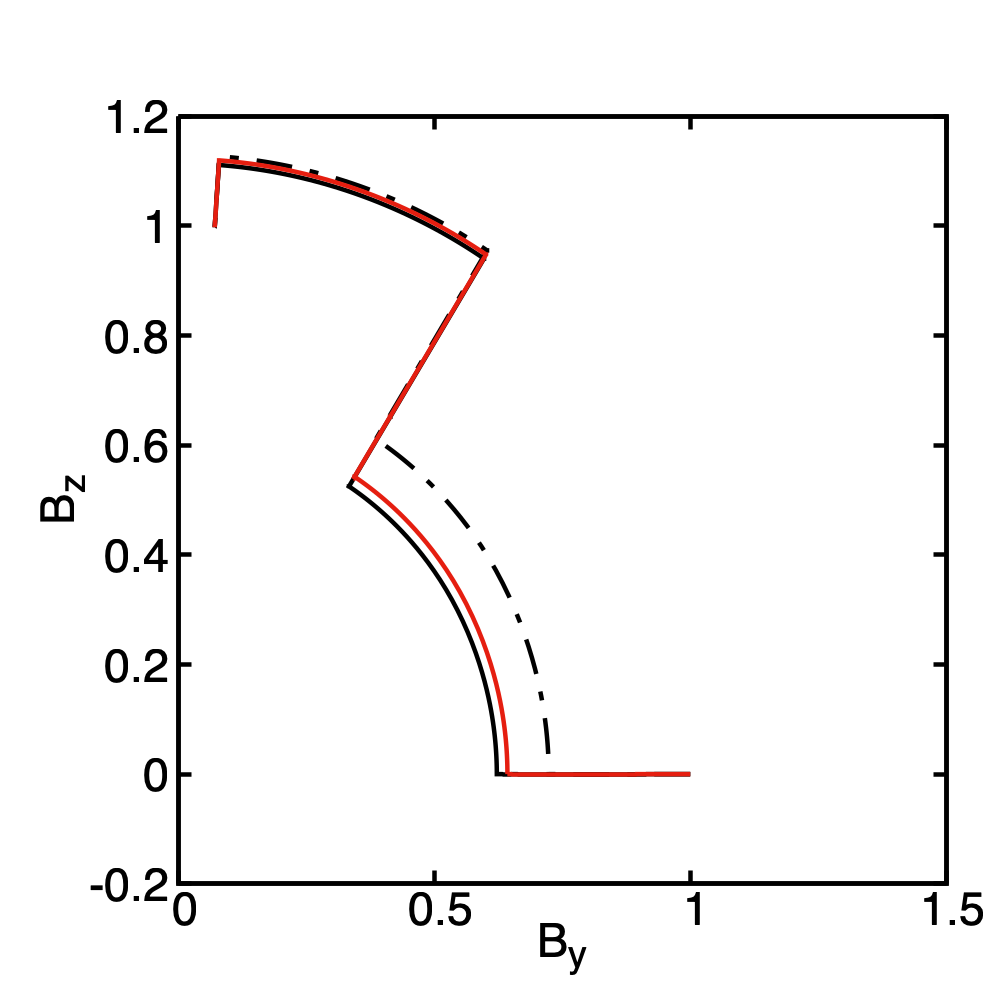}}
   \vspace{-3mm} 
   \caption{The normalized density, $x$-velocity, $y$-velocity, $z$-velocity, plasma pressure, (plasma + radiation) pressure, $y$-magnetic field, $z$-magnetic field, radiation energy \nmn{density and plasma temperature} solutions and the $B_y$ vs $B_z$ plot, at $t = 0.4$ for \citet{torrilhon2003uniqueness} 1.75D Riemann shock-tube problem. Wherever applicable, the corresponding solution for the non-radiative case by \citet{torrilhon2003uniqueness} is also shown as a solid red line. The solutions we show in black are either full FLD (solid) or in diffusion limit (dot-dashed).}
\label{fig:torrilhon}%
\end{figure*}

The solutions at time $t = 4697.96$ s for the radiative Brio-Wu case, equivalent to the non-dimensional value of $t = 0.2$, are shown in Figure~(\ref{fig:briowu}). The solid black line is the radiative solution with full FLD on, in order to allow the flow to switch between the optically thin (freestreaming) and optically thick (diffusion limit) regimes. The dot-dashed line is the radiative solution simulated in the diffusion limit. The non-radiative, ideal MHD solution by \citet{brio1988upwind} is also shown as a solid red line for comparison, wherever applicable. This case consists of 5 waves from left to right: a fast rarefaction, a slow compound wave, a contact discontinuity, a slow shock, and a fast rarefaction wave. For the diffusion limit, the fast rarefaction to the right is replaced by a fast shock. Similarly, solutions for the radiative Torrilhon case, are shown in Figure~(\ref{fig:torrilhon}), for time $t = 103.876$ s, equivalent to the non-dimensional value of $t = 0.4$. As earlier, the solid black, dot-dashed and solid red lines represent the full-FLD, diffusion limit, and non-radiative solutions, respectively. The following seven waves are clearly detectable from left to right: a fast rarefaction, a rotational discontinuity, a slow rarefaction, a contact discontinuity, a slow magnetosonic shock, a rotational discontinuity, and finally a fast magnetosonic shock. Across the rotational discontinuities, the $y$- and $z$- velocities and magnetic fields undergo a jump whereas the density, $x-$velocity, plasma pressure do not jump. The $z-$velocity and $z-$magnetic field do not change across the fast rarefaction wave towards the left. For both these cases, we observe that for the ideal MHD case, only the density jumps across the contact discontinuity whereas for the radiative case in the diffusion limit, the density and the plasma thermal pressure both jump. This major distinction between the non-radiative and radiative cases is due to the contribution of the radiation energy to total pressure. \nmn{
In the freestreaming (e.g. ideal MHD) limit, when $\lambda = 0$, the radiation passes freely through the plasma without exerting any pressure on the plasma and there is therefore no coupling between the radiation pressure and plasma momentum. On the other hand, in the diffusion limit, when $\lambda = 1/3$, the radiation exerts a radiation pressure of $E/3$ on the plasma and is thus strongly coupled with the plasma momentum. In this case, the effective total pressure experienced by the plasma is a sum of these pressure contributions, $(p+E/3)$. Hence, across the contact discontinuity, the equilibrium force balance requires the sum $(p+E/3)$ to remain unchanged. This explains the absence of a jump in $(p+E/3)$ across contact discontinuities, as is observed in the diffusion limit solutions in Figures (\ref{fig:briowu}) and (\ref{fig:torrilhon}), while the individual components $p$ and $E/3$ both undergo jumps. With FLD, the coupling is partial, and the effective radiation pressure experienced by the plasma is less than $E/3$. We therefore see a jump in $(p+E/3)$ across the contact discontinuity in the FLD solutions of Figures (\ref{fig:briowu}) and (\ref{fig:torrilhon}).}

\begin{figure*} 
   \centering
  \FIG{ \includegraphics[width=13cm,clip]{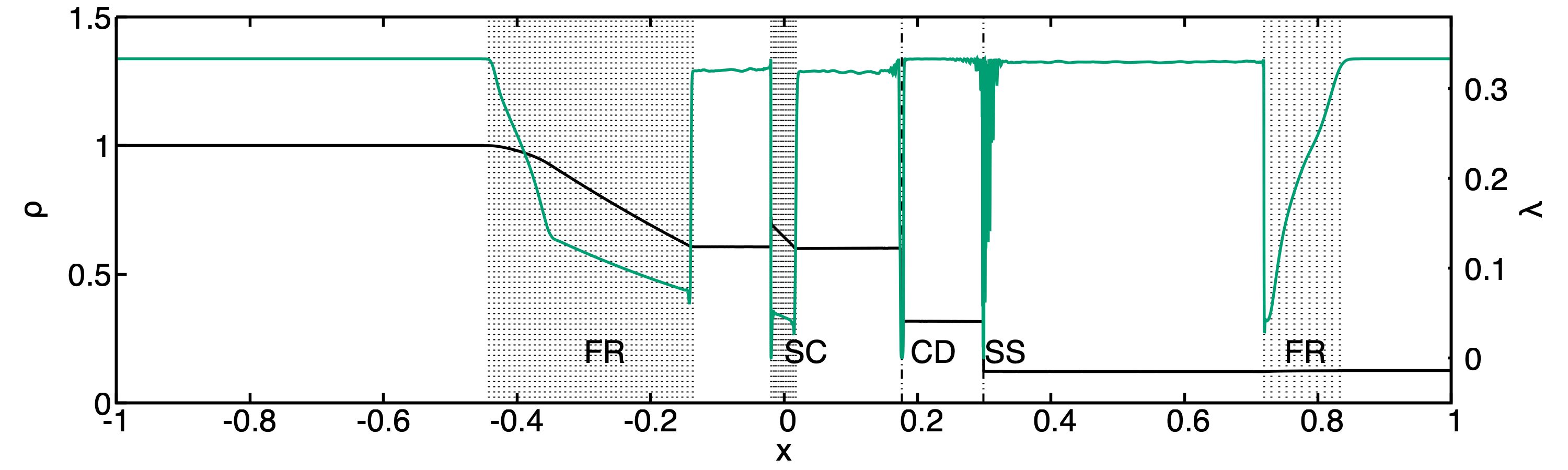}
   \includegraphics[width=13cm,clip]{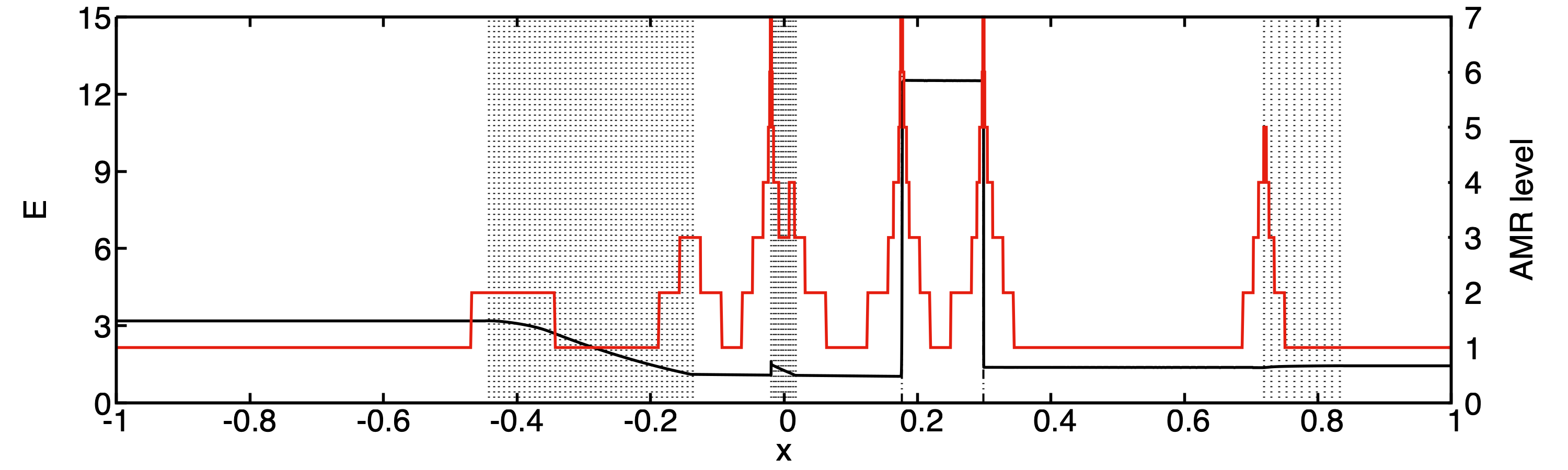}}
   \caption{The density (top) and radiation energy \nmn{density} (bottom) solutions for the full FLD radiative Brio-Wu test. The flux-limiter $\lambda$ solution (green) is superimposed on the density solution\nmn{, whereas the AMR level is superimposed on the radiation energy density solution}. The dotted vertical lines span the rarefaction waves and the dash-dot-dot vertical lines mark discontinuities. \nmn{In the top panel, the 5 waves formed are distinctly marked : fast rarefaction (FR), slow compound wave (SC), contact discontinuity (CD), slow shock (SS), and fast rarefaction wave (FR).}}
    \label{fig:briowu_lambda}%
\end{figure*}

\begin{figure*} 
   \centering
  \FIG{ \includegraphics[width=13cm,clip]{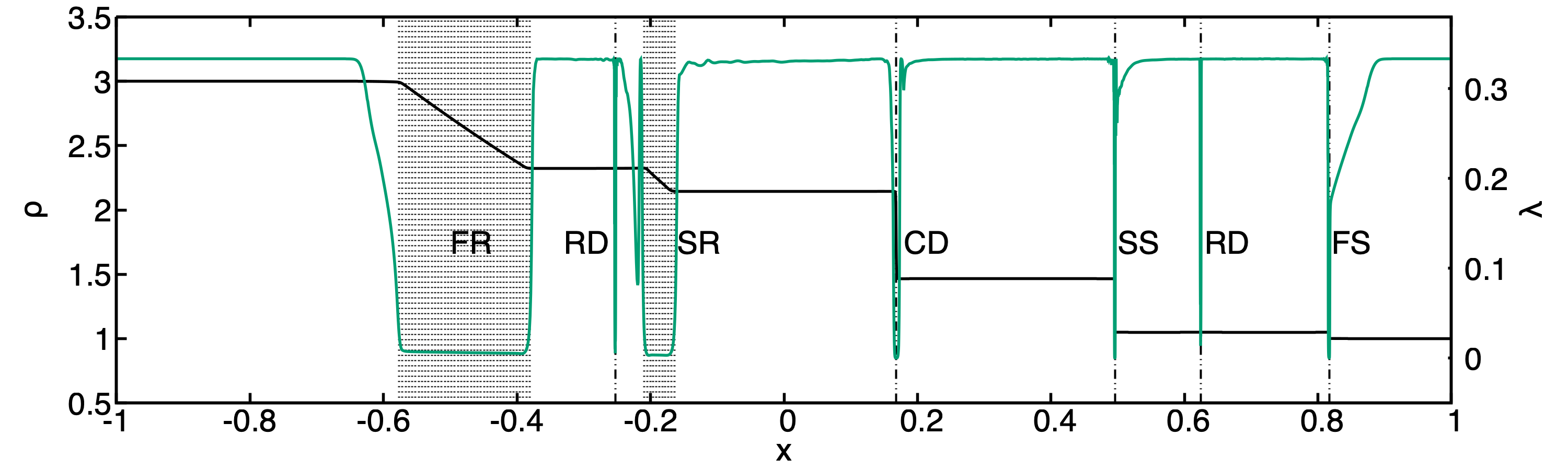}
   \includegraphics[width=13cm,clip]{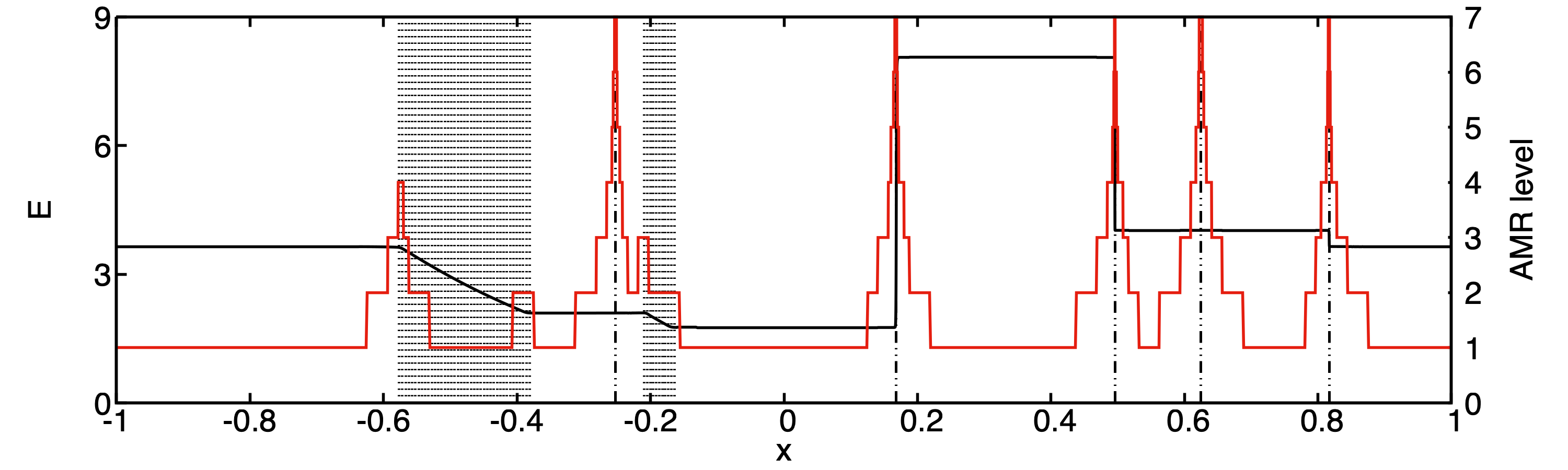}
   \includegraphics[width=13cm,clip]{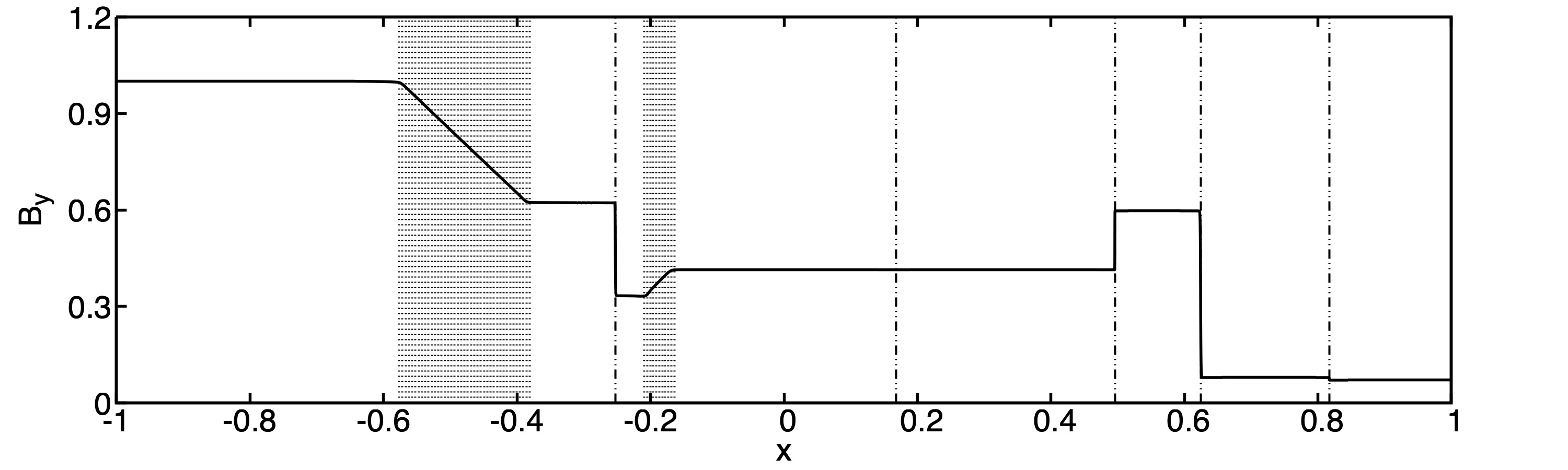}}
   \caption{The density (top), radiation energy \nmn{density} (middle) and $y-$magnetic field solutions for the full FLD radiative Torrilhon test. The flux-limiter $\lambda$ solution (green) is superimposed on the density solution\nmn{, whereas the AMR level is superimposed on the radiation energy density solution}. The dotted vertical lines span the rarefaction waves and the dash-dot-dot vertical lines mark discontinuities. \nmn{In the top panel, the 7 waves formed are distinctly marked : fast rarefaction (FR), rotational discontinuity (RD), slow rarefaction (SR), contact discontinuity (CD), slow shock (SS), rotational discontinuity (RD), fast shock (FS).}}
    \label{fig:torrilhon_lambda}%
\end{figure*}

Magnified versions of the density \nmn{solutions}, superimposed with the flux-limiter $\lambda$, and the radiation energy density solutions\nmn{, superimposed with the AMR level,} for the Brio-Wu test are shown in Figure~(\ref{fig:briowu_lambda}) for the full FLD case. The various wave locations are also marked here (dotted lines across the span of the rarefaction and compound waves, and dash-dot-dot lines for the discontinuities). Similar plots for the Torrilhon test are shown in Figure~(\ref{fig:torrilhon_lambda}), also including the $B_y$ solution. \nmn{For both tests, the AMR does a good job in capturing the various waves created, reaching the highest refinement level at the discontinuities.} For both these tests, we observe that $\lambda$ stays at its diffusion limit value of $1/3$ in regions with zero radiation energy density gradients, while showing clear variation at shocks and contact discontinuities. For the Torrilhon test, $\lambda$ shows a spike also at the two rotational discontinuities. This is due to very small remaining variations in $E$, which are too small to be observed in Figure~(\ref{fig:torrilhon_lambda}). These arise due to the difficulty in the exact capturing of rotational discontinuities in the numerical schemes used. This is a challenge left for future work, as we do not expect any $\lambda$ variation across them, where $E$ should remain constant. This also necessitated the extreme use of AMR on this simple 1D experiment, as too low effective resolutions can show unphysical oscillatory behavior in especially $\lambda$ (i.e. related to the gradient of $E$, which we evaluate numerically), as also impacted by the choice of limiter used in center-to-edge reconstructions.  In the rarefaction regions, $\lambda$ switches to its freestreaming value of $\sim0$ for the Torrilhon case, while showing a smooth transition between the optically thick and thin regimes for the Brio-Wu case. We must note that the $\lambda$ value also depends on the density, opacity and (dimensionalized) length scale of the problem. Higher values of these drive $\lambda$ closer to the optically thick regime, while lower values bring it closer to the freestreaming regime.

\begin{table}
\caption{Initial left and right states for the RMHD Riemann problems}
\label{table:RMHD_Riemann}      
\centering   
\begin{tabular}{c | c c} 
\hline\hline
\multicolumn{3}{c|}{1.5D Brio-Wu test ($\gamma = 2$)}\\
\hline
\multicolumn{3}{c|}{Dimensionalized initial states}\\
\hline
Variable & Left state & Right state \\ 
\hline  
   $\rho$ (g/cm$^3$) & $1.67429\times10^{-14}$ & $2.09287\times10^{-15}$ \\ 
   $p$ (erg/cm$^3$)    & $3.04046\times10^{-3}$ & $3.04046\times10^{-4}$ \\
   $B_x$ (Gauss)     & $1.46601\times10^{-1}$ & $1.46601\times10^{-1}$ \\
   $B_y$ (Gauss)     & $1.95468\times10^{-1}$ & $-1.95468\times10^{-1}$ \\
   $T$ (K)    & $1.1\times10^{3}$ & $8.0\times10^{2}$ \\
   $E$ (erg/cm$^3$)    & $1.10764\times10^{-2}$ & $4.53637\times10^{-3}$ \\
   \hline
   $L$ (cm) &\multicolumn{2}{c|}{$2\times10^{10}$} \\
\hline 
\multicolumn{3}{c|}{Non-dimensionalized initial states}\\
\hline
Variable & Left state & Right state \\ 
\hline  
   $\rho$ & $1.0$ & $0.125$ \\ 
   $p$     & $1.0$ & $0.1$ \\
   $B_x$  & $0.75$ & $0.75$ \\
   $B_y$  & $1.0$ & $-1.0$ \\
   $T$      & $1.0$ & $0.8$ \\
   $E$      & $3.643$ & $1.492$ \\
   \hline
   $L$    &\multicolumn{2}{c|}{$2$} \\
\hline\hline 
\multicolumn{3}{c|}{1.75D Torrilhon test ($\gamma = 5/3$)}\\
\hline
\multicolumn{3}{c|}{Dimensionalized initial states}\\
\hline
Variable & Left state & Right state \\ 
\hline  
   $\rho$ (g/cm$^3$) & $1.67429\times10^{-14}$ & $5.02288\times 10^{-14}$ \\ 
   $p$ (erg/cm$^3$)    & $3.04046\times10^{-3}$ & $9.12139\times10^{-3}$ \\
   $B_x$ (Gauss)     & $2.93201\times10^{-1}$ & $2.93201\times10^{-1}$ \\
   $B_y$ (Gauss)     & $1.95468\times10^{-1}$ & $1.38268\times10^{-2}$ \\
   $B_z$ (Gauss)     & $0.0$    & $1.94978\times10^{-1}$ \\
   $T$ (K)    & $1.1\times10^{3}$ & $1.1\times10^{3}$ \\
   $E$ (erg/cm$^3$)    & $1.10764\times10^{-2}$ & $1.10764\times10^{-2}$ \\
   \hline
   $L$ (cm) &\multicolumn{2}{c|}{$2\times10^8$} \\
\hline 
\multicolumn{3}{c|}{Non-dimensionalized initial states}\\
\hline
Variable & Left state & Right state \\ 
\hline  
   $\rho$ & $1.0$ & $3.0$ \\ 
   $p$      & $1.0$ & $3.0$ \\
   $B_x$  & $1.5$ & $1.5$ \\
   $B_y$  & $1.0$ & $\mathrm {cos}(1.5\mathrm {rad})$ \\
   $B_z$  & $0.0$  & $\mathrm {sin}(1.5\mathrm {rad})$ \\
   $T$      & $1.0$ & $1.0$ \\
   $E$      & $3.643$ & $3.643$ \\
   \hline
   $L$    &\multicolumn{2}{c|}{$2$} \\
\hline\hline 
\end{tabular}
\end{table}

\subsection{2D Magnetoconvection with radiation}\label{sec:convection}

We simulate the case of two-dimensional compressible convection in the presence of a magnetic field, originally studied by \citet{hurlburt1988magnetic}, but here we also account for the effects of radiation. This scenario is relevant for the plasma motion inside the convective zone of the Sun, and the original study showed how an initial uniform vertical field gets swept into concentrated flux sheets, relevant for explaining the granular appearance of convective cells at the solar photosphere. This is a proof-of-principle simulation of an important multi-D effect that is present in the convective layers of stars. In these convective regions, energy transport through the stellar atmosphere is not necessarily carried by radiation, but also by convective transport. This convective transport effect can be altered or dampened due to the presence of magnetic fields as the fieldlines can constrict the motion of the convective cells. Understanding magnetoconvection is a crucial ingredient in our understanding of stellar envelopes.
This magnetoconvection case requires physical effects such as viscous diffusion, magnetic diffusion (resistivity), anisotropic thermal conduction and external gravity.

Accounting for these effects, the governing equations are
\begin{equation}\label{eq:mhd_mass_convection}
\frac{\partial {\rho}}{\partial {t}} + {\nabla} \cdot \left({\rho} {\mbox{\bf v}}\right) = 0,
\end{equation}
\begin{equation}\label{eq:mhd_mom_convection}
\frac{\partial (\rho{\mbox{\bf v}})}{\partial {t}} + {\nabla}\cdot\left(\rho \mbox{\bf v} \mbox{\bf v} - \mbox{\bf B} \mbox{\bf B} + {\left(p + \frac{\mbox{\bf B} \cdot \mbox{\bf B}}{2}\right)}\mbox{\bf I}\right) = \bf f_r + \rho {\bf g} + {\nabla}\cdot \mathbf{\boldsymbol \tau},
\end{equation}
\begin{equation}\label{eq:mhd_energy_convection}
\begin{split}
\frac{\partial e}{\partial t} + 
\nabla\cdot\left( \left(e + p + \frac{\mbox{\bf B}\cdot\mbox{\bf B}}{2}\right)\mbox{\bf v} - (\mbox{\bf B}\cdot \mbox{\bf v})\mbox{\bf B} \right) = \mbox{\bf v}\cdot {\bf f_r} + \dot{q} + \rho \mbox{\bf v}\cdot{\bf g} \\ - \nabla\cdot \left( \left(\eta\nabla\times \mbox{\bf B}\right)\times \mbox{\bf B} - \mathcal{K} \hat{\mathbf{b}}\hat{\mathbf{b}}\cdot \nabla T_g - \mbox{\bf v}\cdot \mathbf{\boldsymbol \tau} \right),
\end{split}
\end{equation}
\begin{equation}\label{eq:mhd_mag_convection}
\frac{\partial{\mbox{\bf B}}}{\partial t} + {\nabla}\cdot \left(\mbox{\bf v} \mbox{\bf B} - \mbox{\bf B} \mbox{\bf v}\right) = - \nabla\times\left(\eta \nabla \times \mbox{\bf B} \right).
\end{equation}

Here, ${\bf g}$ is the gravitational acceleration vector, $\eta$ is the magnetic diffusivity, $\mathcal{K}$ is the thermal conductivity, and $\mathbf{\boldsymbol \tau}$ is the viscous stress tensor given by
\begin{equation}\label{eq:viscous_stress_tensor}
\mathbf{\boldsymbol \tau} = \mu_g \left(\nabla\mbox{\bf v} + (\nabla\mbox{\bf v})^{\bf T} - \frac{2}{3} (\nabla\cdot\mbox{\bf v})\right),
\end{equation}
where $\mu_g$ is the plasma's dynamic viscosity. Also, ${\hat{\bf b}} = \mbox{\bf B}/|\mbox{\bf B}|$ is the unit vector in direction of the magnetic field. For this case, we consider a 2D rectangular $(x,y)$ domain of dimensionless length, $L_x$ = 3, and depth $L_y$ = 1. The depth is represented using a dimensionless depth variable $d$, where $d=-y$, that ranges from $d = d_0$ at the top to $d = d_0 + 1$ at the bottom, where $d_0 = 0.1$. Initially the plasma is taken to be stagnant i.e. $\mbox{\bf v}(t=0) = {\bf 0}$ everywhere. The magnetic field is initially vertical and uniform with a dimensionless value of $B_{y,0} = 6.573\times10^{-2}$. The non-dimensional initial profiles for temperature, density and pressure are given by $T_g(d) = d$, $\rho(d) = d/d_0$, and $p(d) = d^2/d_0$, respectively. Therefore, at the top, $(T_g,\rho,p)_{top} = (0.1,1,0.1)$, and at the bottom, $(T_g,\rho,p)_{bot} = (1.1,11,12.1)$. The Prandtl number, given by
\begin{equation}\label{eq:Prandtl}
\sigma = \mu_g C_p/\mathcal{K},
\end{equation}
is taken to be unity, resulting in a constant anisotropic thermal conduction coefficient. The adiabatic index is $\gamma = 5/3$. The Chandrasekhar number given by
\begin{equation}\label{eq:Chandrasekhar}
Q = \frac{B^2}{\mu_0} \frac{d^2}{\mu_g \eta}\,,
\end{equation}
is taken to be 72. The magnetic Prandtl number given by
\begin{equation}\label{eq:mag_Prandtl}
\zeta_0 = \frac{\eta \rho_0 C_p}{\mathcal{K}}
\end{equation}
is taken to be 0.25. The nondimensional gravitational acceleration is given by
\begin{equation}\label{eq:grav_acc}
g\frac{m_p\mu}{k_B \nabla T_g} = -1\,,
\end{equation}
where $\nabla T_g$ is the initial non-dimensionalized vertical temperature gradient. 

The dimensionalized length and depth of the domain are $L_x = 3 \times 10^8$ cm and $L_y = 10^8$ cm. The dimensionalized values for the plasma properties are $(T_g,\rho,p)_{top} = (6000$K, $3.61286\times10^{-10}$ g/cm$^2$, $3.57864\times 10^{2}$~erg$/$cm$^3)$ at the top and $(T_g,\rho,p)_{bot} = (66000$ K, $3.97415\times10^{-9}$ g/cm$^2$, $4.33016\times 10^{4}$ erg$/$cm$^3)$ at the bottom. The dimensionalized magnetic field becomes $B_{y,0} = 13.93887$ Gauss. The externally applied gravitational field is $g = -989.53838$ m/s$^2$. Initially, radiative equilibrium is assumed everywhere i.e. the plasma and radiation temperatures are equal. Accordingly, the radiation energy density values at the top and bottom are $E_{top} = 9.80519$ erg$/$cm$^3$ and $E_{bottom} = 1.43558\times 10^{5}$ erg$/$cm$^3$, respectively. The corresponding non-dimensional values are $E_{top} = 2.73992\times10^{-3}$ and $E_{bottom} = 40.11518$, respectively. An opacity of $\kappa = 0.4\times10^{-3}$ cm$^2$/g is used for this case. Apart from the plasma density, the orders of magnitude of these values are representative of those in the convection zone of the Sun. The plasma density is approximately 10$^3$ times lower than the typical values in the convection zone of the Sun. Using density values similar to those in the Sun would proportionally increase the plasma pressure and reduce the significance of the radiation pressure. As the current work focuses on radiation-dominant cases, we stick to these density values.

The temperature is held fixed at the top boundary $d = d_0$, and its vertical derivative is held fixed to unity at the bottom boundary, $d = d_0+1$. The radiation temperature is held equal to the plasma temperature at both top and bottom boundaries. The horizontal magnetic field is held fixed at zero at the top and bottom boundaries. 
The magnetic field lines are forced to stay vertical there, but are free to move laterally and can exhibit compression or expansion. 
The vertical velocity, the vertical gradient of the horizontal velocity and the horizontal magnetic field all vanish at the top and bottom boundaries. Therefore, the horizontal components of the viscous and magnetic stresses also vanish at these boundaries. 
These boundary conditions can be summarized as:
\begin{equation}\label{eq:convection_BCs}
\begin{split}
d = d_0: v_y = 0, \frac{\partial v_x}{\partial y} = 0, T_g = d_0, T_r = T_g, B_x = 0, \frac{\partial B_y}{\partial y} = 0. 
\\
d = d_0+1: v_y = 0, \frac{\partial v_x}{\partial y} = 0, \frac{\partial T_g}{\partial y} = 1, T_r = T_g, B_x = 0, \frac{\partial B_y}{\partial y} = 0.
\end{split}
\end{equation}
Periodic boundary conditions are used at the left and right boundaries for all variables. To initiate the convective instability, small-amplitude velocity perturbations are introduced into the stagnant plasma. The above initial and boundary conditions, apart from those for the radiation energy, have been described in detail by \citet{hurlburt1988magnetic}.

\begin{figure}
   \centering
   \includegraphics[trim={100 50 100 250},width=\hsize,clip]{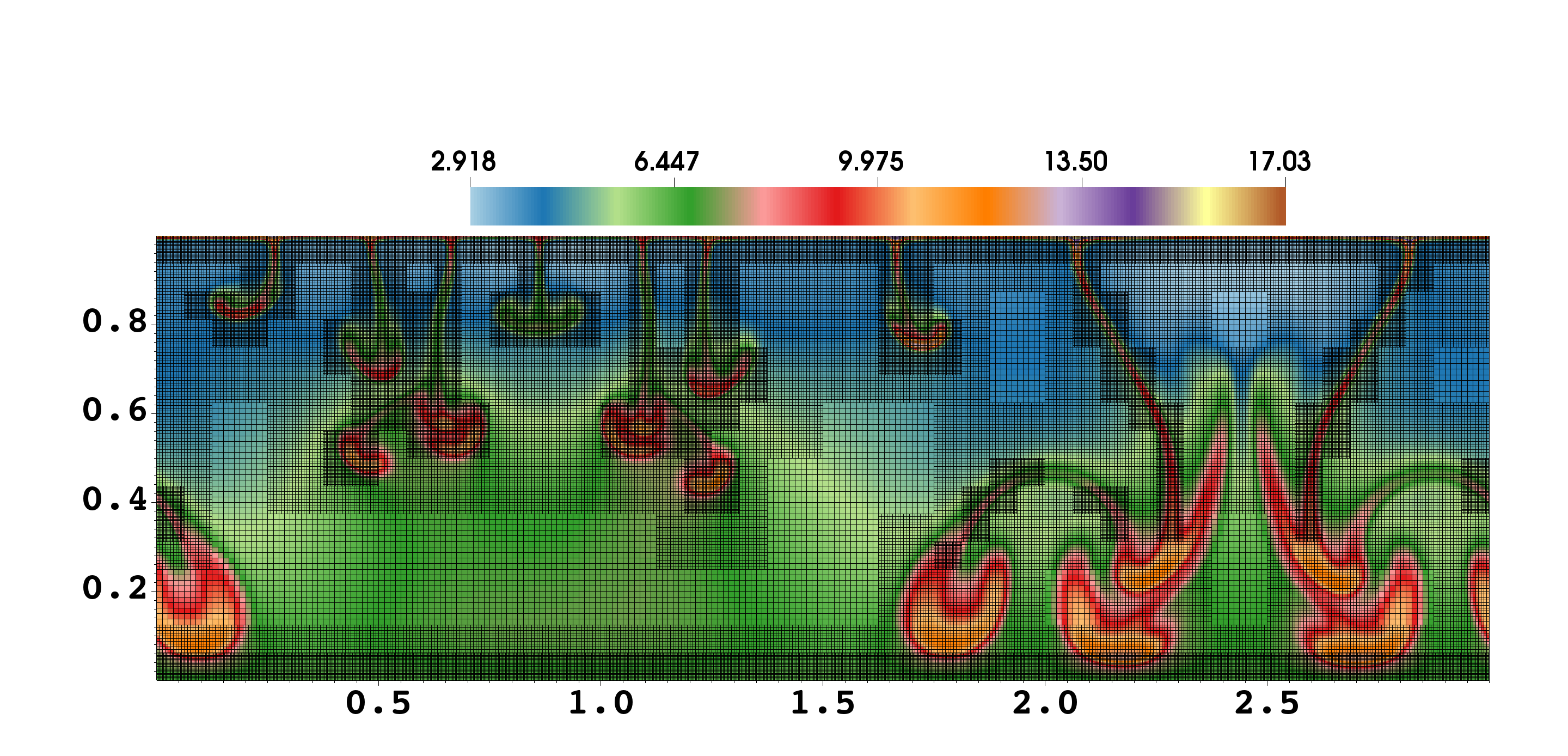}
   \includegraphics[trim={100 50 100 250},width=\hsize,clip]{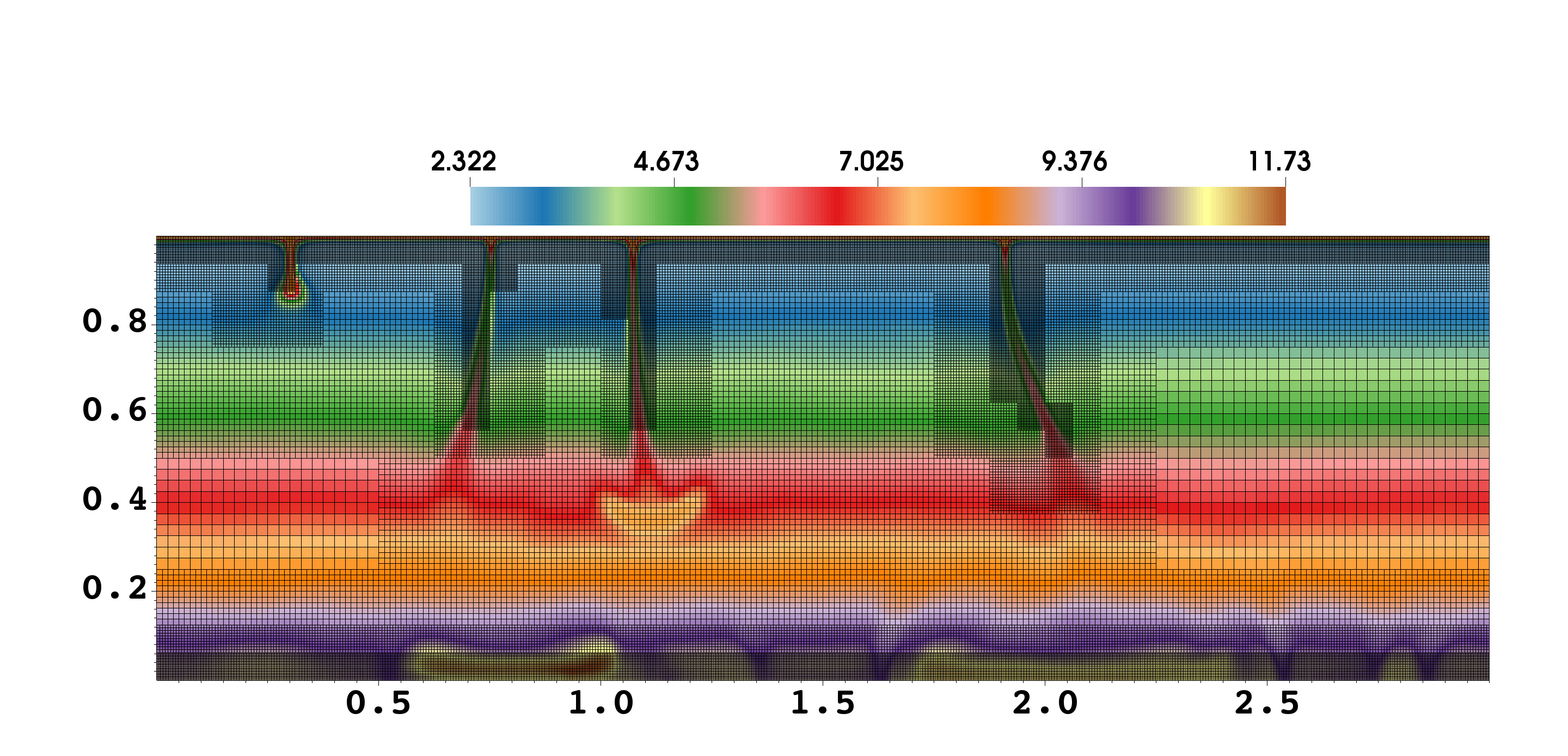}
      \caption{The density for the radiative magnetoconvection case at $t = 9$ (top) and at $t = 429$ (bottom).}
         \label{fig:CONV_AMR}
\end{figure}

\begin{figure}
   \centering
   \includegraphics[trim={100 50 100 350},width=\hsize,clip]{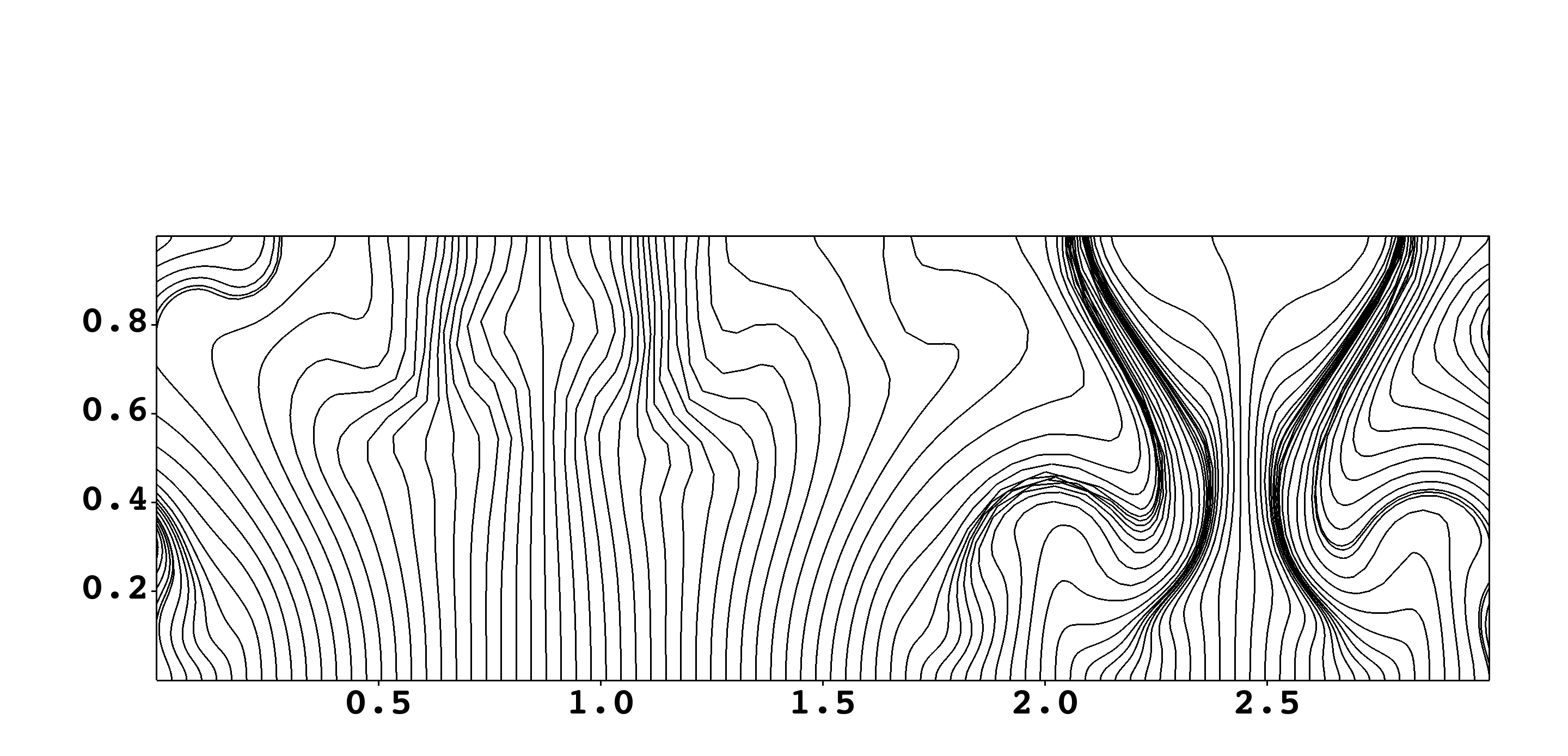}
   \includegraphics[trim={100 50 100 350},width=\hsize,clip]{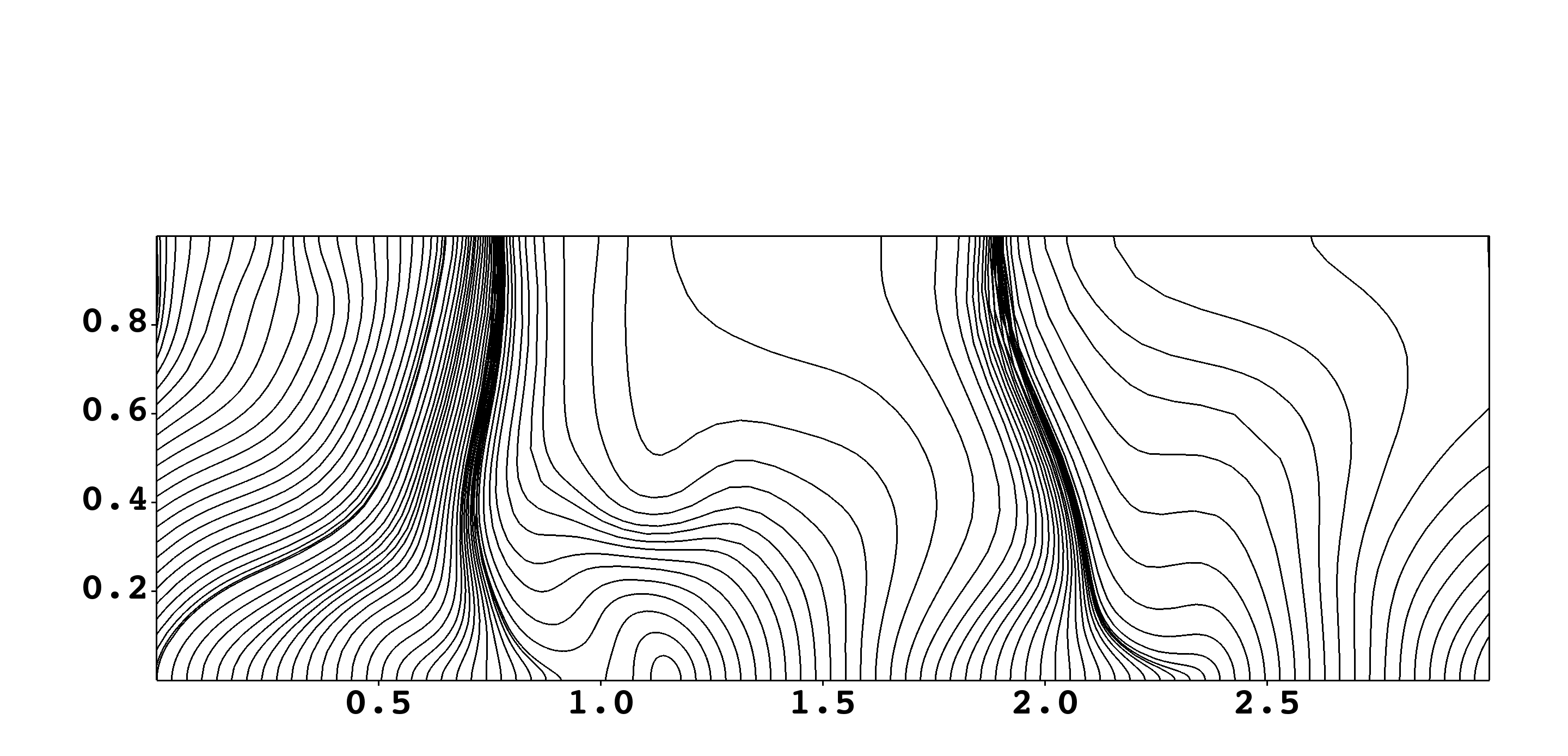}
      \caption{The magnetic field lines for the radiative magnetoconvection case \nmn{at $t = 9$ (top) and} at $t = 429$ \nmn{(bottom)}.}
         \label{fig:CONV_Blines}
\end{figure}

\begin{figure}
   \centering
   \includegraphics[trim={100 50 100 250},width=\hsize,clip]{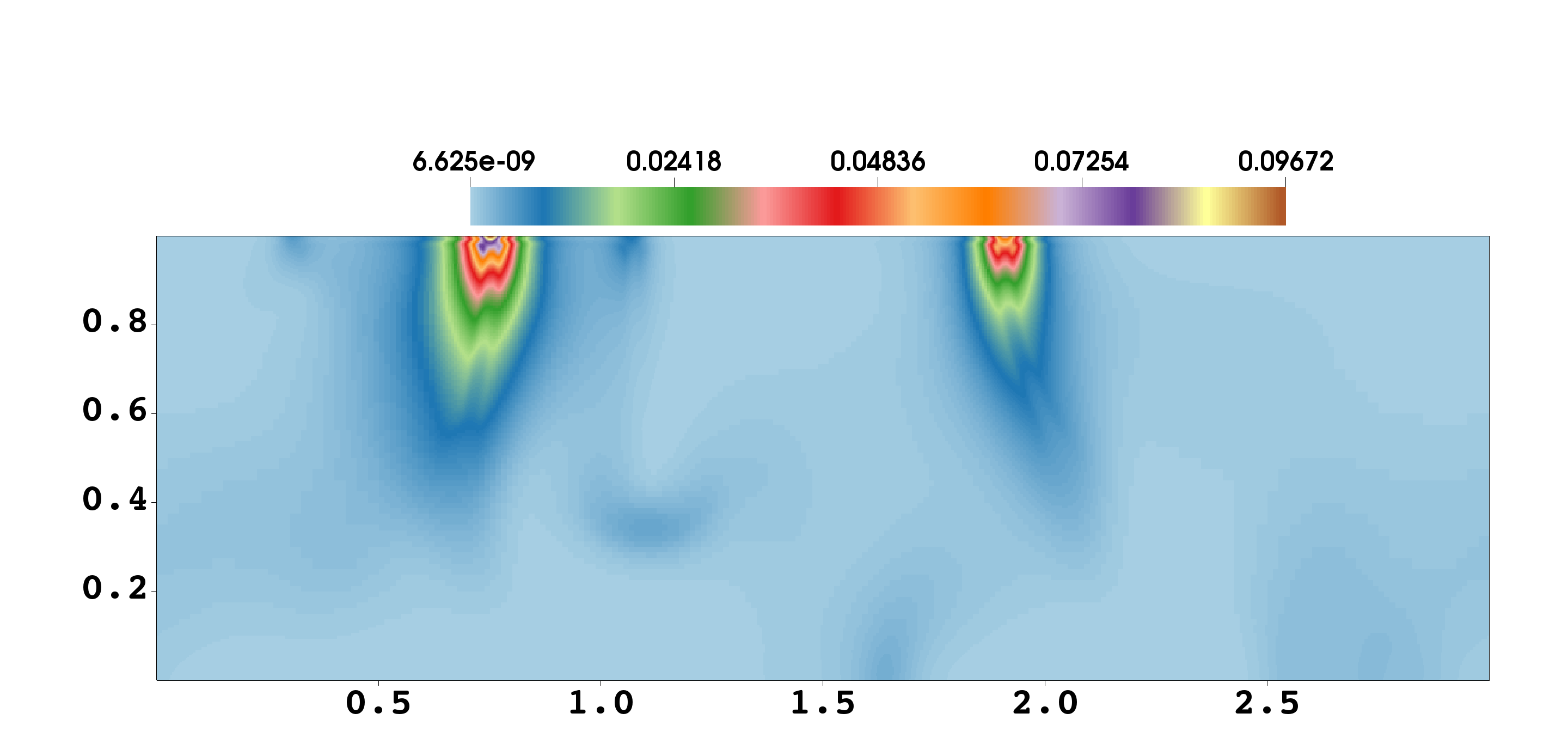}
   \includegraphics[trim={100 50 100 250},width=\hsize,clip]{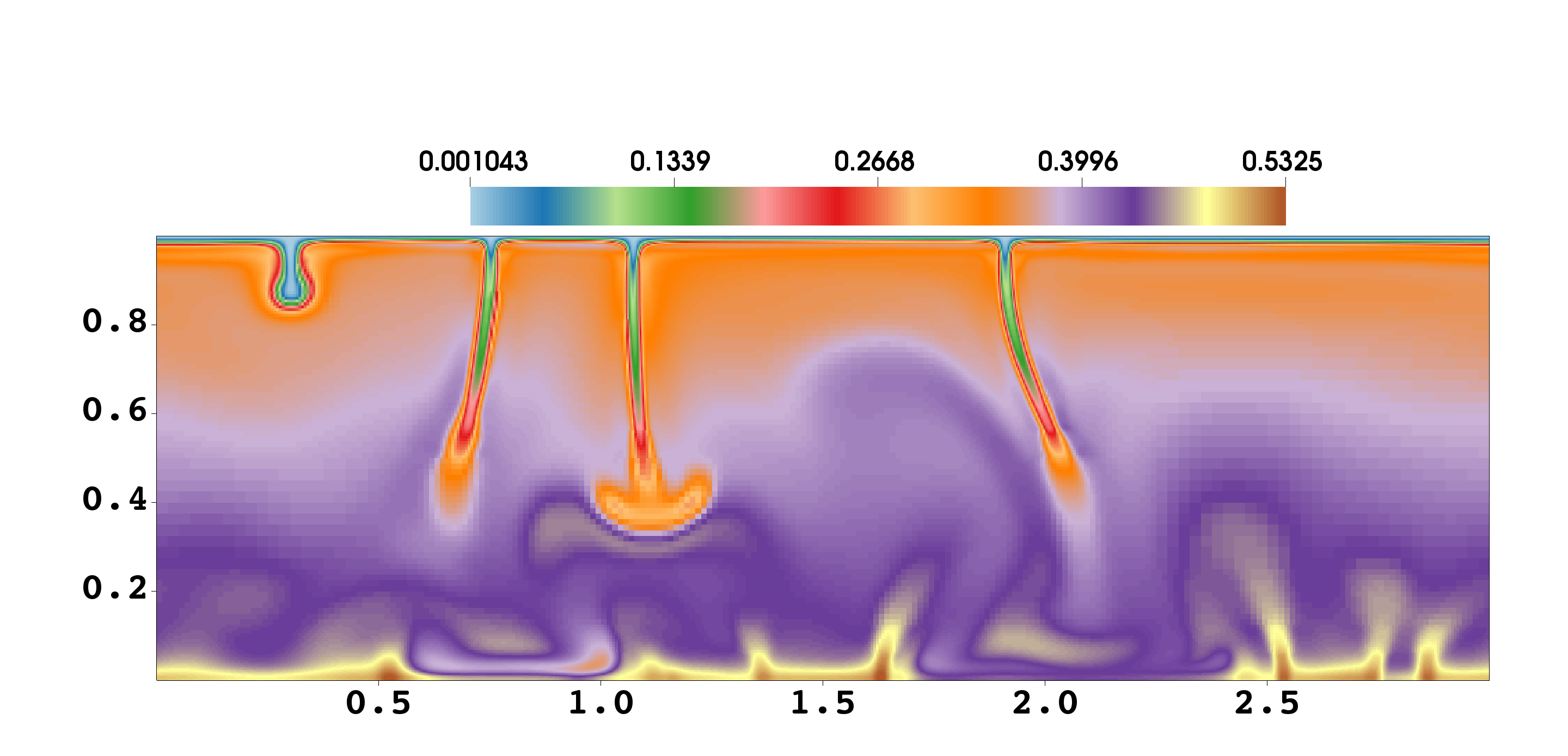}
      \caption{The ratio of magnetic pressure (top) and radiation pressure (bottom) to plasma pressure at $t = 429$.}
         \label{fig:CONV_magEpbyp}
\end{figure}

\begin{figure} 
   \centering
   \includegraphics[width=9cm,clip]{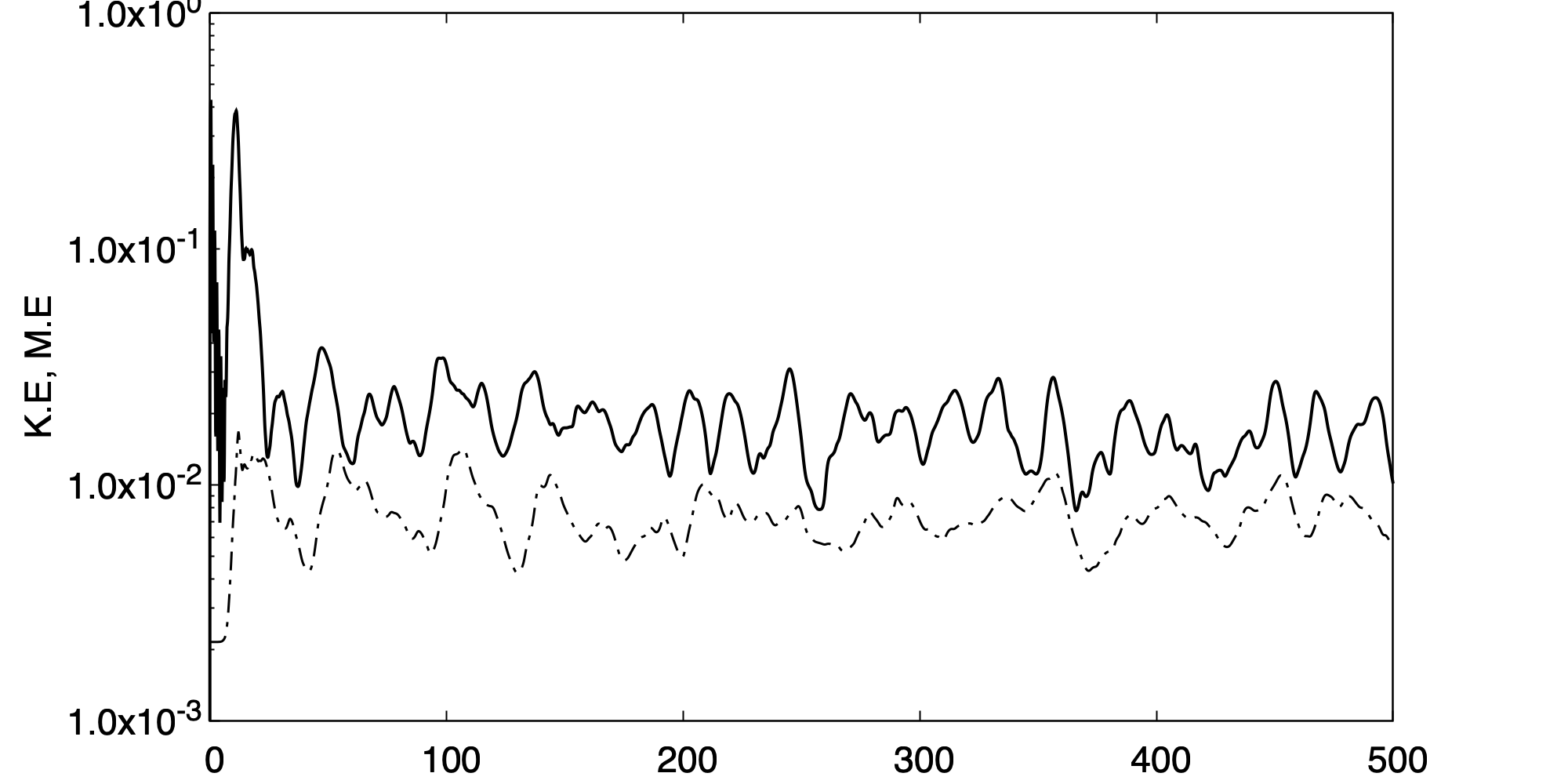}
   \includegraphics[width=9cm,clip]{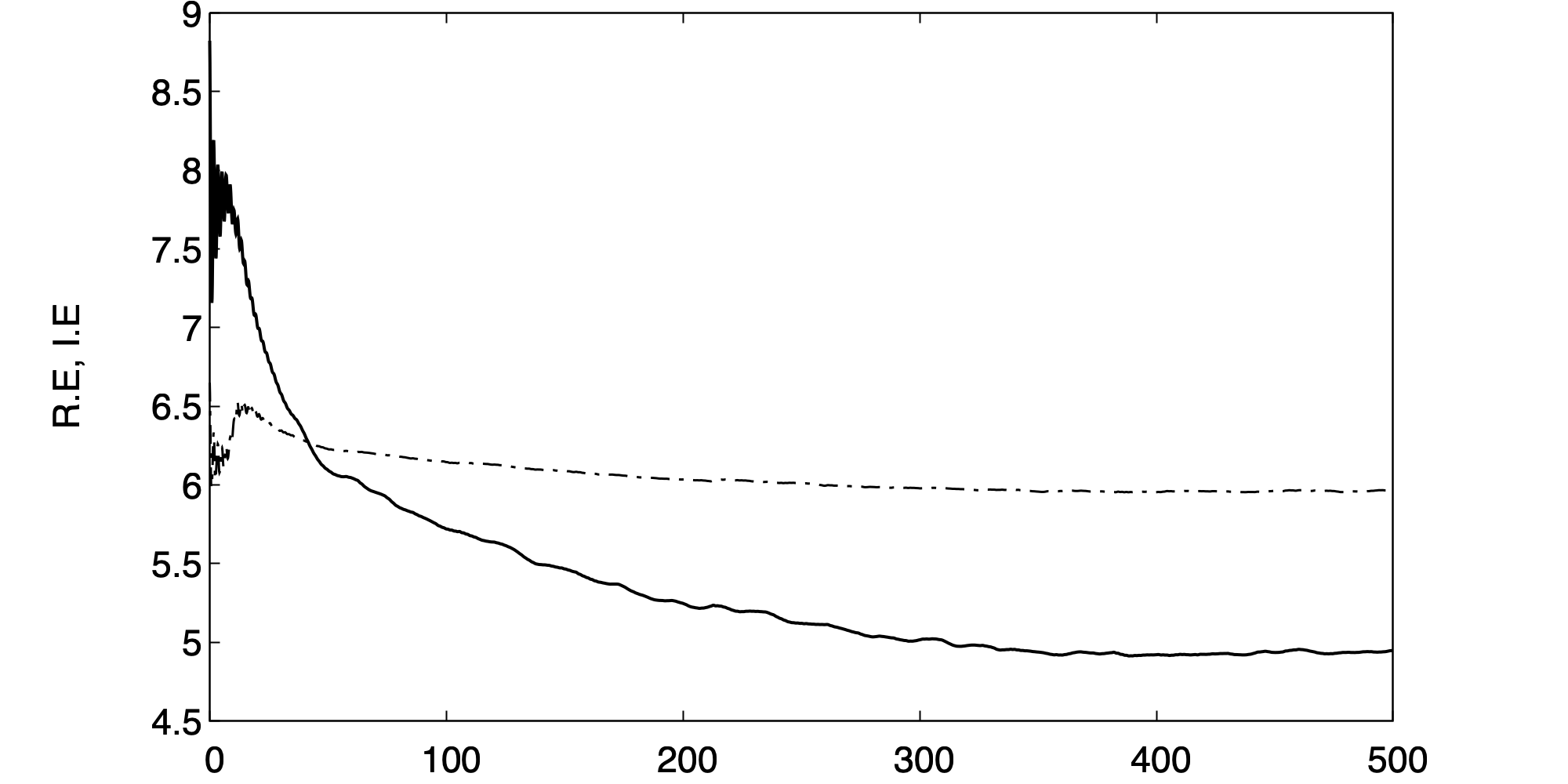}
   \includegraphics[width=9cm,clip]{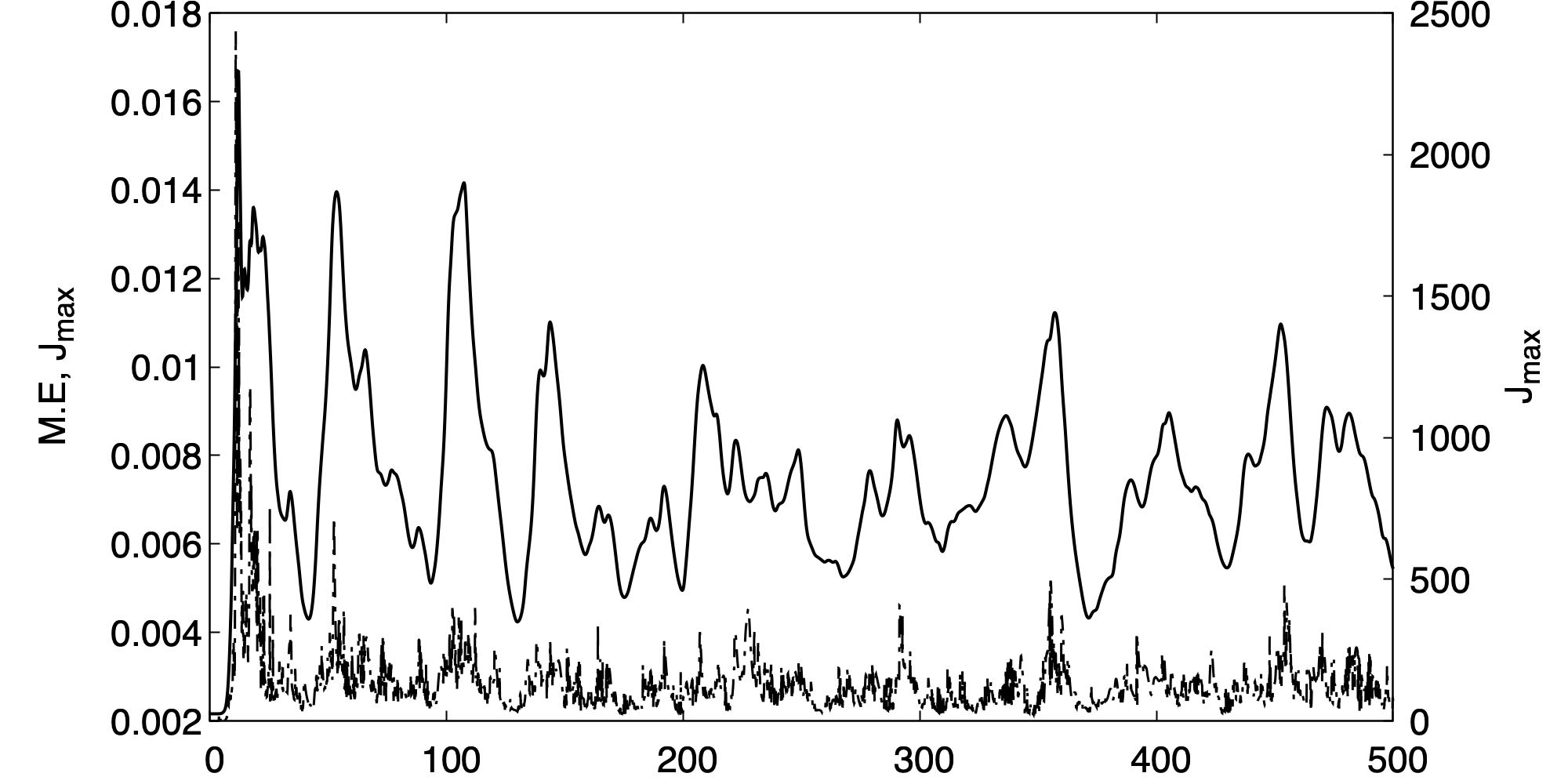}
   \caption{Time evolutions for the diffusion limit radiative magnetoconvection test. \emph{Top}: Total kinetic energy (solid) and total magnetic energy (dash-dot). \emph{Middle}: Total radiation energy (solid) and total plasma internal energy (dash-dot). \emph{Bottom}: Total magnetic energy (solid) and maximum current density (dash-dot).}
    \label{fig:CONV_energy}%
\end{figure}

\begin{figure}
   \centering
   \includegraphics[trim={30 0 150 0},width=\hsize,clip]{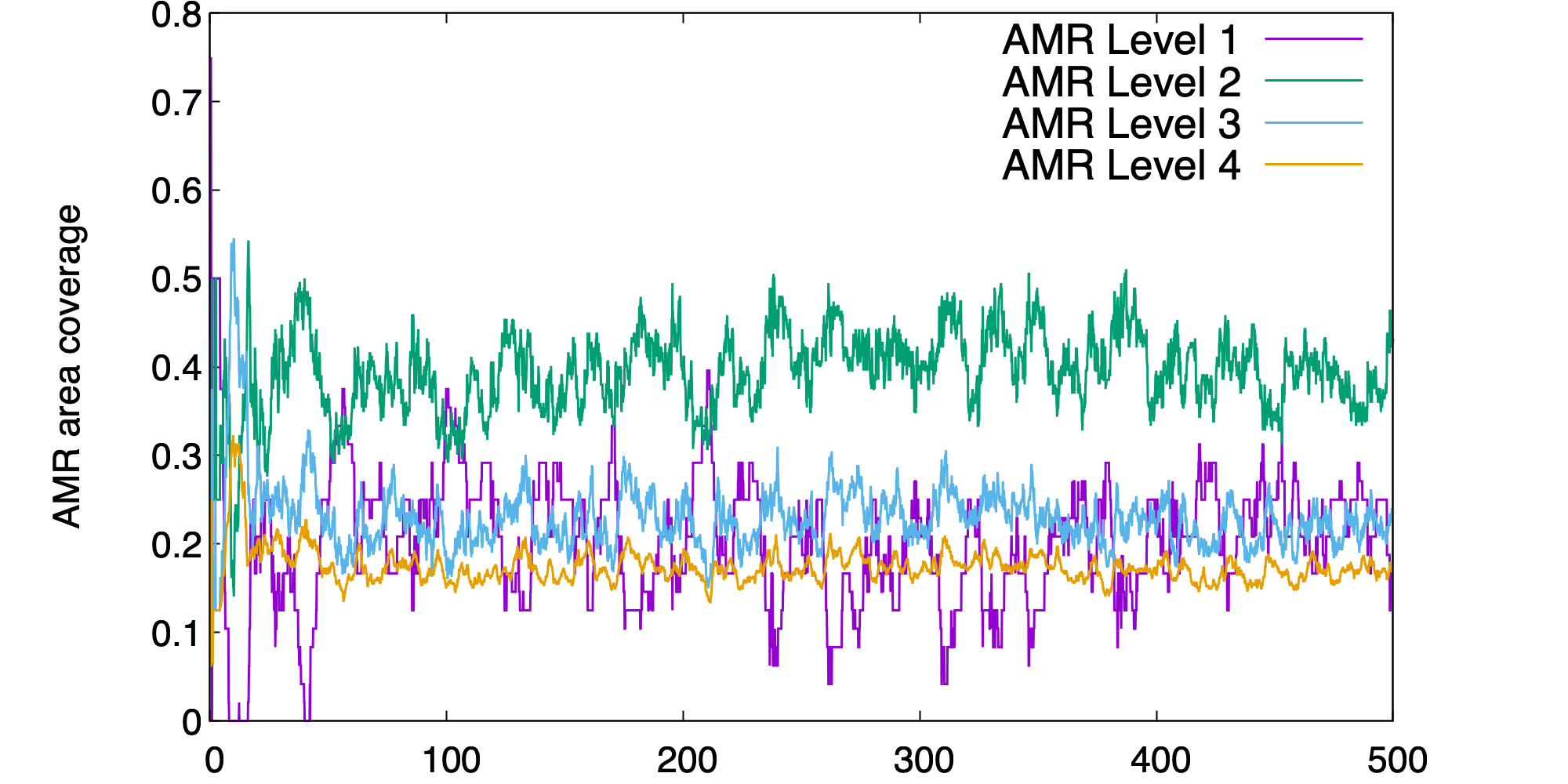}
      \caption{Grid coverage (proportion of total area) for the various AMR levels.}
         \label{fig:AMR_coverage}
\end{figure}

The simulation is run for a non-dimensional time $t = 500$, corresponding to a dimensionalized physical time of 4.417 hours. AMR is driven by the L{\"o}hner's criterion using gradients of density, pressure and radiation energy density. The very first high density downflows occurring due to convection are observed at around $t = 9$. Figure~(\ref{fig:CONV_AMR}) shows the density profile at this time, with the adaptive mesh. The AMR captures these plumes very well. After these relatively high number of plumes, the flow settles into a dynamically steady, oscillatory state, marked by intermittent downflows and periodic compressions of the magnetic field. A typical solution at such a later time is also shown in Figure~(\ref{fig:CONV_AMR}) along with the mesh. Figure~(\ref{fig:CONV_Blines}) shows a snapshot of the magnetic field lines at \nmn{both these time instances
}, clearly showing how the initially uniform vertical magnetic field is swept into concentrated flux sheets by the convective flow. Figure~(\ref{fig:CONV_magEpbyp}) shows the ratios of the magnetic and radiation pressures to the plasma pressure at this later time. Figure~(\ref{fig:CONV_energy}) shows the time-evolution of the total volume-integrated kinetic, magnetic, internal and radiation energies. In the bottom subfigure, it also shows the maximum current density, $J_{\mathrm{max}}$, superimposed along with the magnetic energy evolution. Before the onset of convection, the magnetic energy shows a flat profile until about $t = 8$. During this time, the kinetic energy shows an almost instantaneous increase at $t = 0$, due to the initial flow profile not being in vertical pressure equilibrium. This is because of the added radiation pressure on top of the original equilibrium initial conditions specified by \citet{hurlburt1988magnetic}. The vertical motions allow the plasma to equilibriate without affecting the magnetic field. After these initial transients, the magnetic field lines start to first compress, causing a sharp increase in the total magnetic energy. This increase coincides with an increase in kinetic energy, corresponding to the onset of convection. The magnetic energy then oscillates about an average value in a dynamically steady state. Loss of magnetic energy occurs due to decompression and due to the magnetic field being wound up by the flow and slowly getting dissipated away from the interior by reconnection. The internal and radiation energies settle down at about $t = 400$ to steady values and reach radiative equilibrium. Figure~(\ref{fig:AMR_coverage}) shows the evolution of the grid coverage as a function of time, showing the proportion of the total area of the computational domain covered by the various AMR levels.

\begin{figure}
   \centering
   \includegraphics[trim={100 50 100 250},width=\hsize,clip]{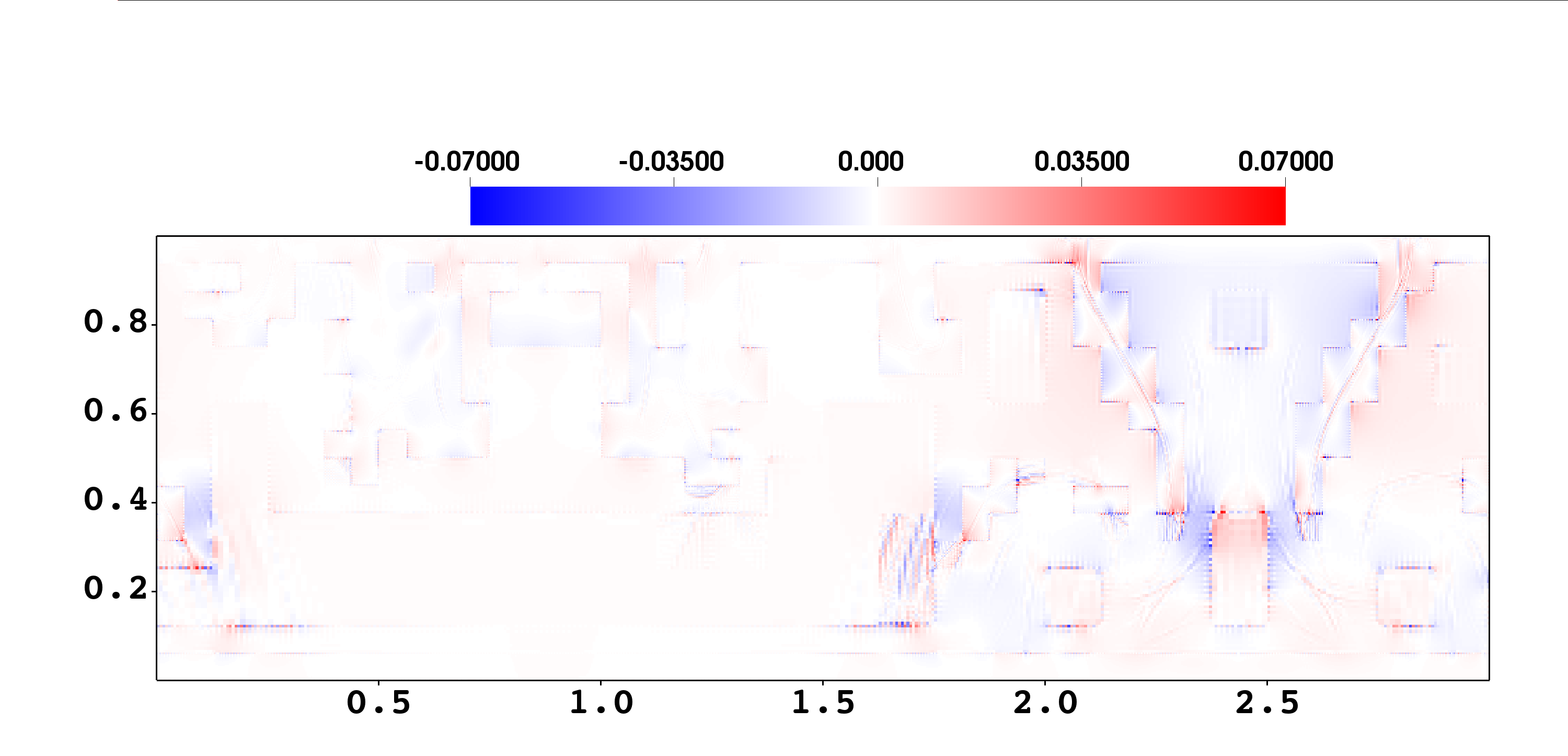}
   \includegraphics[trim={100 50 100 250},width=\hsize,clip]{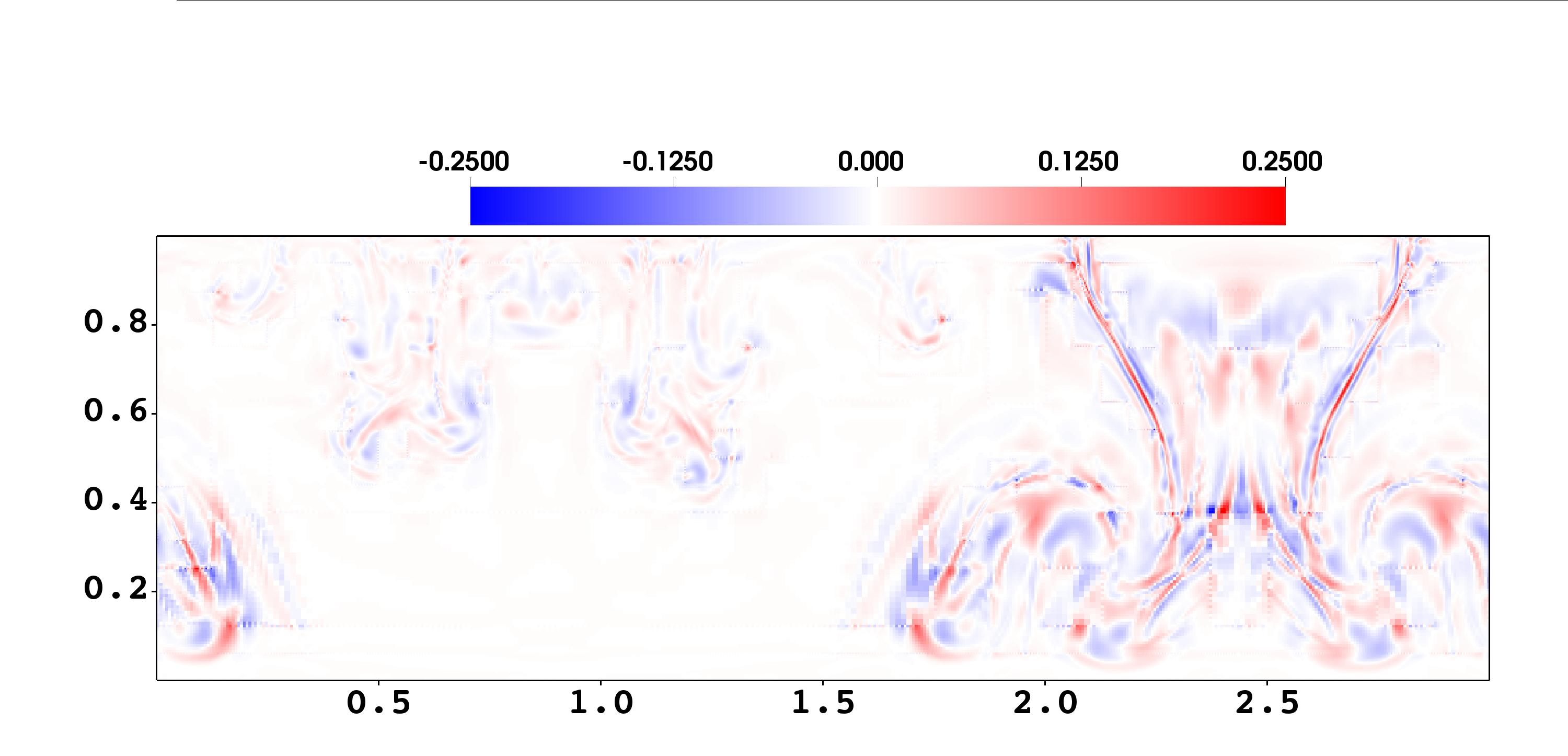}
   \includegraphics[trim={100 50 100 250},width=\hsize,clip]{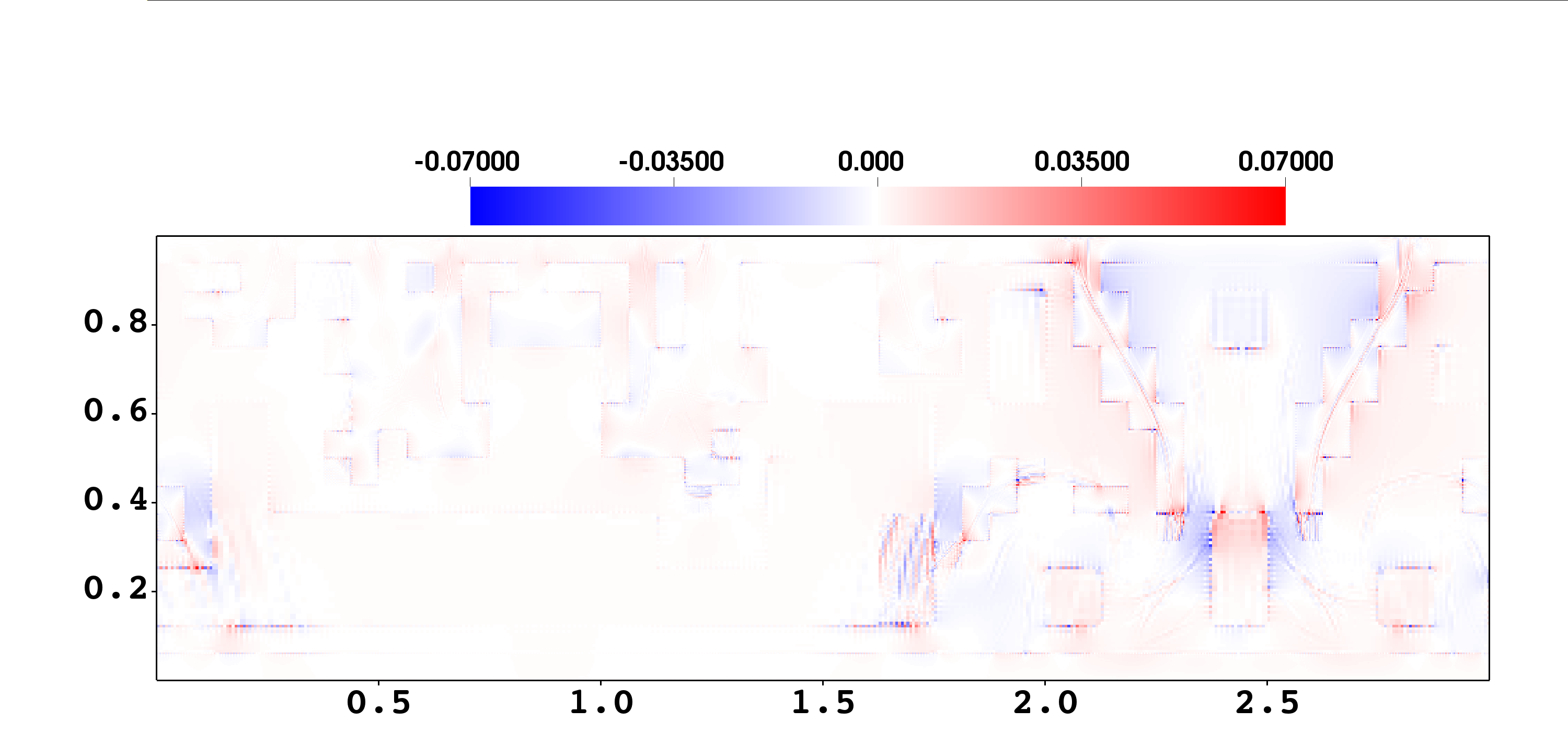}
   \includegraphics[trim={100 50 100 250},width=\hsize,clip]{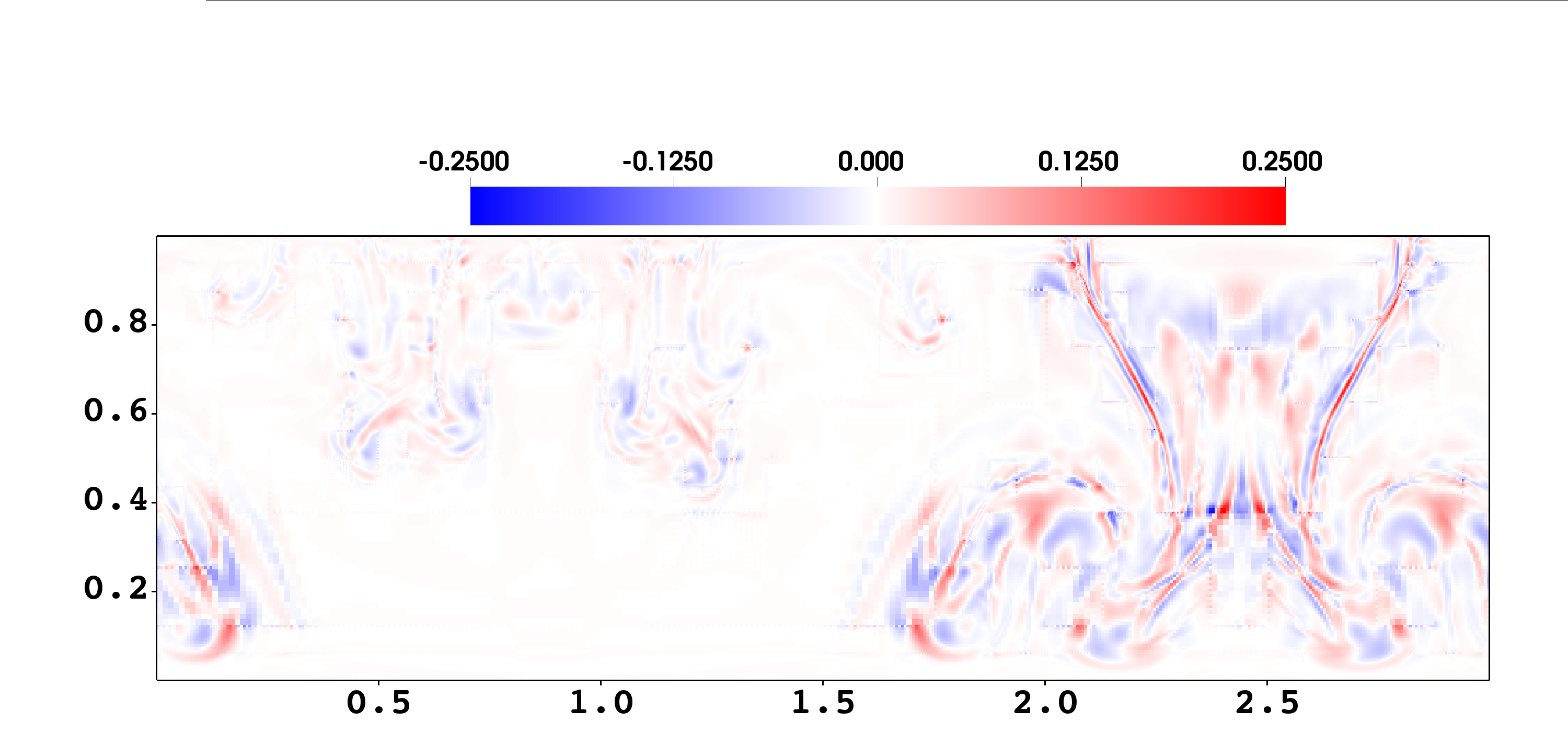}
      \caption{\nmn{The magnetic field discrete divergence error for the radiative magnetoconvection case at $t = 9$ for the (top to bottom) Linde, Janhunen, Linde-Janhunen and Powell methods.}}
         \label{fig:CONV_divB}
\end{figure}

\nmn{As this case involves the winding of the magnetic field by the convective plumes and the resultant magnetic reconnection, the use of the Linde method for divergence control can create significant errors in the discrete divergence in regions where such phenomena occur. As mentioned earlier, such monopole effects can lead to errors in the magnetic energy and thus the total energy, leading to artificial heating or cooling effects. It is therefore in our interest to examine such errors. The top panel in Figure (\ref{fig:CONV_divB}) shows the cell-integrated divergence at $t = 9$, obtained using the Linde method. This corresponds to the density flow solution and magnetic field lines shown in the top panels of the Figures (\ref{fig:CONV_AMR}) and (\ref{fig:CONV_Blines}), respectively. For comparisons, this case was also run using several other divergence control methods, namely the Powell source term method \citep{powell1999}, the constrained transport (CT) method \citep{olivares2019}, the Janhunen method \citep{janhunen2000} and the Linde-Janhunen method. The discrete divergence for the Powell, Janhunen and Linde-Janhunen methods is also shown in the rest of the panels in Figure (\ref{fig:CONV_divB}). The Linde and Linde-Janhunen methods are observed to produce errors of somewhat lower magnitude as compared to the Powell and Janhunen methods. In this particular discretization, the CT method produces divergence errors of the order of machine precision zero. However, the CT method produces monopole errors similar to other schemes, when a different discretization is used for the evaluation of the divergence, and still contributes to artificial heating like other schemes.}

\section{Discussion and Conclusions}\label{sec:conclusions}
This paper describes the recently added radiation-magnetohydrodynamics (RMHD) capabilities of the \texttt{MPI-AMRVAC} computational framework. To test this RMHD module, a variety of benchmark tests were designed. Several cases of radiation-dominated MHD shocks were simulated. We extend on previous works~\citep{mihalas1984foundations,coggeshall1986lie,bouquet2000analytical,lowrie2007radiative,lowrie2008radiative} discussing the radiation-modified Rankine-Hugoniot jump conditions for hydrodynamical shocks in the diffusion limit, to corresponding relations for RMHD shocks. We highlighted the various types of classical MHD shocks (fast, slow, intermediate and switch-on) to modified versions in RMHD. These steady shocks are testing the solver's robustness, its AMR capability of capturing strong shocks and discontinuities and the handling of conserved quantities. Next, we performed a linear perturbation analysis to derive the RMHD dispersion relations for the damping of magnetoacoustic waves in a radiative, magnetized, static background plasma. The analytical damping rates for each of the characteristic waves (slow, fast and undamped Alfv\'en), obtained from the dispersion relation also showed an excellent match with the rates obtained from numerical simulations. The radiation-modified versions of MHD Riemann shock-tube problems are simulated and used to \nmn{present 
example cases involving 
} transitions between optically thick to thin radiative conditions. Including a 1.75D problem from \citet{torrilhon2003uniqueness} well demonstrated the effect of radiation on the various waves generated in an MHD system, where especially the rotational discontinuities represent a challenge to capture numerically. Although several researchers performed RMHD simulations in the past, most of these were targeting specific applications occurring in realistic astrophysical environments, which are not always easily reproduced. Our benchmark cases made deliberate simplifications (such as assuming constant opacities) to highlight the numerical challenges in RMHD settings, even in essentially 1D scenarios, requiring AMR. This paper also seeks to address the scarcity of available literature on standard test cases for RMHD. All of these tests can be used by future researchers to validate and test their codes. We also showed the potential to study the detailed interaction between convective and diffusive radiative flux transport in multi-dimensional setups, in an idealized setting motivated by solar magnetoconvection. 

\texttt{MPI-AMRVAC}'s newly developed RMHD module is now ready to be used to study more realistic scientific applications, such as the envelopes, atmospheres, and winds of massive stars with significant magnetic fields. Since the tests in this paper were quite basic, a constant value of the opacity, $\kappa$, was assumed. However, for realistic applications, more physically accurate models of plasma opacities would have to be used, accounting for opacity dependence on local plasma properties. For instance, the variation of opacities is crucial to simulate phenomena such as certain radiation-driven instabilities occurring in stellar atmospheres and luminous accretion discs~\citep{blaes2003local}. As such, future work would involve relaxation of the constant opacity assumption and using opacity tables. For example, \citet{moens2022first} and \citet{debnath20242d} have already demonstrated the use of opacity tables in \texttt{MPI-AMRVAC}'s RHD module, for their study of Wolf-Rayet stars and O-stars, respectively. \nmn{More realistic scientific applications could involve studying regions of large radiation pressures and transition regions between optically thin and optically thick regimes. For such cases, numerical techniques such as the modified Godunov method of \citet{miniati2007} would help with better capturing of the coupling between the radiation and plasma momentum and energy terms. Future work would also entail the implementation of such techniques.}

For solar applications, the RMHD module demonstrated here can bridge the gap between coronal-only simulations, which have been the main research application for \texttt{MPI-AMRVAC} to date, and advanced RMHD simulations including optically thick, convective layers. There, it will be of interest to clarify the connections between coronal condensations driven by optically thin prescriptions (as the entropy mode can turn into a thermal instability), and the more complete RMHD description with a radiative field incorporated and dynamically coupled to the plasma evolution. It may be necessary to develop a solar-atmosphere tailored variation, where the net heat-loss $\dot{q}$ prescription can be varied depending on temperature and optical thickness, to smoothly connect the typical $\dot{q}=-\rho^2\Lambda(T)$ tabulated cooling curve $\Lambda(T)$ variations applicable in the corona (see \citet{Hermans2021}) with those obtained through the FLD prescriptions. This is left for follow-up studies.

\nmn{Although \texttt{MPI-AMRVAC}'s RMHD module is now ready to be used for scientific solar and astrophysical applications, the use of the FLD module must be approached with caution. FLD has several potentially critical limitations \citep{mihalas1984foundations, hayes2003beyond, davis2012radiation, jiang2012godunov} and its use must be carefully considered on a case-by-case basis. For example, FLD has been known to produce spurious solutions in transition regions, and this has been observed in the accretion disk simulations of \citet{boley2007three}. Inaccuracies of FLD in accretion disk simulations have also been highlighted in the work of \citet{davis2012radiation}. FLD is a monochromatic, frequency-independent approximation which can fail to accurately model radiative flows with a strong dependence on wave frequencies. A key aspect of the FLD model is that the radiation flux is assumed to be in the direction of the radiation energy density gradient, rather than computed from the radiative transfer equation. \citet{jiang2012godunov} demonstrate a photon bubble instability simulation where the radiation flux direction is not aligned with the direction of the radiation energy density gradient at the generated shock fronts. The FLD assumption would clearly fail in such cases. This assumption can also introduce serious errors in optically thin regions that have highly anisotropic radiation fields. Therefore, the FLD method fails in capturing shadows and beams, as has been clearly demonstrated in the work of \citet{hayes2003beyond} and \citet{davis2012radiation}. 

FLD also forces the Eddington factor to strictly lie between $1/3$ and $1$, whereas \citet{jiang2012godunov} have demonstrated cases of radiative shocks where the Eddington factor can fall below $1/3$. The FLD approximation retains only the terms shown in the box 
in the radiation momentum equation, i.e. Equation (\ref{eq:mhd_r_e_mom_CMF}) from Appendix B. This relation between the divergence of the pressure tensor and the radiative flux is the basis for using the Fick's diffusion law as the closure relation. This removes the effect of radiation inertia from radiation flow solutions. Finally, the radiation pressure tensor is taken to be a diagonal tensor in the optically thick limit of the FLD model, with the off-diagonal terms being neglected. This eliminates effects such as radiation viscosity  \citep{mihalas1984foundations, castor2004radiation, jiang2012godunov}. Both these effects, radiation inertia and radiation viscosity, can play a significant role in the dynamics of accretion disk flows. More sophisticated radiation models could be used to improve upon these limitations. For example, \citet{jiang2012godunov} clearly demonstrated the ability of the Variable Eddington Tensor (VET) method in capturing beams and shadows, as opposed to FLD. However, increasingly sophisticated models would come with increased complexity and higher computational cost. The possible incorporation of such models into \texttt{MPI-AMRVAC}'s RMHD module would be the subject of follow-up efforts. These efforts would then also involve more rigorous tests such as the shadow test, beam test and the photon bubble instability test as well as applications such as accretion disk simulations.}

\begin{acknowledgements}
   RK is supported by an FWO project G0B9923N 
      and received funding from the European Research Council (ERC) under the European Union Horizon 2020 research and innovation program (grant agreement No. 833251 PROMINENT ERC-ADG 2018). RK acknowledges KU Leuven C1 project C16/24/010 UnderRadioSun. A.u.-D. acknowledges NASA ATP grant number 80NSSC22K0628 and support by NASA through Chandra Award number TM4-25001A issued by the Chandra X-ray Observatory 27 Center, which is operated by the Smithsonian Astrophysical Observatory for and on behalf of NASA under contract NAS8-03060. JS and NM acknowledge the support of the European Research Council (ERC) Horizon Europe grant under grant agreement number 101044048. A.u.-D. and JS acknowledge support from the Belgian Research Foundation Flanders (FWO) Odysseus program under grant number G0H9218N. The authors thank the referee, Dr. Lizhong Zhang, for valuable feedback that greatly improved the manuscript.
\end{acknowledgements}

%
%
\bibliographystyle{aa}
\bibliography{nmn}{}

\nmn{\section*{Appendix A: Frame transformation for the RMHD equations}\label{appendix:transformation}}

\nmn{
Following \citet{jiang2012godunov}, the radiation MHD equations, including the equations of the radiation energy density and radiation flux, with the radiative terms evaluated in the inertial lab frame, are given by
\begin{equation}\label{eq:mhd_mass_LF}
\frac{\partial {\rho}}{\partial {t}} + {\nabla} \cdot \left({\rho} {\mbox{\bf v}}\right) = 0,
\end{equation}
\begin{equation}\label{eq:mhd_mom_LF}
\begin{split}
\frac{\partial (\rho{\mbox{\bf v}})}{\partial {t}} + {\nabla}\cdot\left(\rho \mbox{\bf v} \mbox{\bf v} - \mbox{\bf B} \mbox{\bf B} + {\left(p + \frac{\mbox{\bf B} \cdot \mbox{\bf B}}{2}\right)}\mbox{\bf I}\right) = -\bf S_r(\bf P),
\end{split}
\end{equation}
\begin{equation}\label{eq:mhd_energy_LF}
\begin{split}
\frac{\partial e}{\partial t} + 
\nabla\cdot\left(\left(e + p + \frac{\mbox{\bf B}\cdot\mbox{\bf B}}{2}\right)\mbox{\bf v} - (\mbox{\bf B}\cdot \mbox{\bf v})\mbox{\bf B} \right) = -c{\bf{S}}_r(E),
\end{split}
\end{equation}
\begin{equation}\label{eq:mhd_mag_LF}
\frac{\partial{\mbox{\bf B}}}{\partial t} + {\nabla}\cdot \left(\mbox{\bf v} \mbox{\bf B} - \mbox{\bf B} \mbox{\bf v}\right) = \bf 0,
\end{equation}
\begin{equation}\label{eq:mhd_r_e_LF}
\frac{\partial {E}_{LF}}{\partial {t}} + {\nabla} \cdot \mbox{\bf F}_{LF} = c{\bf{S}}_r(E),
\end{equation}
\begin{equation}\label{eq:mhd_Fr_LF}
\frac{1}{c^2}\frac{\partial \mbox{\bf F}_{LF}}{\partial {t}} + {\nabla} \cdot \mbox{\bf P}_{LF} = {\bf S}_r(\mbox{\bf P}).
\end{equation}
Here, ${E}_{LF}$, ${\bf F}_{LF}$ and $\mbox{\bf P}_{LF}$ are the frequency-integrated radiation energy density, flux vector and pressure tensor in the inertial lab frame. The source terms in the momentum and plasma energy equations are given by
\begin{equation}\label{eq:SrP_LF}
\begin{split}
{\bf{S}}_r(\mbox{\bf P}) = -\rho (\kappa_F + \kappa_S) (\mbox{\bf F}_{LF} - (\mbox{\bf v} E_{LF} + \mbox{\bf v}\cdot \mbox{\bf P}_{LF}))/c \\ + \rho \mbox{\bf v} (\kappa_P a_r T_g^4 - \kappa_E E_{LF})/c,
\end{split}
\end{equation}
\begin{equation}\label{eq:SrE_LF}
\begin{split}
{\bf{S}}_r(E) = \rho (\kappa_F - \kappa_S) \frac{\mbox{\bf v}}{c^2} \cdot(\mbox{\bf F}_{LF} - (\mbox{\bf v} E_{LF} + \mbox{\bf v}\cdot \mbox{\bf P}_{LF})) \\ + \rho(\kappa_P a_r T_g^4 - \kappa_E E_{LF}).
\end{split}
\end{equation}
Here, $\kappa_P$, $\kappa_E$, $\kappa_F$ and $\kappa_S$, are the Planck, energy density, flux mean absorption and flux mean scattering opacities, respectively. The relations between the radiation energy density, flux vector and pressure tensor in the inertial and co-moving frames are
\begin{equation}\label{eq:E_CMF_LF}
E_{LF} = E_{CMF} + 2\frac{\mbox{\bf v}}{c^2} \cdot \mbox{\bf F}_{CMF}\,,
\end{equation}
\begin{equation}\label{eq:F_CMF_LF}
\mbox{\bf F}_{LF} = \mbox{\bf F}_{CMF} + \mbox{\bf v}E_{CMF} + \mbox{\bf v} \cdot \mbox{\bf P}_{CMF}\,,
\end{equation}
\begin{equation}\label{eq:P_CMF_LF}
\mbox{\bf P}_{LF} = \mbox{\bf P}_{CMF} + 2\frac{\mbox{\bf v}}{c^2} \mbox{\bf F}_{CMF}\,,
\end{equation}
where ${E}_{CMF}$, ${\bf F}_{CMF}$ and $\mbox{\bf P}_{CMF}$ are the frequency-integrated radiation energy density, flux vector and pressure tensor in the co-moving frame.

Equations (\ref{eq:mhd_mass_LF}) and (\ref{eq:mhd_mag_LF}) are identical to Equations (\ref{eq:mhd_mass}) and (\ref{eq:mhd_mag}), respectively. Substituting Equations (\ref{eq:SrP_LF})--(\ref{eq:P_CMF_LF}) in Equations (\ref{eq:mhd_mom_LF}), (\ref{eq:mhd_energy_LF}), (\ref{eq:mhd_r_e_LF}) and (\ref{eq:mhd_Fr_LF}), assuming $\kappa_S = 0$, and substituting
\begin{equation}\label{eq:heating}
\dot{q}_{CMF} = c \kappa_E \rho E_{CMF} - 4 \kappa_P \rho \sigma T_g^4\,,
\end{equation}
and
\begin{equation}\label{eq:rad_force}
{\bf f_{r,CMF}} = \frac{\rho \kappa_F \mbox{\bf F}_{CMF}}{c}\,,
\end{equation}
we get the following equations:
\begin{equation}\label{eq:mhd_mom_CMF}
\begin{split}
\frac{\partial (\rho{\mbox{\bf v}})}{\partial {t}} + {\nabla}\cdot\left(\rho \mbox{\bf v} \mbox{\bf v} - \mbox{\bf B} \mbox{\bf B} + {\left(p + \frac{\mbox{\bf B} \cdot \mbox{\bf B}}{2}\right)}\mbox{\bf I}\right) = {\bf f_{r,CMF}} + \mathcal{O}(v/c),
\end{split}
\end{equation}
\begin{equation}\label{eq:mhd_energy_CMF}
\begin{split}
\frac{\partial e}{\partial t} + 
\nabla\cdot\left(\left(e + p + \frac{\mbox{\bf B}\cdot\mbox{\bf B}}{2}\right)\mbox{\bf v} - (\mbox{\bf B}\cdot \mbox{\bf v})\mbox{\bf B} \right) = \mbox{\bf v}\cdot {\bf f_{r,CMF}} \\ + \dot{q}_{CMF} + \mathcal{O}(v/c),
\end{split}
\end{equation}
\begin{equation}\label{eq:mhd_r_e_CMF}
\frac{\partial {E}_{CMF}}{\partial {t}} + {\nabla} \cdot \left({E}_{CMF} {\mbox{\bf v}}\right) + {\nabla} \cdot \mbox{\bf F}_{CMF} + \mbox{\bf P}_{CMF} : {\nabla} \mbox{\bf v}  = -\dot{q}_{CMF} + \mathcal{O}(v/c),
\end{equation}
\begin{equation}\label{eq:mhd_r_e_mom_CMF}
\frac{1}{c^2}\left(\frac{\partial \mbox{\bf F}_{CMF}}{\partial {t}} + {\nabla} \cdot \left(\mbox{\bf F}_{CMF} {\mbox{\bf v}}\right) \right) + \boxed{{\nabla} \cdot \mbox{\bf P}_{CMF} = -{\bf f_{r,CMF}}} + \mathcal{O}(v/c).
\end{equation}
Here, $\mathcal{O}(v/c)$ comprises terms proportional to $(v/c)$ or higher order powers. For non-relativistic velocities, these terms are very small and can be conveniently neglected. For non-relativistic velocities, and dropping the subscript $CMF$ for simplicity, the Equations (\ref{eq:mhd_mom_CMF}), (\ref{eq:mhd_energy_CMF}) and (\ref{eq:mhd_r_e_CMF}) reduce to Equations (\ref{eq:mhd_mom}), (\ref{eq:mhd_energy}) and (\ref{eq:mhd_r_e}), respectively. 
In Equation (\ref{eq:mhd_r_e_mom_CMF}), the terms with the $(1/c^2)$ factor on the left-hand side and the higher order terms are ignored, resulting in the relation between the divergence of the pressure tensor and the radiative flux as shown in the box drawn.
}


\nmn{\section*{Appendix B: Dispersion relation for damping of MHD waves in a radiative medium}\label{appendix:dispersion}}

To obtain the dispersion relation, we first rewrite the Equations~(\ref{eq:mhd_mass})--(\ref{eq:mhd_mag}) in non-conservative form, and reformulate the energy equation to a pressure evolution equation:
\begin{equation}\label{eq:mhd_mass_noncon}
\frac{\partial {\rho}}{\partial {t}} + \rho {\nabla} \cdot {\mbox{\bf v}} + {\mbox{\bf v}} \cdot {\nabla} {\rho} = 0,
\end{equation}
\begin{equation}\label{eq:mhd_mom_noncon}
{\rho}\frac{\partial {\mbox{\bf v}}}{\partial {t}} +  {\rho}\mbox {\bf v} \cdot {\nabla} \mbox {\bf v} + {\nabla} p + \left({\nabla} \mbox {\bf B}\right) \cdot \mbox {\bf B} - \left(\mbox {\bf B} \cdot {\nabla}\right) \mbox {\bf B} = \bf f_r,
\end{equation}
\begin{equation}\label{eq:mhd_energy_noncon}
\frac{\partial p}{\partial t} + \mbox {\bf v} \cdot {\nabla}p + \gamma p {\nabla} \cdot \mbox {\bf v} = (\gamma-1)\dot{q},
\end{equation}
\begin{equation}\label{eq:mhd_mag_noncon}
\frac{\partial {\mbox{\bf B}}}{\partial t} + \left(\mbox {\bf v} \cdot {\nabla}\right) \mbox {\bf B} + \mbox {\bf B} {\nabla} \cdot \mbox {\bf v} - \left(\mbox {\bf B} \cdot {\nabla}\right) \mbox {\bf v} = \bf 0.
\end{equation}
These above equations will be linearized assuming an infinite, static, time-independent ($\partial /\partial t = 0$), uniform background for the plasma properties. In this equilibrium state, the heating and cooling function $\dot{q}$ is also assumed to vanish, and therefore the equilibrium radiation temperature equals the plasma temperature. The equilibrium radiation force ${\bf f_r}$ also vanishes due to the constant radiation energy. The equilibrium state is hereby denoted by $\rho_0$, $\mbox{\bf v}_0 = \bf 0$, $p_0$, $\mbox{\bf B}_0$, $T_{g,0}$, $E_0$ and $T_{r,0} = T_{g,0}$ for the equilibrium density, velocity, plasma pressure, magnetic field, plasma temperature, radiation energy density and radiation temperature, respectively. Small linear deviations from this equilibrium are denoted by the subscript `1'. The linearized versions of Equations~(\ref{eq:mhd_mass_noncon})--(\ref{eq:mhd_mag_noncon}) for $\lambda = 1/3$ (the optically thick limit) are given by
\begin{equation}\label{eq:mhd_mass_lin}
\frac{\partial {\rho_1}}{\partial {t}} + {\rho}_0{\nabla} \cdot  {\mbox{\bf v}_1} = 0,
\end{equation}
\begin{equation}\label{eq:mhd_mom_lin}
{\rho}_0\frac{\partial {\mbox{\bf v}_1}}{\partial {t}} + {\nabla} p_1 + \left({\nabla} \mbox {\bf B}_1\right) \cdot \mbox {\bf B}_0 - \left(\mbox {\bf B}_0 \cdot {\nabla}\right) \mbox {\bf B}_1 = -\frac{1}{3} {\nabla} E_1,
\end{equation}
\begin{equation}\label{eq:mhd_energy_lin}
\frac{\partial p_1}{\partial t} + \gamma p_0  {\nabla} \cdot \mbox {\bf v}_1 = (\gamma-1)(\dot{q}_{\rho} \rho_1 + \dot{q}_{T_g} T_{g,1} + \dot{q}_{T_r} T_{r,1}),
\end{equation}
\begin{equation}\label{eq:mhd_mag_lin}
\frac{\partial {\mbox{\bf B}_1}}{\partial t} + \mbox {\bf B}_0 {\nabla} \cdot \mbox {\bf v}_1 - \left(\mbox {\bf B}_0 \cdot {\nabla} \right) \mbox {\bf v}_1 = \bf 0.
\end{equation}
The solenoidality condition, Equation~(\ref{eq:mhd_mag}), reduces to 
\begin{equation}\label{eq:div_B_lin}
{\nabla} \cdot {\mbox{\bf B}_1} = 0.
\end{equation}
and the ideal gas law, Equation~(\ref{eq:ideal_gas_law}), neglecting variations in $\mu$, reduces to
\begin{equation}\label{eq:ideal_gas_law_lin}
\frac{p_1}{p_0} = \frac{\rho_1}{\rho_0} + \frac{T_{g,1}}{T_{g,0}} \,.
\end{equation}
The linearized version of Equation~(\ref{eq:mhd_r_e}) for radiation energy is
\begin{equation}\label{eq:mhd_r_e_lin}
\frac{\partial {E}_1}{\partial {t}} + \frac{4}{3}{E}_0{\nabla} \cdot {\mbox{\bf v}_1} = -(\dot{q}_{\rho} \rho_1 + \dot{q}_{T_g} T_{g,1} + \dot{q}_{T_r} T_{r,1}) + \nabla \cdot \left(\frac{c}{3 \rho_0 \kappa} \nabla E_1 \right).
\end{equation}
Here, $\dot{q}_\rho$, $\dot{q}_{T_r}$ and $\dot{q}_{T_g}$ are partial derivatives of the $\dot{q}$ function with respect to $\rho$, $T_r$ and $T_g$, respectively, and are given by
\begin{equation}\label{eq:qdot_rho}
\dot{q}_{\rho} = c \kappa a_r (T_{r,0}^4 - T_{g,0}^4),
\end{equation}
\begin{equation}\label{eq:qdot_Tr}
\dot{q}_{T_r} = 4 c\rho_0 \kappa a_r T_{r,0}^3,
\end{equation}
\begin{equation}\label{eq:qdot_Tg}
\dot{q}_{T_g} = -4 c\rho_0 \kappa a_r T_{g,0}^3.
\end{equation}
These derivatives are to be evaluated for the background state, such that $\dot{q}_{\rho}=0$ as we have adopted radiative equilibrium. Note that we here get rather simple expressions for these derivatives, as we assumed constant opacities $\kappa$, while many of the physically relevant unstable wave mode solutions in stellar interiors or atmospheres relate to detailed variations of opacities with density and temperatures. This complication is e.g. handled in \citet{blaes2003local}.
The relation between radiation energy density and radiation temperature can be linearized to give 
\begin{equation}\label{eq:rad_temp_lin}
\frac{E_1}{E_0} = 4 \frac{T_{r,1}}{T_{r,0}}.
\end{equation}
To obtain dispersion relations for linear stability analysis, plane-wave perturbations of the form
\begin{equation}\label{eq:density_plane_wave}
\rho_1 = \hat{\rho}e^{i(\bf k \cdot \bf x - \omega t)}
\end{equation}
can be assumed, where $\bf k$ is the wavenumber and $\omega$ is the angular frequency for the plane wave perturbation. For the 1.75D cases studied numerically further on, we will choose $\bf k$ to be along the $x$-axis
${\bf k} = (k,0,0)$. 

Applying such perturbations to $\rho$, $\bf v$, $p$, $\bf B$ and $E$ and substituting them in Equations~(\ref{eq:mhd_mass_lin})--(\ref{eq:mhd_mag_lin}) and (\ref{eq:mhd_r_e_lin}) gives the following
\begin{equation}\label{eq:eigen_mass}
\omega \hat{\rho} = \rho_0 (\bf k \cdot \bf \hat{v}),
\end{equation}
\begin{equation}\label{eq:eigen_mom}
\omega \rho_0 {\bf \hat{v}} = {\bf k} (\hat{p} + \hat{E}/3) + {\bf k} ({\bf \hat{B}} \cdot {\bf B}_0) - ({\bf B}_0 \cdot {\bf k}) {\bf \hat{B}},
\end{equation}
\begin{equation}\label{eq:eigen_energy}
\omega \hat{p} = \gamma p_0 ({\bf k} \cdot {\bf \hat{v}}) + i (\gamma-1)\left(\dot{q}_{\rho} \hat{\rho} + \dot{q}_{T_g} \hat{T}_g + \dot{q}_{T_r} \hat{T}_r\right),
\end{equation}
\begin{equation}\label{eq:eigen_mag}
\omega {\bf \hat{B}} = {\bf B}_0 ({\bf k} \cdot {\bf \hat{v}}) - ({\bf B}_0 \cdot {\bf k}) {\bf \hat{v}},
\end{equation}
\begin{equation}\label{eq:eigen_r_e}
\omega \hat{E} = \frac{4}{3} E_0 ({\bf k} \cdot {\bf \hat{v}}) - i \left(\dot{q}_{\rho} \hat{\rho} + \dot{q}_{T_g} \hat{T}_g + \dot{q}_{T_r} \hat{T}_r\right) - i \frac{k^2 c \hat{E}} {3 \rho_0 \kappa}.
\end{equation}
Equations~(\ref{eq:eigen_mass})--(\ref{eq:eigen_r_e}) originally constitute a set of 9 equations, accounting for each of the 3 dimensions of the momentum and magnetic field eigenvalue equations. However, for ${\bf k} = (k,0,0)$, the equation for the $x$-component of the magnetic field reduces to $\hat{B}_x = 0$, reducing it to a set of 8 equations. This is the solenoidality condition for the perturbed magnetic field, and it makes the number of physical eigenfrequencies $\omega$ to be 8 in total, one more than the familiar ideal MHD result. In the above equations, $\hat{T}_g$ can be substituted in terms of density and plasma pressure perturbation amplitudes using Equation~(\ref{eq:ideal_gas_law_lin}).
%
%
Similarly, $\hat{E}$ can be substituted using
%
%
Equation~(\ref{eq:rad_temp_lin}).

The above equations then lead to the dispersion relation, where we use certain characteristic speeds of the flow: 
\begin{equation}\label{eq:char_speeds}
c_{g,0}^2 = \frac{\gamma p_0}{\rho_0}, c_{r,0}^2 = \frac{4 E_0}{9\rho_0}, v_{A,0}^2 = \frac{B_{x,0}^2 + B_{y,0}^2 + B_{z,0}^2}{\rho_0}, v_{A,x,0}^2 = \frac{B_{x,0}^2}{\rho_0}
\end{equation}
where $c_{g,0}$, $c_{r,0}$, $v_{A,0}$ and $v_{A,x,0}$ are the adiabatic plasma sound speed, the radiation sound speed, the total Alfv\'en speed and the $x$-Alfv\'en speed, respectively. The final dispersion relation can be written as:
\begin{equation}\label{eq:dispersion_relation_simplified}
\begin{split}
\Biggl(\omega^2 - k^2 v_{A,x,0}^2\Biggr) \Biggl(\omega^6 + i\Biggl(\frac{ck^2}{3\rho_0\kappa} + \frac{\dot{q}_{T_r} T_{r,0}}{4 E_0} - (\gamma-1)\frac{\dot{q}_{T_g} T_{g,0}}{p_0}\Biggr)\omega^5 \\ - k^2\Biggl(c_{r,0}^2 + c_{g,0}^2 + v_{A,0}^2 - (\gamma-1)\dot{q}_{T_g} \frac{T_{g,0}}{p_0} \frac{c}{3\rho_0\kappa}\Biggr)\omega^4 \\ + ik^2 \Biggl(-\frac{ck^2}{3\rho_0\kappa}\biggl(c_{g,0}^2 + v_{A,0}^2\biggr) + \\(\gamma-1)\dot{q}_{T_g} \frac{T_{g,0}}{\rho_0} \biggl(\frac{4}{3} + \frac{\gamma c_{r,0}^2}{c_{g,0}^2} + \frac{\gamma v_{A,0}^2}{c_{g,0}^2} \biggr) \\ - \dot{q}_{T_r} \frac{T_{r,0}}{\rho_0} \biggl(\frac{c_{g,0}^2}{9 c_{r,0}^2} + \frac{\gamma-1}{3} + \frac{v_{A,0}^2}{9 c_{r,0}^2} \biggr)\Biggr)\omega^3 \\ + k^4\Biggl((\gamma-1)\biggl(-\dot{q}_{T_g} \frac{T_{g,0}}{\rho_0}\frac{c}{3\rho_0\kappa}\biggr)\biggl(1+\frac{\gamma v_{A,0}^2}{c_{g,0}^2}\biggr) \\ + v_{A,x,0}^2\biggl(c_{r,0}^2 + c_{g,0}^2\biggr)\Biggr)\omega^2 \\ + ik^4 \biggl(v_{A,x,0}^2\biggr)\Biggl(k^2 c_{g,0}^2\frac{c}{3\rho_0\kappa} - \dot{q}_{T_g} \frac{T_{g,0}}{\rho_0}(\gamma-1)\biggl(\frac{\gamma c_{r,0}^2}{c_{g,0}^2} + \frac{4}{3}\biggr) \\ + \dot{q}_{T_g} \frac{T_{g,0}}{\rho_0} \biggl(\frac{\gamma-1}{3} + \frac{c_{g,0}^2}{9 c_{r,0}^2}\biggr)\Biggr)\omega \\ + k^6(\gamma-1)\biggl(v_{A,x,0}^2\biggr)\dot{q}_{T_g} \frac{T_{g,0}}{\rho_0}\frac{c}{3\rho_0\kappa} \Biggr) = 0 \,.
\end{split}
\end{equation}

\end{document}